\documentclass{aa}  

\usepackage{graphicx}
\usepackage{txfonts}
\usepackage{xcolor}
\usepackage[flushleft]{threeparttable}

\begin{document}

   \title{HARMONI view of the host galaxies of active galactic nuclei around cosmic noon:}

   \subtitle{Resolved stellar morpho-kinematics and the M$_{BH}$-$\sigma_{\star}$ relation}

   \author{B. Garc\'ia-Lorenzo
          \inst{1,2}
          \and
          A. Monreal-Ibero\inst{3,1,2}
          \and
          M. Pereira-Santaella\inst{4}
          \and
          N. Thatte\inst{5}
          \and
          C. Ramos Almeida\inst{1,2}
          \and
          L. Galbany\inst{6,7}
          \and
          E. Mediavilla\inst{1,2}
          }

   \institute{Instituto de Astrof\'isica de Canarias, C/ V\'ia L\'actea s/n, E-38205 La Laguna, Tenerife, Spain\\
              \email{bgarcia@iac.es}
         \and
             Departamento de Astrof\'isica, Universidad de La Laguna, E-38200 La Laguna, Tenerife, Spain
         \and
             Leiden Observatory, Leiden University, PO Box 9513, 2300 RA, Leiden, The Netherlands
         \and
             Centro de Astrobiolog\'ia (CSIC-INTA), Ctra. de Ajalvir, Km 4, 28850, Torrej\'on de Ardoz, Madrid, Spain
         \and
             Department of Physics, Denys Wilkinson Building, Keble Road, University of Oxford, OX1 3RH, UK
         \and
             Institute of Space Sciences (ICE, CSIC), Campus UAB, Carrer de Can Magrans, s/n, E-08193 Barcelona, Spain.
         \and
             Institut d'Estudis Espacials de Catalunya (IEEC), E-08034 Barcelona, Spain.
             }

   \date{Received May 27, 2021; accepted November 23, 2021}

 
  \abstract
  {The formation and evolution of galaxies appear linked to the growth of supermassive black holes, as evidenced by empirical scaling relations in nearby galaxies. Understanding this co-evolution over cosmic time requires the revelation of the dynamical state of galaxies and the measurement of the mass of their central black holes (M$_{BH}$) at a range of cosmic distances. Bright active galactic nuclei (AGNs) 
are ideal for this purpose.}
  { The High Angular Resolution Monolithic Optical and Near-infrared Integral field spectrograph (HARMONI), the first light integral-field spectrograph for the Extremely Large Telescope (ELT), will transform visible and near-infrared ground-based astrophysics thanks to its advances in sensitivity and angular resolution. We aim to analyse the capabilities of HARMONI to reveal the stellar morpho-kinematic properties of the host galaxies of AGNs at about cosmic noon.}
  {We made use of the simulation pipeline for HARMONI (HSIM) to create mock observations of representative AGN host galaxies at redshifts around cosmic noon. We used observations taken with the Multi Unit Spectroscopic Explorer (MUSE) of nearby galaxies showing different morphologies and dynamical stages combined with theoretical AGN spectra to create the target inputs for HSIM.
  }
  {According to our simulations, an on-source integration time of three hours should be enough to measure the M$_{BH}$ and to trace the morphology and stellar kinematics of the brightest host galaxies of AGNs beyond cosmic noon. For host galaxies with stellar masses $<$10$^{11}$ M$_{\sun,}$ longer exposure times are mandatory to spatially resolve the stellar kinematics.}
  {}
  

   \maketitle
%

\section{Introduction}

   The formation and evolution of galaxies result from the hierarchical assembly across cosmic time. Besides, the empirical relations between large-scale properties of galaxies (e.g. bulge mass, stellar velocity dispersion, etc.) and the mass of their central supermassive black holes (SMBH) suggest a co-growth scenario (see e.g. \citealt{kormendy2013} for a review). To constrain models and provide a comprehensive picture of the formation and evolution of galaxies, it is necessary to study the change of the scaling relations between SMBHs and their host galaxies over cosmic time. A lack of change will suggest feedback mechanisms controlling the co-growth (e.g. \citealt{dimatteo2005}), while an evolution will point to a non-causal origin that will accidentally result after several interactions and mergers (e.g. \citealt{jahnke2011}).
   
   Both star formation and nuclear activity peak at a redshift in the 1.5 $\le$ $z$ $\le$ 2 range (e.g. \citealt{madau2014, aird2015}), a period often referred to as 'cosmic noon'. A proper cosmic period to study the evolution of the scaling relations supporting the co-growth scenario should include this highest activity ridge in the Universe's timeline. Such study requires an estimation of both the mass of central SMBHs (M$_{BH}$) and the properties of the host galaxies (e.g. luminosity, size, morphology, stellar velocity dispersion, etc.) for a representative sample of objects at different redshifts. 
   
   Estimating M$_{BH}$ over cosmic time can be done by observing active galactic nuclei (AGNs) spectra and applying the single epoch virial method (e.g. \citealt{kaspi2000, mclure2002}). It assumes that the motions of the gas surrounding the black hole (i.e. the broad-line region, BLR) are virialised (e.g. \citealt{2013shen}). 
   The procedure uses the empirical relation between the BLR size and the AGN continuum luminosity (e.g. \citealt{peterson2004}) to approach M$_{BH}$ from parameters easily measured from AGN type 1 spectra. For example, using the H$\beta$ emission line (see \citealt{vestergaard2006}):
   
   \begin{multline}
    log \left(\frac{M_{BH}}{M_{\sun}}\right) = ( 6.67 \pm 0.03 )\ + \\
    + log \left( \left[ \frac{FWHM(H\beta)}{1000\ km\ s^{-1}} \right]^{2} \left[ \frac{L(H\beta)}{10^{42} erg\ s^{-1} } \right]^{0.63} \right),
    \label{relation}
    \end{multline}
    
    \noindent where FWHM(H$\beta$) and L(H$\beta$) are the full width at half maximum and luminosity, respectively, of the broad component of H$\beta$ (see e.g. \citealt{shen2013} for other simple relations connecting emission lines and 
continuum luminosities and M$_{BH}$). The empirical correlation found between the BLR kinematics and the ionisation state of the narrow-line region (NLR) has extended the M$_{BH}$ estimation to type 2 AGNs. This approach only requires us to measure the narrow line luminosities of [\ion{O}{iii}] and H$\beta$ (\citealt{baron2019}). We note that around cosmic noon, the H$\beta$, H$\alpha$, [\ion{O}{iii}]$\lambda\lambda4959,5007,$ and [NII]$\lambda\lambda6548,6584$ emission lines (i.e. the traditional NLR and BLR tracers) are redshifted into the near-infrared range.
    
    The single epoch virial method requires high signal-to-noise ratio (S/N)
    spectra (e.g. \citealt{marziani2012}). The most luminous quasi-stellar objects (QSOs) are then ideal to determine M$_{BH}$ across cosmic time. However, bright QSOs hide their underlying galaxies with contrasts ranging from 10$^{-1}$ to 10$^{-3}$ (e.g. \citealt{floyd2004}). Hence, the simultaneous determination of M$_{BH}$ and galaxy parameters requires a compromise between the S/N of the AGN spectrum and its host galaxy spectra to control uncertainties. An efficient procedure for a proper de-blending will benefit that compromise (see e.g. \citealt{garcialorenzo2005}, \citealt{vayner2016}, \citealt{husemann2016}, \citealt{rupke2017}, and \citealt{varisco2018}). 
    
    The deviations of nearby mergers and bulgeless galaxies from the host-scaling relations with M$_{BH}$ (see e.g. \citealt{kormendy2013} and references therein) indicate that controlling the morphology of the AGN hosts is essential when studying the evolution of scaling relations with redshift. Besides this, the stellar velocity dispersion ($\sigma_{\star}$) is the most fundamental parameter linking galaxies and SMBHs \citep{Shankar2016, Marsden2020}. Moreover, only the M$_{BH}$-$\sigma_{\star}$ relation holds to all galaxy types, from dwarfs to the most massive galaxies (e.g. \citealt{2016Bosch}), although with deviations for evolved interactions \citep[e.g.][]{kormendy2013, 2016Saglia}. Then, unveiling the stellar dynamical state of galaxies over cosmic time is crucial to a complete understanding of the evolution of galaxies in the framework of the co-growth scenario.
   
   Integral field spectroscopic (IFS) surveys have extensively characterised the stellar \citep[e.g.][]{falcon2017, 2018Stark, 2021Fraser} and ionised gas \citep[e.g.][]{2015GarciaLorenzo, 2018Stark, 2020denBrok} kinematics of nearby galaxies ($z<$0.1) for a broad range of masses (M$_{\star}\sim10^{8}-10^{12}$ M$\odot$) and morphological types. Galaxies over $z>$0.1 have scarcely been studied using IFS so far, 
   focusing on the gaseous component to trace their kinematic evolution \citep{2015Wisnioski, 2016Contini, 2018ForsterSchreiber, 2020Foster, 2021Sharma}. These studies reveal that rotation dominates the gas kinematics in many galaxies from cosmic noon. However, gas kinematics (even more in AGN hosts) is complex, with features of inflows, outflows, turbulence, clumps, and so on. Instead, the stellar component is a good tracer of the gravitational potential of galaxies. Beyond the local Universe (z$>$0.1), spatially resolved stellar kinematics has barely been explored through IFS, reporting velocity fields consistent with rotating stellar disks \citep{guerou2017}. Close to or beyond the cosmic noon, there are only $\sigma_{\star}$ measurements from integrated or stacked spectra of massive quiescent galaxies \citep[e.g.][and references therein]{2020Mendel, 2021Esdaile, 2021vanderWel}. The limitation on collecting power and spatial resolution of the current IFS instruments (e.g. Multi-Espectrógrafo en Gran Telescopio Canarias (GTC) de Alta Resolución para Astronomía (MEGARA\footnote{http://www.gtc.iac.es/instruments/megara/megara.php}) at the 10-metre GTC, the OH-Suppressing Infrared Integral Field Spectrograph (OSIRIS\footnote{https://www2.keck.hawaii.edu/inst/osiris/}) at the 10-metre Keck telescope, the K-band Multi Object Spectrograph (KMOS\footnote{https://www.eso.org/sci/facilities/paranal/instruments/kmos.html}) or the Multi Unit Spectroscopic Explorer (MUSE\footnote{https://www.eso.org/sci/facilities/develop/instruments/muse.html}) at the 8-metre Very Large Telescope (VLT), etc.) are the causes behind this lack of resolved stellar dynamics of galaxies around cosmic noon. Only the new generation of the 30-40 metre-class telescopes will allow the resolving of the stellar kinematics of active and non-active galaxies around cosmic noon.
    
    
    In this work, we explored the capabilities of the High Angular Resolution Monolithic Optical and Near-infrared Integral field spectrograph \citep[HARMONI,][]{2016Thatte, 2020Thatte} on the Extremely Large Telescope\footnote{https://elt.eso.org/} (ELT) to reveal the morphological and stellar dynamical state of AGN host galaxies and to trace the cosmic evolution of the scaling relations connecting the host parameters with the mass of their central SMBHs. In \S \ref{instrument}, we describe the instrument and its different configurations to identify an optimal setup for observing AGNs and their hosts. In \S\ref{mock}, we describe the procedure of generating inputs for the simulation pipeline and the mock HARMONI observations. The analysis and discussion are presented in \S\ref{analysis}. Section \S\ref{conclusions} summarises the main conclusions.
In the following, we assume a standard $\Lambda$CDM model with H$_0$ = 67.9 km s$^{-1}$ Mpc$^{-1}$, $\Omega_{M}$ = 0.31 and $\Omega_{\Lambda}$ = 0.69 \citep{Planck16}.


\section{HARMONI}
\label{instrument}

HARMONI\footnote{https://harmoni-elt.physics.ox.ac.uk/} will provide the first light spectroscopic capability to the ELT (see e.g. \citealt{2016Thatte,2020Thatte}). HARMONI was conceived as a workhorse instrument to support a wide variety of science programmes (see e.g. \citealt{2010Arribas,2016Thatte,2020Thatte}) and address many of the key science cases for the ELT\footnote{http://www.eso.org/sci/facilities/eelt/science/}. 

HARMONI includes single conjugated and laser tomography adaptive optics (SCAO and LTAO, respectively) systems to compensate the atmospheric turbulence, although seeing-limited observations are also available. SCAO will provide an excellent image quality (< 20\% degradation of the telescope point-spread function; PSF) using a single, bright (J<14) natural star reference located up to 15 arcsec from the science target. LTAO will compensate the atmospheric turbulence ($\sim$30\% Strehl in K band), with 50\%-90\% sky coverage. This will be done by using six laser guide stars and a reference natural star (J$<19$) within up to 1 arcmin radius of the science field. HARMONI will have a similar angular resolution to the Atacama Large Millimeter/submillimeter Array (ALMA\footnote{https://www.almaobservatory.org/} ) and comparable sensitivity to the James Webb Space Telescope (JWST\footnote{https://jwst.nasa.gov/}) at wavelengths $<$2.5 $\mu$m, which perfectly complement each other due to many synergies. At the time of writing, the development of HARMONI is in the final design phase, with the first light scheduled for the end of 2026. In this section, we go through the multiple configurations of HARMONI with the aim of identifying the optimal setup to observe AGNs with a signal to noise and resolution that permit us to study galaxy black hole co-evolution over cosmic time.

\subsection{Spatial scales and field of view}
\label{spatial}

\begin{table}
        \centering
        \caption{HARMONI spatial configurations. }
        \label{spatial-scales}
        \begin{tabular}{cccc} 
                \hline
Spaxel scale  & FoV & Spaxel size & FoV \\ 
(mas$^{2}$) & (arcsec$^{2}$)   & (pc$^{2}$ at $z$=1.5) & (kpc$^{2}$ at $z$=1.5)\\ \hline
$30\times 60$ & $6.12\times 9.12$ & $\sim260\times 520$ & $\sim53\times 79$  \\
$20\times 20$ & $4.08\times 3.04$ & $\sim173\times$ 173 & $\sim35\times 26$ \\
$10\times 10$ & $2.04\times 1.52$ & $\sim87\times 87$ & $\sim18\times 13$\\
$4\times 4$ & $0.82\times 0.61$ & $\sim35\times 35$ & $\sim7.1\times 5.3$ \\ \hline
\end{tabular}
   \tablefoot{ Columns (1) and (2) correspond to angular sizes on the sky of the spaxel scale and field of view (FoV), respectively. Columns (3) and (4) indicate the corresponding physical sizes at redshift $z$=1.5.  }

\end{table}

The size of the galaxies over cosmic time varies: R$_{eff}$ (kpc) = B$_{z}$ ( 1 + $z$ )$^ { \beta_z }$, where B$_{z}$ and $\beta_z$ parameters depend on stellar mass and galaxy type (\citealt{2014vanderWel}), and R$_{eff}$ is the effective radius. For redshift 2.75 > $z$ > 0.75, the average sizes range between 1.3 $\leq$ R$_{eff}$ $\leq$ 2.9 kpc and 3 $\leq$ R$_{eff}$ $\leq$ 5 kpc for early- and late-type objects, respectively (see Table 2 in \citealt{2014vanderWel}), with AGN hosts lying in-between (\citealt{Silverman2019}). These numbers correspond to full (i.e. 2$\times$R$_{eff}$) angular sizes in the $\sim$0.3-1.3 arcsec range. Moreover, many AGN hosts show gas outflows, typically extending over R$_{eff}$ at $z$>0.6 (e.g. \citealt{2016Brusa, 2017Vayner,Schreiber_2019}), sometimes reaching even a few R$_{eff}$ (e.g. \citealt{2006Nesvadba, 2012Harrison, 2019Leung}). Considering this, we need a field of view (FoV) of a few times the largest angular size expected for these high-redshift AGN host+outflows, that is 
a few arcsec.

HARMONI offers four different spatial configurations (see Table \ref{spatial-scales}) providing spectra of more than 31000 spatial positions arranged in a rectangular array of 152 $\times$ 204 spaxels. The two finer spatial scales will provide high spatial resolution, but their FoVs are smaller than some of the most powerful outflows observed in AGNs (e.g. \citealt{2006Nesvadba, 2012Harrison, 2019Leung}). On the contrary, the FoV of the two coarser scales fits the expected size of the most extended outflows, with physical resolutions comparable to surveys of nearby galaxies like the Calar Alto Legacy Integral Field Area (CALIFA) survey (\citealt{2016Sanchez}), the Mapping Nearby Galaxies at APO (MANGA) survey (\citealt{2015Bundy}), or the Sydney-Australian-Astronomical-Observatory Multi-object Integral-Field Spectrograph (SAMI) survey (\citealt{2012Croom}). Moreover, the coarser scales also simultaneously sample the background allowing on-frame sky subtraction. We thus selected the 20$\times$20 mas$^{2}$ scale for our HARMONI simulations of AGN+host galaxies. This scale provides the best sensitivity among the four HARMONI spatial scales (\citealt{2015Zieleniewski}), and it preserves, at a fixed S/N threshold, better spatial resolution than the 30$\times$60 mas$^{2}$ scale. Some results on mock HARMONI observations of QSO+host galaxies using the 30$\times$60 mas$^{2}$ spatial scale were presented in \cite{2019Garcia}.


%
%
\subsection{Spectral resolutions and gratings}
\label{spectral}

\begin{figure}
\centering
\includegraphics[trim={0 1.5cm 0cm 1.5cm},clip,width=7cm,angle=270]{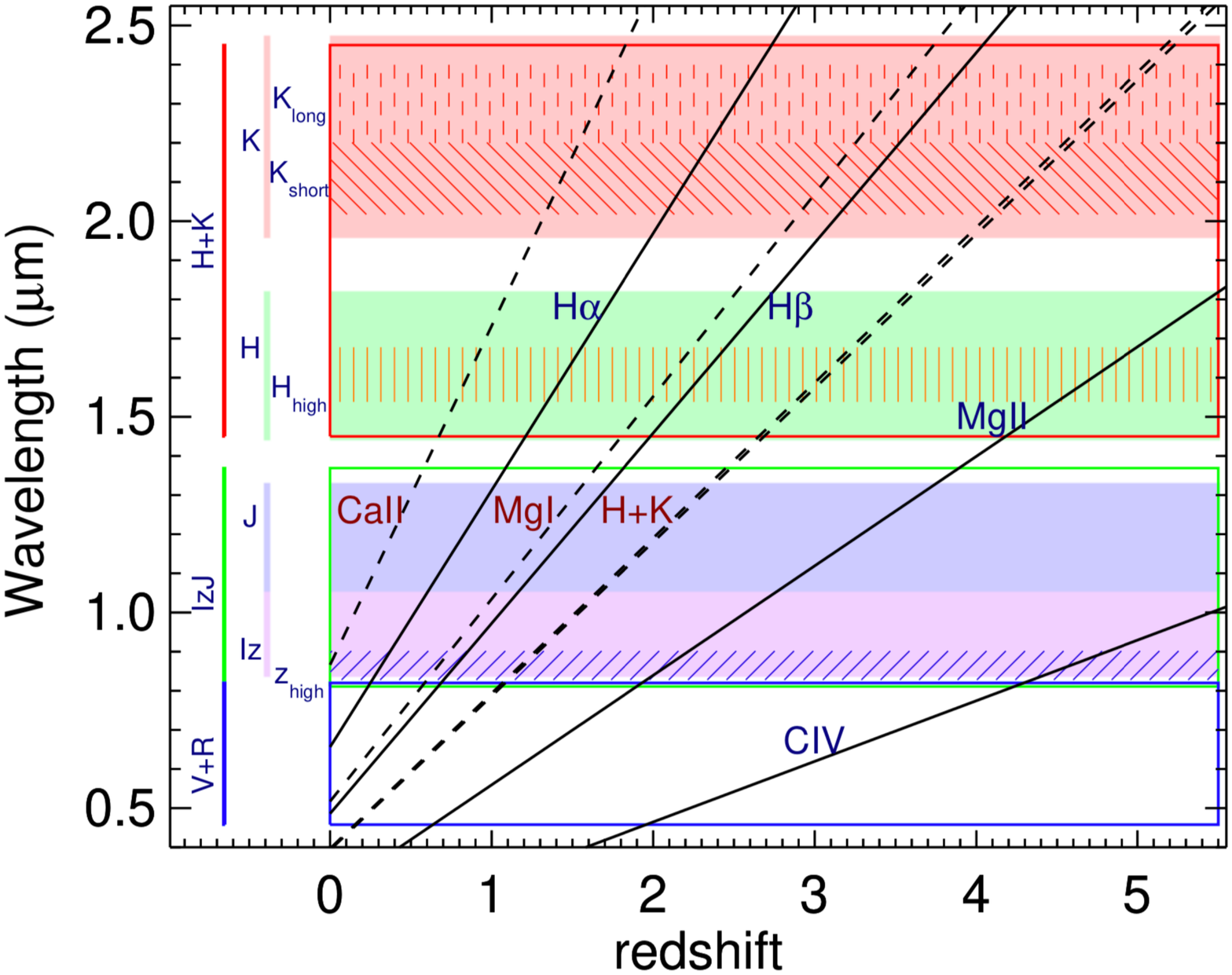}
  \caption{Key spectral features to measure M$_{BH}$ and $\sigma_{\star}$ observable at each HARMONI band as a function of the redshift. Black solid and dashed lines correspond to emission (i.e. CIV, MgII, H$\beta,$ and H$\alpha$) and absorption (i.e. CaIIH+K, MgI, and CaII) features, respectively. The low-resolution bands correspond to the wavelength range between blue (V+R), green (I+z+J), and red (H+K) rectangles. Purple (I+z), blue (J), green (H), and red (K) filled rectangles mark the wavelength range of medium-resolution gratings. Hatch patterns mark the high-resolution bands: purple, striped, inclined: Z$_{high}$; orange striped: H$_{high}$; red, striped; inclined: K$_{short}$; and  red, dashed, striped: K$_{long}$.   }
    \label{fig_gratings}
\end{figure}

To optimise the ELT+HARMONI observing time, it is desirable to have all the required information in a
single shot, that is, a grating whose wavelength coverage includes both emission lines and stellar features to measure the M$_{BH}$ and $\sigma_{\star}$ parameters.
HARMONI will provide three resolving powers (R=$\lambda$/$\Delta\lambda$ of $\sim$3200, $\sim$7000, and $\sim$18000) in different bands (see Fig. \ref{fig_gratings}). Depending on the redshift of the target, the relevant spectral features will be shifted to  different wavelengths. Figure \ref{fig_gratings} includes the wavelength of some emission and absorption lines as a function of redshift.
 
 The HARMONI high spectral resolution bands will provide accurate estimates of the stellar velocity dispersion, with a spectral resolution of $\sim$1 \AA \ (or $\sigma_{\star}$$\sim$7 km s$^{-1}$) at 1.92 $\mu$m. However, only for the K bands (i.e. K$_{short}$, and  K$_{long}$) could stellar and broad ionised gas features be simultaneously observed (i.e. from H$\beta$ to MgI lines) for objects at $z \gtrapprox $3.2. The HARMONI medium-resolution bands can sample the H$\beta$-MgI region for AGNs in the $\sim$0.6 to $\sim$3.7 redshift range with a spectral resolution of about 2.7 \AA \ (or $\sigma_{\star}$$\sim$18 km s$^{-1}$). Nevertheless, the simultaneous observation of the MgI stellar feature and H$\beta$ emission line is unfeasible using these medium-resolution gratings for objects in two redshift gaps (i.e. $\sim$[1.5-2.0], and $\sim$[2.5-3.1]). The HARMONI low-resolution gratings enable us to observe a wide spectral range including several emission and absorption features from nearby to high-redshift objects (see Fig. \ref{fig_gratings}). Indeed, the estimation of M$_{BH}$ and $\sigma_{\star}$ is possible with HARMONI low-resolution gratings for AGNs in the 0 $\leq z \leq$ 3.7 range observing the H$\beta$-MgI region (there is only a gap for objects in the $\sim$[1.7-2.0] redshift range).
 
 We note that a resolution of $\sim$3200 corresponds to a $\Delta\lambda$ of $\sim$6 \AA \ (or $\sigma_{\star}$$\sim$40 km s$^{-1}$) at 1.92 $\mu$m. Current optical spectroscopic surveys of nearby galaxies measure stellar velocity dispersions with spectral resolutions that range from 30 to 140 km s$^{-1}$ (see e.g. The Baryon Oscillation Spectroscopic Survey (BOSS), \citealt{Thomas2013}; MANGA,  \citealt{Law2016}; SAMI, \citealt{Sande2017}; CALIFA, \citealt{GarciaBenito2015}). For intermediate redshift (0.2 $\leq$ $z$ $\leq$ 0.7) galaxies, the MUSE at the VLT provides spatially resolved stellar kinematics with spectral resolution ranging from 40 to 70 km s$^{-1}$  \citep{guerou2017}. Therefore, HARMONI low-resolution bands could potentially resolve, using a homogeneous procedure, the stellar kinematics of galaxies at around cosmic noon with comparable velocity resolution to current measurements for nearby and intermediate redshift galaxies. We note that if the instrumental resolution is similar to that of the galaxy (i.e. $\sigma_{\mathrm{instrumental}}$$\sim$$\sigma_{\star}$$\sim$ 40 km s$^{-1}$), the net effect will be to add in quadrature, which can be detected with enough S/N. Pioneer works on stellar kinematics focused on the H$\beta$-MgI spectral region, proving this range to trace both the ionised gas and stellar kinematics in nearby galaxies (e.g. The Spectroscopic Areal Unit for Research on Optical Nebulae (SAURON) project, \citealt{2001Bacon}). Hence, we focused our HARMONI simulations of AGNs and host galaxies on observing the H$\beta$-MgI range through HARMONI low-resolution gratings.

\section{Mock HARMONI observations}
\label{mock}

During the design phases of instrumental developments, simulations permit us to check capabilities and limitations. HSIM\footnote{https://github.com/HARMONI-ELT/HSIM} \citep{2015Zieleniewski} is a dedicated tool for simulating observations with HARMONI. It incorporates detailed models of the sky emission and transmission spectra,
telescope, and instrument responses to produce realistic mock data for a given input target. Running a simulation requires the selection of observing conditions (such as seeing and airmass), telescope and instrument configuration (observing mode, number of exposures, exposure time, grating, spaxel scale, etc.). HSIM needs input data cubes characterising the target physics to compute the signal and noise of the mock HARMONI observations. We used version 300 of HSIM to explore the potential of HARMONI to study AGNs and their host galaxies at three redshifts around cosmic noon, in particular, 0.76, 1.50, and 2.30. To generate input data cubes for HSIM representative of the targets, we combined 3D information of nearby galaxies with theoretical QSOs, as explained in the following sections (i.e. \S \ref{host}, \S \ref{QSO}, and \S \ref{targets}).

\subsection{The host galaxies}
\label{host}

\begin{figure}
\includegraphics[trim={0cm 0cm 0cm 0cm},clip,width=6.25cm,angle=270]{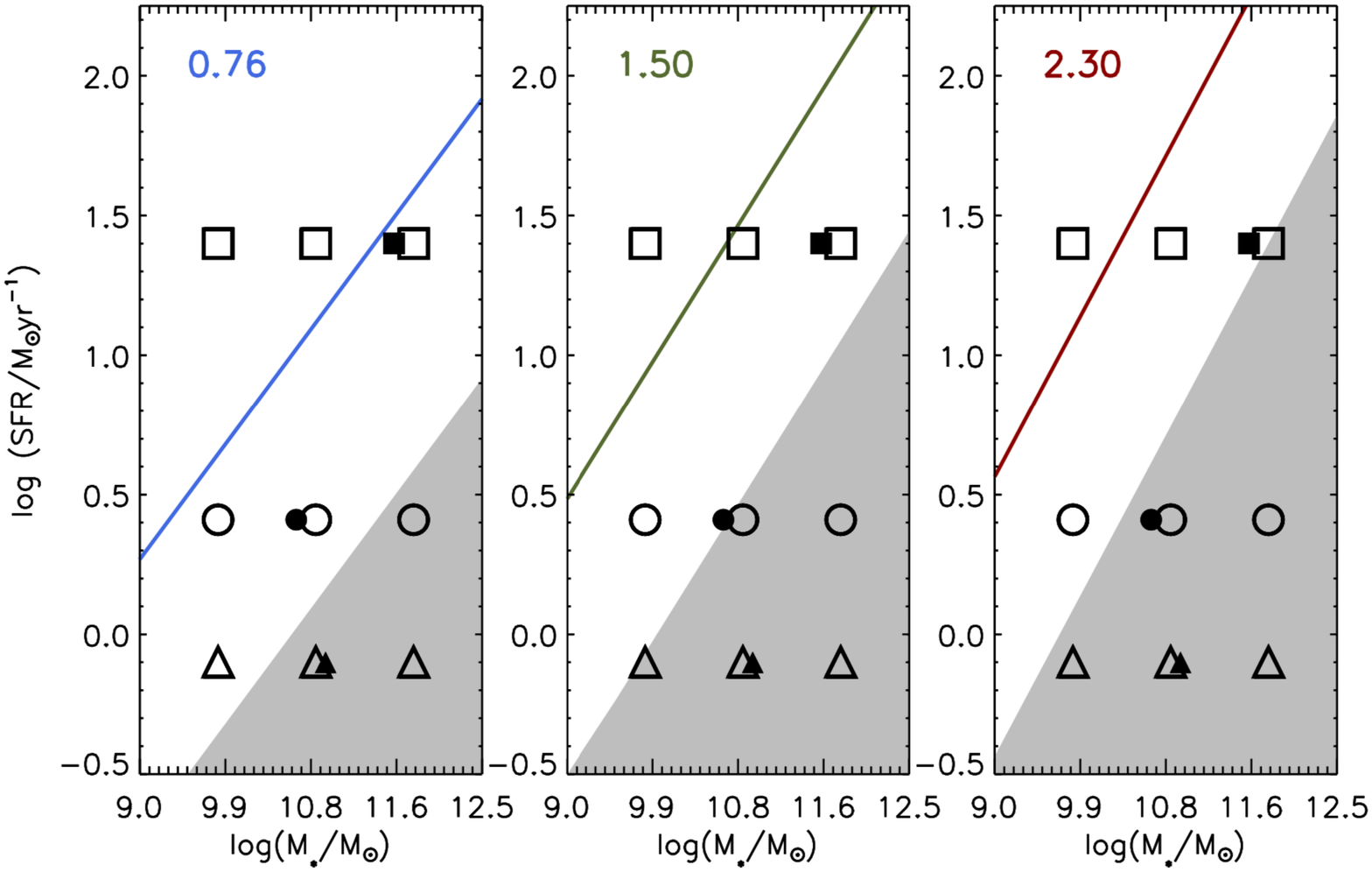}
  \caption{ Location in SFR stellar-mass diagram of the three galaxies selected as AGN hosts scaled to 10$^{12}$L$_{v,\odot}$ ($\log(M_{\star}/M_{\sun}) \sim$ 11.8), 10$^{11}$L$_{v,\odot}$ ($\log(M_{\star}/M_{\sun}) \sim$ 10.8), and 10$^{10}$L$_{v,\odot}$ ($\log(M_{\star}/M_{\sun}) \sim$ 9.8) luminosities. Symbols indicate the distinct hosts: circles$\rightarrow$NGC\,809, triangles$\rightarrow$PGC\,055442, and squares$\rightarrow$NGC\,7119A. The filled symbols correspond to the actual location of the three nearby galaxies (see Table \ref{tabgalaxies}). At each redshift (top left labels), the solid line traces the main sequence (SFR$_{\mathrm{MS}}$) fitted by
 \cite{2018Pearson} (see also Table \ref{SFR_table}). Objects in the grey region correspond to quenched host galaxies.  }
    \label{mass-sfr}
\end{figure}

At $z$ $\lessapprox$0.4, many different galaxy types host AGNs (e.g. \citealt{ 2006Guyon, 2012Bessiere, 2019Marina}). Nonetheless, the most luminous AGNs are typically hosted by massive (i.e. M$_{\star}> 10^{11}$ M$_{\odot}$) early-type galaxies (e.g. \citealt{2003MNRAS.346.1055K}) or triggered by interactions and mergers (see e.g. \citealt{Cris_2011, Chiaberge_2015}). The picture could be similar around cosmic noon, with the AGN host population including spheroids, disk-like galaxies, interactions, and mergers (e.g. \citealt{2006Guyon, Treister2012, Floyd2013, Zakamska2019, Silverman2019, 2020Pensabene, 2020Rizzo}). As already mentioned, assessing the morphological and dynamical stage of AGN hosts is relevant for tracing the M$_{BH}$-$\sigma_{\star}$ relation over cosmic time because a few morphological types deviate from the scaling relations (e.g. \citealt{2008Graham, 2016Saglia}). Thus, we chose a set of three nearby (distances between $\sim$70 and 145 Mpc) galaxies showing distinct morphologies (i.e. lenticular, spiral, and interacting) as possible AGN hosts for our simulations. We used MUSE observations obtained in the framework of the All-weather MUse Supernova Integral field Nearby Galaxies (AMUSING) project \citep{Galbany16}. The spatial and spectral sampling of the MUSE data are 0\farcs2 spa$^{-1}$, and 1.25\,\AA~pix$^{-1}$, respectively, aptly suiting the needs of HSIM. The MUSE instrument covers almost the whole optical spectral range (4800-9300 \AA ) in its standard mode, including the spectral features needed for our scientific goals.
 Appendix \ref{appen} presents the fundamental parameters and properties of these galaxies. The selected spiral galaxy matches the SFR
 and stellar mass connection defining the main sequence (MS) of nearby star-forming galaxies well. The lenticular and interacting hosts are more than 3 sigma above the MS, assuming a typical scatter of $\sim$0.3 dex in SFRs \citep[e.g.][]{2015Schreiber,2018Circosta,2020Sebastian}. Nonetheless, the SFR depends on redshift and peaks at cosmic noon,  and the MS follows a simple two-parameter power law of the following form: log(SFR)=$\alpha(z)\times$[log(M$_{\star})-$10.5]+$\beta(z)$, where $\alpha(z)$=0.38+0.12$\times z$, and $\beta(z)$=1.10+[0.53$\times$ln(0.03 + $z$)] \citep{2018Pearson}. Then, the SFR of these galaxies at the three selected redshifts would be more than 3 sigma below the MS of high redshift star-forming galaxies in approximately half of the cases (see Table \ref{SFR_table}). Their location in the SFR-M$_{\star}$ diagram (see Fig. \ref{mass-sfr}) corresponds to the transition between star-forming and quenched galaxies, which is the so-called green valley. Observations show that many AGN host galaxies populate the green valley, suggesting AGNs as a transition phase between star-forming and retired galaxies \cite[see e.g.][]{2020Sebastian, 2020Ishino}. Indeed, AGN feedback seems to be an efficient quenching mechanism to control galaxies' growth over cosmic time in numerical simulations \cite[see e.g.][for a review]{2017Naab}. However, observations indicate that galaxies hosting an AGN may have the SFR quenched \citep[e.g.][]{2020Smith} or enhanced \citep[e.g.][]{2017Mahoro} as a consequence of the AGN feedback. There are different criteria to separate quenched from star-forming galaxies \cite[see Table 2 in][]{2019Donnari}. Taking the distance from the main sequence as the reference, the cases where $\Delta$SFR$_{\mathrm{0}-\mathrm{MS}}$ are less than -1 would correspond to a quenched host galaxy in our simulations (see Table \ref{SFR_table}).
 
  \begin{table}
        \centering
        \caption{ Estimated star formation rates (SFRs) of high-redshift galaxies at the main sequence (SFR$_{\mathrm{MS}}$) compared with the SFR of the nearby galaxies selected to act as AGN hosts in the mock HARMONI simulations (see Table \ref{tabgalaxies} for SFR$_{0}$ values).  }  
        \label{SFR_table}
        \begin{tabular}{ccccc} 
                \hline
         &                    &                  & {\bf Hosts}           &                    \\ 
         &                                            & {\bf Lenticular} & {\bf Spiral} & {\bf Interaction} \\ \hline
               &  SFR$_{\mathrm{MS}}$                        &    1.02          &     1.16          &  1.49 \\
{\bf z=0.76}   &  $\Delta$SFR$_{\mathrm{0}-\mathrm{MS}}$ &    -0.61         &    -1.26          &  -0.09 \\
               &  AGN host                       &   S-F               &    Q               &        S-F \\\hline 
               &  SFR$_{\mathrm{MS}}$                        &    1.38          &     1.55          &  1.94 \\
{\bf z=1.50}   &  $\Delta$SFR$_{\mathrm{0}-\mathrm{MS}}$ &    -0.97         &    -1.65          &  -0.54 \\
               &  AGN host                               &    S-F           &      Q             &  S-F       \\\hline 
               &  SFR$_{\mathrm{MS}}$                        &    1.68          &     1.89          &  2.37 \\
{\bf z=2.30}   &  $\Delta$SFR$_{\mathrm{0}-\mathrm{MS}}$ &    -1.27         &    -1.99          &  -0.97 \\
               &  AGN host                              &      Q            &       Q            &   S-F      \\\hline 
\end{tabular}
\tablefoot{ Units of SFR and $\Delta$SFR$_{\mathrm{0}-\mathrm{MS}}$ are in log M$_{\odot}$/yr and 1 dex, respectively. S-F and Q classify the galaxy as star-forming or quenched, respectively, based on the distance to the MS criteria.}
\end{table}

To cover three different ranges of luminosity and stellar mass, the MUSE data cubes of the three galaxies were scaled to sample three absolute $V$-band luminosities ($\log(L_V / L_{V,\sun})$ =10, 11, 12), corresponding to $M_V = -20.19, -22.69, -25.19$ \footnote{$V_{\sun}$=+4.81, \citep{Willmer18}} , and $\log(M_{\star}/M_{\sun})$ > 9.8, 10.8, 11.8 using the mass-luminosity ratio in \cite{2001Bell}. Since the total absolute magnitude in the $V$-band was not available in Hyperleda\footnote{http://leda.univ-lyon1.fr/\label{ledacat}} \citep{2014makarov} for the three galaxies, we used the $B$ magnitude (see Table \ref{tabgalaxies}) and applied an average colour correction for their morphological types ($B-V$$\sim$$0.88$, \citealt{Mannucci01}) to estimate the $V$-band magnitudes.
These scaled galaxies are representative of the population of host galaxies of type 1 AGNs at least up to redshift 1 \citep{2020Ishino}. Their locations in the SFR stellar-mass diagram (see Fig. \ref{mass-sfr}) indicate that nearly half of the mock targets would correspond to star-forming hosts
and the other half to quenched galaxies.

Before using these galaxies to create the inputs for HSIM, we brought their data cubes to rest-frame wavelength using their systemic velocities (see Table \ref{tabgalaxies}). In the case of NGC\,7119A, we used the systemic velocity of the NGC7119 system (i.e. 9666 km s$^{-1}$ from Hyperleda). 

\subsection{The AGN}
\label{QSO}

To simulate the emission of the central AGN, we used a theoretical QSO spectrum assuming a continuum modelled as a power law of the following form: F$_{\lambda} =$ A $\times$ $\lambda ^ {\alpha}$, with $\alpha$ = $-1.72$, and A is a constant (in arbitrary units at this step) related to the luminosity density \citep{Neugebauer87}. 

\begin{figure}
\includegraphics[trim={1cm 2cm 1cm 1cm},clip,width=9cm]{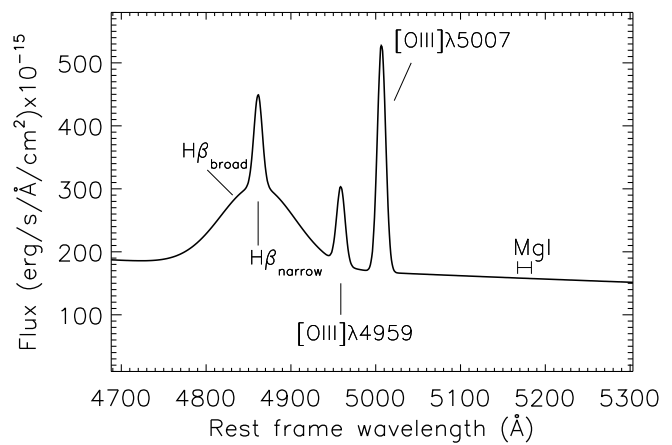}
  \caption{Theoretical QSO spectrum scaled to L$_{\mathrm{[OIII]}}$ = 10$^{43}$ erg s$^{-1}$ in the wavelength range of interest. Emission features in the range are labelled. For reference, we also marked the location of the MgI features indicative of the host's stellar component.   }
    \label{QSO_model}
\end{figure}

On top of this continuum, we included the brightest emission lines from H$\beta$ to [\ion{S}{ii}]$\lambda$6717,6731 adopting Gaussian profiles for them. Using the catalogue of spectral properties of type 1 AGNs by \cite{2017MNRAS.472.4051C}\footnote{https://qsfit.inaf.it/ \label{catalogue}} (version 1.2.0), we fixed the parameters of these Gaussians by taking the mean values for objects at z < 0.7. In practice, we assumed Gaussian widths of $\sigma$=4.8 \AA \ and $\sigma$=44 \AA \ (equivalent to a FWHM of $\sim$$700$, and $\sim$$6360$ km s$^{-1}$ at H$\beta$ rest-frame wavelength), for the narrow and broad components of the emission lines. 
We also considered equivalent widths of $\sim$7 and $\sim$58 \AA \ for the narrow and broad component of H$\beta$, respectively, and $\sim$19 \AA \ for [OIII]$\lambda$5007. From the same catalogue and redshift range, the median and brightest [\ion{O}{iii}] luminosities of AGNs are roughly 10$^{42}$ and 10$^{43}$ erg s$^{-1}$, corresponding to bolometric luminosities of 10$^{45.5}$ and 10$^{46.5}$ erg s$^{-1}$ \citep{2004heckman}, respectively. Then, the resulting theoretical spectrum in arbitrary units was scaled to these two [\ion{O}{iii}] luminosities (see Fig. \ref{QSO_model}). According to the characteristic X-ray luminosity of AGNs \citep{aird2015} and the correlation between X-ray and [\ion{O}{iii}] luminosities for type 1 AGNs \citep{2015A&A...578A..28B}, these two synthetic spectra should be representative of type 1 AGNs at around cosmic noon. 


Using the simulated spectra and equation \ref{relation}, we find that the faintest and brightest AGNs account for a black hole mass of $\sim$3.8$\times$10$^{8}$ M$_{\sun}$, and $\sim$1.6$\times$10$^{9}$ M$_{\sun}$, respectively. These M$_{BH}$ match the central black holes of galaxies around the cosmic noon according to observations \citep{2017Kozowski} and the average growth history of SMBHs \citep{2006Marconi}. Taking into account the M$_{BH}$ stellar-mass relation (e.g. Equation 9 from \citealt{2016Bosch}) and assuming that is valid for any $z$, galaxies with stellar masses in the $\sim$1.3-4.6$\times$10$^{11}$M$_{\sun}$ range should host black holes of that M$_{BH}$ at the redshifts of interest. Moreover, AGN luminosities appear related to the stellar mass and the SFR of their host galaxies. Taking into account the SFR of the selected hosts (see \S\ref{host}) and the AGN luminosities adopted in this work, our mock AGN+host would reside in evolutionary tracks closer to quenching  \cite[e.g.][]{2016Mancuso}. We note that a small fraction of quenched host galaxies at cosmic noon will host luminous AGNs \cite[e.g.][]{2020Florez}.

\begin{table*}
        \centering
        \caption{Summary of the input parameters to simulate HARMONI observations of AGNs and their host.}
        \label{HSIM_par}
        \begin{tabular}{lcl} 
        \hline
HSIM parameter & Input value & Comments \\
                \hline
Exposure time & 900 & Integration time of each exposure in seconds \\
N$_{exp}$ & 12 & Number of exposures \\
Grating & Iz+J \& H+K & Selection according to target redshift \\
Spaxel Scale & 20$\times$20 & in mas$\times$mas \\
Zenith seeing & 0.64 arcsec & Average seeing at Cerro Armazones \citep{2009Skidmore} \\
Air mass & 1.1 & Average air mass during the HARMONI observation \\
Moon illumination & 0 & Fraction of the Moon illuminated in dark nights \\
Jitter & 3 & Telescope PSF blur in mas \\
Telescope temperature & 280 & Typical temperature of the site and telescope in K \\
ADR on or off & True & Simulate the atmospheric differential refraction is simulated \\
AO mode & LTAO & Adaptive optics (AO) mode of the observations \\
Detector systematics & None & Detector systematics not included \\
Noise seed & 1234  & Random number generator \\
\hline
\end{tabular}
\end{table*}

\subsection{Creating the input targets and running HSIM}
\label{targets}

The first step to create the input data cubes for HSIM was to build 11 rest-frame data cubes, two associated with the two simulated AGN spectra (see \S\ref{QSO}) and nine with the distinct luminosity-scaled host galaxies (see \S\ref{host}). The FoV of these data cubes were cropped to the size of the galaxies to save space and simulation computing time. Then, we added the AGN and host galaxy data cubes by placing the QSO at the nucleus of each host, limiting the spectral range to rest-wavelengths from 4770 to 5400 \AA \ (i.e. H$\beta$-MgI region). 
We note that for the interacting system NG\,7119, we placed the QSO at the centre of the northern galaxy (i.e. NGC\,7119A; see Appendix \ref{appen}). All these data cubes were moved to three redshifts ($z = 0.76, 1.50,$ and $  2.30$), applying the corresponding spatial resampling, spectral shifting, and brightness dimming. We fixed the spatial and spectral scales of these data cubes to 10mas$\times$10mas per spaxel and 2 \AA \ per pixel, respectively, suiting the oversampling recommendations of HSIM but saving disk space. We note that to minimise effects when convolving the line-spread function and the PSF, HSIM internally performs spatial and spectral flux-conserving interpolations to the nominal scales for the selected setup \citep{2015Zieleniewski}.  In summary, we obtained mock HARMONI observations for three redshifts, three host galaxy morphologies, three host galaxy luminosities, and two QSO bolometric luminosities, corresponding to 54 targets. In addition, we simulated  27 targets without an AGN (3 redshift $\times$ 3 morphologies $\times$ galaxy luminosities) and six targets with pure-QSO spectra (3 redshift $\times$ 2 QSO bolometric luminosities). These data cubes are the inputs for HSIM (HSIM$_{\mathrm{Input}}$ hereafter).


Table \ref{HSIM_par} summarises the HSIM input parameters for the 87 generated targets.
In all the cases, we fixed the total on-source observing time to 3 hours, split into 12 exposures of 900 seconds each.
We checked the dependency of the average S/N 
with the number of exposures (N$_{exp}$) for a fixed on-source integration time of 3 hours, taking 3$\times$3600 seconds as reference. We found a S/N percentage variation of $\Delta$S/N(\%)$\sim$2.4$-$0.83$\times$N$_{exp}$ and $\Delta$S/N(\%)$\sim$1.4$-$0.49$\times$N$_{exp}$ for IzJ and H+K gratings, respectively. Therefore, the mock HARMONI observations analysed in this work have a $\sim$7.5\% in IzJ and $\sim$4.5\% lower S/N than simulations performed splitting the 3 hours of the total integration time in only three exposures of 3600 seconds each. \cite{2019Augustin} reported a 7\% S/N gain when choosing 5$\times$3600 seconds over 20$\times$900 seconds for V+R grating.
 However, we note that in AO-assisted instruments, AO performances may vary during the exposure time. These variations result from changes in the atmospheric turbulence conditions and the airmass evolution (both functions of the wavelength). Thus, shorter individual exposures will be wiser to preserve the quality of the AO correction.



\begin{figure}
\includegraphics[trim={0cm 0cm 0cm 2cm},clip,width=9.5cm,angle=180]{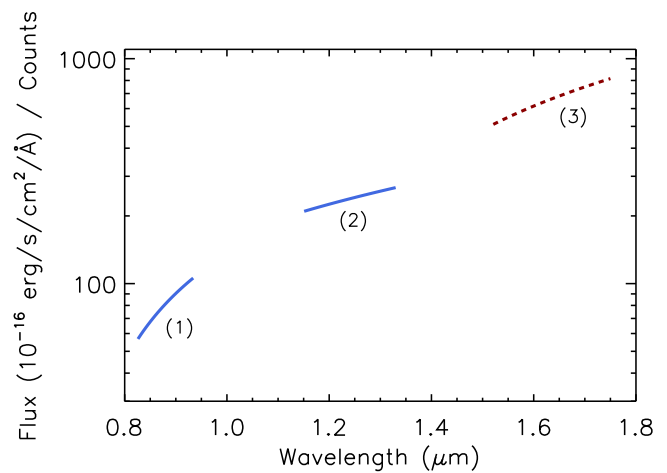}
  \caption{Flux-density-to-counts ratio for the gratings (IzJ-blue and HK-dashed red lines) used for the mock HARMONI observations of AGNs+hosts. We note that the derived functions only cover the H$\beta$+MgI spectral range (4770-5400 \AA \ rest-frame wavelengths) at redshifts $0.76$ (1), $1.50$ (2), and $2.30$ (3) due to the design of our simulations.}
    \label{calibrafunction}
\end{figure}

\subsection{Flux calibration of the mock HARMONI observations}
\label{fluxcalibra}

The HARMONI simulator (HSIM-version 300) can provide a flux-calibrated data cube. This requires the definition of a large enough aperture to determine the correction factors to translate detector counts into flux units with minimal flux uncertainties. However, we had to perform an independent flux calibration because the FoV of the HSIM$_{\mathrm{input}}$ data cubes was matched to the size of the host galaxies. We performed the flux calibration for IzJ and HK gratings after running HSIM and only in the spectral range corresponding to the H$\beta$+MgI region at redshifts $0.76$, $1.50$, and $2.30$ (see \S\ref{spectral}). To calibrate the mock observations, one should mock observe a spectro-photometric standard star. Instead, as we applied HSIM to two point-like sources (QSOs of [OIII] luminosities 10$^{42}$ and 10$^{43}$ erg s$^{-1}$ ) by each configuration, we used these data cubes to calibrate in flux. To do so, we integrated the flux using a circular aperture of 1.2 arcsec in radius to maximize the covered flux and S/N. We then compared the extracted spectra of each mock QSO observation at each luminosity and spectral range with the corresponding flux-calibrated spectra obtained from the input target. Such comparison provides the following linear calibration functions with lambda (see also Fig. \ref{calibrafunction}) in the fitted spectral ranges: \\
\\
  f$_{\mathrm{IzJ}}$($\lambda$) = ($-3.19 + 4.55 \lambda$) $\times$ 10$^{-18}$, if $\lambda < $ 0.94 $\mu$m, \\
  f$_{\mathrm{IzJ}}$($\lambda$) = ($-1.60 + 3.21 \lambda$) $\times$ 10$^{-18}$, if 1.15 < $\lambda < $ 1.33 $\mu$m, \\
  f$_{\mathrm{HK}}$($\lambda$) = ($-1.51 + 1.33 \lambda$) $\times$ 10$^{-19}$, if 1.54 < $\lambda < $ 1.73 $\mu$m. \\
  
\noindent Finally, we applied these functions to each spaxel in our mock HARMONI observations to transform counts into flux. Comparing the resulting data cubes for the mock HARMONI pure AGNs with their corresponding HSIM inputs, we found uncertainties in the flux calibration that range between 1\% and 5\%.

\begin{figure*}[ht]
 \centering
\includegraphics[trim={0cm 1cm 2cm 0cm},width=12.5cm,angle=270]{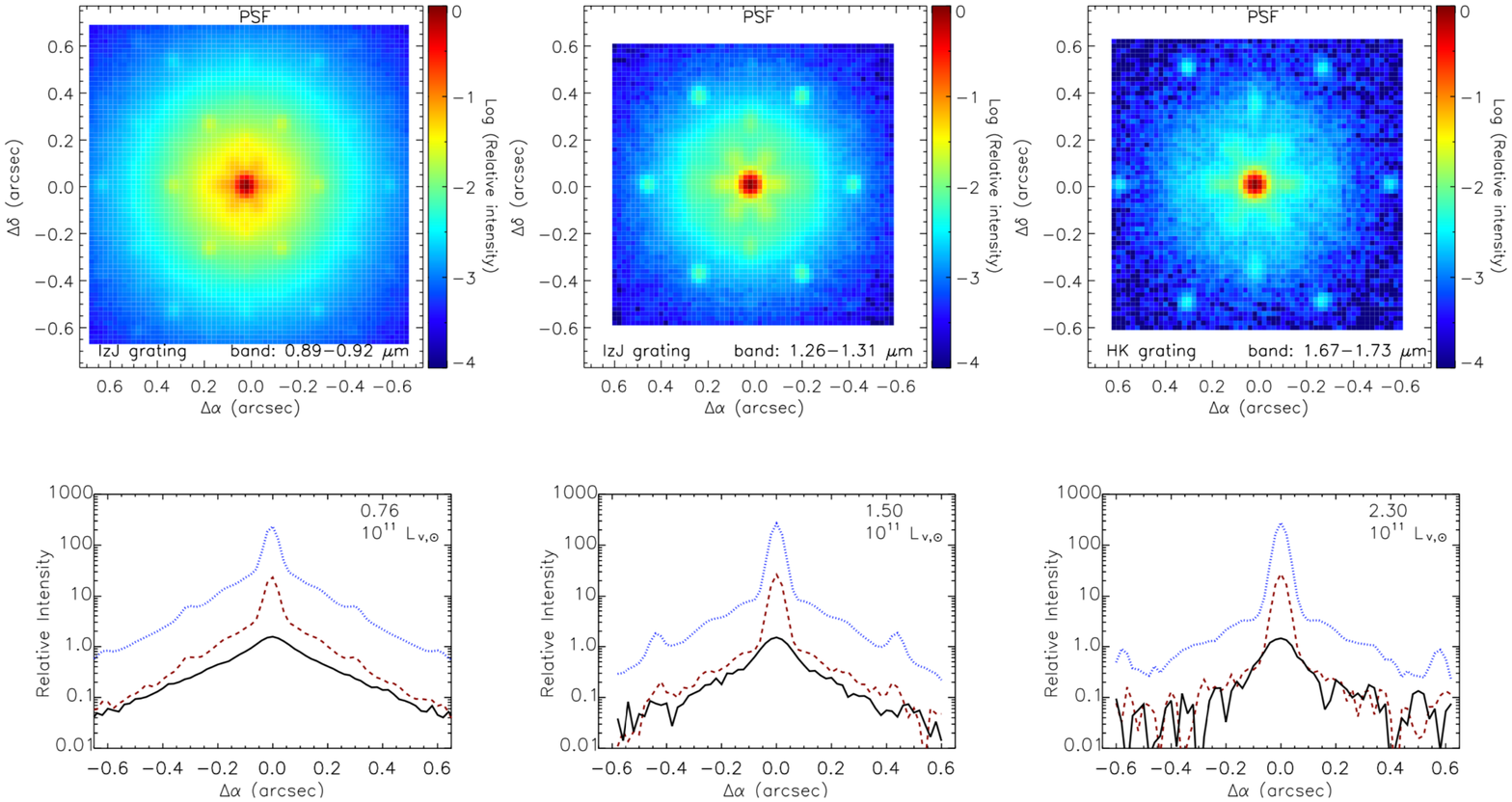}

  \caption{
  The PSF for HARMONI in the instrument configuration summaries in Table \ref{HSIM_par} at three wavelengths ranges. (Upper panels): Filter-band images (0.89-0.92 $\mu$m (left), 1.26-1.31 $\mu$m (centre), and 1.67-1.73 $\mu$m (right)). The 20$\times$20 mas scale of our mock HARMONI observations undersamples the potential FWHM of the PSF, which varies from $\sim$6 mas at the bluest filter to $\sim$11 mas at the reddest one. Intensities are in logarithmic scale and normalised to the peak in each case. Lower panels: Radial profiles along the right ascension axis and through the centre obtained from the filter-band image of the lenticular host galaxy scaled to 10$^{11}$L$_{v,\odot}$ without QSO (black profile), the QSO of L$_{[OIII]}$ = 10$^{42}$ (red dashed-profile), and the QSO of L$_{[OIII]}$ = 10$^{43}$ (blue dotted-line). Intensities are normalised to the peak of the host in each case. }
    \label{cut_PSF}
\end{figure*}
 
\begin{figure*}
\centering
\includegraphics[trim={1cm 1.5cm 5cm 0cm},width=5.5cm]{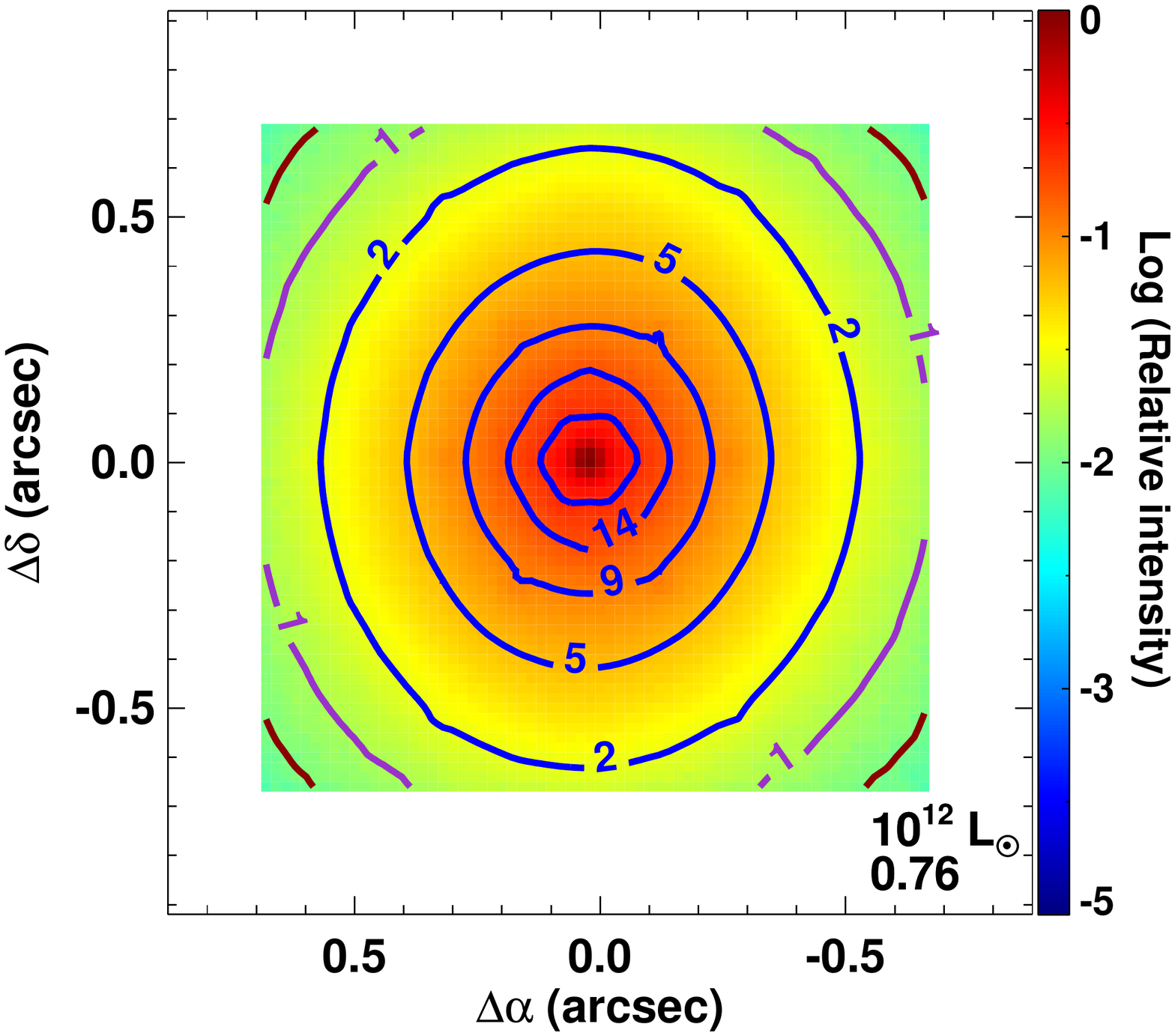}
\includegraphics[trim={1cm 1.5cm 5cm 0cm},width=5.5cm]{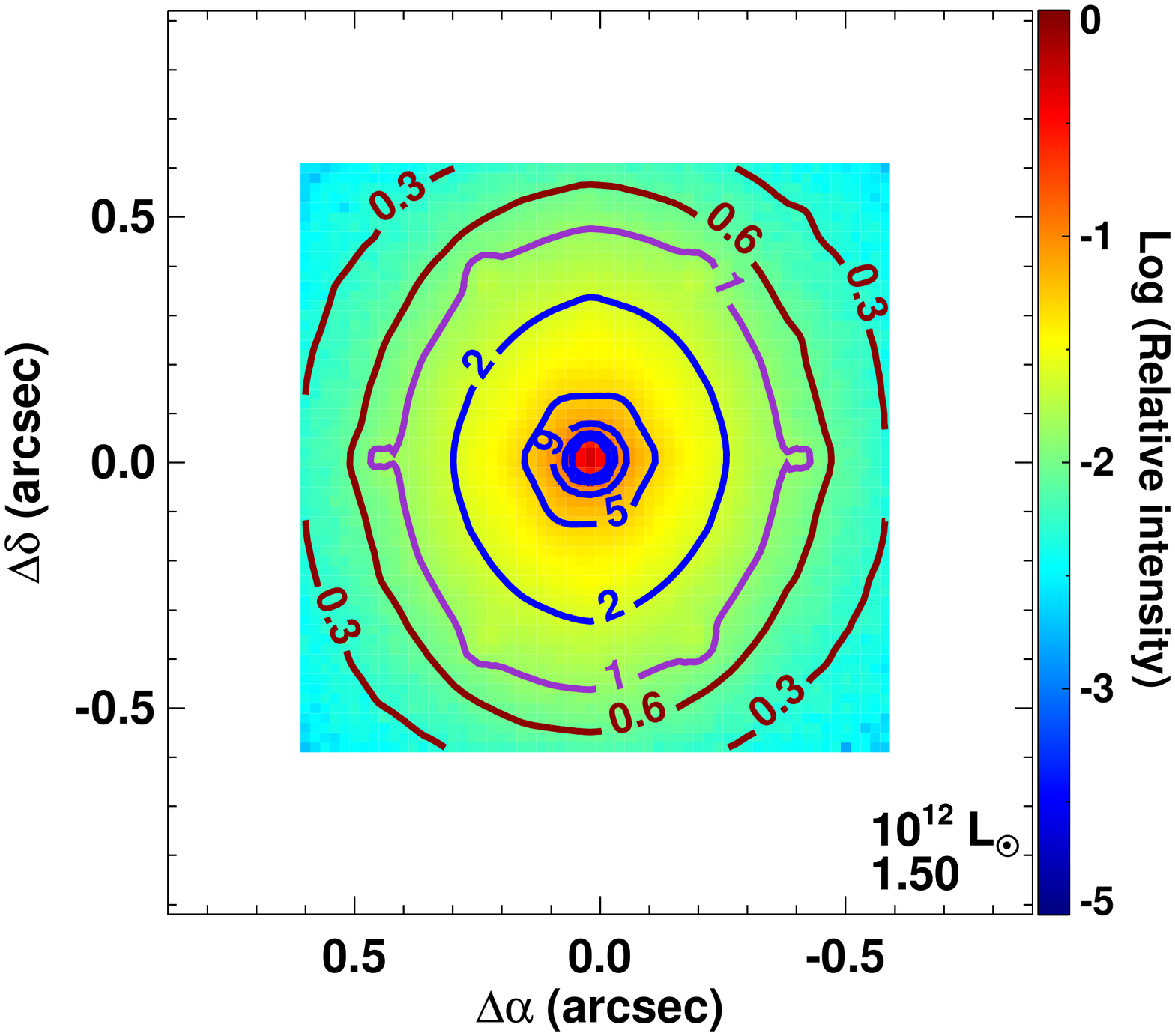}
\includegraphics[trim={1cm 1.5cm 5cm 0cm},width=5.5cm]{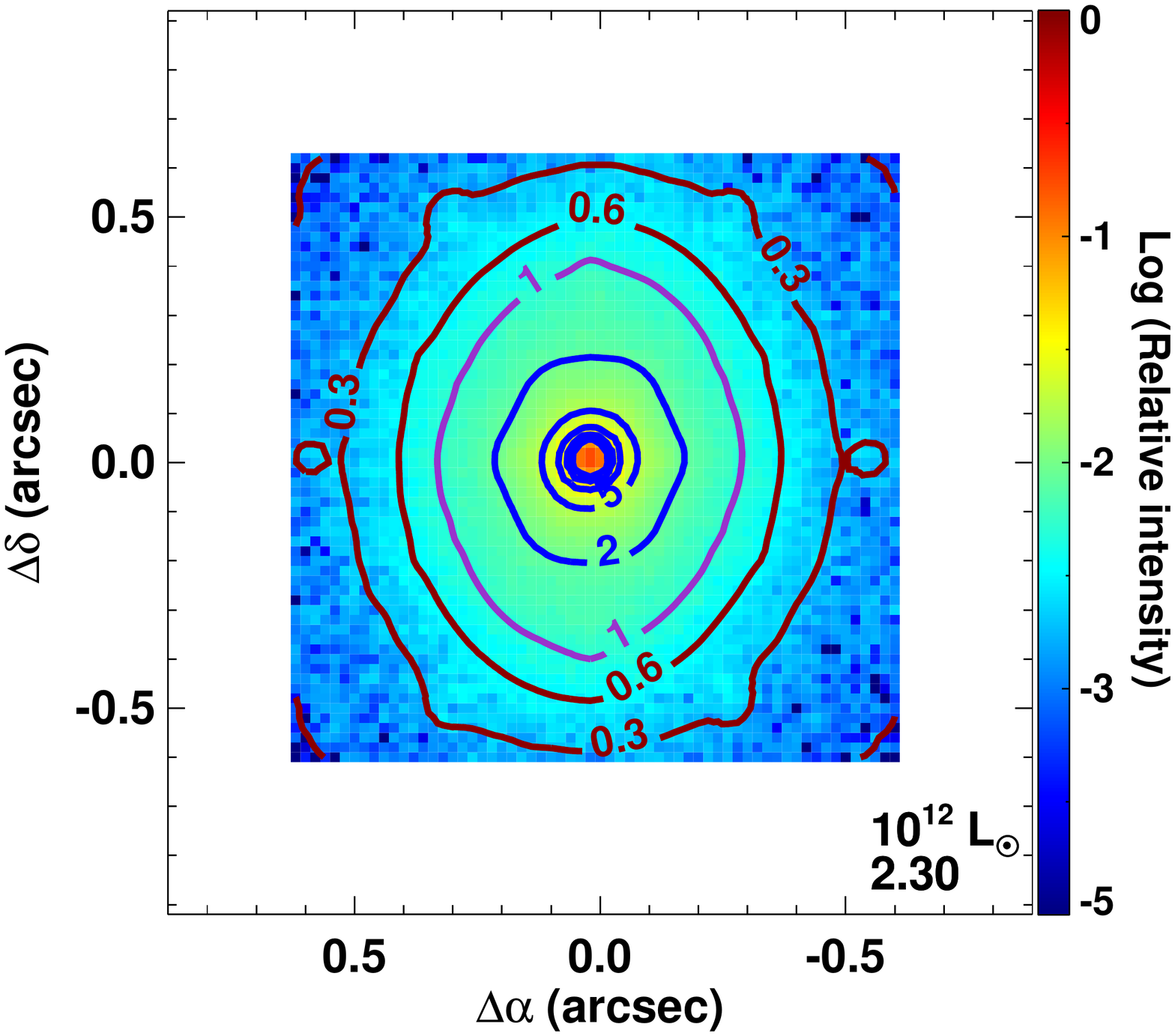}

 \includegraphics[trim={1cm 1.5cm 5cm 0cm},width=5.5cm]{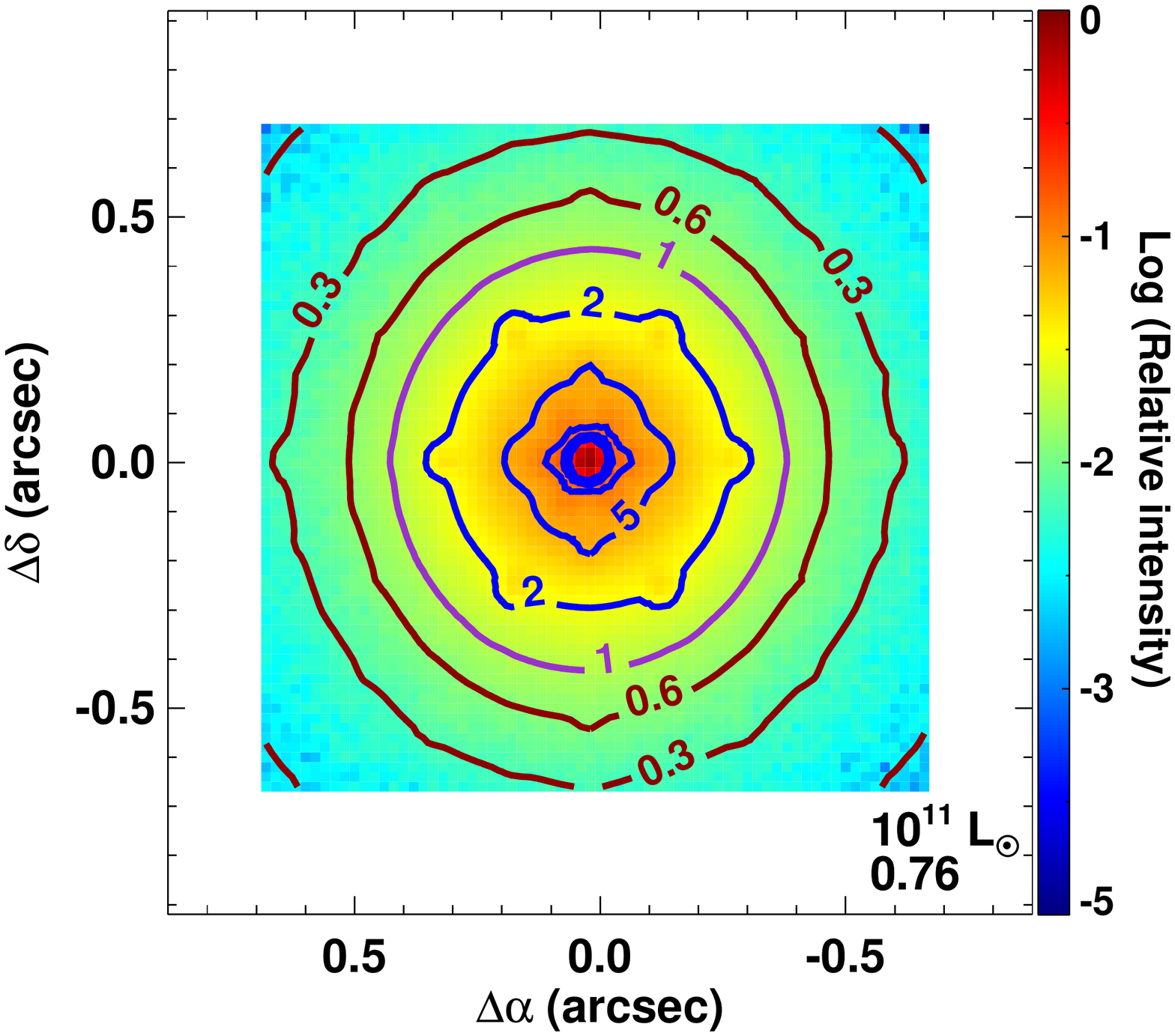}
 \includegraphics[trim={1cm 1.5cm 5cm 0cm},width=5.5cm]{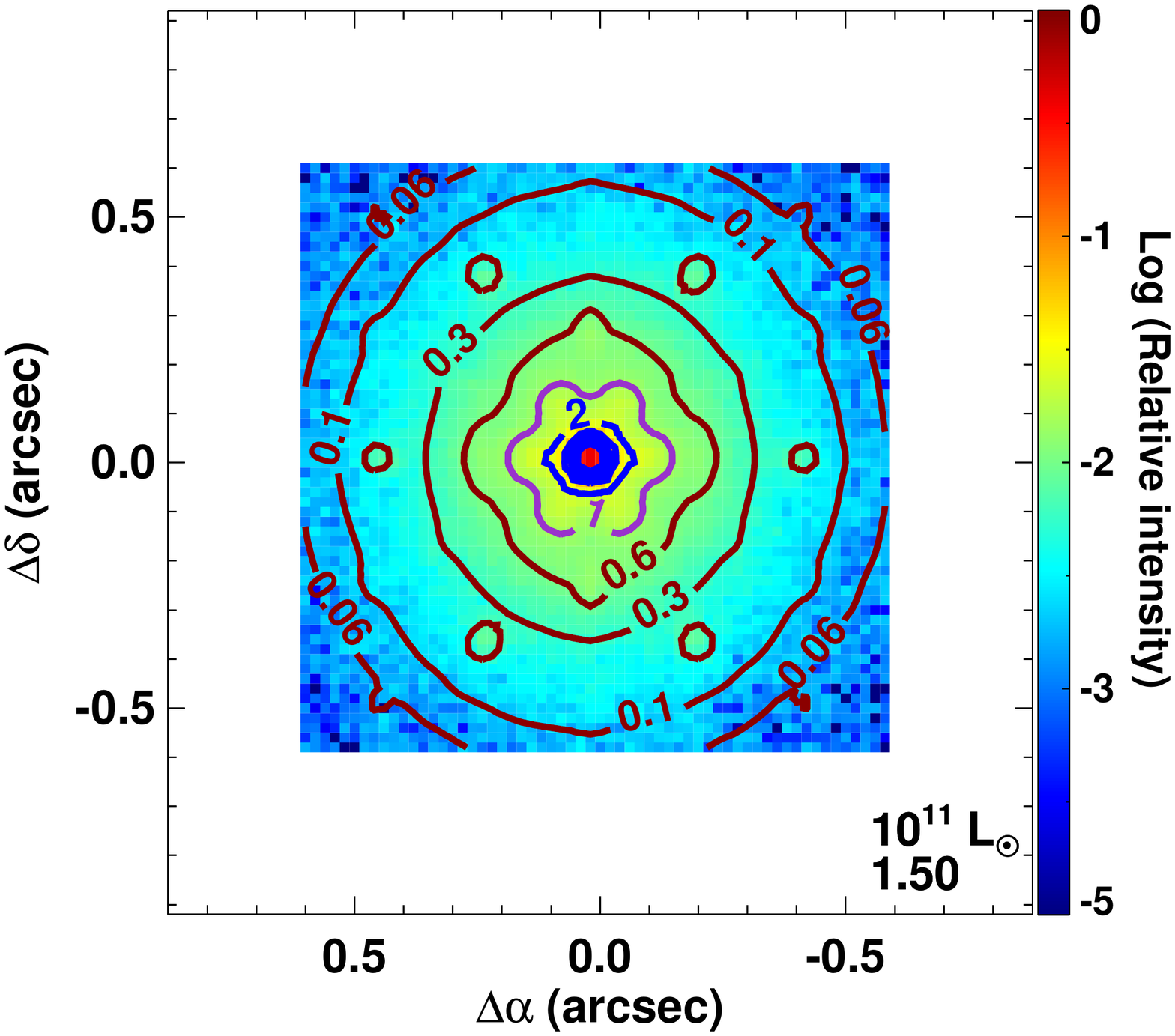}
 \includegraphics[trim={1cm 1.5cm 5cm 0cm},width=5.5cm]{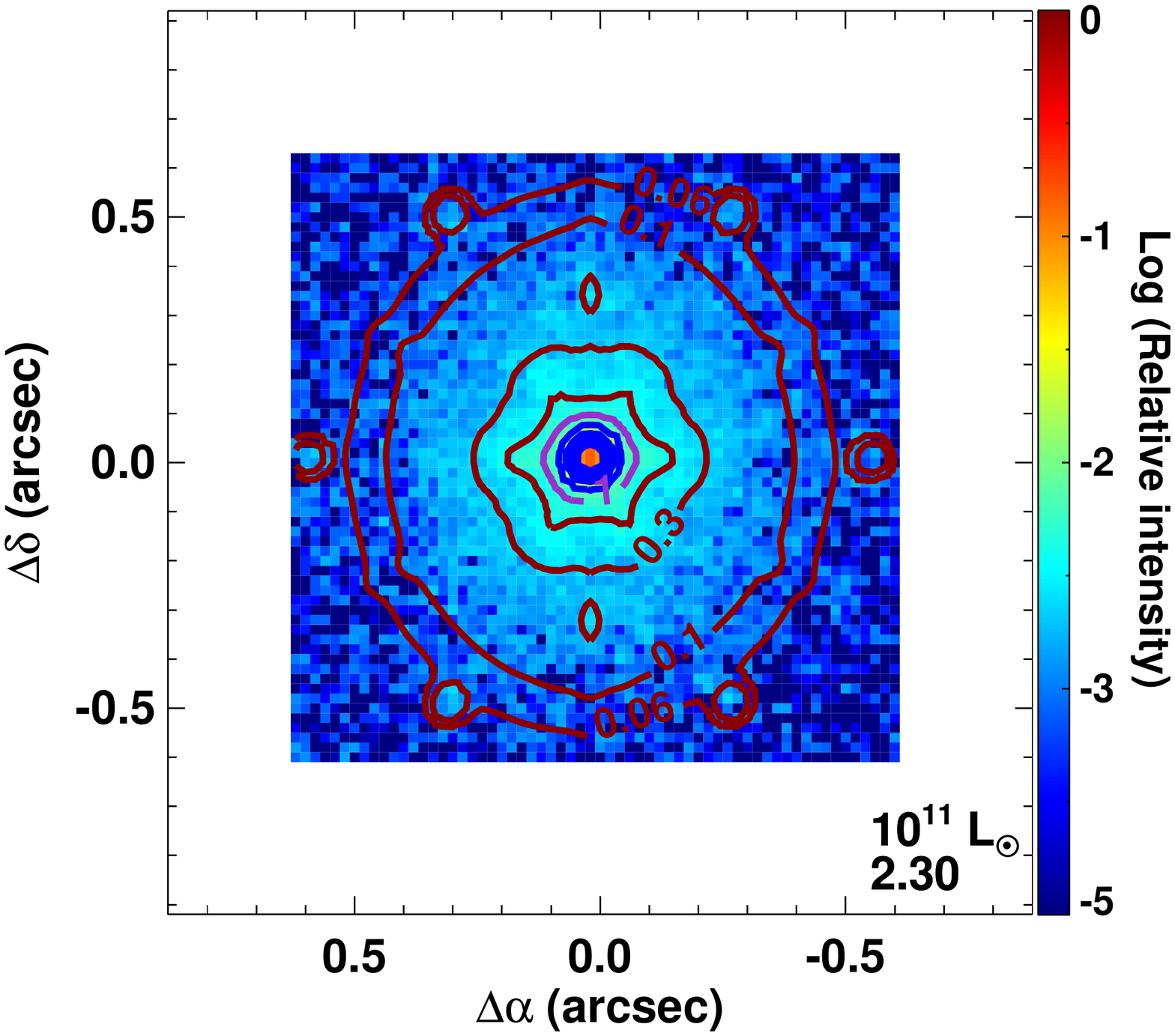}
 
  \includegraphics[trim={1cm 1.5cm 5cm 0cm},width=5.5cm]{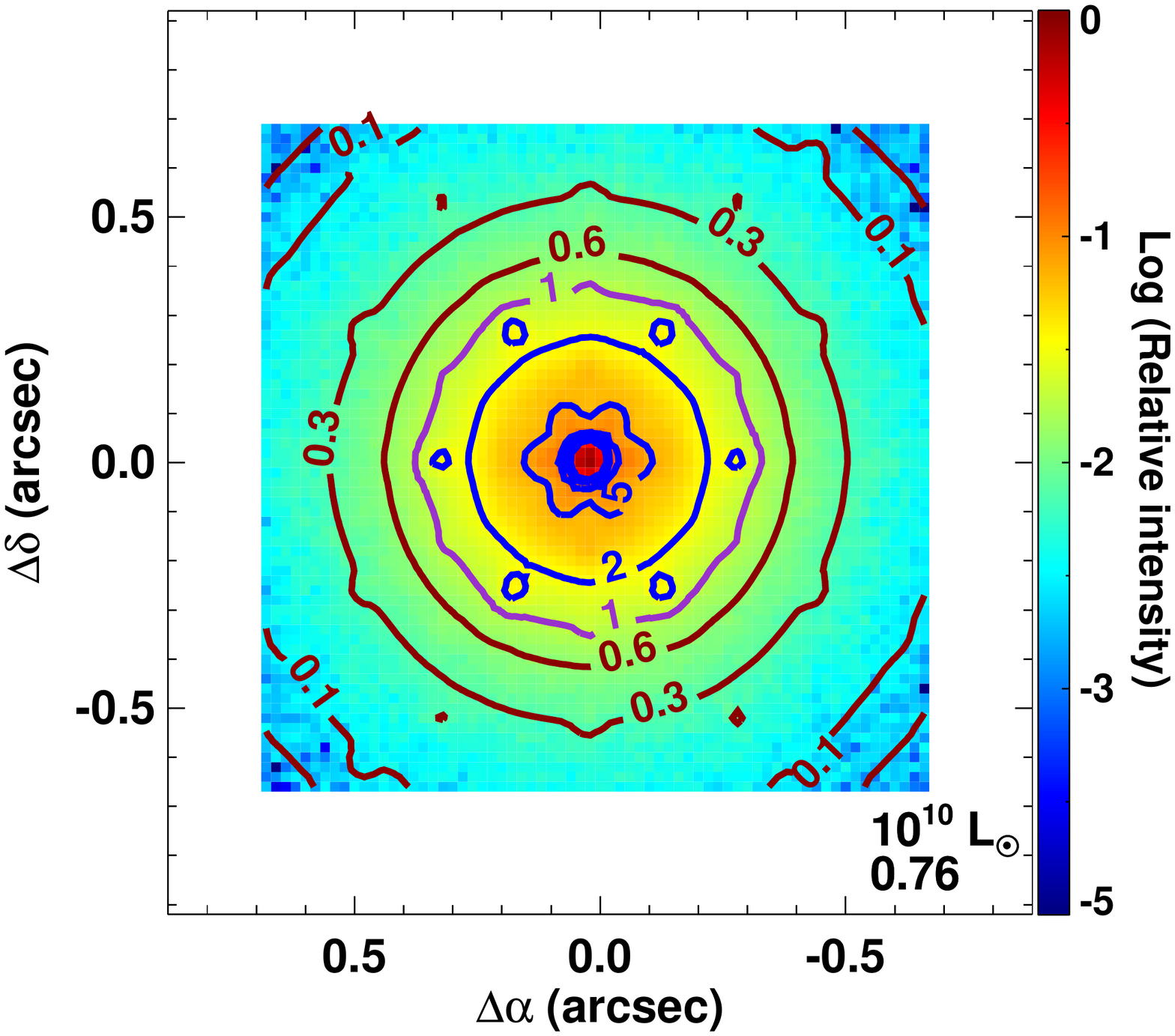}
 \includegraphics[trim={1cm 1.5cm 5cm 0cm},width=5.5cm]{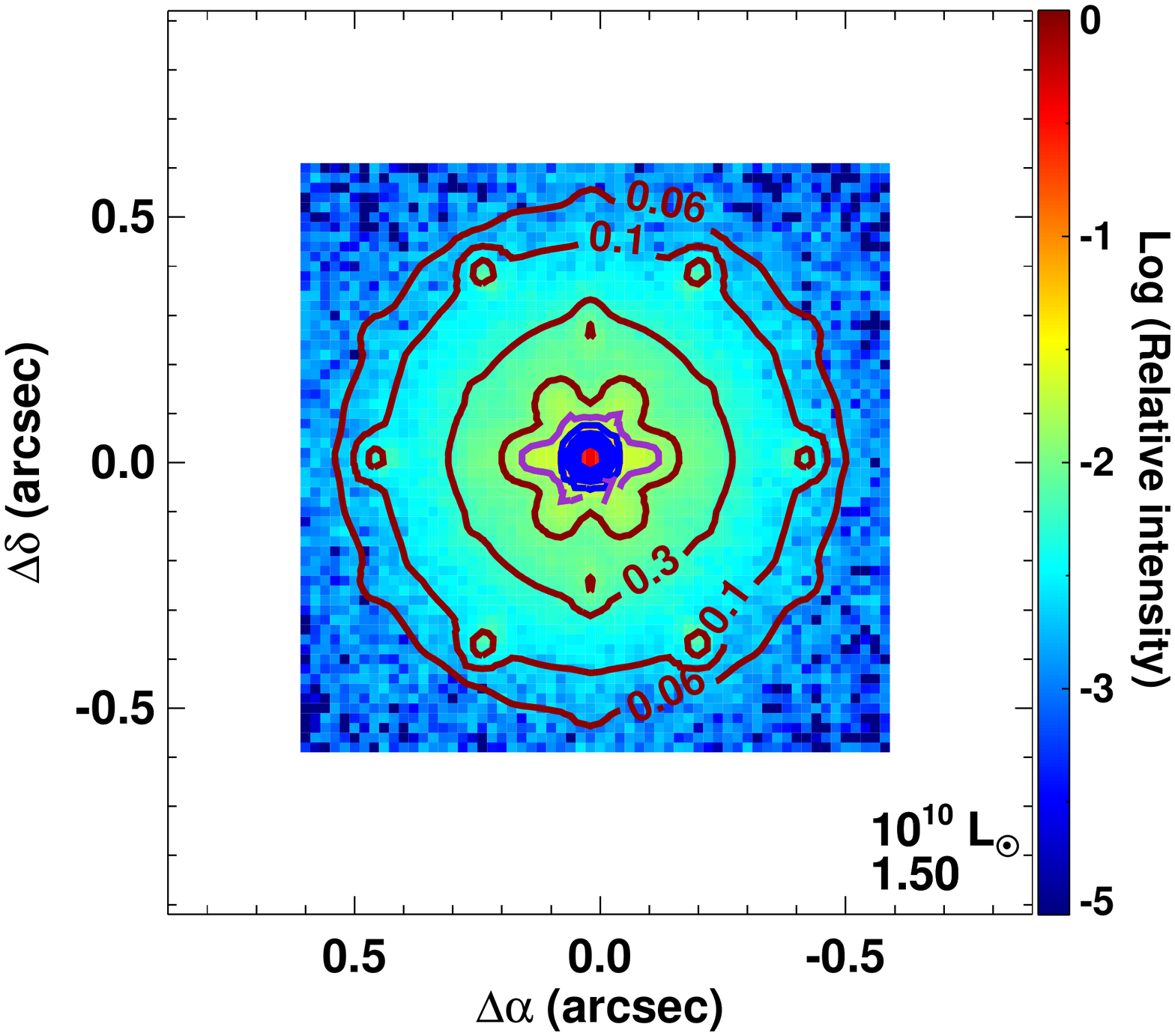}
 \includegraphics[trim={1cm 1.5cm 5cm 0cm},width=5.5cm]{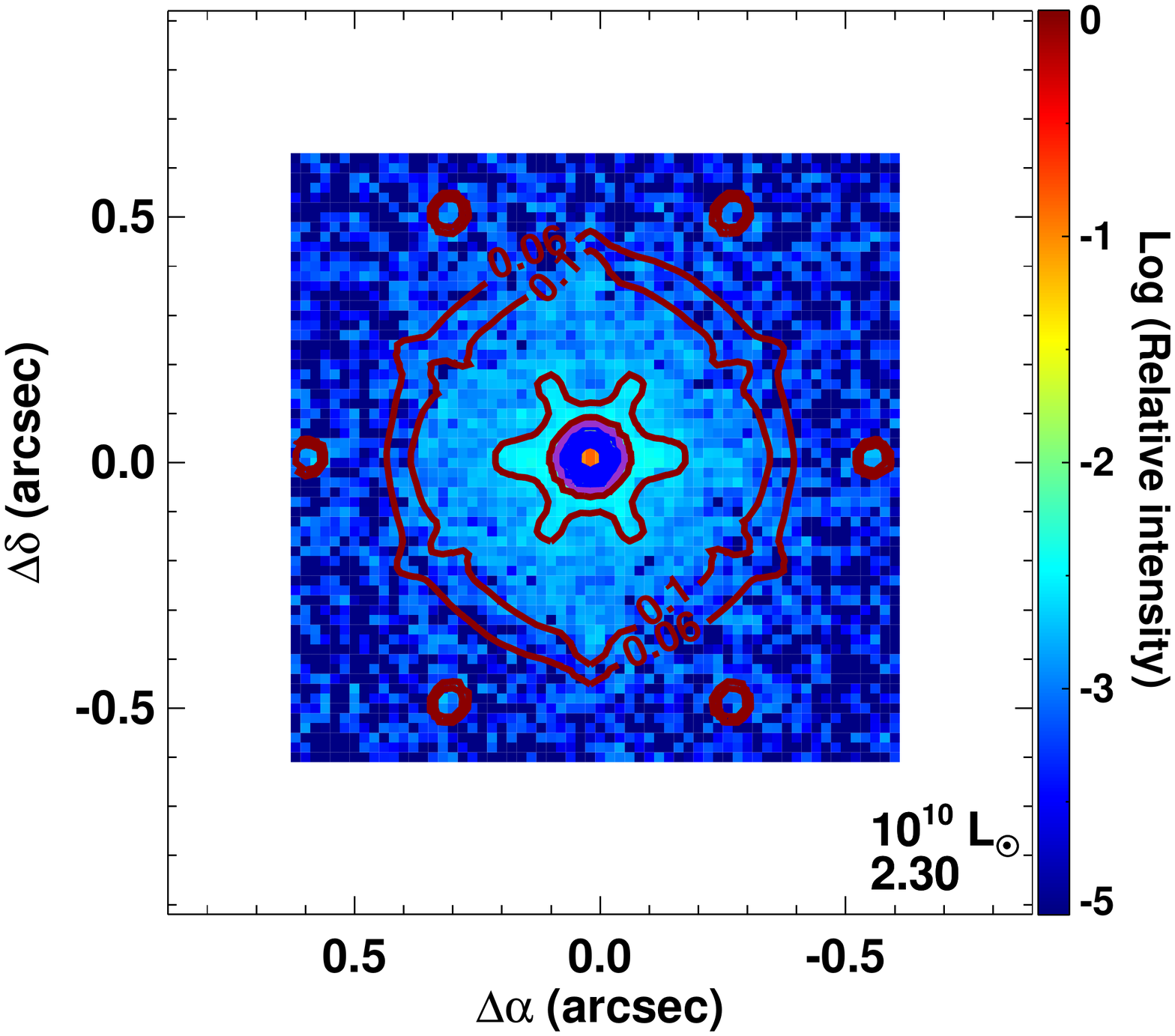}
  \caption{Filter-band images obtained by integrating the signal of the HSIM output cubes in the spectral bands (H$\beta$+MgI region) of the IzJ-grating (0.82-0.93 $\mu$m (left) and 1.17-1.33 $\mu$m (centre)) and the H+K-grating (1.55-1.75 $\mu$m (right)) when considering a QSO of L$_{[OIII]}\sim$10$^{42}$ erg s$^{-1}$ at the nucleus of the lenticular host (i.e. NGC\,809) at redshifts 0.76 (left), 1.50 (middle), and 2.30 (right). Host luminosities are scaled to 10$^{12}$L$_{v,\odot}$ (top), 10$^{11}$L$_{v,\odot}$ (middle) and 10$^{10}$L$_{v,\odot}$ (bottom). Intensity is in logarithmic scale and normalised to the peak of the brightest case (top left panel). Contours correspond to the average S/N in the rest-frame 5050-5250 \AA \ corresponding to a stellar continuum region including the MgI features (brown contours are S/Ns of 0.06, 0.1, 0.3, and 0.6; S/Ns values of 2, 5, 9, 14, and 20 are drawn as blue contours; the purple contour indicates a S/Ns of 1). }
 \label{NGC809_HbMgI_range_42}
\end{figure*}

\begin{figure*}
\centering
\includegraphics[trim={1cm 1.5cm 5cm 0cm},width=5.5cm]{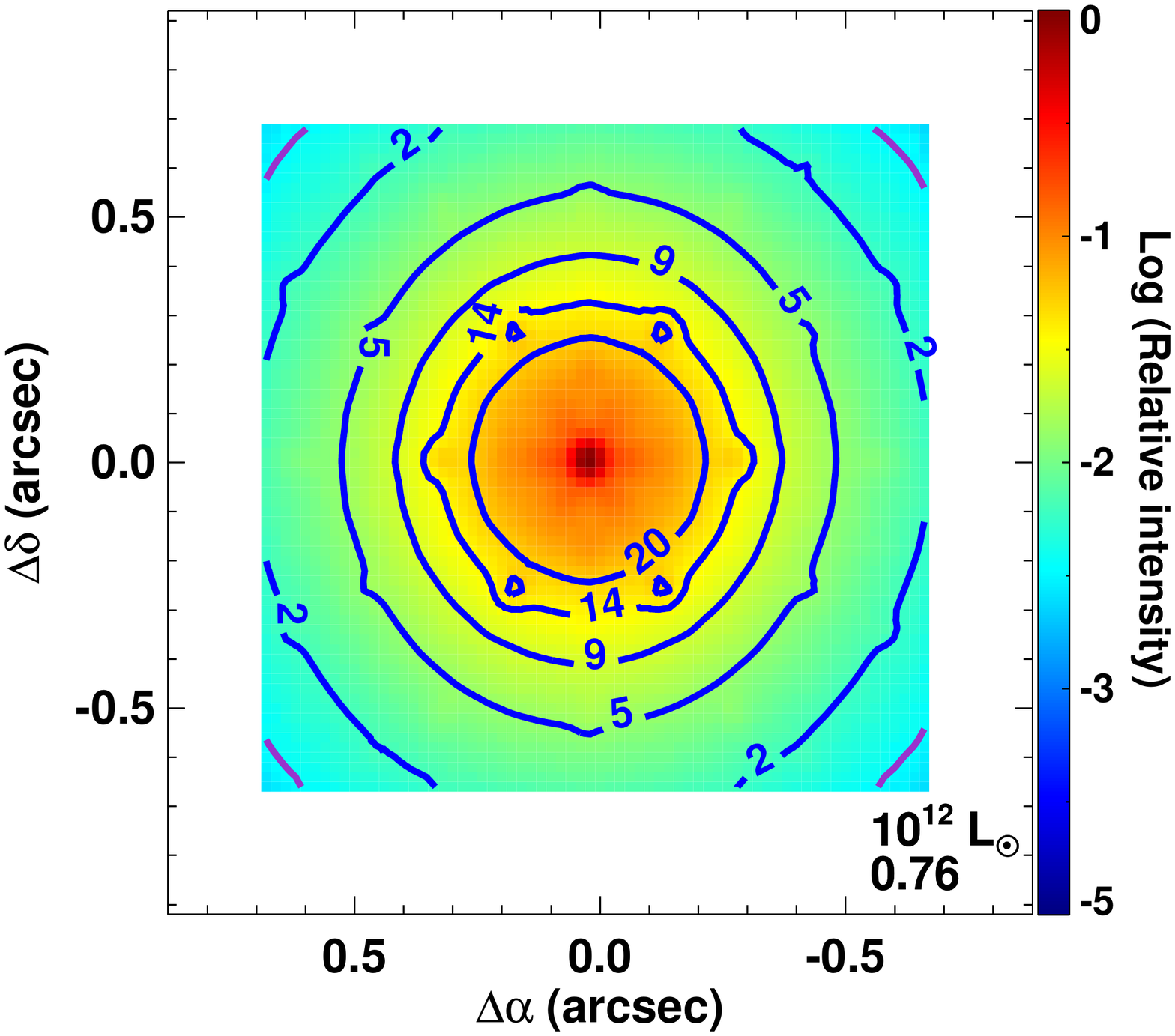}
\includegraphics[trim={1cm 1.5cm 5cm 0cm},width=5.5cm]{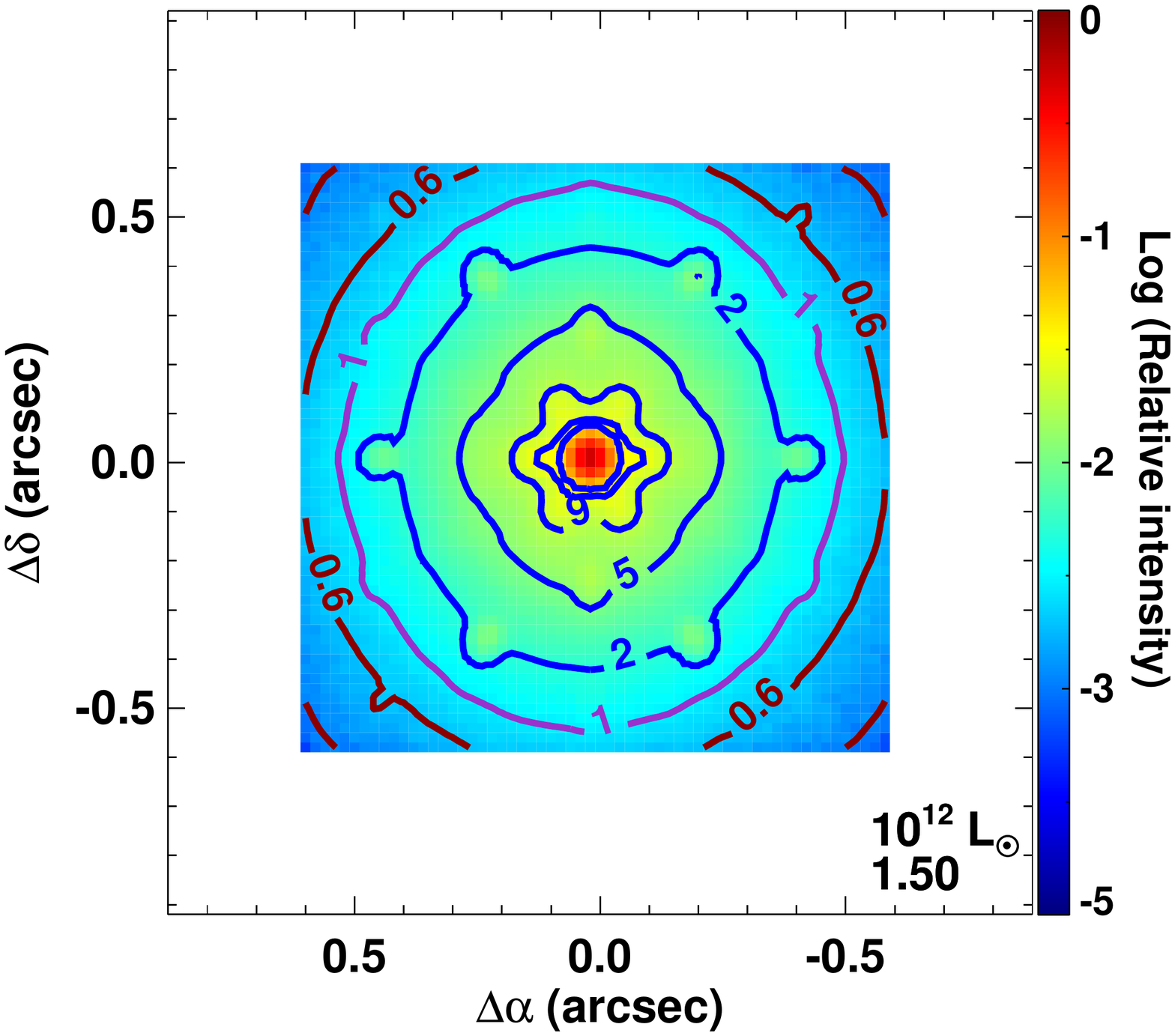}
\includegraphics[trim={1cm 1.5cm 5cm 0cm},width=5.5cm]{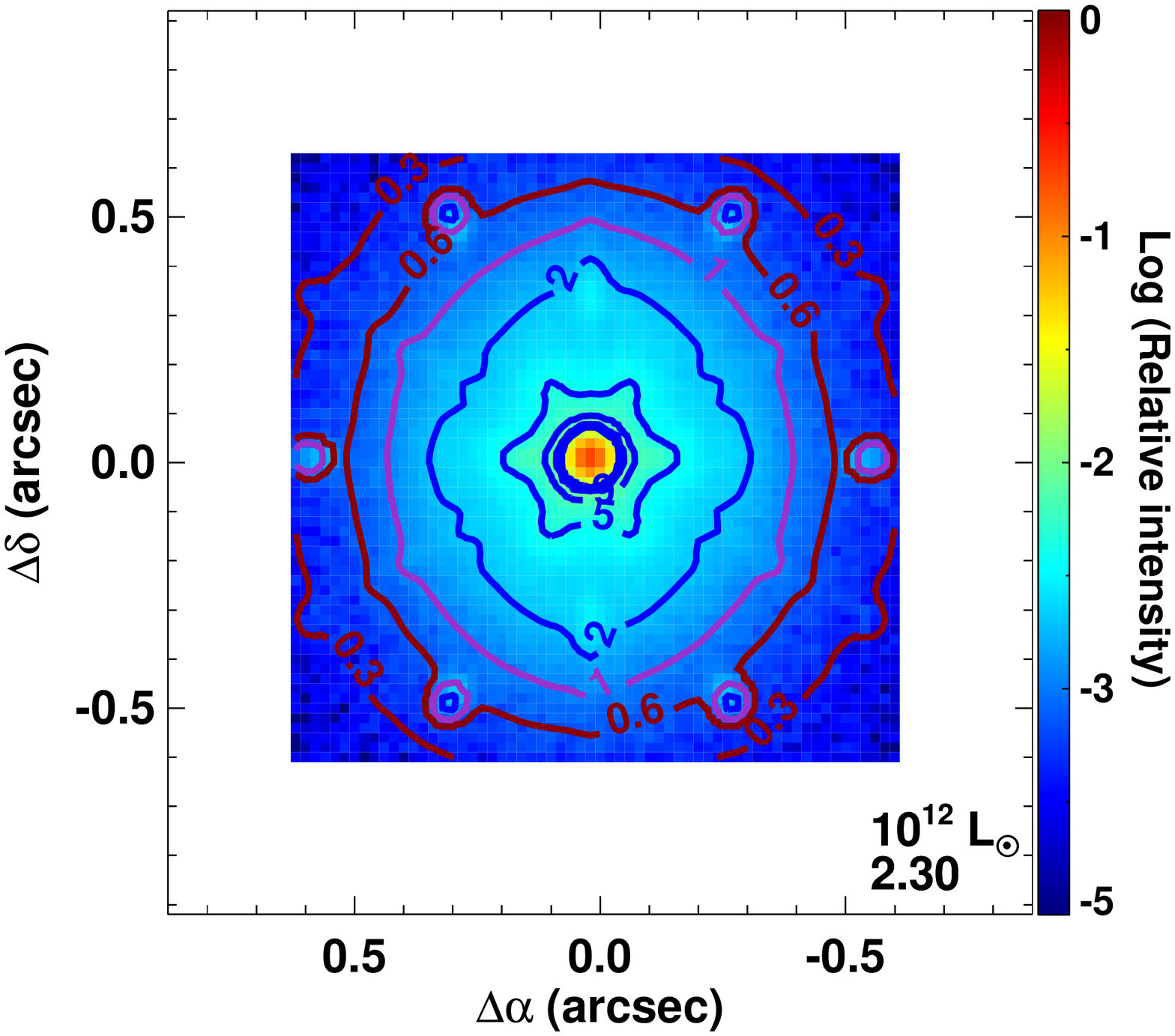}

 \includegraphics[trim={1cm 1.5cm 5cm 0cm},width=5.5cm]{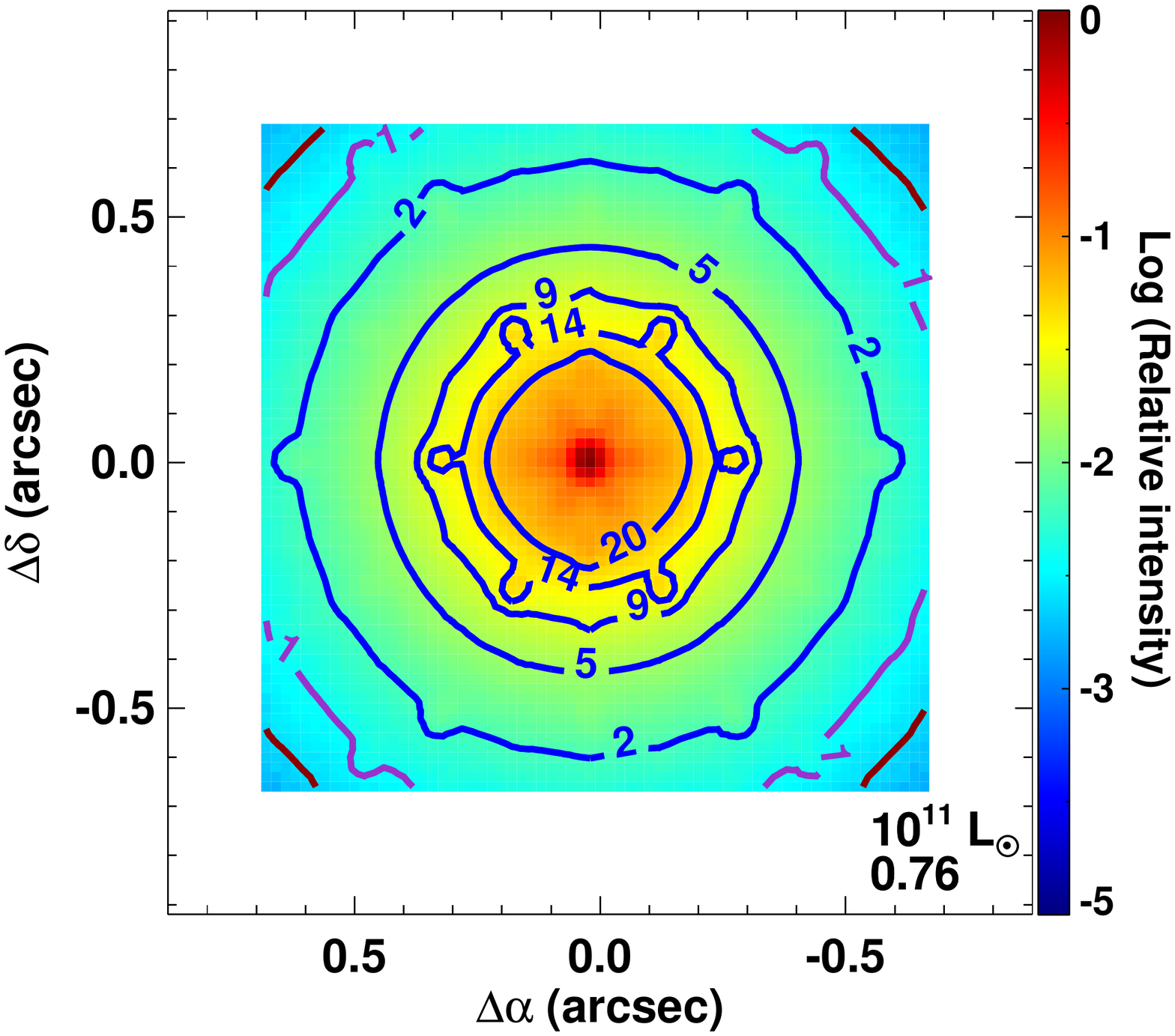}
 \includegraphics[trim={1cm 1.5cm 5cm 0cm},width=5.5cm]{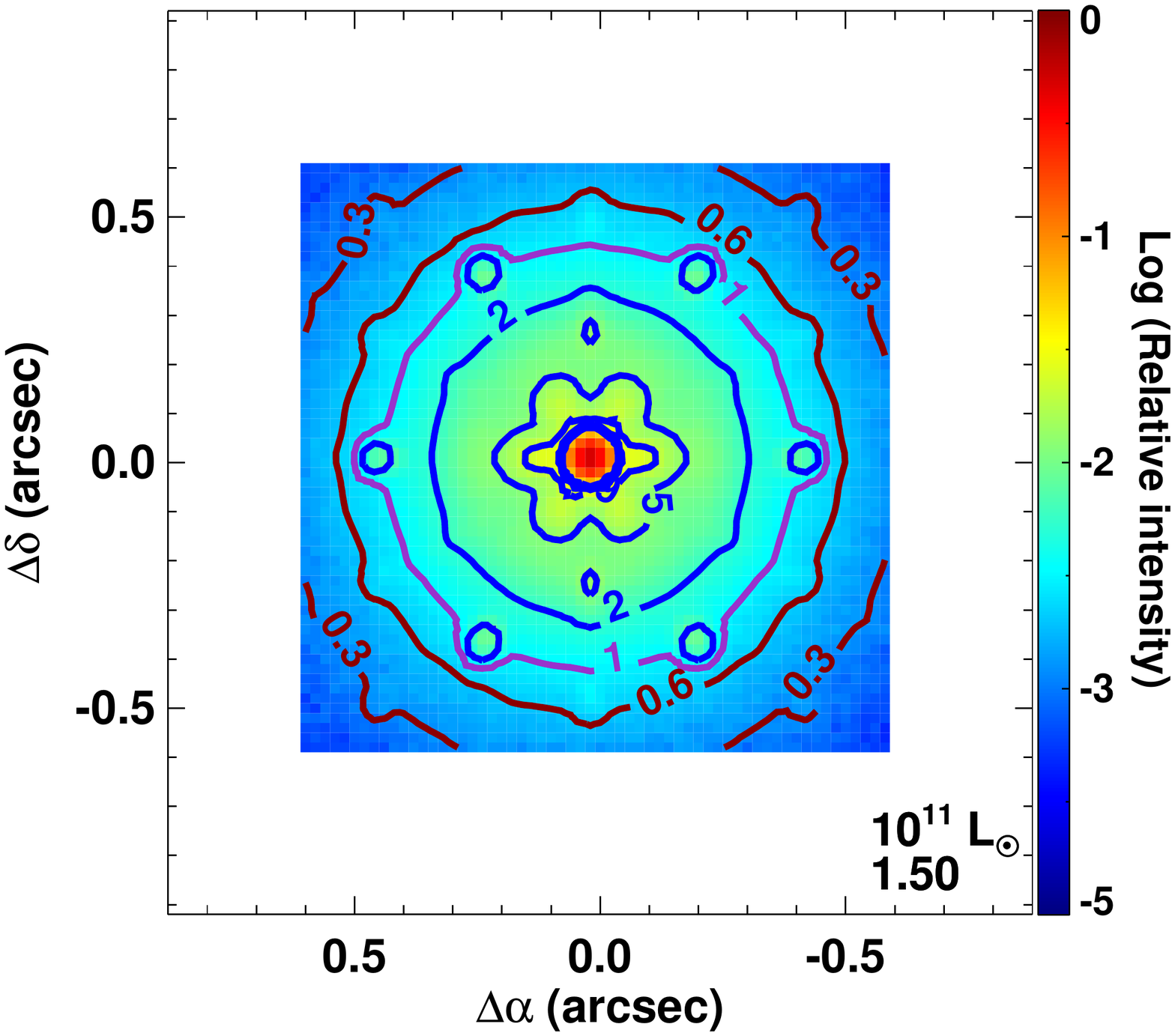}
 \includegraphics[trim={1cm 1.5cm 5cm 0cm},width=5.5cm]{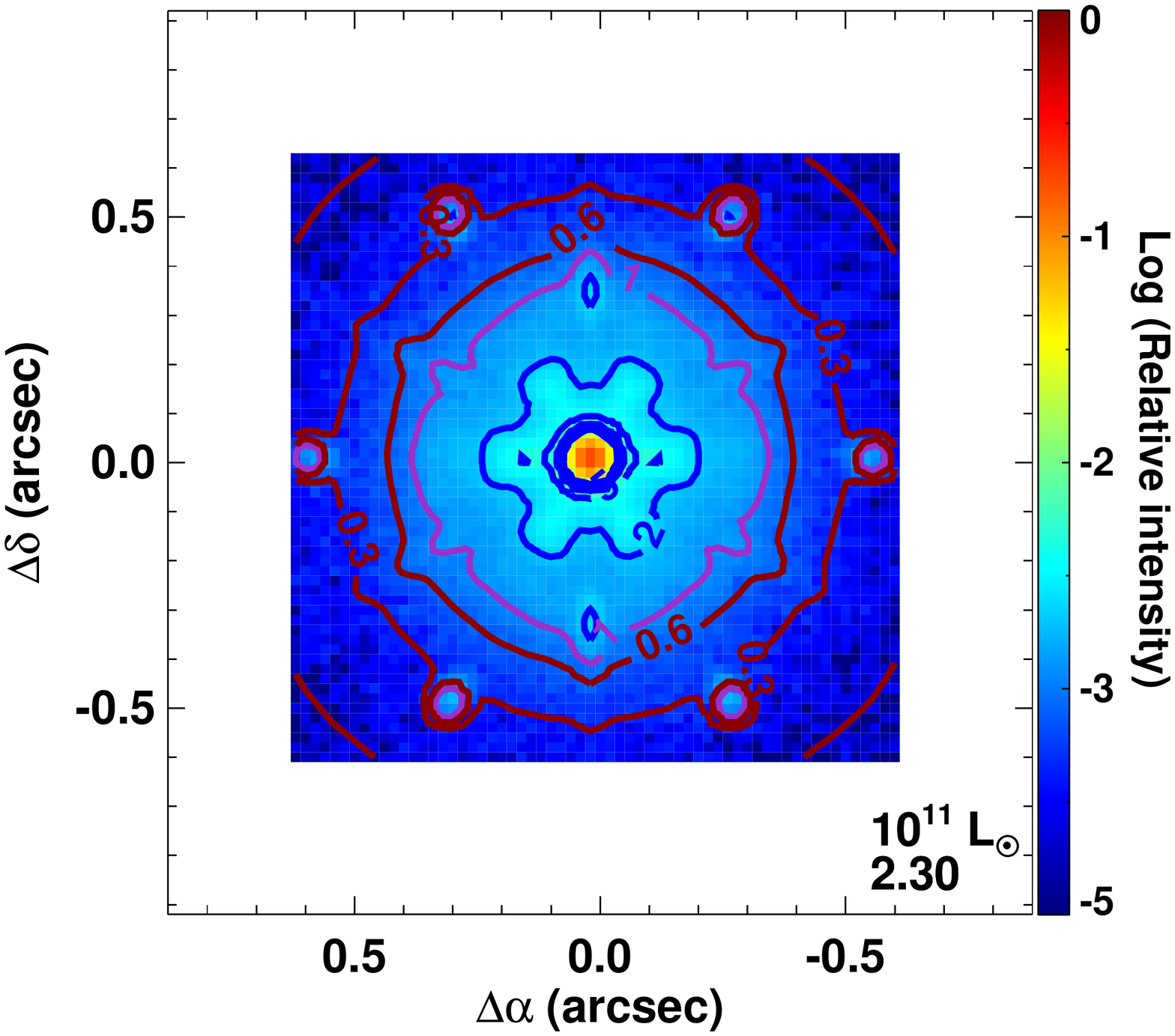}
 
  \includegraphics[trim={1cm 1.5cm 5cm 0cm},width=5.5cm]{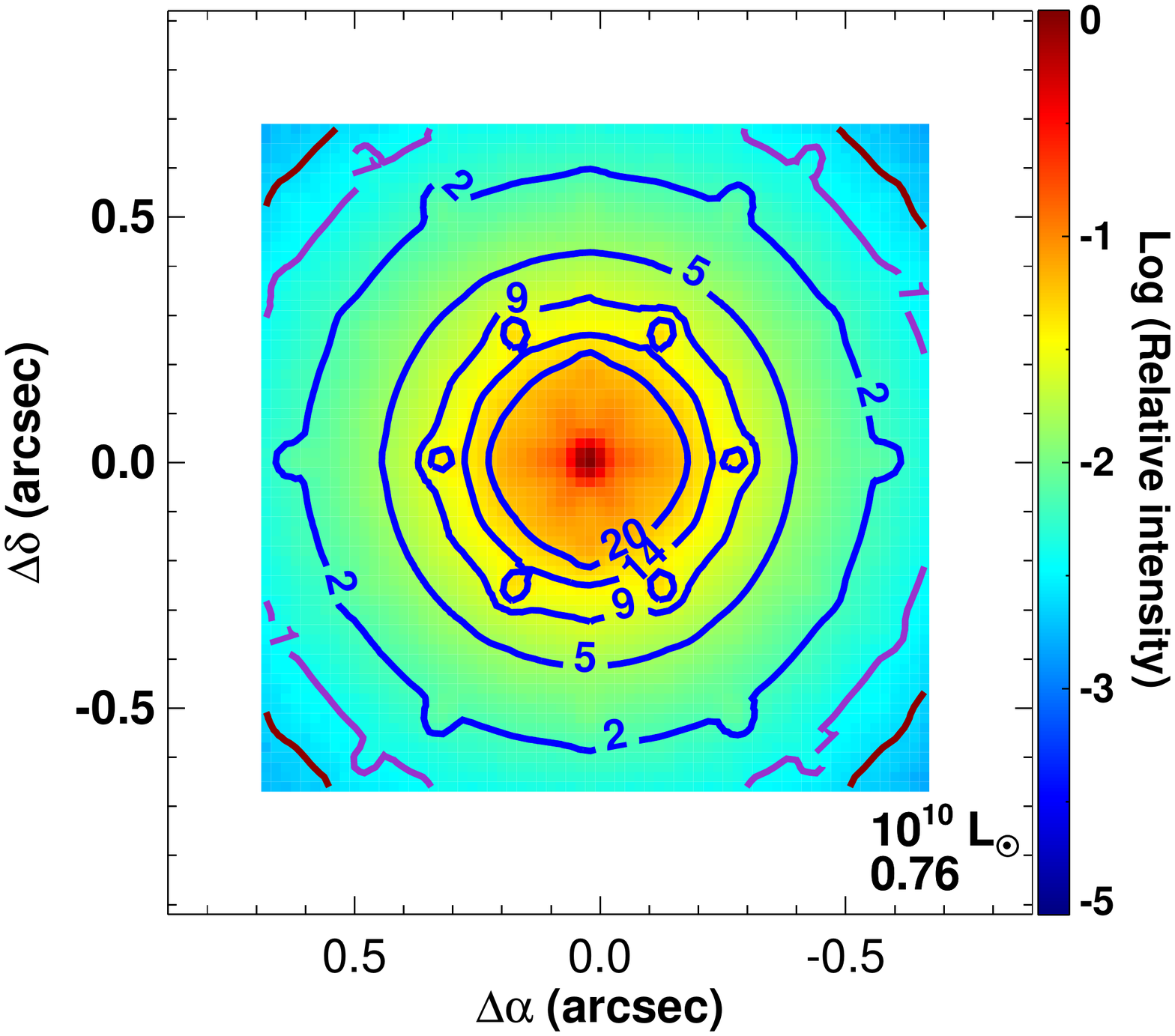}
 \includegraphics[trim={1cm 1.5cm 5cm 0cm},width=5.5cm]{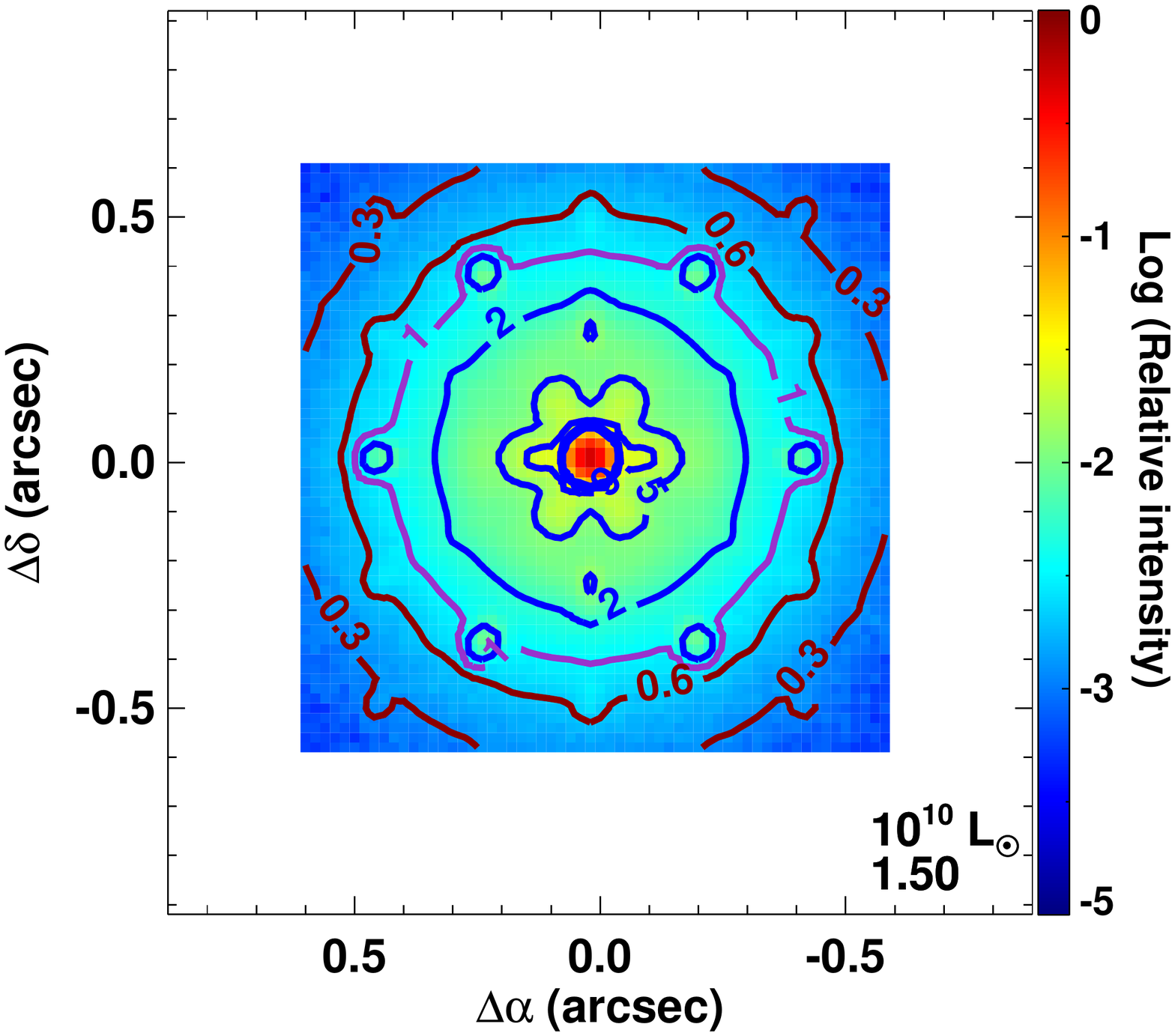}
 \includegraphics[trim={1cm 1.5cm 5cm 0cm},width=5.5cm]{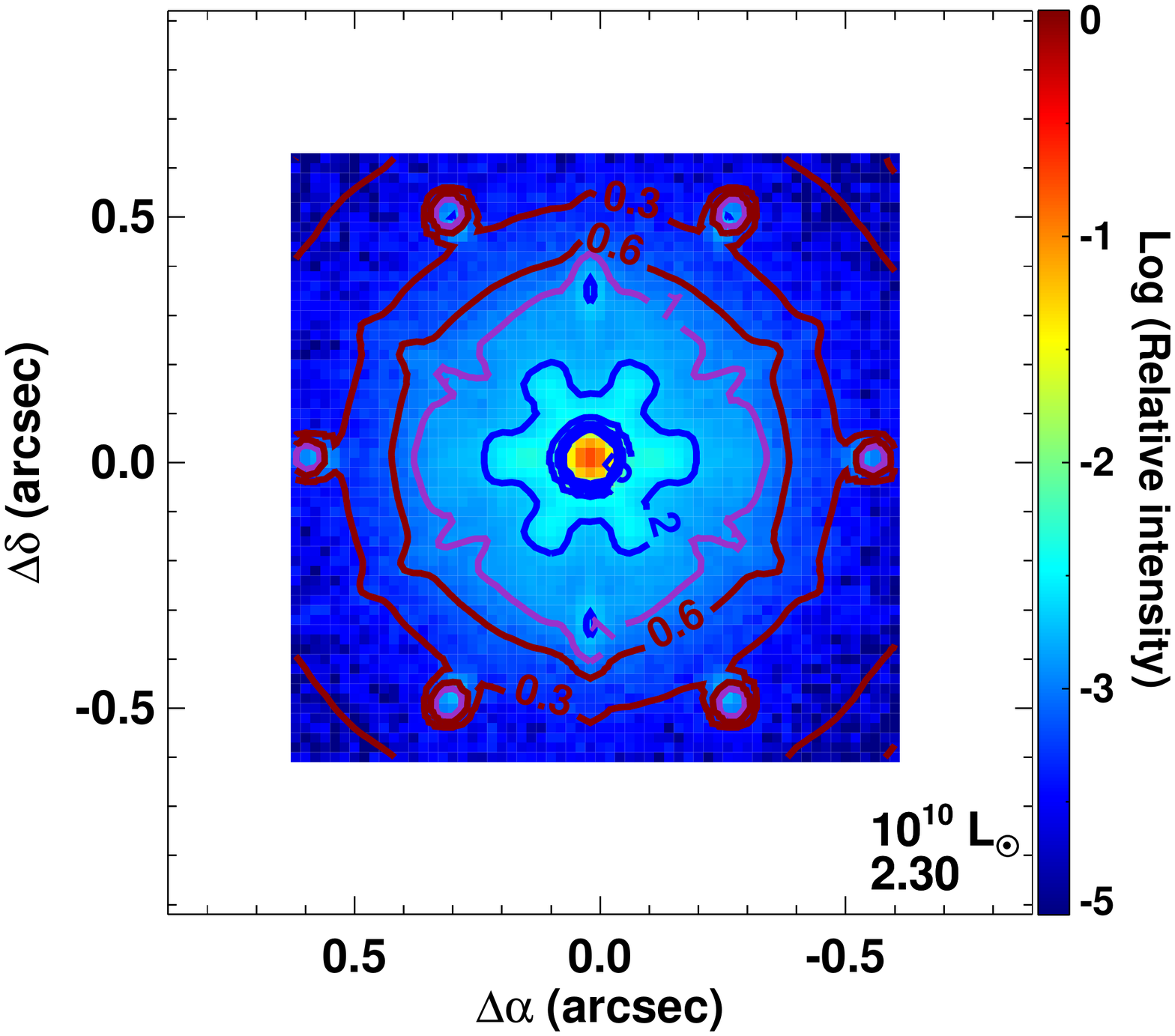}
  \caption{Same as Fig. \ref{NGC809_HbMgI_range_42}, but considering a QSO of L$_{[OIII]}\sim$10$^{43}$ erg s$^{-1}$ at the nucleus of the lenticular host (i.e. NGC\,809). }
 \label{NGC809_HbMgI_range_43}
\end{figure*}


\begin{figure*}
\centering
\includegraphics[trim={1cm 1.5cm 5cm 0cm},width=5.5cm]{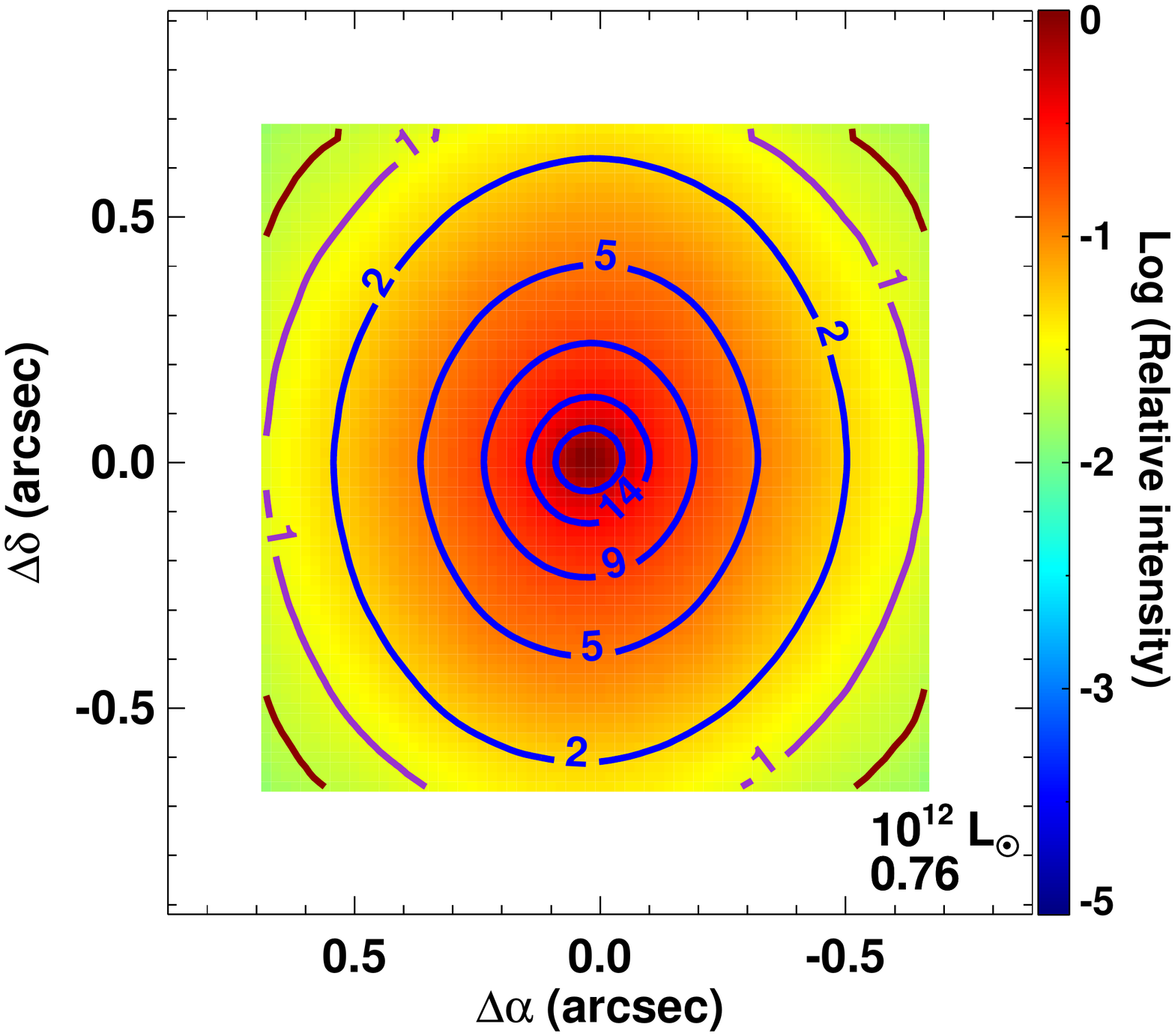}
\includegraphics[trim={1cm 1.5cm 5cm 0cm},width=5.5cm]{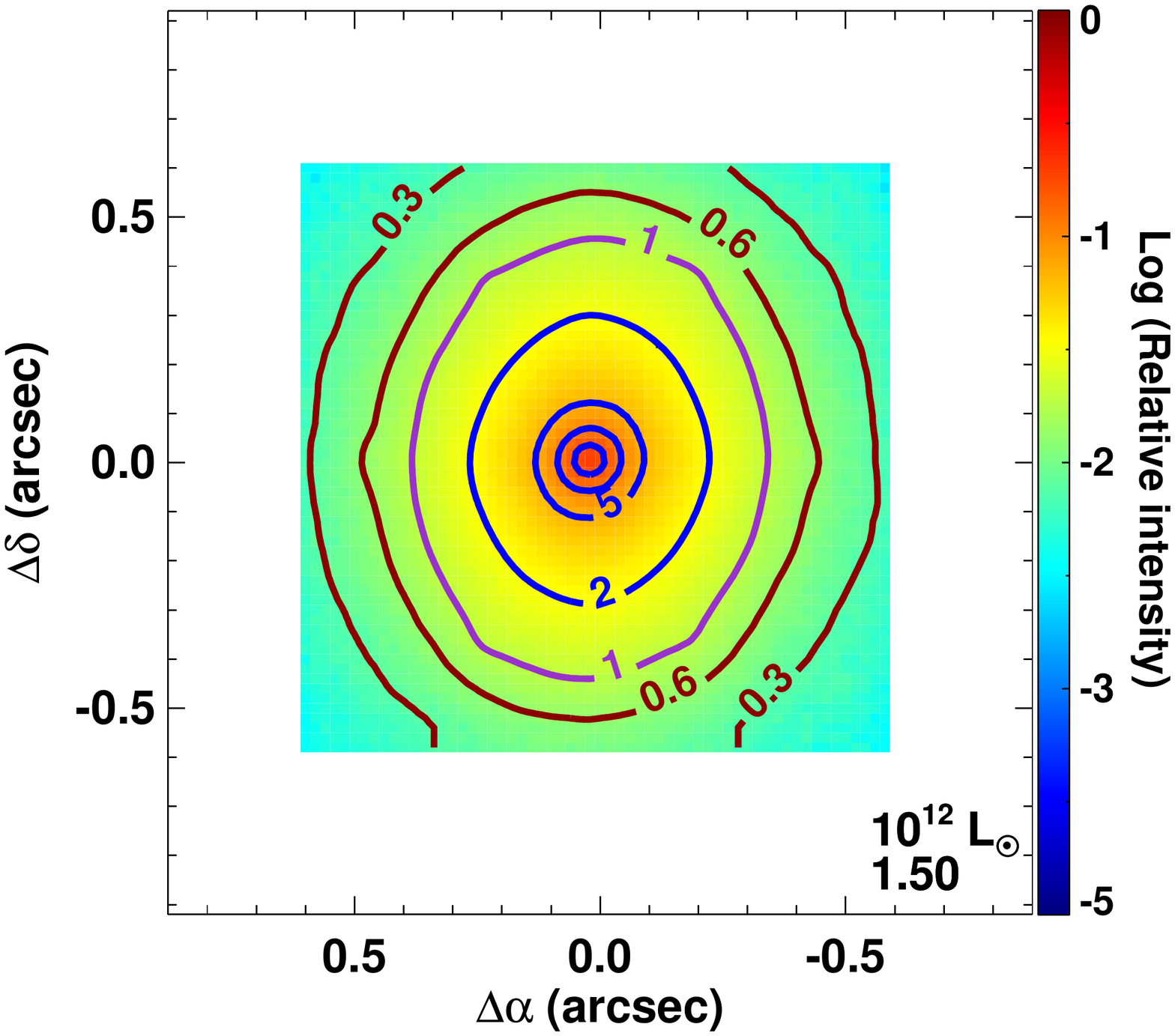}
\includegraphics[trim={1cm 1.5cm 5cm 0cm},width=5.5cm]{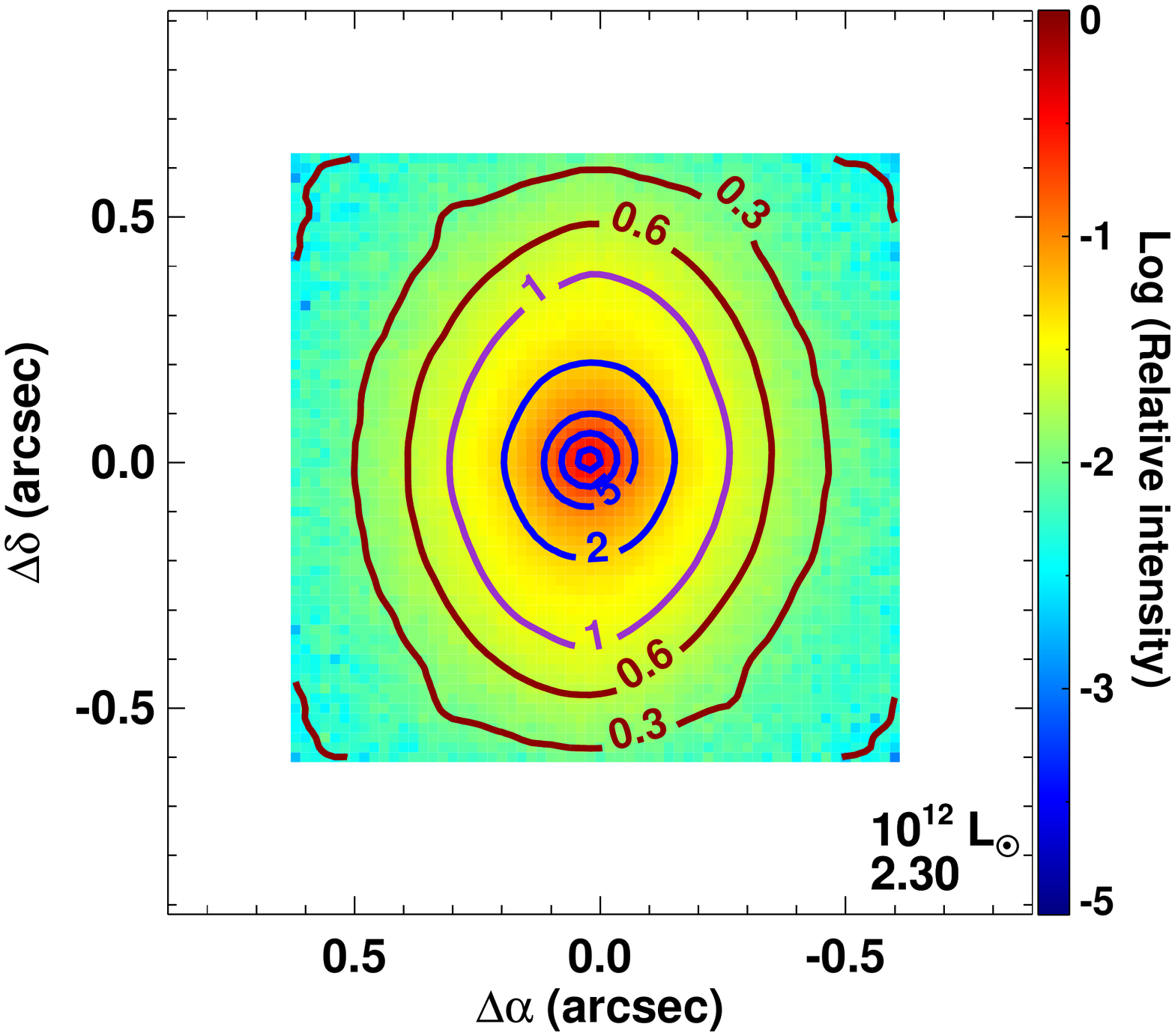}

 \includegraphics[trim={1cm 1.5cm 5cm 0cm},width=5.5cm]{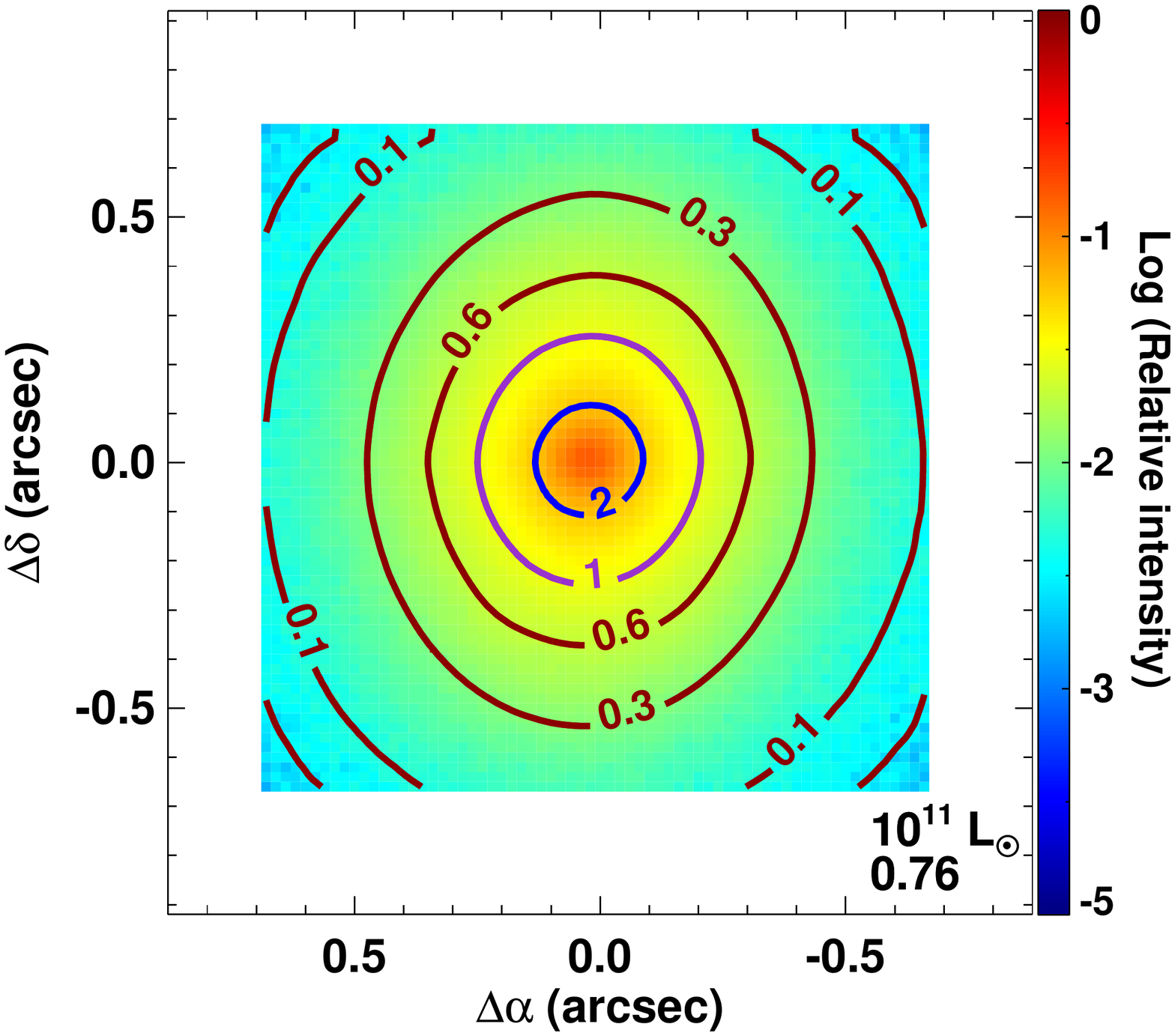}
 \includegraphics[trim={1cm 1.5cm 5cm 0cm},width=5.5cm]{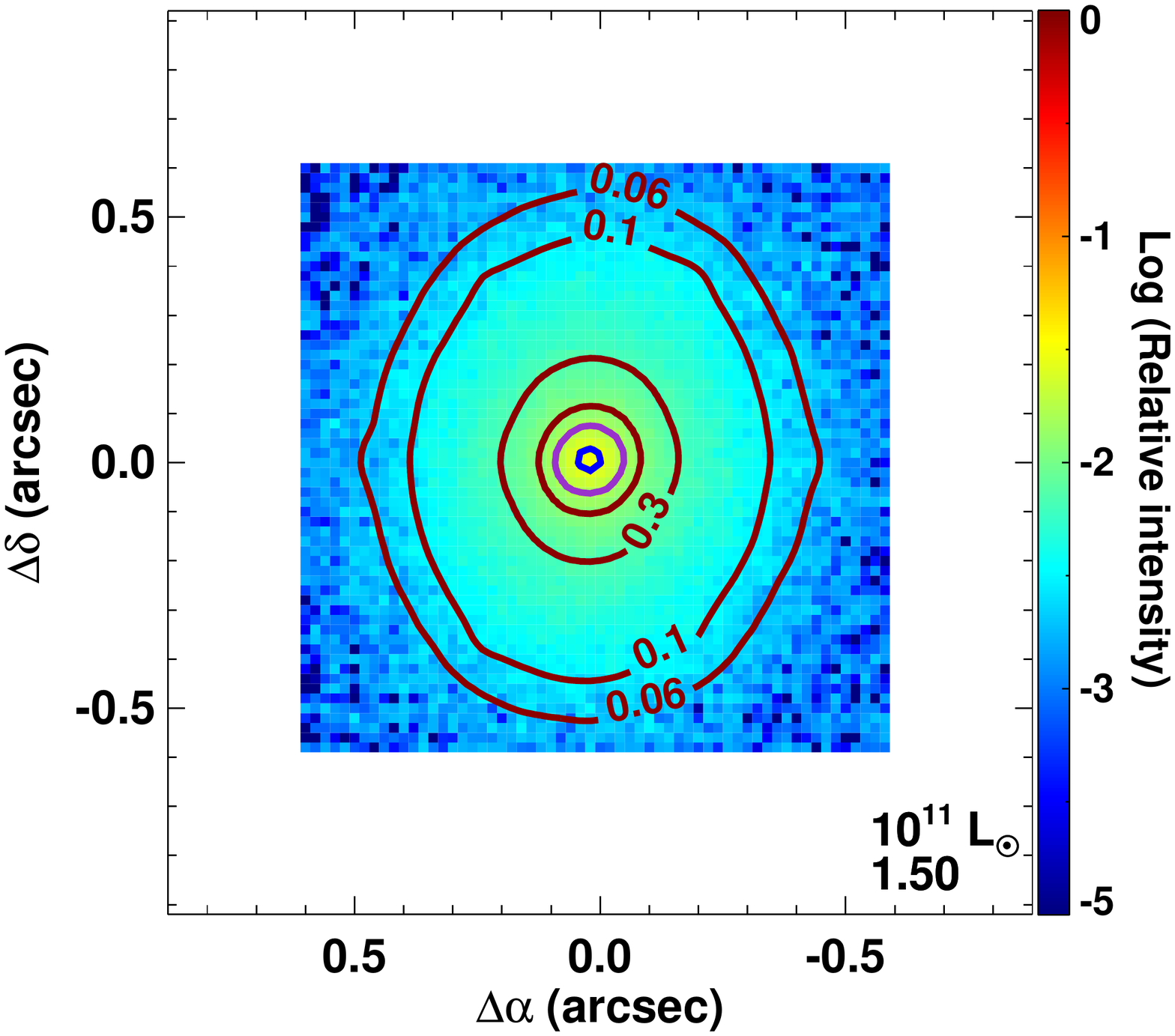}
 \includegraphics[trim={1cm 1.5cm 5cm 0cm},width=5.5cm]{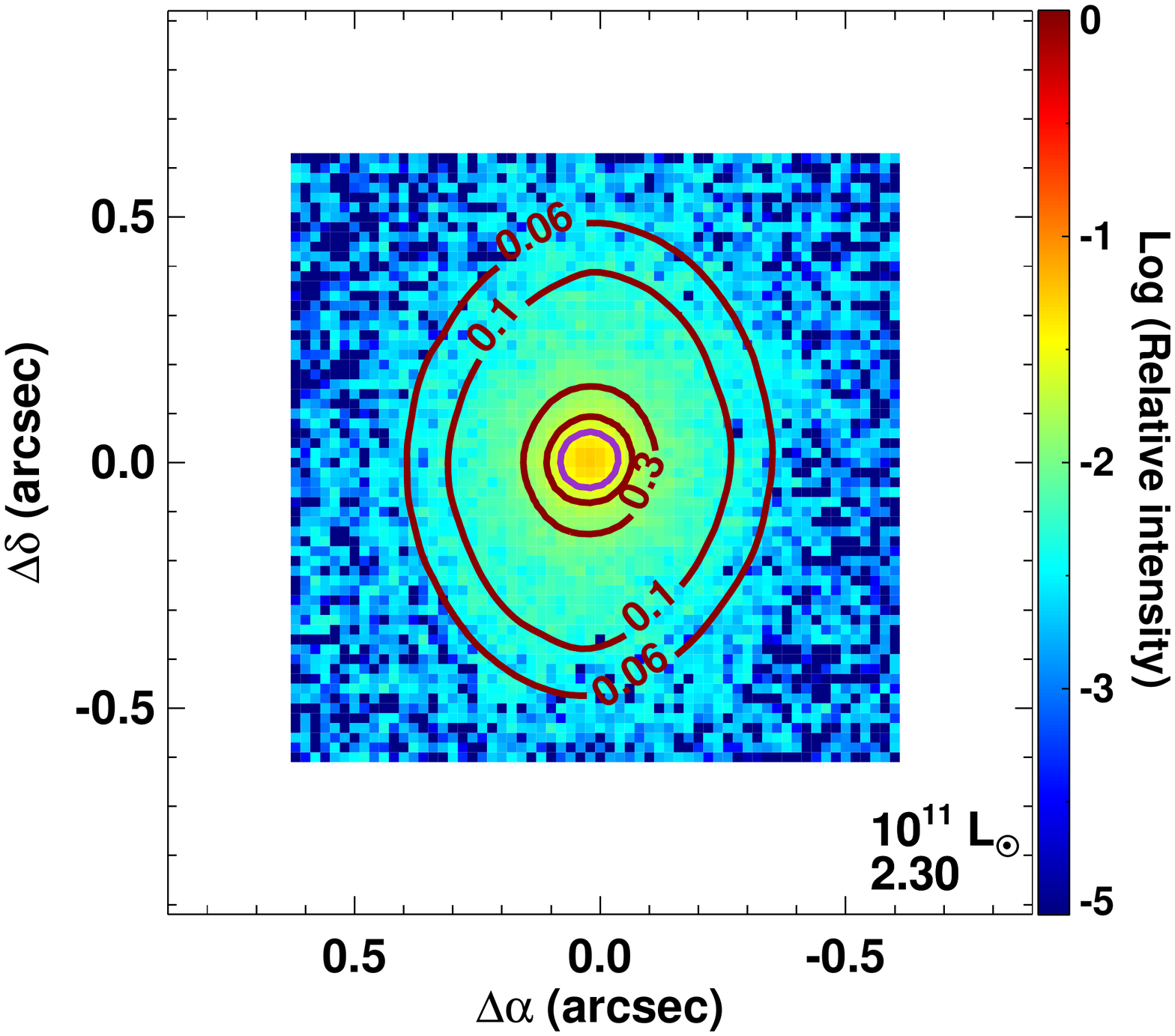}
 
  \includegraphics[trim={1cm 1.5cm 5cm 0cm},width=5.5cm]{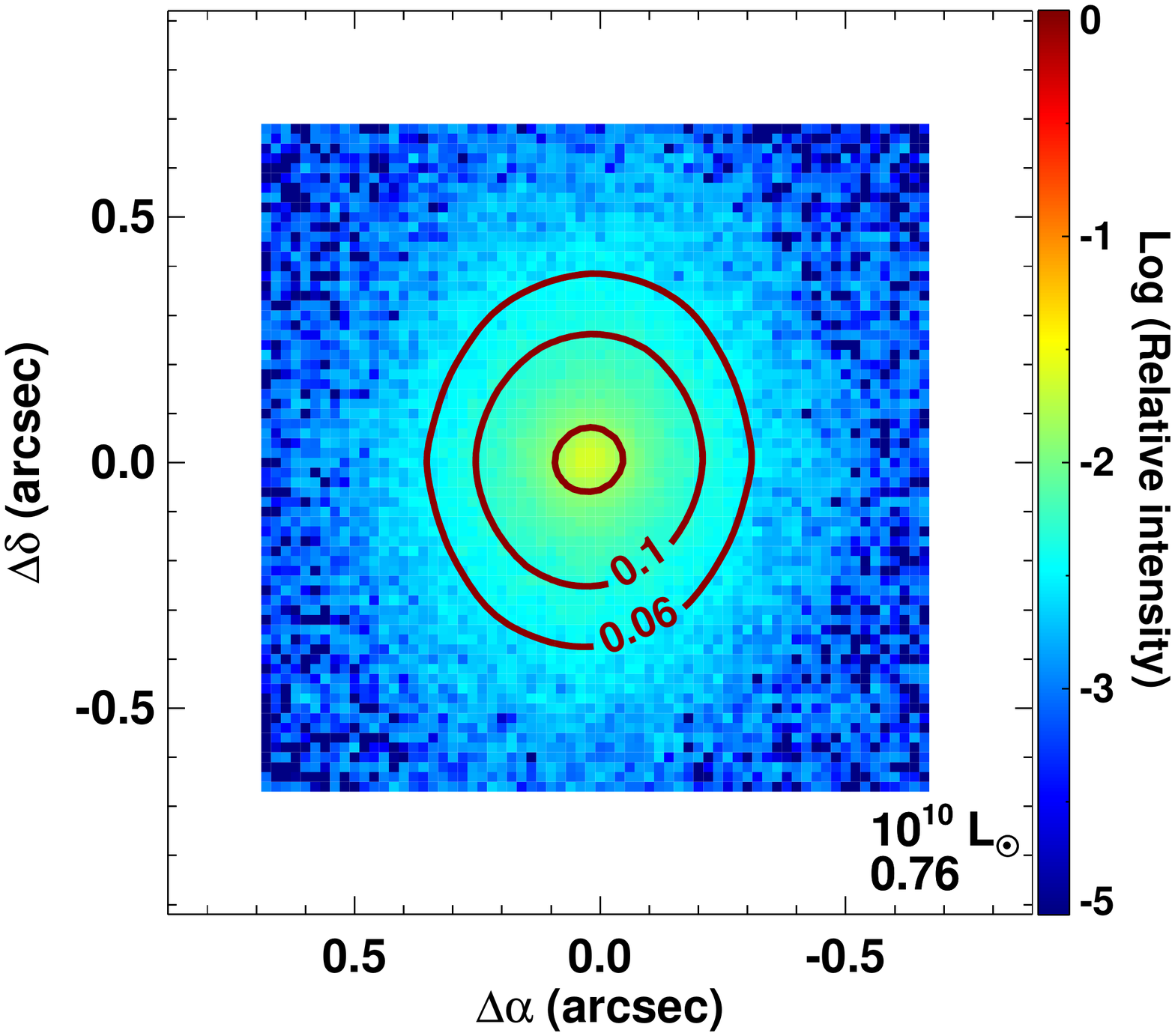}
 \includegraphics[trim={1cm 1.5cm 5cm 0cm},width=5.5cm]{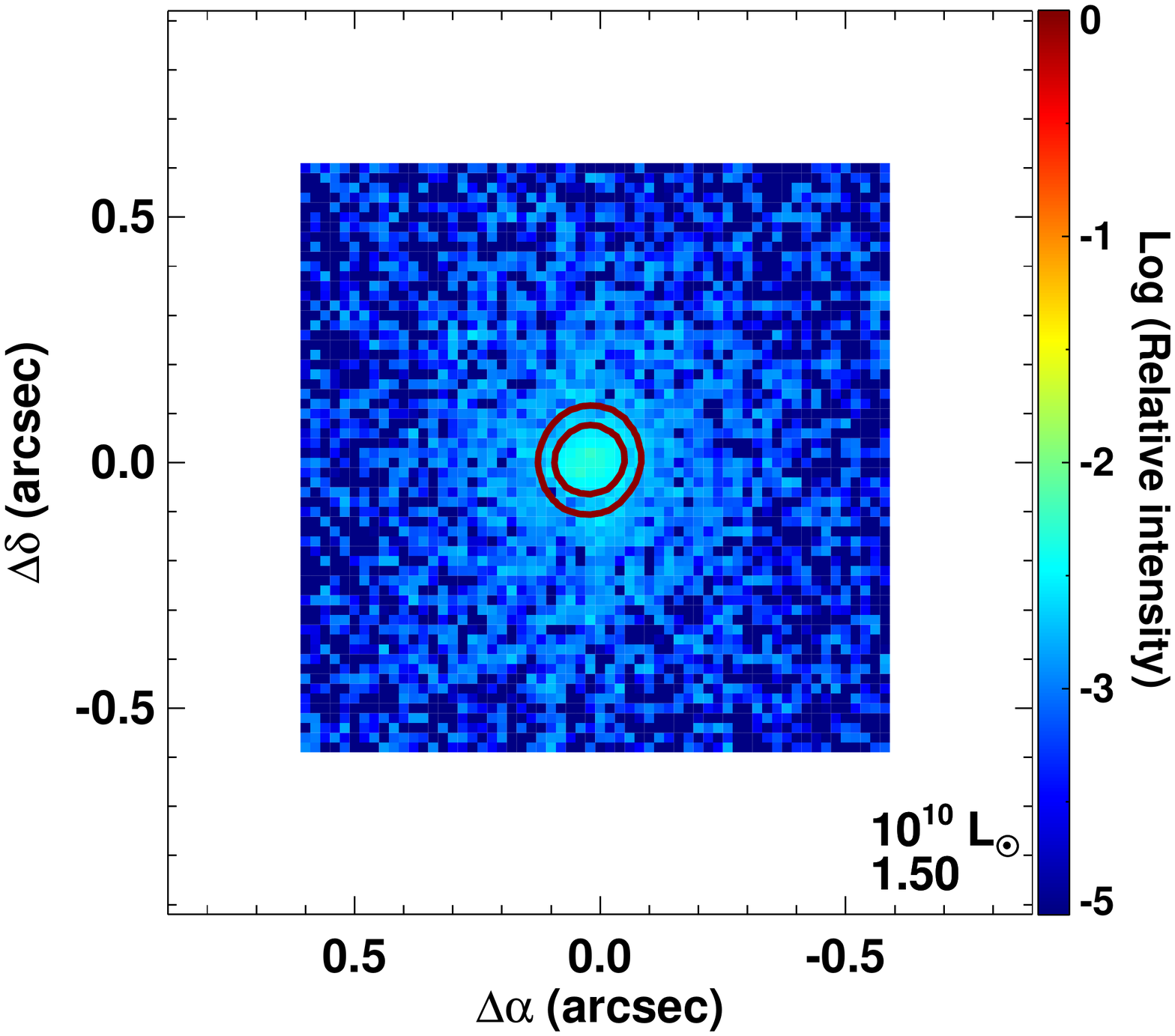}
 \includegraphics[trim={1cm 1.5cm 5cm 0cm},width=5.5cm]{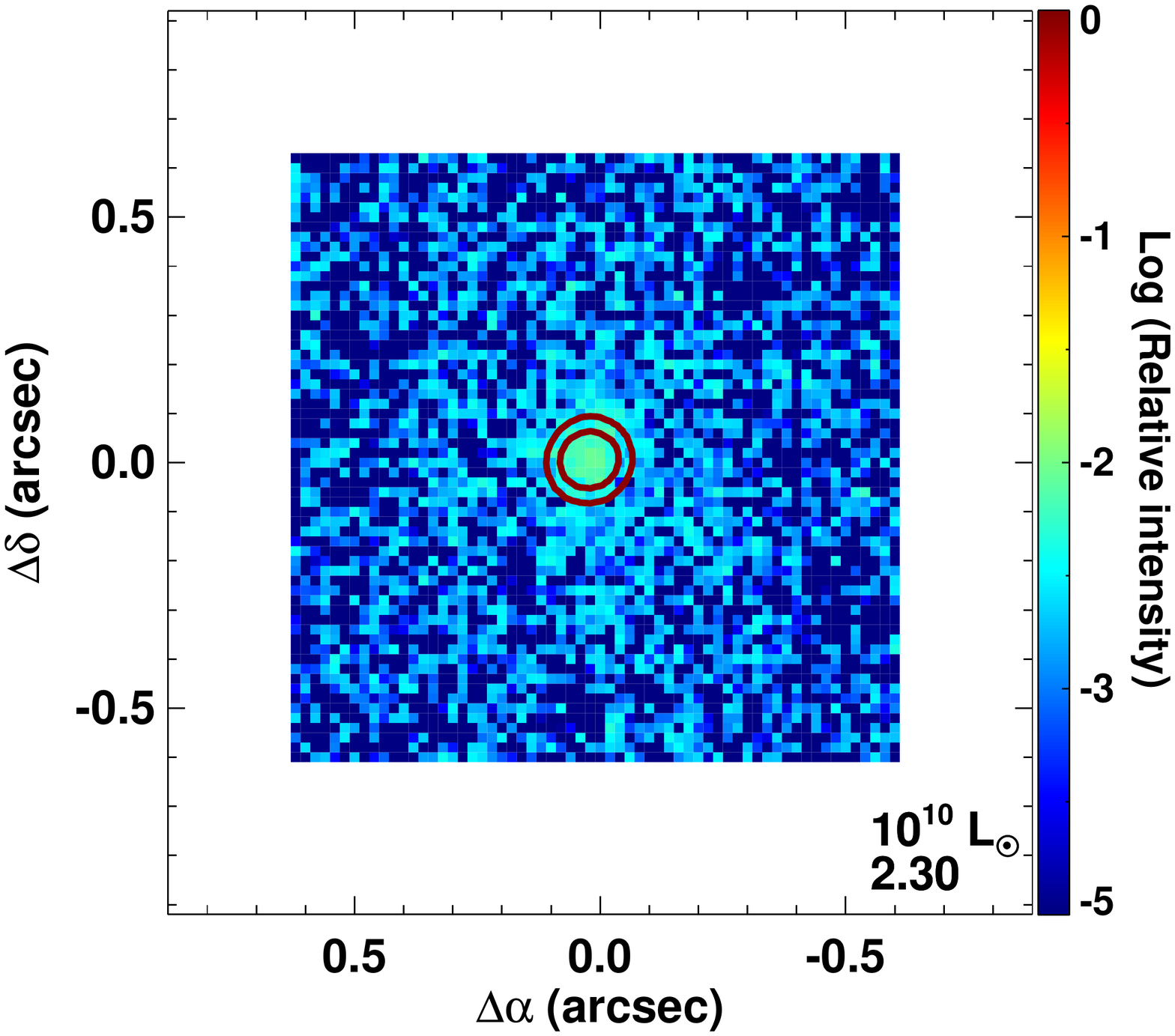}
  \caption{Full-band images obtained by integrating the signal of the HSIM output cubes in the spectral band of the IzJ (left and central; 0.81-1.34 $\mu$m) and the H+K (right; 1.45-2.40 $\mu$m) gratings when considering the lenticular host (i.e. NGC\,809) as a galaxy at redshifts 0.76 (left), 1.50 (middle), and 2.30 (right) scaled to 10$^{12}$L$_{v,\odot}$ (top), 10$^{11}$L$_{v,\odot}$ (middle), and 10$^{10}$L$_{v,\odot}$ (bottom) as also labelled in each panel. Intensity is in logarithmic scale and normalised to the peak of the brightest case (top left panel). As in Figs. \ref{NGC809_HbMgI_range_42} and \ref{NGC809_HbMgI_range_43}, contours correspond to the average S/N in the rest-frame 5050-5250 \AA \ (black contours are S/Ns of 0.06, 0.1, 0.3, and 0.6; S/N values of 2, 5, 9, 14, and 20 are drawn as blue contours; the purple contour indicates a S/N of 1).}
 \label{NGC809_fullrange}
\end{figure*}

\section{Analysis and discussion of results}
\label{analysis}

Through this section, we briefly describe the steps and methodology to analyse our mock HARMONI observations. We note that our simulations are constrained to the size of the objects (see \S\ref{targets}), but the actual FoV of the HARMONI 20x20 mas$^{2}$ scale is wider (see Table \ref{spatial-scales}). For the analysis, we employed popular tools frequently used to analyse data of nearby galaxies. In this section, we only include the figures resulting from the analysis of the lenticular host (i.e. NGC809). Appendix \ref{appen_images} shows the same figures for the spiral and interacting galaxies.

\subsection{Band-filter images and host galaxy parameters}

Adaptive optics 
systems greatly compensate for the atmospheric turbulence impact on data observed at ground-based telescopes. However, the resulting PSFs are complex, showing temporal and spatial variations in addition to their intrinsic wavelength dependence. Furthermore, turbulence residuals will produce a halo around the PSF peak; the worse the quality of the AO correction, the wider that PSF halo. Figure \ref{cut_PSF} shows images of the PSF for our mock HARMONI observations at three wavelength bands. The sizes of these complex PSFs are comparable to the angular size of the host galaxies at around cosmic noon ($\sim$1 arcsec at $z$$\sim$1.5). Then, the actual AGN light spread halo may be brighter than the surface brightness of the host, especially in the higher contrast cases (see Fig. \ref{cut_PSF}). Hence, the bright AGN light spreads all over, outshining the host and resulting in a blend of host and AGN spectra across the whole extent of the host galaxy (see Figs. \ref{NGC809_HbMgI_range_42} and \ref{NGC809_HbMgI_range_43}).
Thus, to study the underlying galaxies of distant AGNs, it is necessary to subtract the AGN contribution.

 Some specific tools have been developed to de-blend (on 2D or 3D data) the emission from the AGN and its host galaxy (see e.g. \citealt{garcialorenzo2005}, \citealt{vayner2016}, \citealt{husemann2016}, \citealt{rupke2017}, \citealt{varisco2018}). These algorithms generally require a careful characterisation of the system PSF, which can be done following distinct reconstruction approaches (see e.g. \citealt{Beltramo2020}). That de-blending will inevitably introduce residuals that impact the S/N of the host spectrum and images. Procedures to optimally de-blend the AGN spectrum and the host galaxy data cube are beyond the scope of this paper and will be the focus of a forthcoming work considering distinct AO performances. In the meantime, we assume an optimal removal of the light from the central AGN, and we focus hereafter on the analysis of the mock HARMONI observations of the host galaxies only, without AGNs. This represents the ideal case of a perfect de-blending of the host and the central AGN light while simultaneously assessing the detection of non-active galaxies. However, even in this case, the shot noise introduced by the AGN light will always be added in quadrature to the detector noise sources, limiting the actual S/N. Detector systematics were not considered in our simulations (see Table \ref{HSIM_par}). Nevertheless, we include some basic estimations to obtain an overview of the de-blending impact.

Hence, we integrated the mock HARMONI observations of the galaxies without AGN in the full spectral range of the corresponding gratings to obtain band-filter images. Figure \ref{NGC809_fullrange} shows these images for the luminosity-scaled lenticular host morphology at the different redshifts. These images indicate a marginal detection (2-sigma $\leq$ signal $\leq$ 4-sigma) of the galaxies for the lowest host luminosity at $z$>0.76. As a reference for redshift dimming, Table \ref{SNR_re_z} includes the average S/N estimated within an aperture of 60 mas in radius centred on the nucleus of the hosts scaled to 10$^{11}$L$_{v,\odot}$.
We used the elliptical isophote analysis within the {\it \emph{photutils Astropy}} package\footnote{https://photutils.readthedocs.io/en/stable/isophote.html} to estimate the photometric position angle (PA$_{p}$) and the ellipticity ($\epsilon$) from these recovered images (see Table \ref{photo_par}). We were not able to perform good fits for the faintest cases. We found a good agreement (within uncertainties) in the position angles obtained for the HSIM$_{\mathrm{Input}}$ and HSIM outputs (see Table \ref{photo_par}). Ellipticities present some deviations, especially in the case of the interacting host. Deviations are larger when comparing with the actual photometric parameters of the three nearby galaxies (see Table \ref{tabgalaxies}). The impact of spatial resolution on defining the morphological structures in the redshifted galaxies is the foremost driver of these deviations \citep[see e.g.][]{2014Mast}. 

 Based on filter-band images, \cite{2009Jahnke} established the detection limit for faint underlying host galaxies to be $\sim$3\% of the total flux of the AGN in a defined aperture. To assess the potential impact of AGN subtraction on our results, we assume that particular 3\% of the AGN flux as the host detection limit. Using the narrow-band flux-calibrated data cubes for the AGN and the host galaxies separately, we note that host galaxies are brighter than 3\% of the AGN fluxes at the selected radii to estimate the photometric parameters. Only the hosts scaled to 10$^{10}$ L$_{V,\sun}$ combined with the brightest AGN (i.e. L$_{[OIII]}$ = 10$^{43}$ erg s$^{-1}$) host fluxes are under that threshold. Thus, the AGN removal will impact the estimation of the photometric parameters mainly in the highest contrast AGNs with faint hosts.

\begin{table*}
        \centering
        \caption{Estimated effective radius at different redshifts of the galaxies acting as AGN hosts in the mock HARMONI simulations. }
        \label{SNR_re_z}
        \begin{tabular}{l|ccc|ccc|ccc|c} 

                     &  \multicolumn{3}{c}{\bf $z$=0.76} &  \multicolumn{3}{c}{\bf $z$=1.50} & \multicolumn{3}{c}{\bf $z$=2.30} &   \\ 
                                                      \cline{2-4} \cline{5-7} \cline{8-10}

{\bf Prototype host}    & {\bf S/N} & {\bf R$_{eff}$} & {\bf R$_{eff}$} & {\bf S/N} & {\bf R$_{eff}$} & {\bf R$_{eff}$} & {\bf S/N} & {\bf R$_{eff}$} & {\bf R$_{eff}$} & {\bf $\beta_{z}$} \\ 

             & 60 mas & (kpc)     & (mas)     & 60 mas & (kpc)     & (mas)     & 60 mas & (kpc) & (mas) & \\ \hline

Lenticular (NGC\,809)     & 7.4 &  1.77 & 234 & 1.2 & 1.15 & 133 & 0.9 & 0.81 & 97 & $-1.24$ \\
Spiral (PGC\,055442)  & 2.0 &  2.57 & 340 & 0.8 & 2.00 & 231 & 0.6 & 1.64 & 195 & $-0.72$ \\
Interaction (NGC\,7119A)   & 1.3 &  6.46 & 854 & 0.2 & 4.88 & 564 & 0.1 & 3.91 & 466 & $-0.80$ \\ \hline
\end{tabular}
\tablefoot{ Columns correspond to: (1) Host morphology (Galaxy name); (2),(5), and (8) Average S/N of the integrated image corresponding to the host scaled to 10$^{11}$L$_{v,\odot}$ (see Figs. \ref{NGC809_fullrange}, \ref{PGC055442_fullrange}, and \ref{NGC7119N_fullrange}) within a circular aperture of 60 mas ($\sim$454 pc, $\sim$520 pc, and $\sim$503 pc at $z$=0.76, $z$=1.50, and $z$=2.30, respectively) in radius centred on the nucleus. (3), (6), and (9) effective radius (in kpc) of the galaxies at the corresponding redshift obtained through the analytic dependence of galaxy sizes in \citealt{2014vanderWel}; (4), (7), and (10) effective radius (in milli-arcsec) derived from (3), (6), and (9) using the appropriate cosmological scale factor \citep{2006Wright}; and (11) Exponent value of the cosmic evolution of galaxy sizes (i.e. R$_{eff}$ $\propto$ (1 + $z$)$^{\beta_{z}}$, \citealt{2014vanderWel}).}
\end{table*}

\begin{table*}
        \begin{threeparttable}

\caption{Photometric position angle (PA$_{p}$ in degrees) and ellipticity ($\epsilon$) measured at 0.6 arcsec from the peak of the full-band images in Figs. \ref{NGC809_fullrange}, \ref{PGC055442_fullrange}, and \ref{NGC7119N_fullrange}. Uncertainties correspond to the standard deviation of the PA$_{p}$ between 0.5 and 0.7 arcsec from the object centre\tablefootmark{a}. }

\centering
\begin{tabular}{lcccccccccc}
   &  & \multicolumn{3}{|c|}{\bf Z=0.76}  & \multicolumn{3}{|c|}{\bf Z=1.50} &  \multicolumn{3}{|c|} {\bf Z=2.30}  \\
\hline
 {\bf Host} & {\bf Host}        & {\bf PA$_{p}$}      & {\bf $\epsilon$} & {\bf PA$_{k}$} & {\bf PA$_{p}$}      & {\bf $\epsilon$} & {\bf PA$_{k}$} & {\bf PA$_{p}$} & {\bf $\epsilon$} & {\bf PA$_{k}$} \\
  {\bf Morph.}            & {\bf Lum}  & {\bf (deg)} &                  &  {\bf (deg)} &  {\bf (deg)} &  & {\bf (deg)} & {\bf (deg)} & & {\bf (deg)}   \\
\hline
                &    {\bf HSIM$^{11}_{\mathrm{Input}}$} & $-$4.3 $\pm$ 0.8 & 0.29  &  $-$6.2 & $-$4.1 $\pm$ 0.6 &   0.28  & $-$6.2 & $-$4.1 $\pm$ 0.6 &   0.29  & $-$6.2 \\
{\bf lenticular} & {\bf 10$^{12}$L$_{v,\odot}$} & $-$4.2 $\pm$ 0.2 & 0.17  & $-$6.2 & $-$4.3 $\pm$ 0.3 & 0.16 & 0.0    & $-$4.2 $\pm$ 0.3 &   0.21 & $-$6.2 \\
{\bf (NGC\,809)}     & {\bf 10$^{11}$L$_{v,\odot}$} & $-$3.9 $\pm$ 0.8 & 0.17  & $-$12.4 & $-$3.1 $\pm$ 1.6  & 0.16 & 49.7 & $-$7.2 $\pm$ 7.1 & 0.20 & ---  \\
     & {\bf 10$^{10}$L$_{v,\odot}$} & $-$5.6 $\pm$ 4.4 &  0.19  & --- & --- & --- & --- & --- & --- & --- \\
\hline
                         &    {\bf HSIM$^{11}_{\mathrm{Input}}$} & 24.4 $\pm$ 9.8 &   0.29 & 12.4 & 19.7 $\pm$  10.4 &   0.26 & 12.4 &  20.6 $\pm$   9.7 &   0.27 & 12.4   \\
{\bf Spiral} & {\bf 10$^{12}$L$_{v,\odot}$} & 24.5 $\pm$   6.8 &   0.20 & 12.4 & 20.2 $\pm$ 6.5 &   0.20 & 18.6 & 21.1 $\pm$   8.1 &   0.23 & 18.6  \\
{\bf (PGC\,055442)}            & {\bf 10$^{11}$L$_{v,\odot}$} & 24.3 $\pm$ 6.3 & 0.20  & 6.2 & 19.9 $\pm$ 8.0 & 0.20 & 12.4 & 20.8 $\pm$ 10.8 & 0.26 & ---  \\
& {\bf 10$^{10}$L$_{v,\odot}$} &  22.3 $\pm$ 7.4 &   0.19  & --- & --- & --- & --- & --- & --- & ---   \\
\hline
                         &    {\bf HSIM$^{11}_{\mathrm{Input}}$} & 121.1 $\pm$   0.1 &   0.50 & 130.3 & 119.9 $\pm$ 1.7 &   0.58 & 130.3 & 121.1 $\pm$   1.1 &   0.54 & 130.3       \\
  {\bf Interaction} & {\bf 10$^{12}$L$_{v,\odot}$} & 132.4 $\pm$ 11.9 & 0.14 & 124.4 & 119.4 $\pm$ 4.7 & 0.25 & 136.5 & 118.1 $\pm$ 3.4 & 0.34 & 120.8  \\
{\bf (NGC\,7119A)} & {\bf 10$^{11}$L$_{v,\odot}$} &  132.6 $\pm$ 14.7 & 0.17 & 136.5  & 123.9 $\pm$  8.5 &   0.24  & --- &  --- & --- & --- \\
& {\bf 10$^{10}$L$_{v,\odot}$} &  ---              & ---  &  ---   & ---              & ---     & --- & ---  & --- & --- \\

\hline
\end{tabular}
\tablefoot{ Uncertainties in $\epsilon$ are always less than 0.1. The third column at each redshift corresponds to the estimation of the kinematic position angle (PA$_{k}$ in degrees), with errors always < 6.5 degrees. HSIM$^{11}_{\mathrm{Input}}$ corresponds to the HSIM$_{\mathrm{Input}}$ data cube for the host scaled to 10$^{11}$L$_{v,\odot}$ in each case. Same results are obtained for the other two host luminosities. For comparison, the photometric and kinematics parameters for the galaxies at $z\sim$0 are in Appendix \ref{appen}.}\\
\tablefoottext{a}{In the case of the spiral host, there is a significant variation of the PA$_{p}$ between these two radial distances due to the transition from the bar (PA$_{\mathrm{bar}}$$\sim$45$\deg$) to the disc (PA$_{\mathrm{disk}}$$\sim$15$\deg;$ see Appendix \ref{appen}).} 

\label{photo_par}
  \end{threeparttable}

\end{table*}

\begin{table*}
\label{host_table}
\caption{Stellar velocity dispersion measured from the mock HARMONI observations of representative host galaxies of AGNs at three redshifts around cosmic noon. }

\centering
\begin{tabular}{lcccccccccc}
   &  & \multicolumn{3}{|c|}{\bf Z=0.76}  & \multicolumn{3}{|c|}{\bf Z=1.50} &  \multicolumn{3}{|c|} {\bf Z=2.30}  \\
\hline
 {\bf Object} & {\bf Host} & {\bf S/N} & {\bf V$_{*}$} & {\bf $\sigma_{*}$}  & {\bf S/N} & {\bf V$_{*}$} & {\bf $\sigma_{*}$}  & {\bf S/N} & {\bf V$_{*}$} & {\bf $\sigma_{*}$} \\
 & {\bf V Luminosity}  & & {\bf km s$^{-1}$} & {\bf km s$^{-1}$}   & & {\bf km s$^{-1}$} & {\bf km s$^{-1}$}  & & {\bf km s$^{-1}$} & {\bf km s$^{-1}$}  \\
\hline
                         &    {\bf HSIM$^{11}_{\mathrm{Input}}$} &      389 &        3 &      158 $\pm$     1 &      327 &        4 &      161 $\pm$     1 &      327 &        0 &      163 $\pm$     1 \\
   {\bf lenticular} & {\bf 10$^{12}$L$_{\odot}$} &      111 &       26 &      147 $\pm$     3 &       78 &       16 &      142 $\pm$     3 &       39 &        8 &      173 $\pm$     6 \\
{\bf (NGC\,809)} & {\bf 10$^{11}$L$_{\odot}$} &       51 &       25 &      152 $\pm$     7 &       10 &       33 &      139 $\pm$    16 &        4 & --- & --- \\
& {\bf 10$^{10}$L$_{\odot}$} &        6 & --- & --- &        3 & --- & --- &        <1 & --- & --- \\
\hline
                           &    {\bf HSIM$^{11}_{\mathrm{Input}}$} &      360 &      $-$39 &      147 $\pm$     1 &      301 &      $-$39 &      147 $\pm$     0 &      300 &      $-$38 &      147 $\pm$     0 \\
   {\bf Spiral} & {\bf 10$^{12}$L$_{\odot}$} &      123 &      $-$13 &      152 $\pm$     2 &       74 &      $-$19 &      144 $\pm$     3 &       42 &      $-$16 &      153 $\pm$    11 \\
{\bf (PGC\,055442)} & {\bf 10$^{11}$L$_{\odot}$} &       59 &      $-$22 &      147 $\pm$     6 &       10 &       17 &      152 $\pm$    20 &        3 & --- & --- \\
& {\bf 10$^{10}$L$_{\odot}$} &        7 & --- & --- &        <1 & --- & --- &        <1 & --- & --- \\
\hline
                            &    {\bf HSIM$^{11}_{\mathrm{Input}}$} &       84 &      209 &      133 $\pm$     9 &       91 &      201 &      132 $\pm$     7 &       90 &      198 &      122 $\pm$     5 \\
   {\bf Interaction} & {\bf 10$^{12}$L$_{\odot}$} &       72 &      236 &      145 $\pm$    12 &       48 &      238 &      133 $\pm$    12 &       23 &      218 &      197 $\pm$    33 \\
{\bf (NGC\,7119A)} & {\bf 10$^{11}$L$_{\odot}$} &       41 &      244 &      144 $\pm$    19 &        8 &      186 &      121 $\pm$    38 &        2 & --- & --- \\
& {\bf 10$^{10}$L$_{\odot}$} &        5 & --- & --- &      <1 & --- & --- &        <1 & --- & --- \\

\hline
\end{tabular}
\tablefoot{ HSIM$^{11}_{\mathrm{Input}}$ corresponds to the HSIM$_{\mathrm{Input}}$ data cube for the host scaled to 10$^{11}$L$_{v,\odot}$ in each case. The same results are obtained for the other two host luminosities. As a reference, Table \ref{tabgalaxies} lists the stellar velocity dispersion for the nearby galaxies taken as model hosts of AGNs at around cosmic noon.}
\label{table1}
\end{table*}

\begin{table}
        \centering
        \caption{Number of Voronoi bins with S/N $\geq$8 in the 5100-5230 \AA \ rest-wavelength resulting from mock HARMONI observations of the redshifted lenticular host (i.e. NGC\,809) using 6, 9, 12, or 15 hours (as indicated) on-source exposure times. For comparison purposes, the table includes that number for three hours of exposure time.  }
        \label{voxels_vari}
        \begin{tabular}{l|cccc|} 

        {\bf Host} & {\bf $z$}   & {\bf Exposure} & {\bf Number of Voronoi } \\
        {\bf V luminosity} &    & {\bf time (hours)} & {\bf bins ({\bf S/N} $\geq$ 8)} \\
                              &             &                             &                            \\ \hline
                              & {\bf 0.76}  &        3                    &  1323  \\
{\bf 10$^{12}$L$_{\odot}$}& {\bf 1.50}  &        3                    &  292  \\
                              & {\bf 2.30}  &        3                    & 97    \\ \hline
                              & {\bf 0.76}  &        3 / 6                & 32 / 111     \\
{\bf 10$^{11}$L$_{\odot}$}& {\bf 1.50}  &        3 / 9                & 3 / 17   \\
                              & {\bf 2.30}  &        3 / 12                &  0 / 6  \\ \hline
                              & {\bf 0.76}  &        3 / 9               &  0 / 7  \\
{\bf 10$^{10}$L$_{\odot}$}& {\bf 1.50}  &        3 / 12                & 0 / 1    \\
                              & {\bf 2.30}  &        3 / 15                &  0 / 0  \\ \hline

\end{tabular}
\end{table}

\subsection{Integrated spectra of the hosts}
\label{integrated_results}

\begin{figure*}
\centering
 \includegraphics[trim={1.25cm 0.5cm 2cm 0},clip,width=5.75cm]{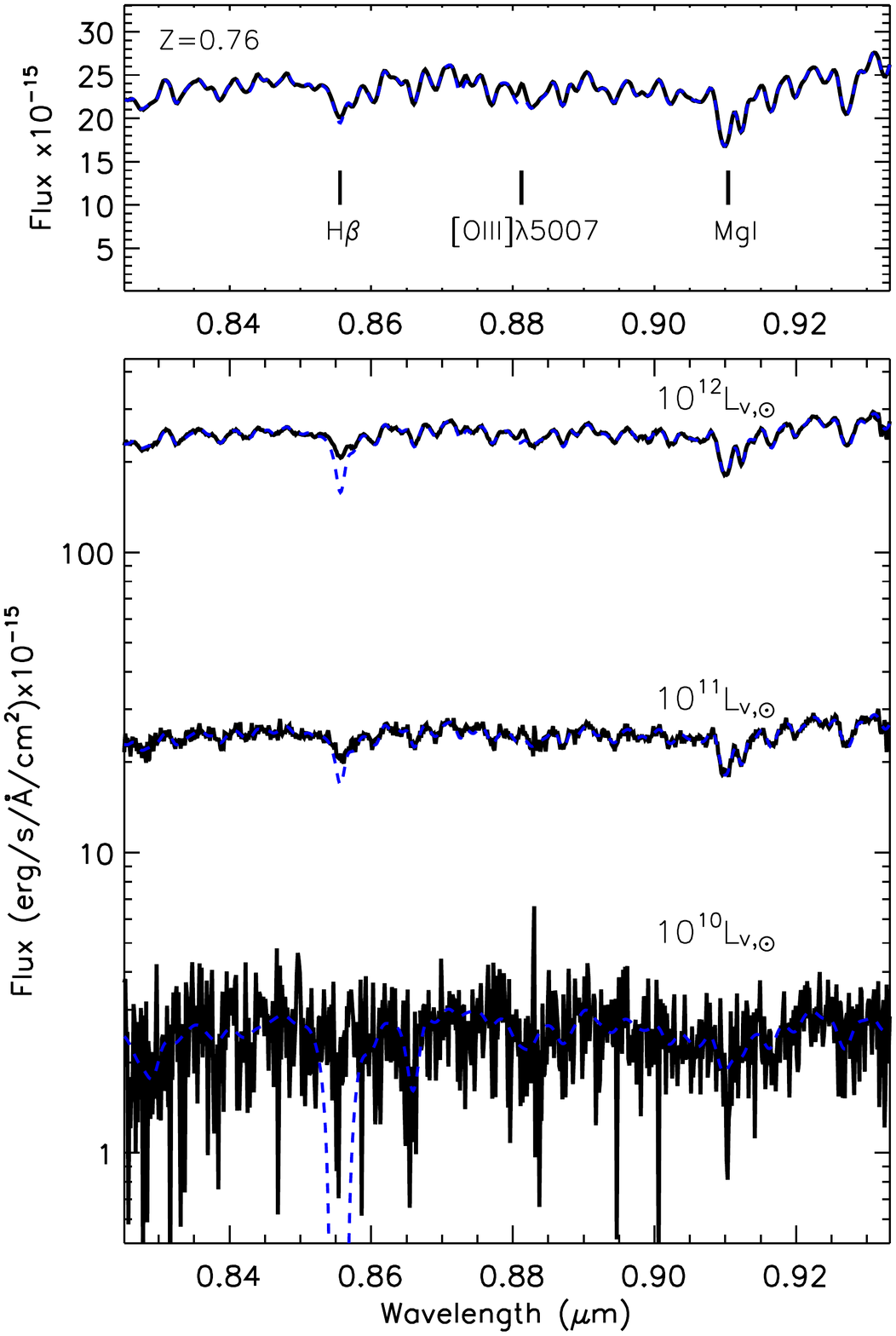}
 \includegraphics[trim={1.25cm 0.5cm 2cm 0},clip,width=5.75cm]{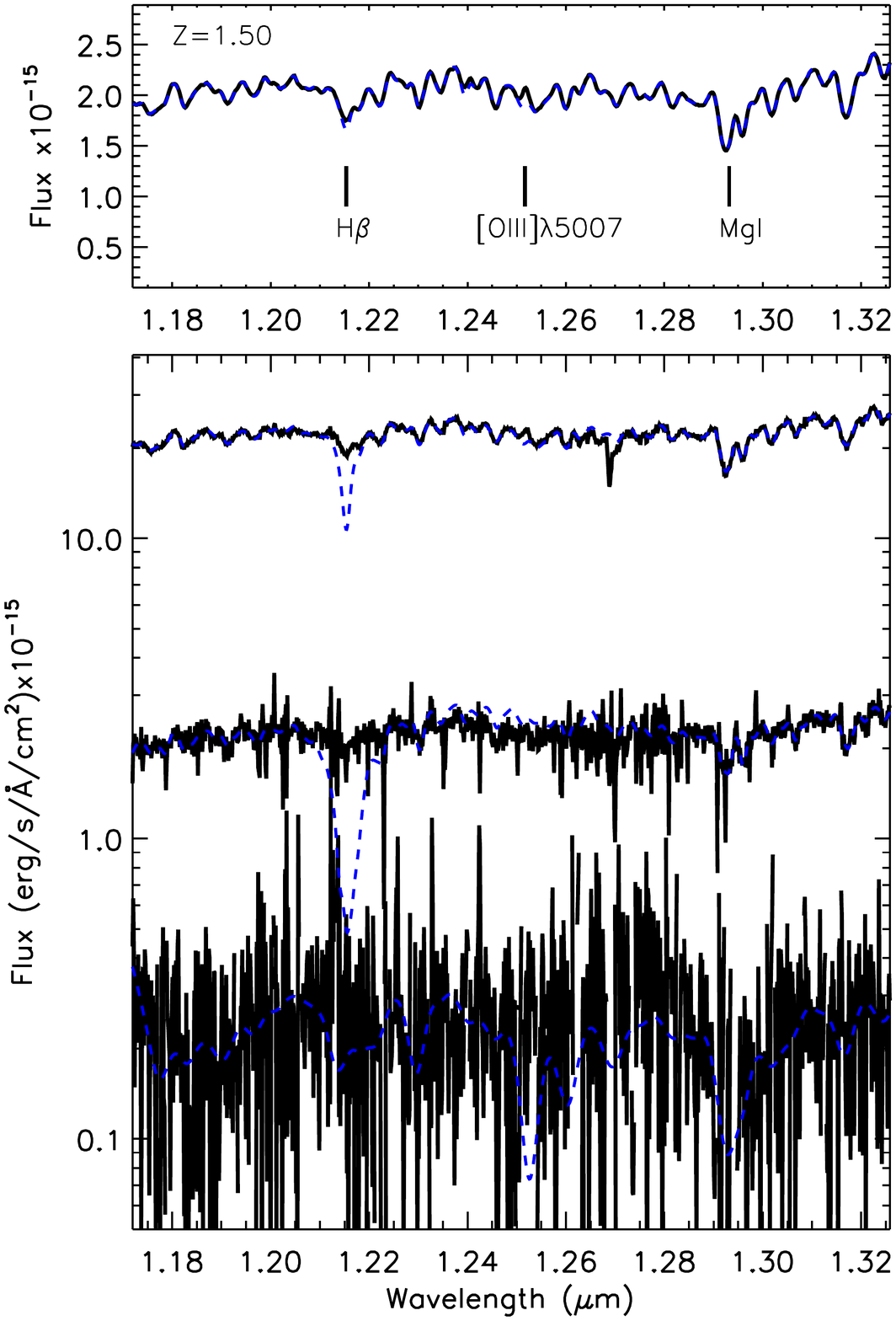}
 \includegraphics[trim={1.25cm 0.5cm 2cm 0},clip,width=5.75cm]{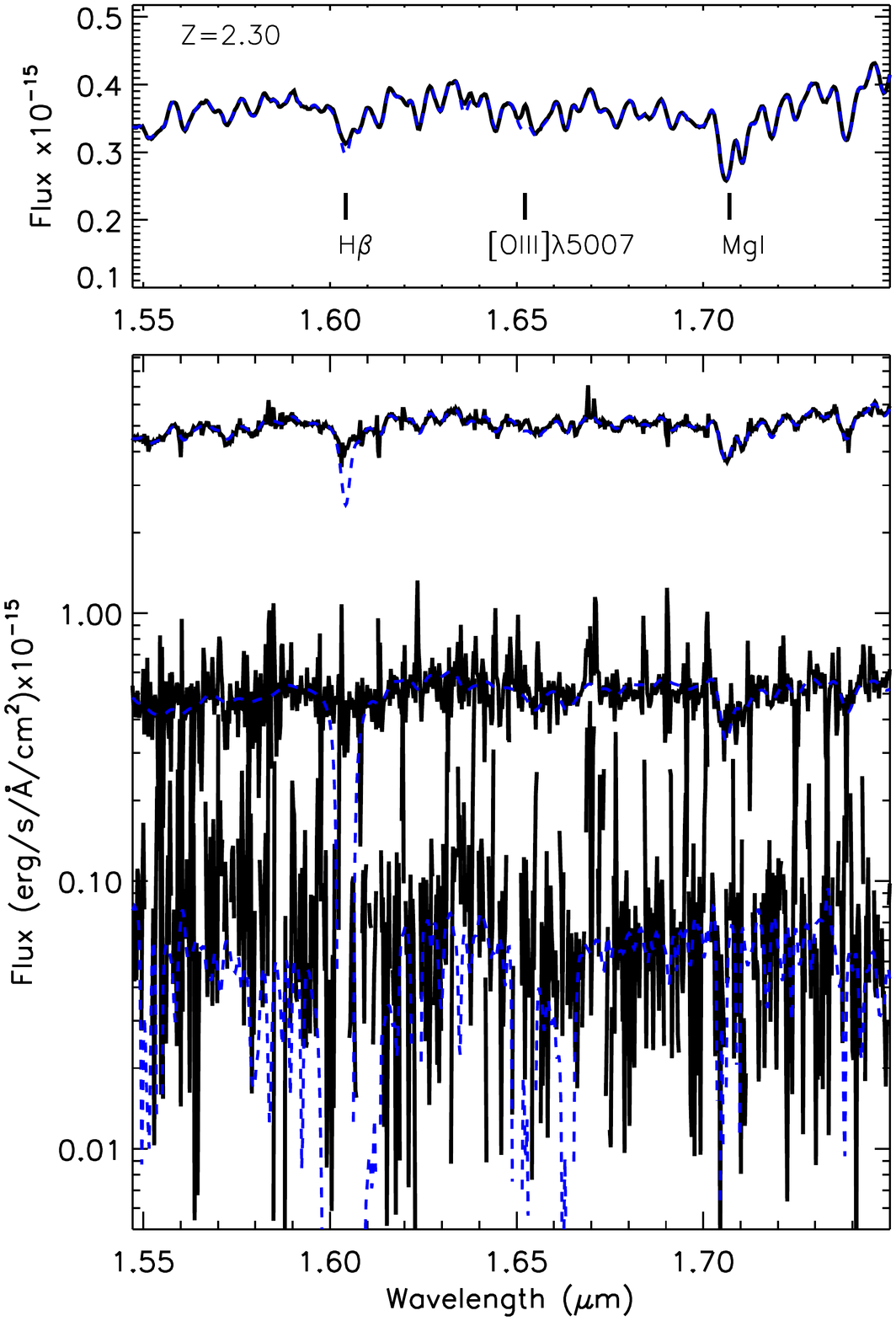}

  \caption{Representative spectra of lenticular host (i.e. NGC\,809) at redshifts 0.76 (left), 1.50 (centre), and 2.30 (right) obtained by adding the signal from the spaxels within the effective radius (see Table \ref{SNR_re_z}). Top panels correspond to the spectra extracted from the HSIM$_{\mathrm{Input}}$s, only for the host galaxy scaled to 10$^{11}$L$_{v,\odot}$. The bottom panels show the spectra from the mock HARMONI observation, after reduction and flux calibration, for the three host luminosities. The blue dotted lines correspond to the {\scshape{ppxf}} best-fit model for each spectrum.}
 \label{NGC809_Reffspectra}
\end{figure*}

For the mock HARMONI observations of the hosts, we obtained the integrated spectra by adding the signal of the spaxels within an aperture of twice the effective radius (R$_{eff}$) in diameter. We estimate the radius of these apertures at each $z$ (see Table \ref{SNR_re_z}) by applying the following: 

\begin{equation} 
\label{beta_z}
R_{eff}(z)= \left[ \frac{1 + z}{1 + z_0} \right]^{\beta_{z}} R_{eff}(z_0) 
,\end{equation}

\noindent where $\beta_{z}$ provides the rate of average size evolution of galaxies as a function of mass and galaxy type (see Table 2 in \citealt{2014vanderWel}). The adopted values for $\beta_{z}$ in each case are listed in Table \ref{SNR_re_z}. Table \ref{tabgalaxies} includes R$_{eff}$($z_{0}$) for the three galaxies selected as hosts. For comparison, we also extracted the integrated spectra within the same apertures from the corresponding HSIM$_{\mathrm{input}}$ data cubes. Figure \ref{NGC809_Reffspectra} shows these aperture-integrated spectra (see also Figs. \ref{PGC055442_Reffspectra} and \ref{NGC7119N_Reffspectra} in Appendix \ref{appen_images}) .

\begin{figure}
\includegraphics[trim={0cm 1cm 0cm 0cm},clip,width=9.25cm, angle=180]{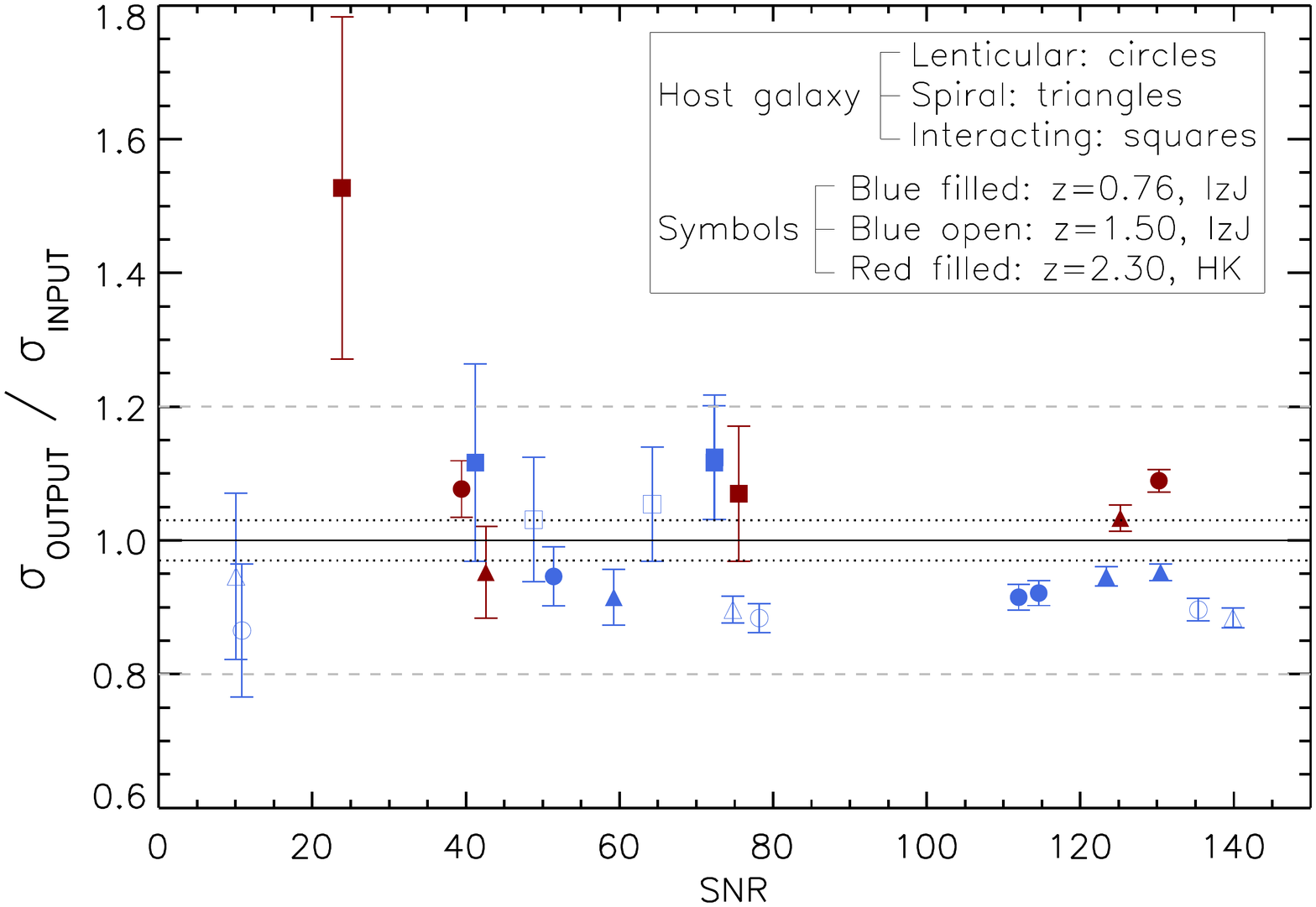}
  \caption{Stellar velocity dispersion of aperture spectra obtained from the mock HARMONI observations relative to the stellar velocity dispersion of the HSIM$_{\mathrm{Input}}$s. Different symbols indicate the distinct hosts: circles$\rightarrow$NGC\,809, triangles$\rightarrow$PGC\,055442, and squares$\rightarrow$NGC\,7119A. Colours correspond to redshift and grating: blue filled symbols $\rightarrow$ $z$=0.76, and IzJ grating; blue open symbols $\rightarrow$ $z$=1.50, and IzJ grating; red filled symbols $\rightarrow$ $z$=2.30, and HK grating. Grey dashed line marks the 20\% level of uncertainty. Black dotted line indicates the typical uncertainty of $\sigma_{\star}$ for the HSIM$_{\mathrm{Input}}$s. Black solid line indicates $\sigma_{\mathrm{OUTPUT}}$ = $\sigma_{\mathrm{INPUT}}$.  }
    \label{aperture_sigma}
\end{figure}

We estimated the S/N of these aperture-integrated spectra in the 5100-5230 \AA \ rest wavelength corresponding to the MgI region (see Table \ref{table1}). To do so, we obtained the best-fit model for the stellar continuum in the 4650-5250 \AA \ rest-frame range using the penalised pixel-fitting ({\scshape{ppxf}}\footnote{https://pypi.org/project/ppxf/}) method \citep{cappellari2004, cappellari2017} masking the emission lines. As templates, we used spectra from the ELODIE stellar library \citep{2001ELODIE} at R$=10000$. This is adequate for the instrumental resolution of the mock HARMONI data adjusted to the rest-frame H$\beta$-MgI range for the three redshifts considered. Similar results are obtained using the MILES stellar library as templates at $z$=0.76 \citep{vazdekis2015}. Figure \ref{NGC809_Reffspectra} includes the best-fit models overplotted on the spectra (see also Figs. \ref{PGC055442_Reffspectra} and \ref{NGC7119N_Reffspectra} in Appendix \ref{appen_images}). The standard deviation of the residuals (spectrum $-$ best-fit) provides an assessment of the noise in the MgI spectral range, to finally compute the S/N. {\scshape{ppxf}} also provides the stellar velocity and velocity dispersion of the spectrum under analysis (Table \ref{table1}). {\scshape{ppxf}} only provides reliable estimations of the stellar kinematics or spectra with moderate-to-high
S/N (e.g. S/N > 8 per spectral bin) \citep{cappellari2004}. The S/N of the integrated spectra for the faintest hosts are under this S/N threshold. As we shifted the spectra to the rest-frame wavelengths before running {\scshape{ppxf}}, the stellar velocities are around 0. The estimation of the stellar velocity dispersion from the aperture-integrated spectra with S/N > 8 for the mock HARMONI observations over the considered $z$ range is in good agreement with the ones estimated for the HSIM$_{\mathrm{Input}}$, with uncertainties < 20\%. Only in the case of the interacting host at the largest $z$, the uncertainty in $\sigma_{\star}$ is larger, reaching 40\% (see Fig. \ref{aperture_sigma}).

 To approach the potential impact on these measurements of the residuals after the AGN-host de-blending, we also obtained the corresponding integrated spectra for the mock HARMONI AGN+host data and fitted the AGN contribution (see \S \ref{fit-AGN}). We used the der\_snr algorithm\footnote{http://www.stecf.org/software/ASTROsoft/DER\_SNR} \citep{2008Stoehr} to estimate the S/N in the MgI region on the residual spectra after subtracting the fitted AGN contribution. On average, we found that the S/N is $\sim$18\% ($\sim$50\%) lower than the S/Ns in Table \ref{host_table} when the host luminosity is 10$^{11}$L$_{\odot}$ (10$^{10}$L$_{\odot}$). For the brightest hosts, the differences are comparable ($\sim$1\%)  to the differences between the S/Ns in Table \ref{host_table} and those estimated using der\_snr ($<$1\%) on the host's integrated spectra.

\subsection{Resolved stellar kinematics}

\begin{figure*}
\centering
 \includegraphics[trim={0.5cm 0.5cm 0.75cm 0},clip,width=4.25cm]{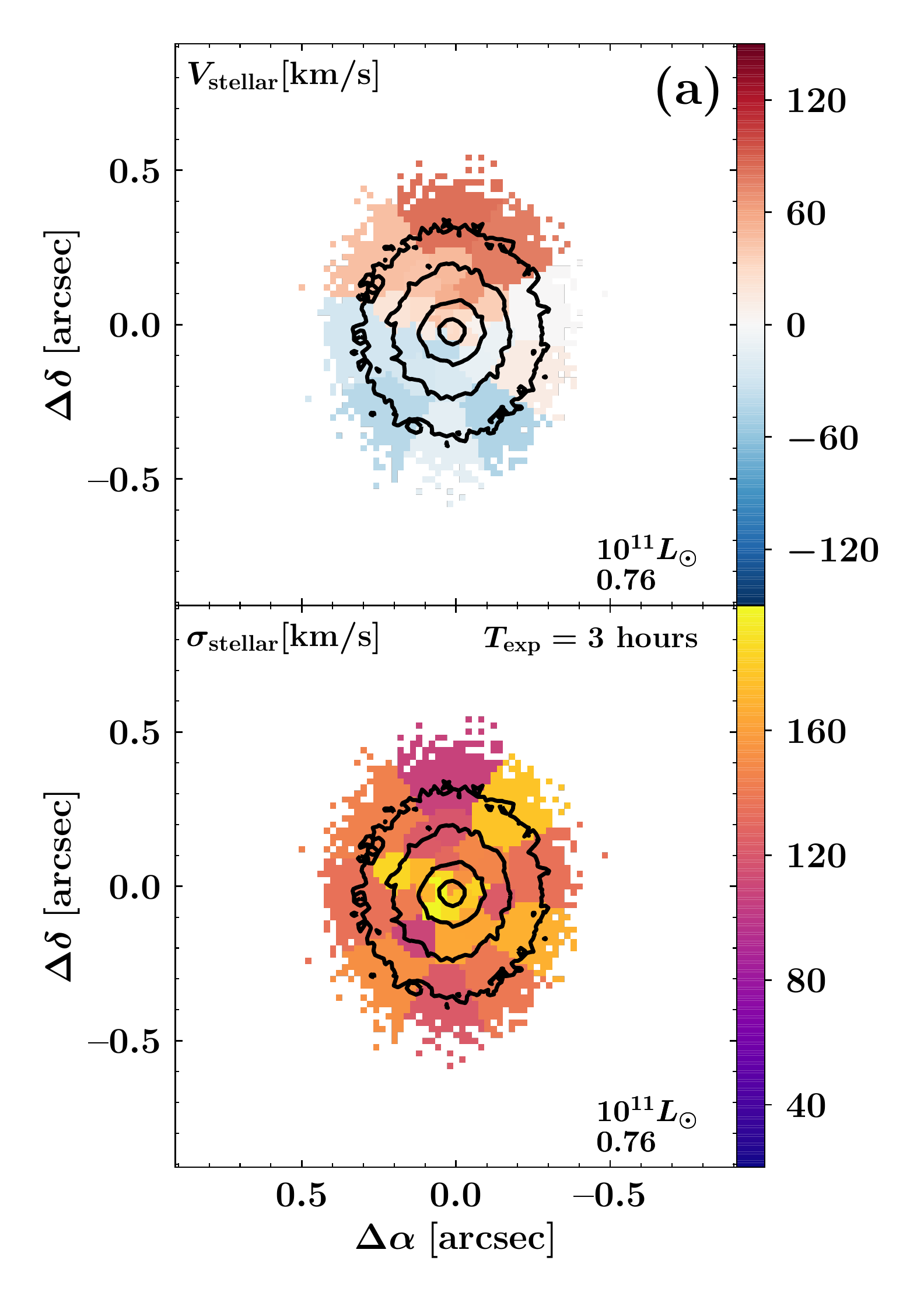}
 \includegraphics[trim={0.5cm 0.5cm 0.75cm 0},clip,width=4.25cm]{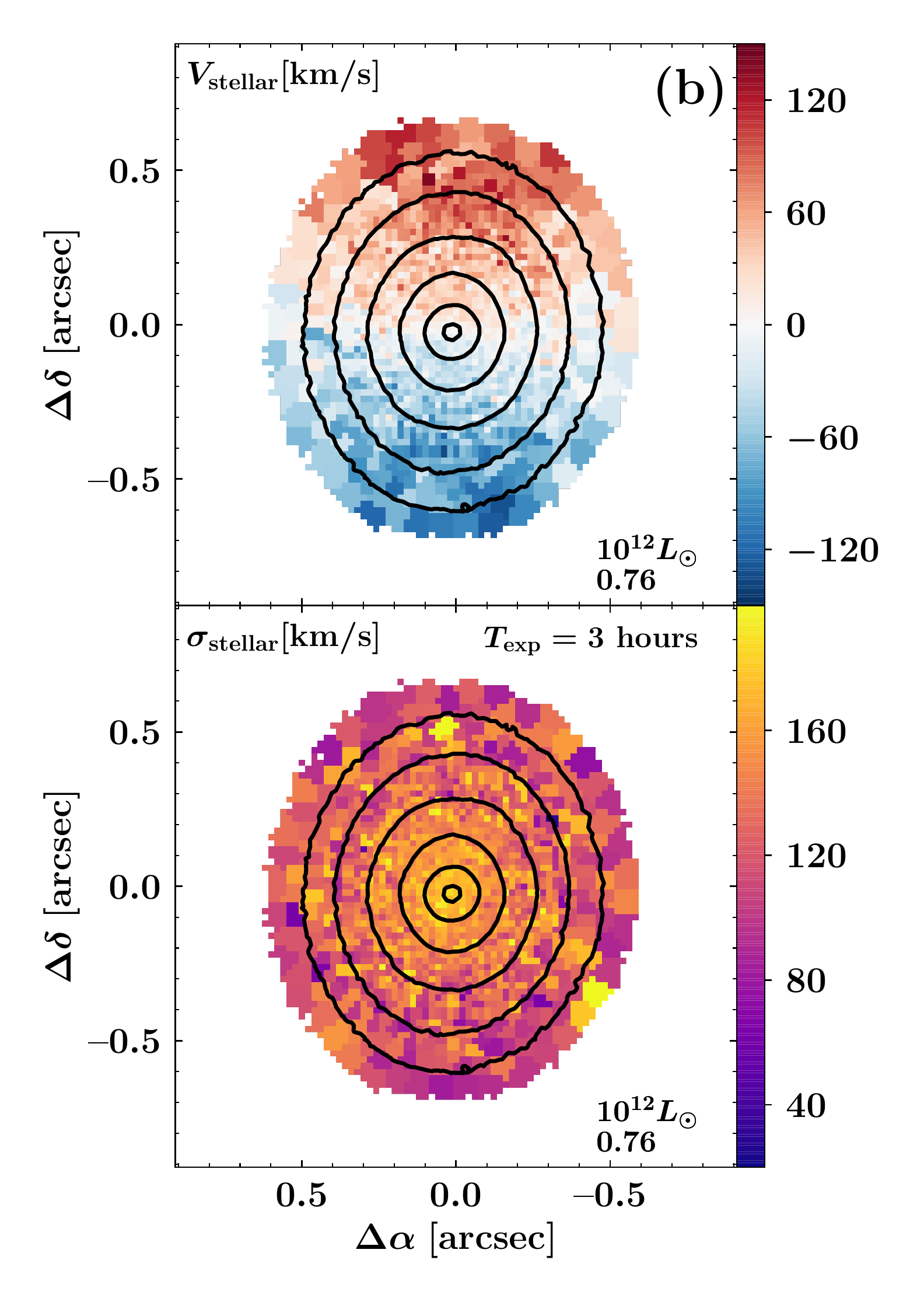}
 \includegraphics[trim={0.5cm 0.5cm 0.75cm 0},clip,width=4.25cm]{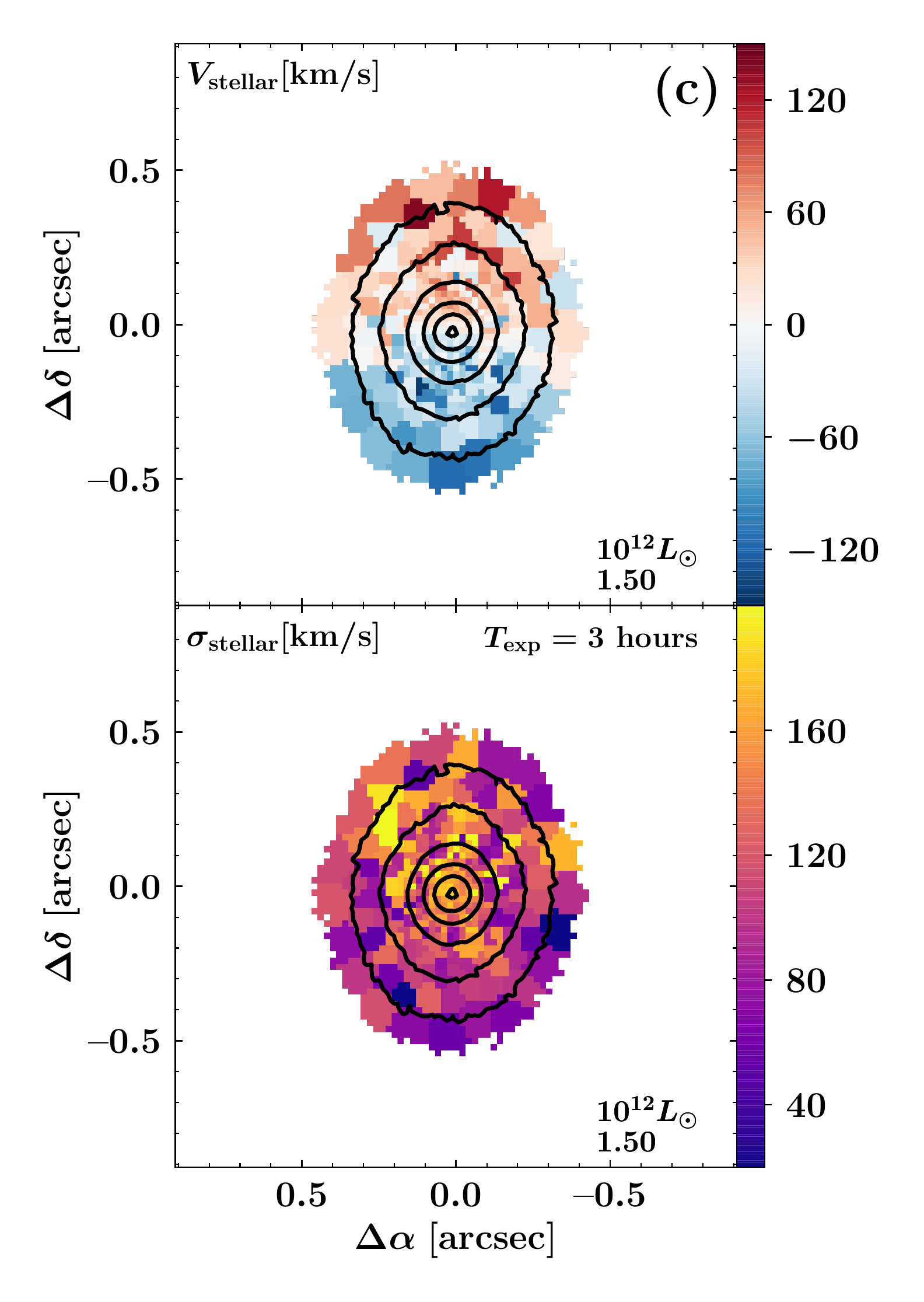}
\includegraphics[trim={0.5cm 0.5cm 0.75cm 0},clip,width=4.25cm]{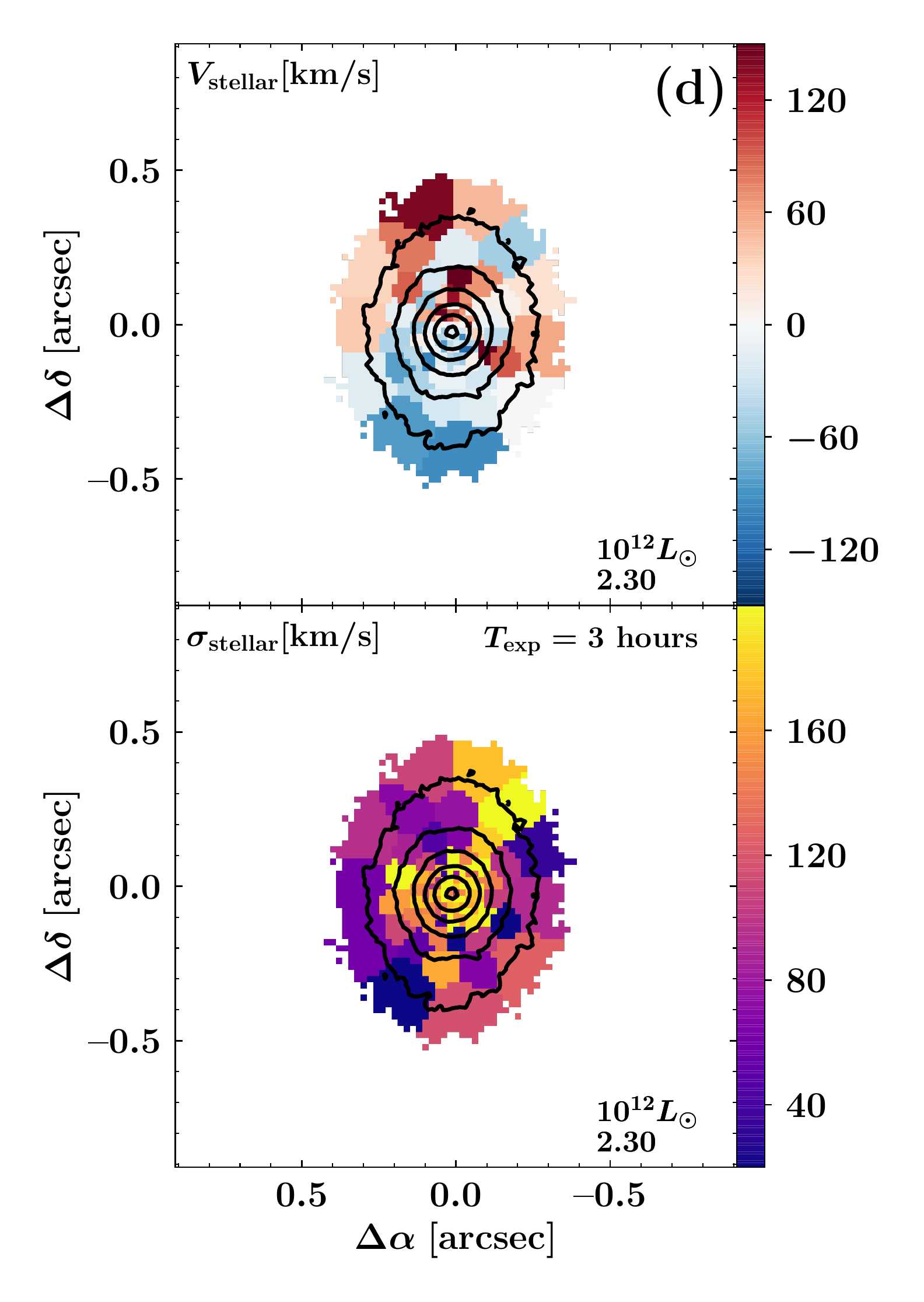}
\includegraphics[trim={0.5cm 0.5cm 0.75cm 0},clip,width=4.25cm]{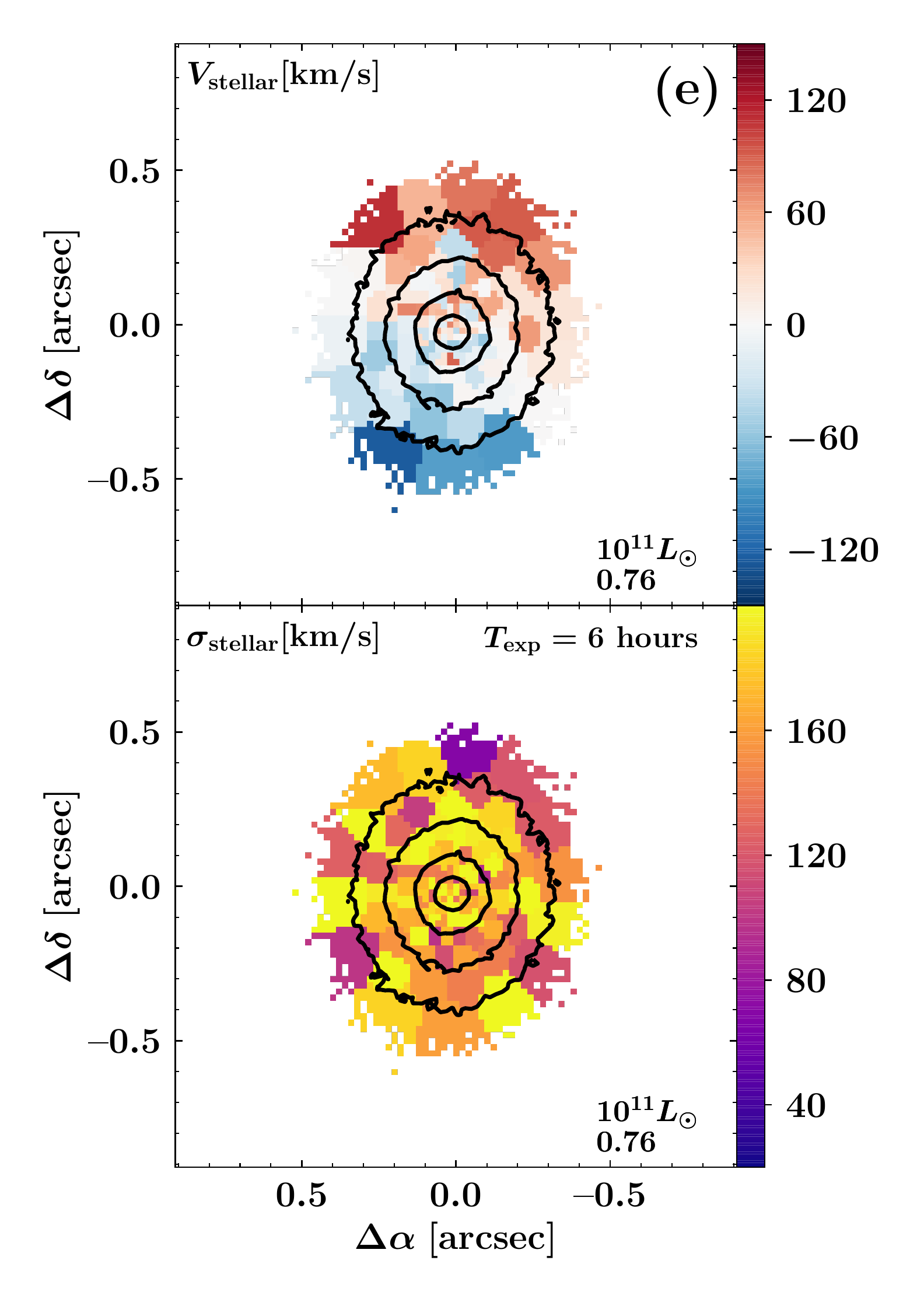}
\includegraphics[trim={0.5cm 0.5cm 0.75cm 0},clip,width=4.25cm]{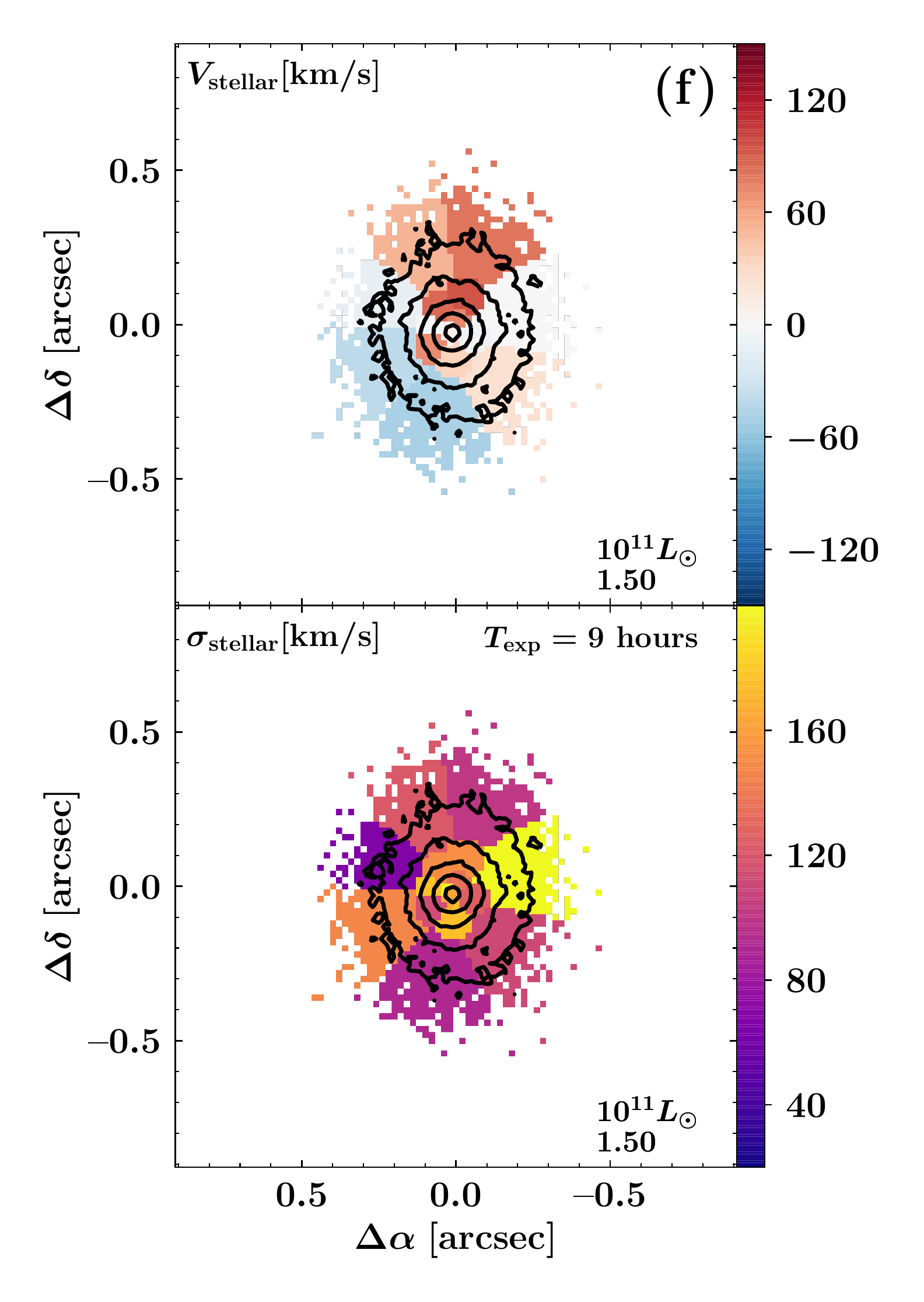}
\includegraphics[trim={0.5cm 0.5cm 0.75cm 0},clip,width=4.25cm]{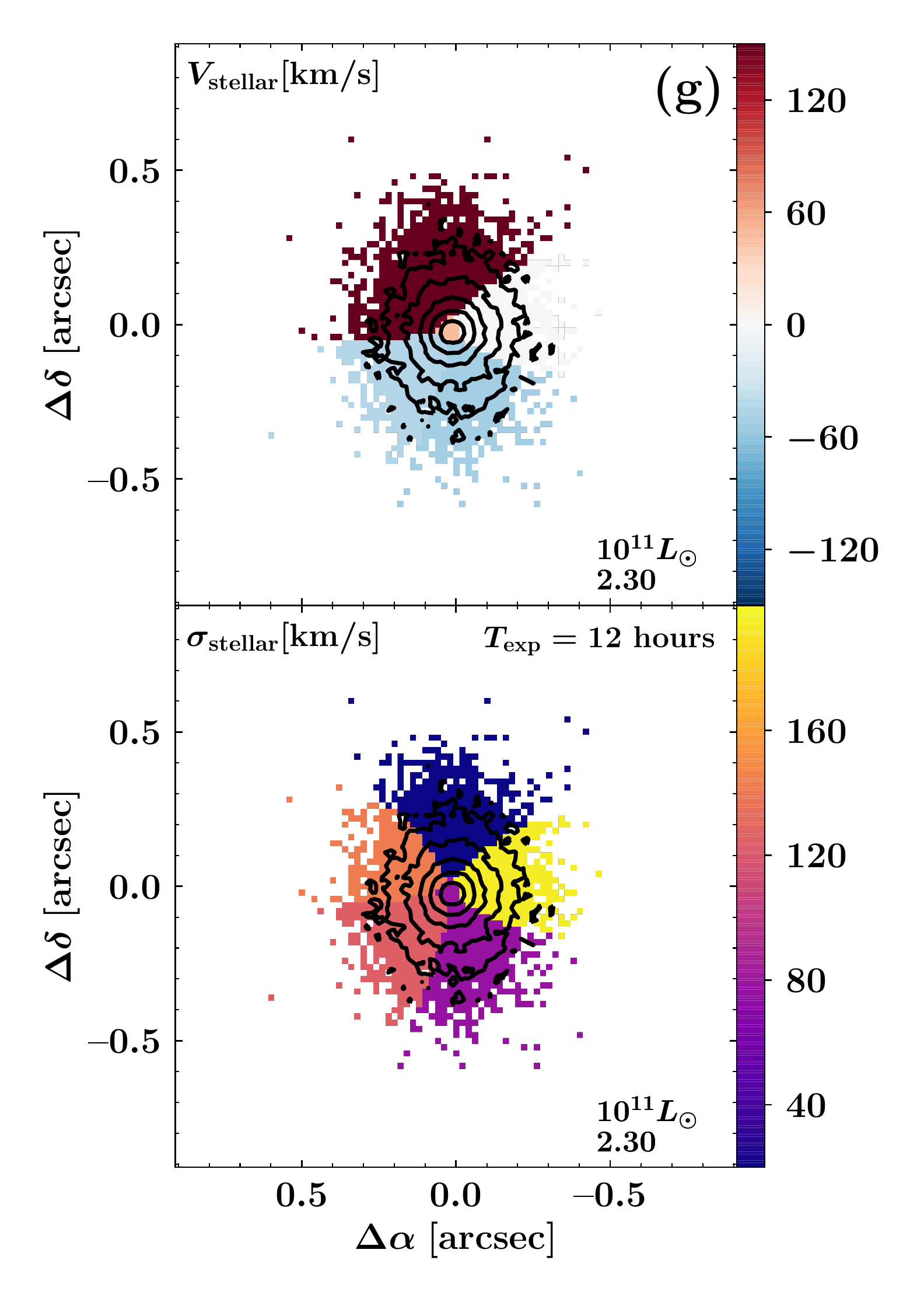}
\includegraphics[trim={0.5cm 0.5cm 0.75cm 0},clip,width=4.25cm]{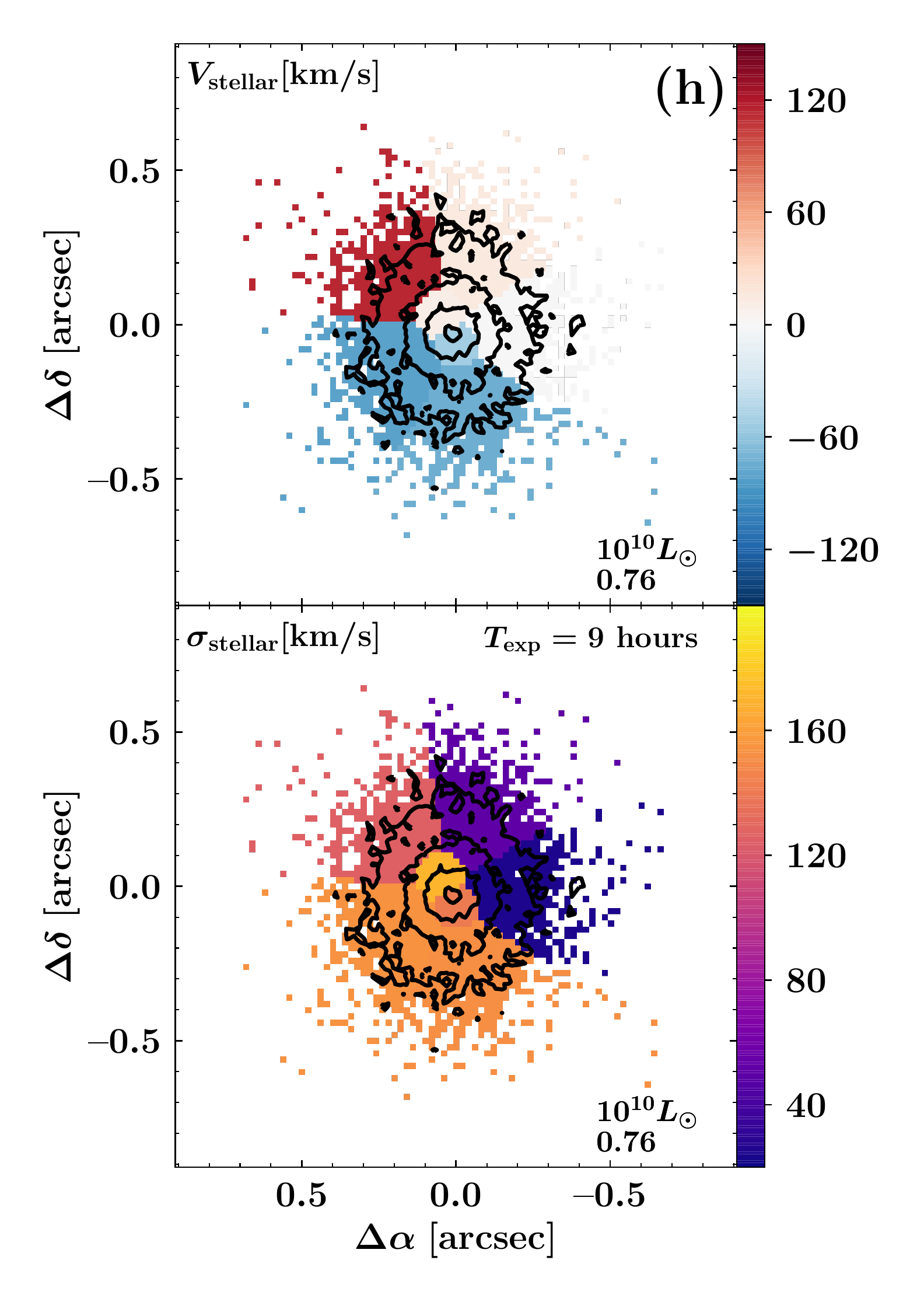}
  \caption{
  Stellar kinematics of galaxies at redshifts around cosmic noon. Top panels: Stellar velocity field (V$_{\star}$) and velocity dispersion distribution ($\sigma_{\star}$) obtained from mock HARMONI observations of three hours of on-source exposure time of lenticular (i.e. NGC\,809) host galaxies scaled to the $V$-band luminosities indicated in each panel at redshifts 0.76 ((a) and (b)), 1.50 (c), and 2.30 (d), as also indicated. Bottom panels: Same 2D-distributions but from mock HARMONI observations of six ((e)), nine ((f) and (h)), and 12 ((g)) hours of total exposure time of a lenticular hosts of the luminosities in the $V$ band, which is labelled (i.e. 10$^{11}$L$_{\odot}$ ((e), (f), and (g)) and 10$^{10}$L$_{\odot}$ ((h)). Contours (in black) from the stellar continuum are overlaid in steps of 0.25 mag. For all panels, the colour bars are in the [-150,150] km s$^{-1}$ and [20,200] km s$^{-1}$ ranges for velocities and velocity dispersions, respectively.}
 \label{resolved_kinematics_NGC809}
\end{figure*}

To spatially resolve the stellar kinematics, we only selected those spaxels within the outermost isophote with S/N$\geq$1,  and spatially binned the spectra to achieve a S/N of about 8. This level provides reliable values of the first two moments of the line-of-sight velocity distribution (see \S \ref{integrated_results}), and it still preserves spatial resolution in the brightest mock HARMONI observations. Using the GIST\footnote{https://abittner.gitlab.io/thegistpipeline/} pipeline \citep{2019Bittner}, we applied the Voronoi binning procedure described in \cite{cappellari2003} to ensure this S/N, and the {\scshape{ppxf}} code to derive the velocity and the velocity dispersion of each Voronoi-bin spectrum. We note that the actual spatial resolution of stellar kinematic maps will depend on the S/N threshold applied and the sizes of the resulting Voronoi-bins, smoothing the potential spatial resolution of the selected HARMONI scale. Figure \ref{resolved_kinematics_NGC809} shows the derived stellar velocity fields and distribution of stellar velocity dispersions for the AGN-lenticular host (see Figs. \ref{resolved_kinematics_PGC055442} and \ref{resolved_kinematics_NGC7119N} in Appendix \ref{appen_images} for the spiral and interacting hosts). The noisy appearance of the stellar kinematics maps is larger when increasing the redshift and decreasing the host luminosity. Residuals resulting after the sky subtraction, mainly for the H+K spectral configuration, strongly impact the derived stellar kinematics. These residuals result from the interpolations inside HSIM to match the spatial and spectral sampling of the input datacube to the HSIM internal samplings used to generate the mock data cubes.

We fit the global kinematic position angle (PA$_{k}$ in Table \ref{photo_par}) of these resolved stellar velocity fields using the PaFit \footnote{https://pypi.org/project/pafit/} code \citep{2006Krajnovi}. In general, we found a good agreement between the position angles of the major kinematic axes derived for the HSIM inputs  and HSIM outputs (Table \ref{photo_par}). Large deviations result for poor, spatially resolved velocity fields (see Table \ref{photo_par}).  From the analysis of these simulations, we infer that three hours of total exposure with HARMONI are enough to resolve the stellar kinematics of the most massive hosts of AGNs at around cosmic noon. 

We checked the dependency of the number of Voronoi bins at S/N$\sim$8 with exposure time to derive a reasonable observing limit to resolve the stellar kinematics of host galaxies with stellar masses in the M$_{\star} \sim 10^{10}-10^{11}$ M$_{\odot}$ range. For that, we performed some additional simulations for the lenticular host (i.e. NGC\,809) increasing the total exposure time to 6, 9, 12, or 15 hours. Table \ref{voxels_vari} summarises the results in terms of number of voxels. Panels from (e) to (h) in Figure \ref{resolved_kinematics_NGC809} include the corresponding 2D 
distributions of velocities and velocity dispersions. Resolving the stellar velocity field of low-mass (i.e. M$_{\star}$ $\leq$ 10$^{10}$ M$_{\odot}$) galaxies beyond cosmic noon is an expensive challenge in terms of observing time, even for the ELT (> 15 hours on-source per target). This estimation is in agreement with previous simulations for galaxies at 2$<z<$4 \citep{2016Kendrew}.

In the case of AGN hosts, the presence of the AGN poses an even bigger challenge. The stellar features will show a certain degree of dilution by the AGN continuum \citep[e.g.][]{2005Garcia-Rissmann}. The AGN removal will impact the S/N of individual spaxels, probably limiting the spatial resolution and coverage of the stellar kinematics. However, the level of residuals might become small when combining spatial and spectral information, to the point that the relevant residuals would only affect the brightest 1 or 2 spaxels that trace the PSF \citep{husemann2016, rupke2017}. Moreover, gas outflows are ubiquitous in luminous AGNs, usually giving rise to double-peaked and multi-component emission-line profiles. The scale and orientation of these AGN-driven outflows might impact the determination of the stellar kinematics \citep[see e.g.][]{2006Gersen}. In particular, at the wavelength range of interest, the MgI features can be strongly contaminated by the complexity of the [NI]$\lambda$5199 emission line doublet resulting from the presence of multiple gaseous components \citep[e.g. Fig. 5e in][]{1999Garcia}. For the considered QSO luminosities (i.e. L$_{[\ion{O}{iii}]}\sim$10$^{42}$,10$^{43}$ erg s$^{-1}$), ionised outflows usually present velocities between 600 and 900 km s$^{-1}$, and mass outflow rates in the range 1.5-50 M$_{\odot}$ year$^{-1}$ \citep[e.g.][]{2017Fiore}. We do not include any feature accounting for such ionised outflows in our simulations. The wide range in the observed outflow properties makes it difficult to define a representative model. 


\subsection{Integrated spectra of the AGN+hosts: AGN parameters }
\label{fit-AGN}

\begin{figure*}
\centering
 \includegraphics[trim={0.25cm 1.cm 1.75cm 1.25cm},clip,width=5cm]{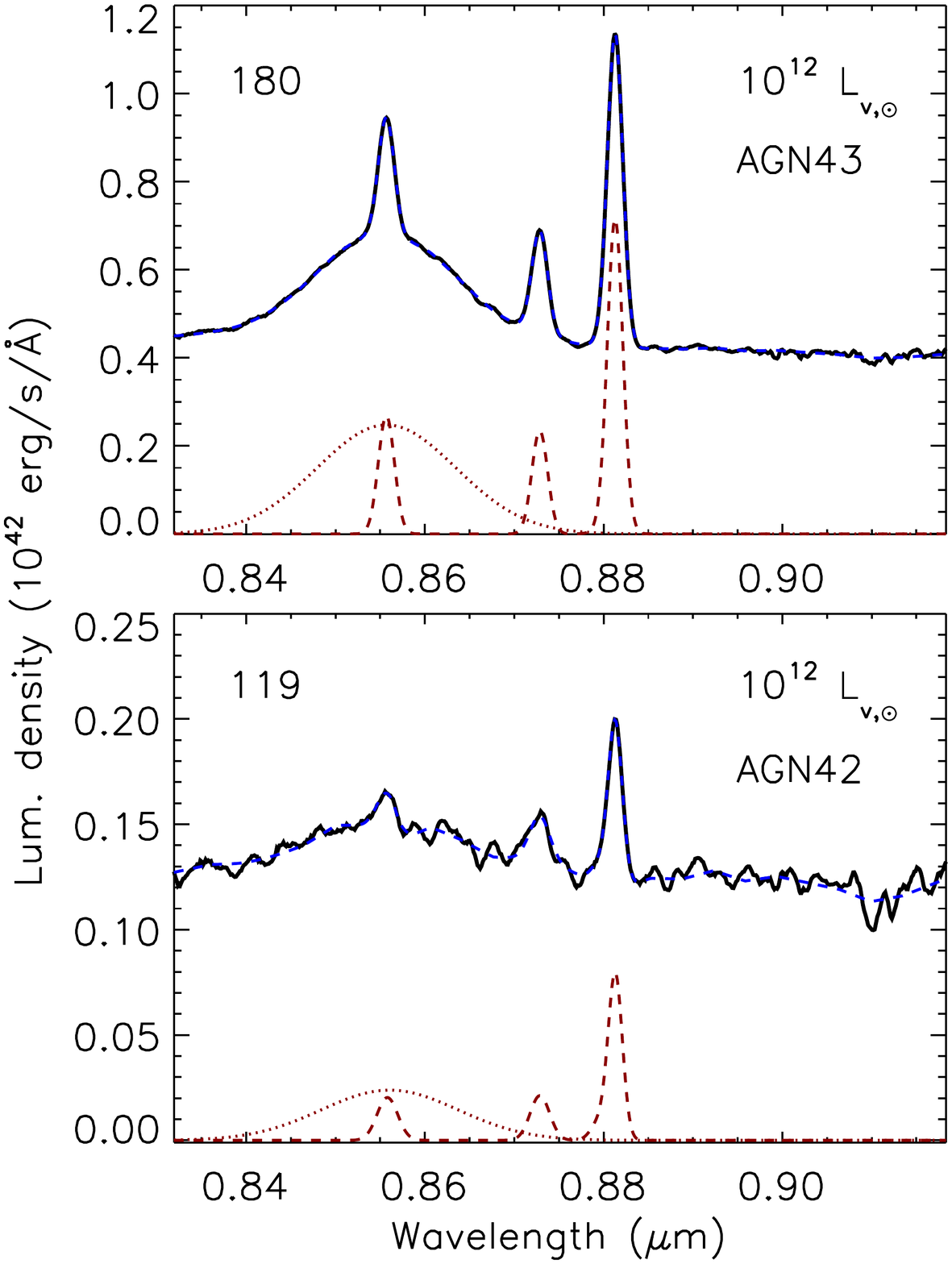}
 \includegraphics[trim={0.25cm 1.cm 1.75cm 1.25cm},clip,width=5cm]{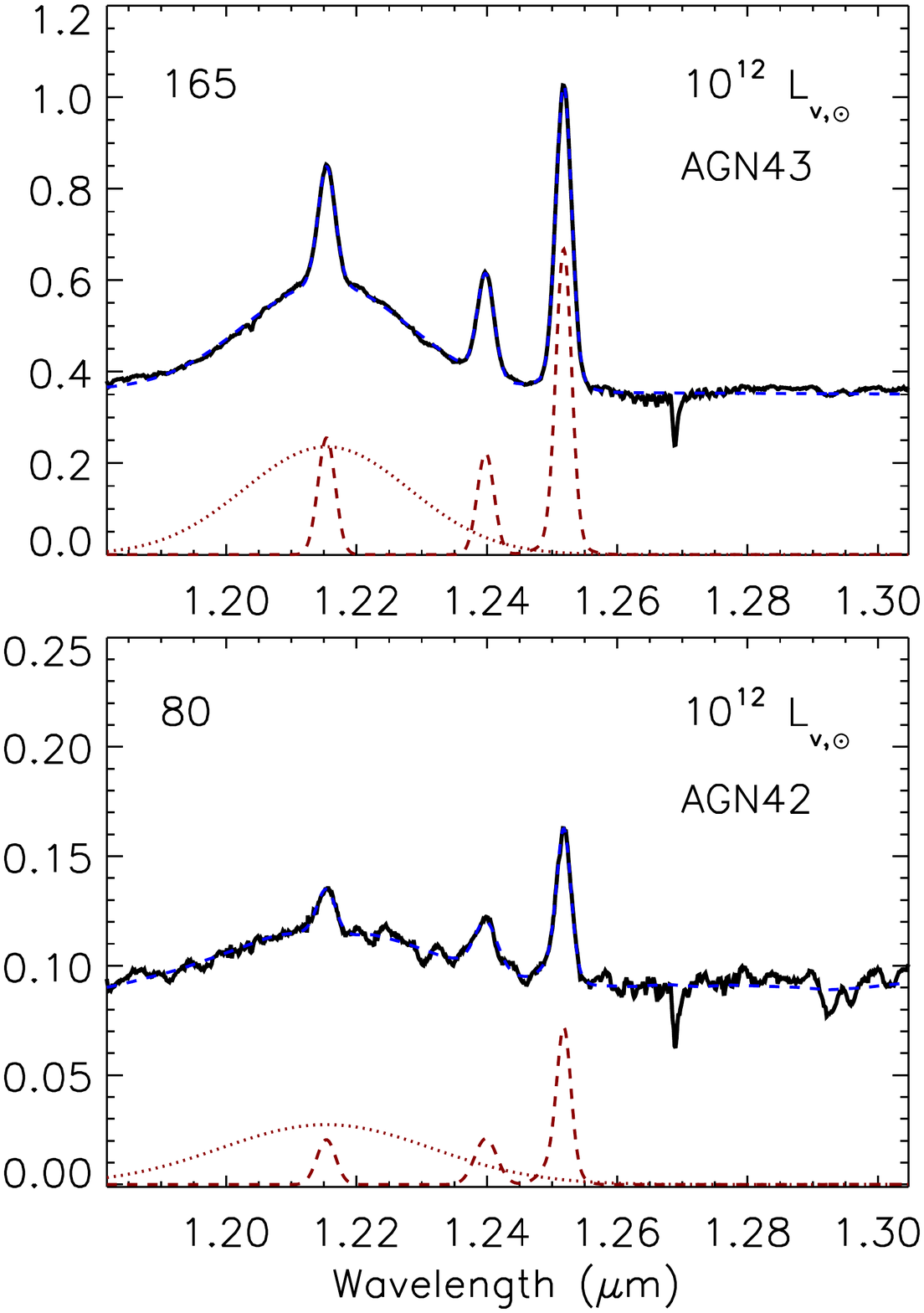}
 \includegraphics[trim={0.25cm 1.cm 1.75cm 1.25cm},clip,width=5cm]{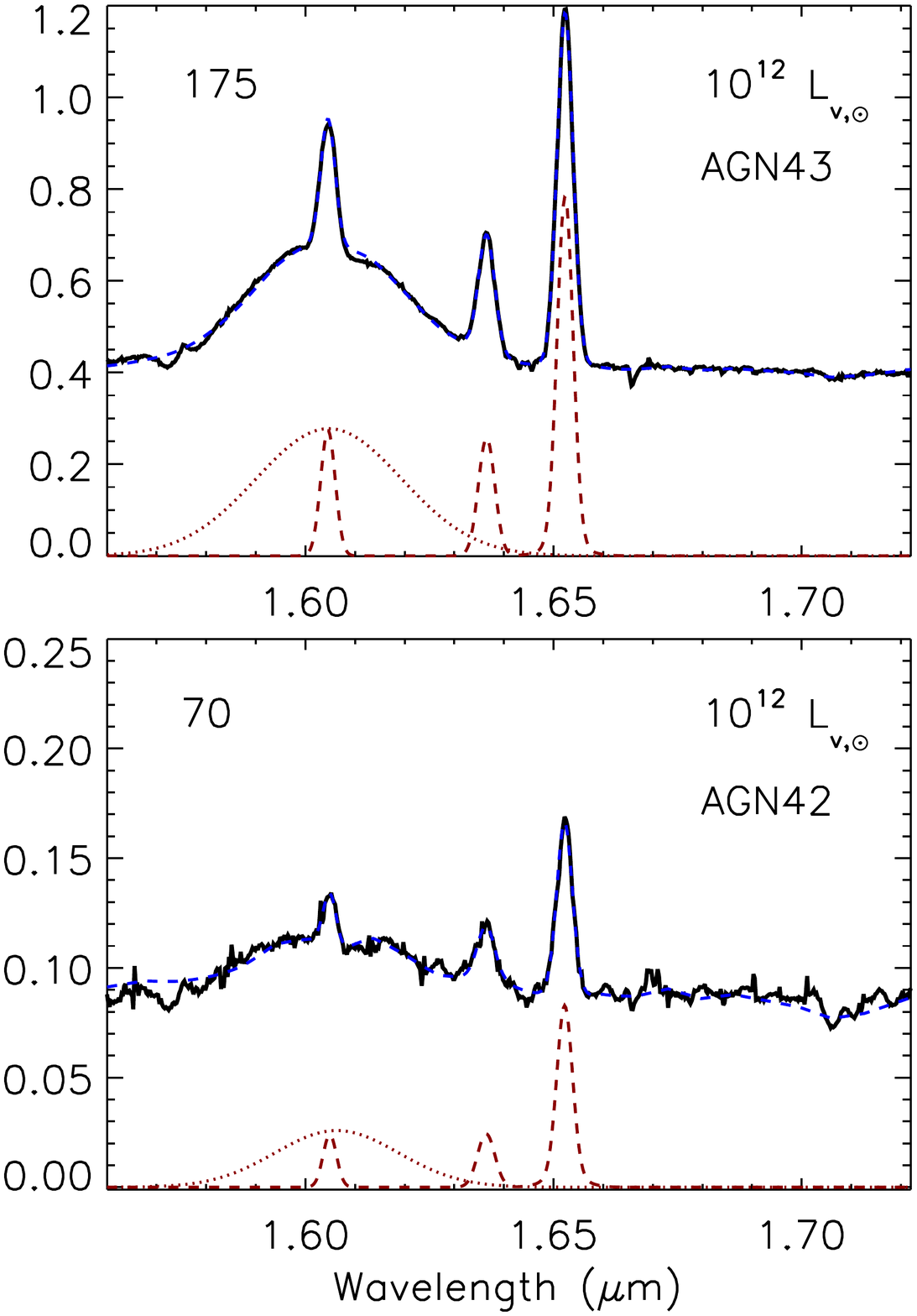}
 
  \includegraphics[trim={0.25cm 1.cm 1.75cm 1.25cm},clip,width=5cm]{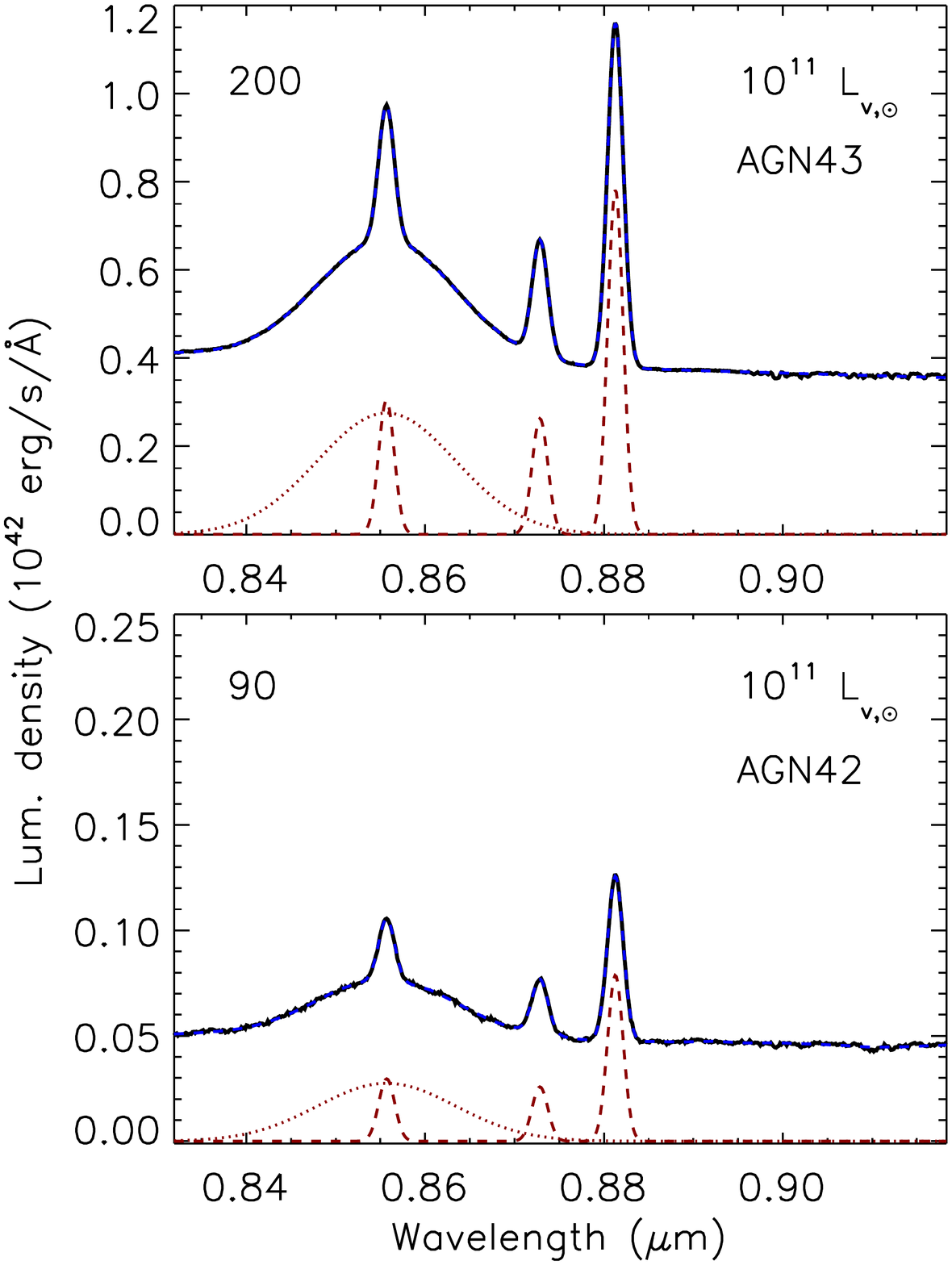}
  \includegraphics[trim={0.25cm 1.cm 1.75cm 1.25cm},clip,width=5cm]{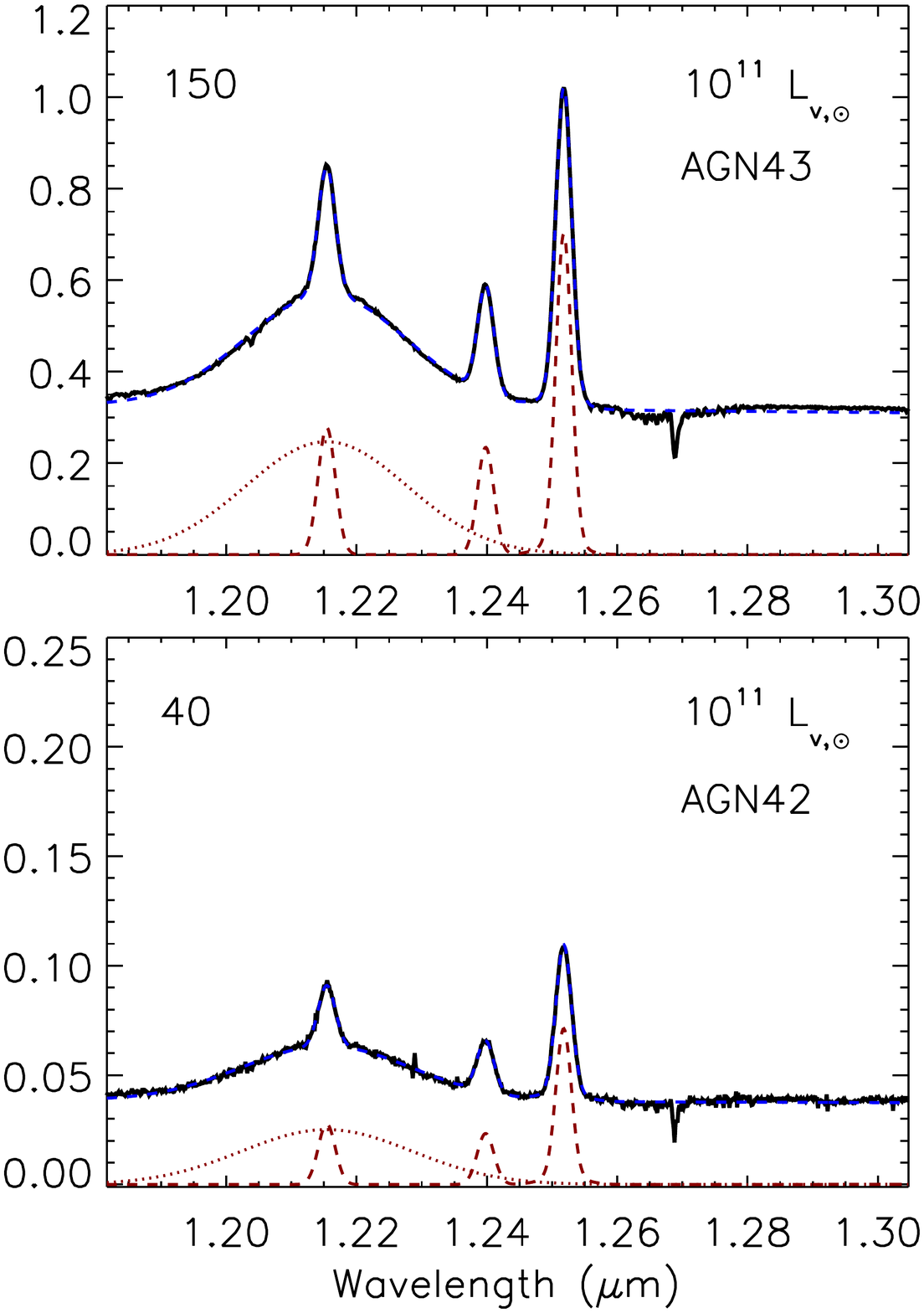}
  \includegraphics[trim={0.25cm 1.cm 1.75cm 1.25cm},clip,width=5cm]{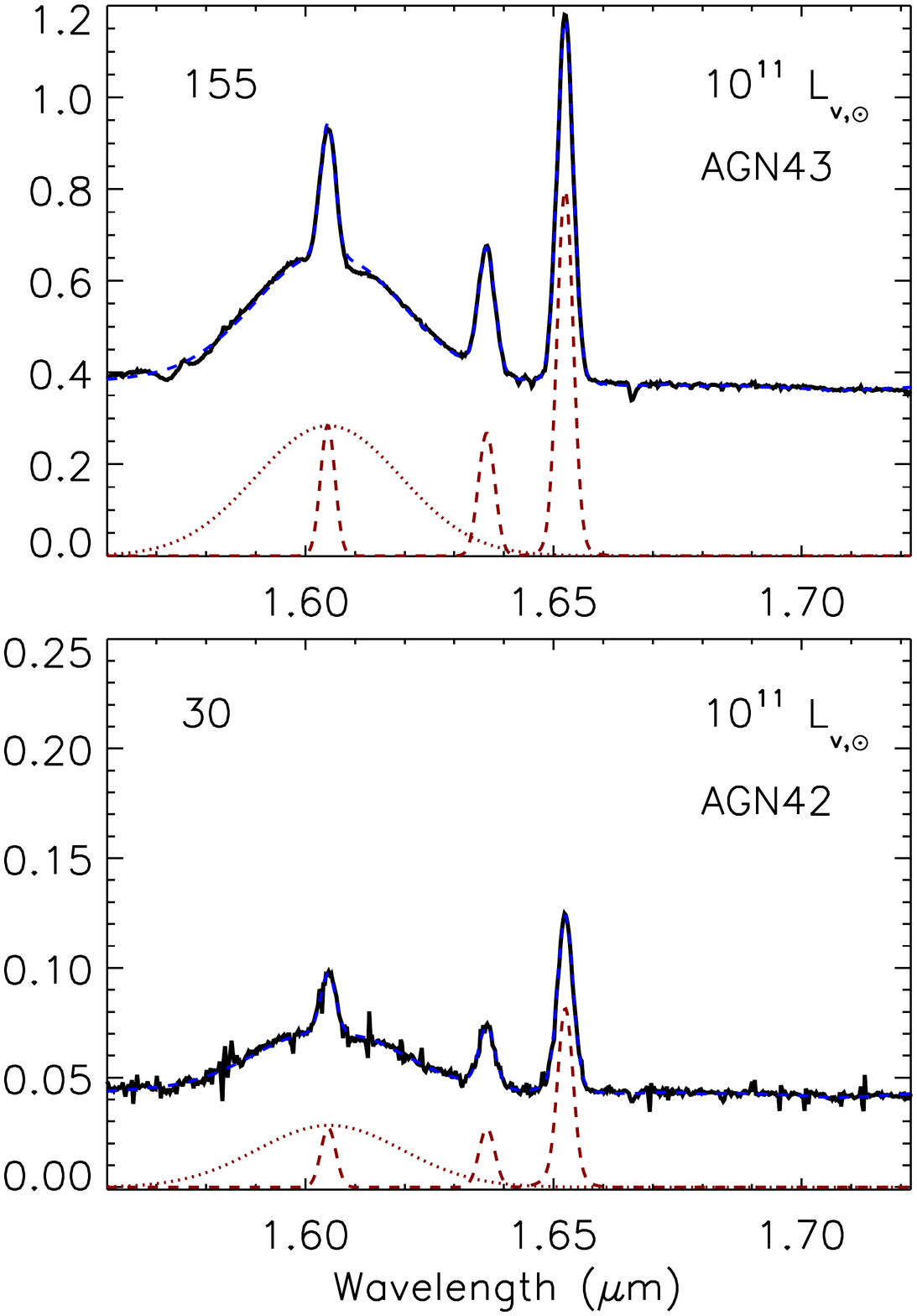}

 \includegraphics[trim={0.25cm 1.cm 1.75cm 1.25cm},clip,width=5cm]{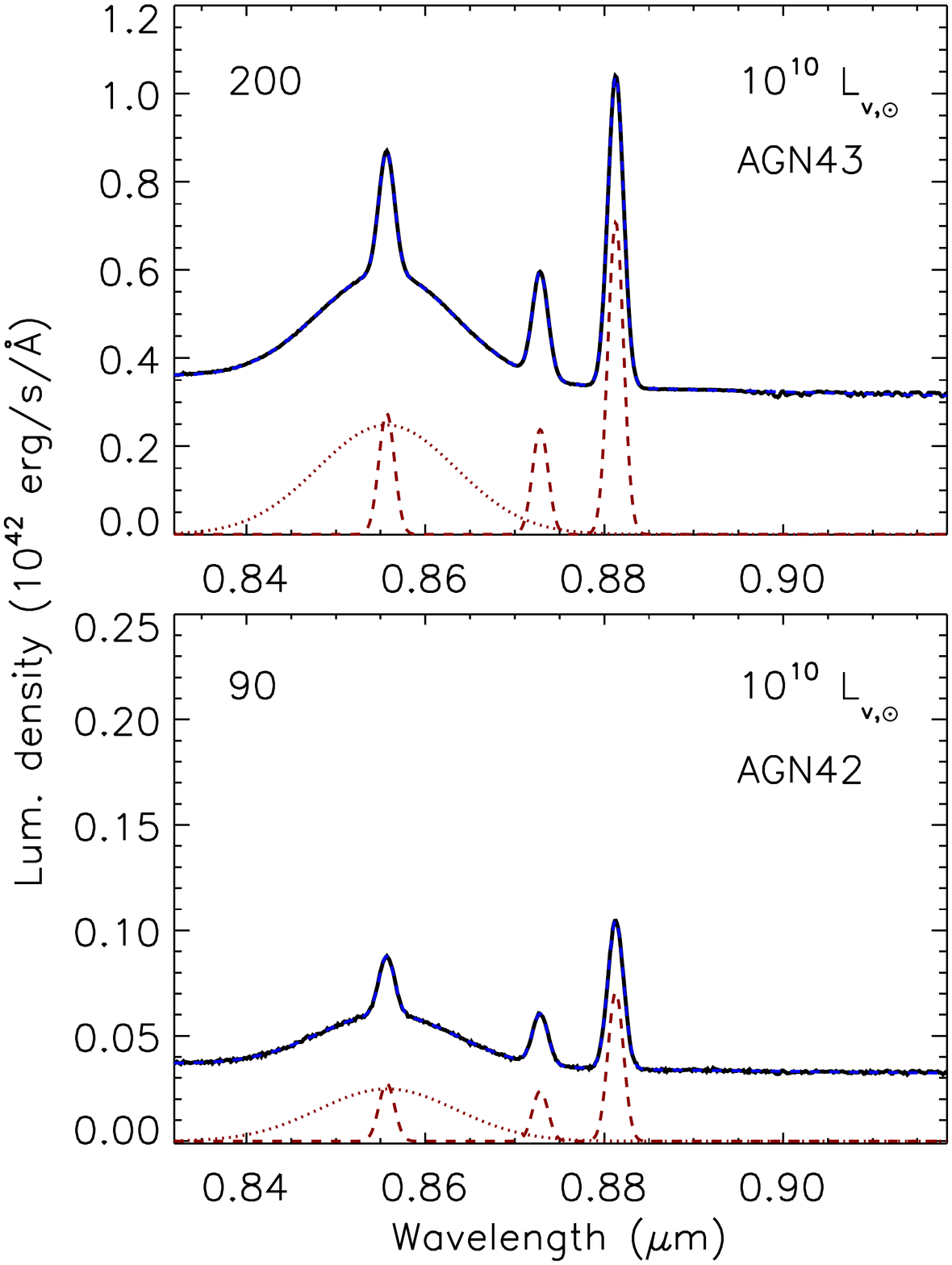}
 \includegraphics[trim={0.25cm 1.cm 1.75cm 1.25cm},clip,width=5cm]{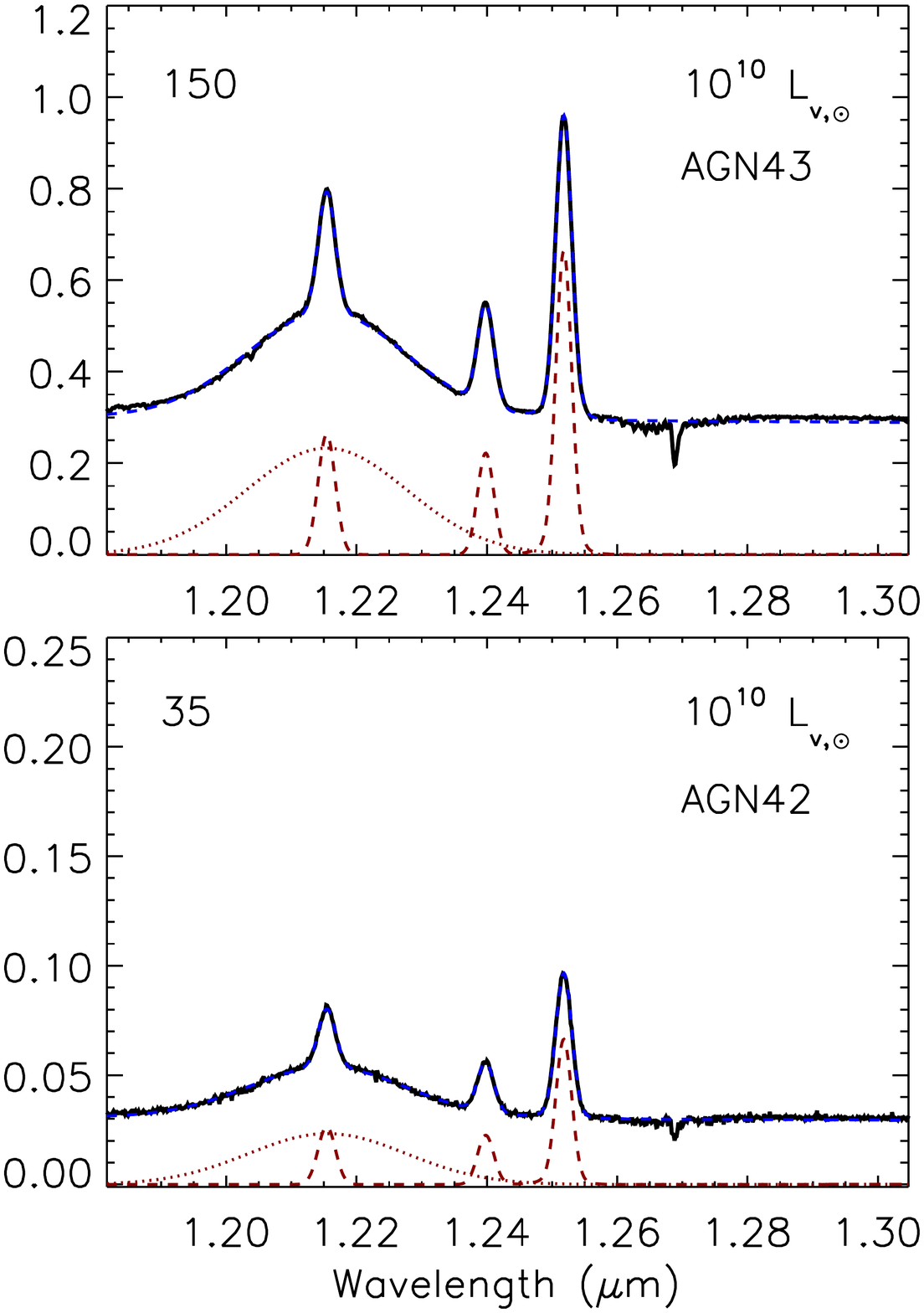}
 \includegraphics[trim={0.25cm 1.cm 1.75cm 1.25cm},clip,width=5cm]{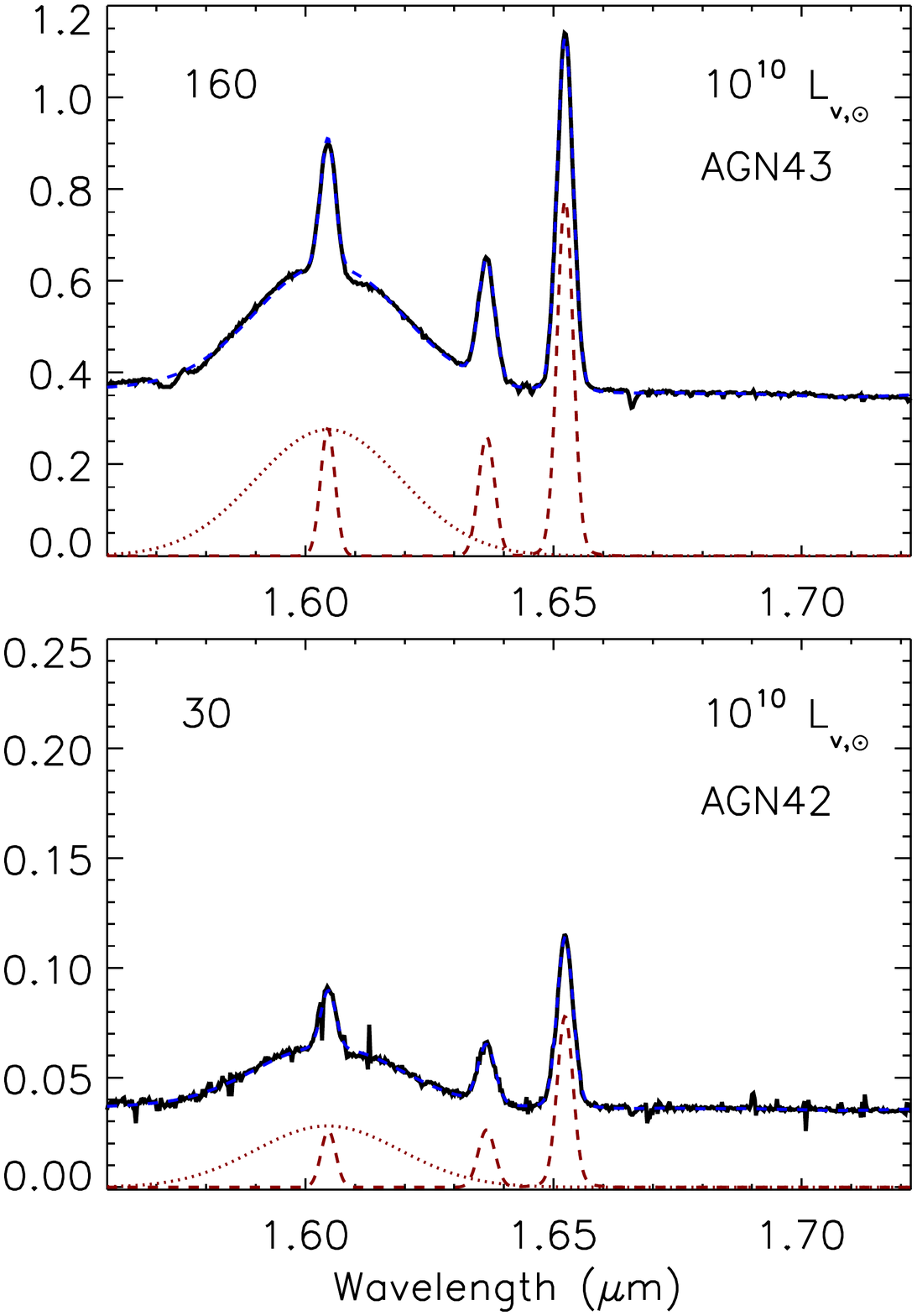}

  \caption{ Integrated spectra for the AGN+host (lenticular galaxy) at redshifts 0.76 (left), 1.50 (centre), and 2.30 (right) obtained by adding the signal from the spaxels within the host's effective radius (see \S\ref{integrated_results}). The labels in the top right corner indicate the host and AGN (AGN42$\rightarrow$10$^{42}$ and AGN43$\rightarrow$10$^{43}$ erg s$^{-1}$) luminosities. The numbers in the top left corners indicate the S/N in the rest-frame 5050-5250 \AA \ obtained using the der\_snr algorithm. The blue-dotted lines correspond to the QSFIT model to each spectrum. Red lines draw the Gaussian profiles fitting the broad (dotted line) and narrow (dashed line) components of the emission lines in the AGN spectra.}
  
 \label{BLR_fit}
\end{figure*}

\begin{figure*}
\includegraphics[trim={0cm 0.0cm 3.5cm 0cm},clip,width=10.75cm, angle=270]{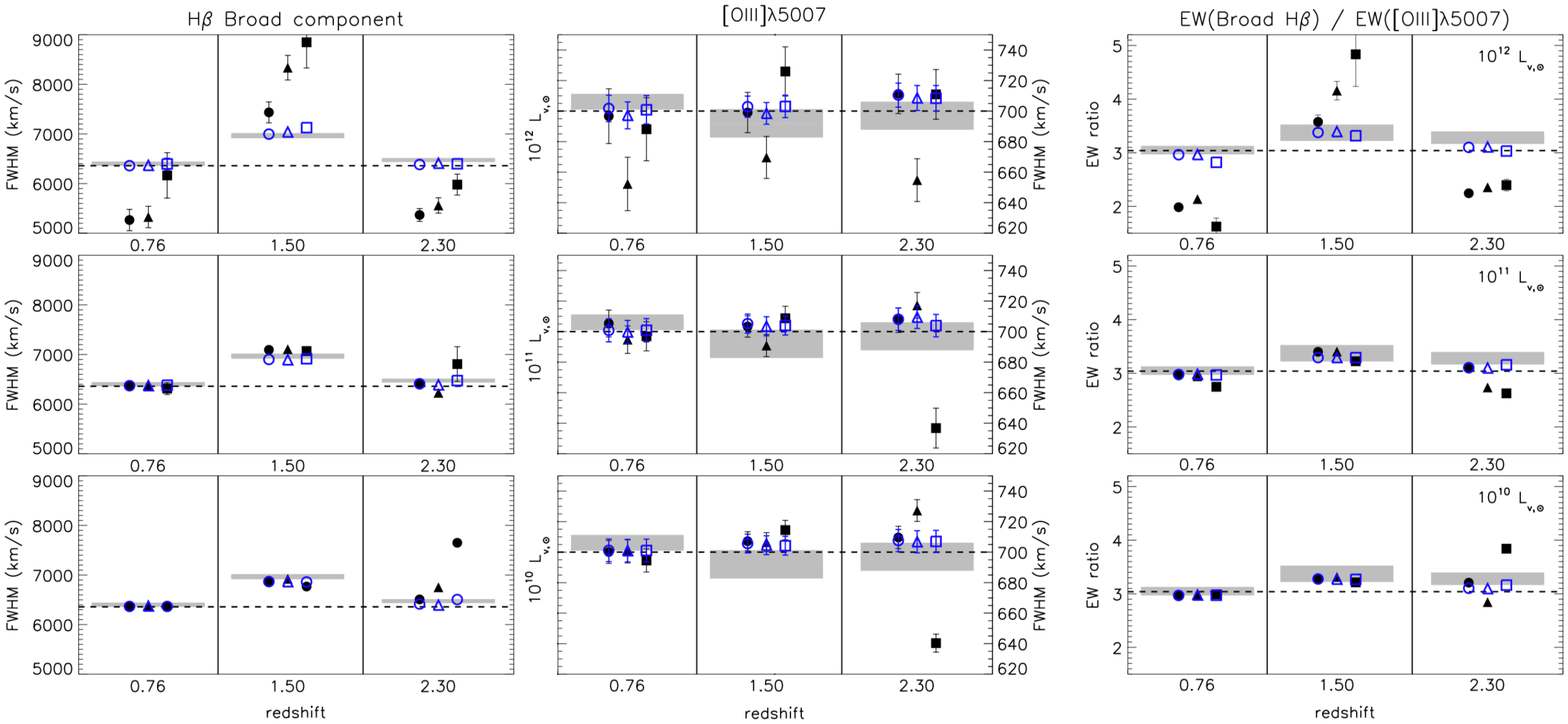}

  \caption{ Full width at half maximum of the broad H$\beta$ (left panels) and narrow [OIII]$\lambda5007$ (central panels) components of the Gaussian profiles modelling the emission lines in the AGN+hosts' integrated-spectra at the different redshifts (see text). Right panel shows the equivalent-width ratio of the broad and narrow components. Different symbols indicate the distinct hosts (circles$\rightarrow$NGC\,809, triangles$\rightarrow$PGC\,055442, and squares$\rightarrow$NGC\,7119A), and their colours indicate the [OIII] luminosities of the AGN (black filled symbols$\rightarrow$10$^{42}$, and blue open symbols$\rightarrow$10$^{43}$ erg s$^{-1}$). Error bars correspond to the uncertainties provided by the QSFIT$^{\ref{catalogue}}$ package if larger than the symbol size. Dashed lines are the adopted values to simulate the emission of the central AGN (see \S\ref{QSO}). Grey bands indicate the parameters obtained when fitting the integrated spectra obtained from the mock HARMONI data for the pure AGN, enclosing the values and uncertainties for the two AGN luminosities. }
    \label{fwhm_emission}
\end{figure*}

  Following the procedure described in \ref{integrated_results}, we also obtained the integrated spectra for the mock HARMONI observations of the AGN+hosts within an aperture of the R$_{eff}$ in radius. It is worth noting that the host galaxy contribution within these apertures is not negligible, resulting in a certain degree of dilution of the AGN features, mainly in the less contrasted cases. We performed a basic spectral analysis of these spectra in the 4725-5220 \AA \ rest-frame spectral range using QSFIT$^{\ref{catalogue}}$ \citep{2017MNRAS.472.4051C}. We did not consider any iron contribution in the QSFIT modelling because it is not present in our AGN model (see \S\ref{QSO}). We note that QSFIT only includes a host galaxy for sources with z < 0.8. Figure \ref{BLR_fit} shows some examples of these integrated QSO+AGN spectra, including the QSFIT models overplotted. The QSFIT model uses Gaussian profiles to fit the broad and narrow components of the emission lines in the AGN+host spectra, providing their parameters. Figure \ref{fwhm_emission} summarises the results for the different mock host's AGN combinations. Differences between values for the input (dashed lines in Fig. \ref{fwhm_emission}) and the mock HARMONI pure AGNs (grey bands in Fig. \ref{fwhm_emission}) are negligible in all the brightest AGNs and many of the faintest AGN cases. Compared to the parameters of the pure AGN, only the faintest AGN shows large deviations when combined with the brightest host, and also the furthest cosmic distance. The design of the simulations, which includes an internal spectral interpolation (see \S\ref{HSIM_par}), probably induces the non-negligible differences.
 
\section{Summary and conclusions}
\label{conclusions}

The new generation of 30-40-metre class telescopes will be a major revolution for astrophysics, opening new frontiers. As a first light instrument for the ELT, HARMONI will be a very versatile integral field spectrograph supporting a broad range of science programmes. Given the multiple instrument settings (e.g. spaxel scale, gratings, AO mode, etc.), it is valuable to quantify in advance the expected performances for specific science cases. 

Focusing on the context of the co-evolution of galaxies and their central black holes, we explored the potential of HARMONI to resolve the morpho-kinematics of the host galaxy of AGNs at redshifts around cosmic noon. We used the dedicated simulation pipeline HSIM (version 300) to perform a set of mock HARMONI observations for the 20$\times$20 mas$^{2}$ spatial scale, a spectral resolution of R=3200, and LTAO working at 0.67 arcsec seeing. To create the input targets for HSIM, we generated synthetic AGN spectra and took advantage of available IFS data of nearby galaxies with distinct morpho-kinematics. We scaled these two ingredients to different AGN and host galaxy luminosities, combined them, and artificially dimmed and redshifted them to the chosen cosmic epoch. Our main results and conclusions are summarised as follows:

   \begin{itemize}
      \item The two coarser scales of HARMONI seem to be the best suited to reveal the host galaxies of distant AGNs. Their FoVs extend beyond the average size of the host galaxies, including the expected size of the outflows, and simultaneously observing the sky background.
      \item The low-resolution gratings of HARMONI provide the opportunity to resolve the stellar kinematics of galaxies using a homogeneous procedure from before cosmic noon to the local Universe, with similar velocity resolution to IFS legacy surveys of nearby and intermediate-redshift galaxies.
      \item In three hours of on-source integration, HARMONI low-spectral resolution observations with the 20$\times$20 mas$^{2}$ spatial scale will reveal the morphology of galaxies brighter than 10$^{10}$ L$_{\sun}$ around cosmic noon, allowing the estimation of basic morphological parameters (type, position angles, or inclination). These observations will also provide reliable estimations of their stellar velocity and velocity dispersions at the effective radius with S/Ns over 8 in the stellar continuum.
      \item ELT+HARMONI will resolve the stellar kinematics of massive galaxies (M$_{\star}$ $\geq$ 10$^{11}$ M$_{\sun}$) up to and beyond cosmic noon  when observing under median atmospheric conditions for three hours.
      \item  HARMONI observations including the ones mock here will allow reliable measurements of the emission line parameters of the central AGN for almost all the considered cases. Only those with the lowest AGN-host contrasts show discrepancies not attributable to the design of the simulations.
      \item The simulations described here show that from the ELT+HARMONI observations we can simultaneously measure the M$_{BH}$ and the resolved stellar kinematics of galaxies (M$_{\star}$ > 10$^{10.5}$ M$_{\sun}$) hosting bright AGN at a redshift around cosmic noon using total exposure times that can vary from three to 15 hours. 
      
   \end{itemize}
   
 As we already discussed, a major challenge to reveal the morpho-kinematics of distant AGN hosts is the de-blending the AGN and host galaxy spectra with negligible residuals. The reconstruction of the complex and variable PSF in IFS data is key to this goal. The next step of this work is to explore the reliability of 3D-PSF reconstruction for AO-assisted IFS observations.

\begin{acknowledgements}
      We are grateful to Laura S\'anchez-Menguiano for kindly sharing the (10), (11), (12), and (13) entries in Table \ref{tabgalaxies} for NGC\,809. We also thank Michele Cappellari, Ignacio Ferreras and Ignacio Mart\'{\i}n-Navarro for their useful comments. BG-L and AM-I acknowledge support from the Spanish Ministry of Science, Innovation and Universities (MCIU), Agencia Estatal de Investigaci\'on (AEI), and the Fondo Europeo de Desarrollo Regional (EU-FEDER) under projects with references AYA20155-68217-P and PID2019-107010GB-100. BG-L, AM-I, CRA and EMG also acknowledge financial support from the State Agency for Research of the Spanish MCIU through the Center of Excellence Severo Ochoa award to the Instituto de Astrof\'{\i}sica de Canarias (SEV-2015-0548 and CEX2019-000920-S). MPS acknowledges support from the Comunidad de Madrid through the Atracci\'on de Talento Investigador Grant 2018-T1/TIC-11035 and PID2019-105423GA-I00 (MCIU/AEI/EU-FEDER).  NT acknowledges support from the Science and Technology Facilities Council (grant ST/N002717/1), as part of the UK E-ELT Programme at the University of Oxford. CRA acknowledges financial support from the Spanish MCIU under grant with reference RYC-2014-15779, from the European Union Horizon 2020 research and innovation programme under Marie Sklodowska-Curie grant agreement No 860744 (BiD4BESt), from the State Research Agency (AEI-MCINN) of the Spanish MCIU under grants PID2019-106027GBC42 and EUR2020-112266. CRA also acknowledges support from the Consejer\'{\i}a de Econom\'{\i}a, Conocimiento y Empleo del Gobierno de Canarias and the EU-FEDER under grant with reference ProID2020010105. LG acknowledges financial support from the Spanish Ministry of Science, Innovation and Universities (MCIU) under the 2019 Ram\'on y Cajal program RYC2019-027683 and from the Spanish MCIU project HOSTFLOWS PID2020-115253GA-I00. We also acknowledge the usage of the HyperLeda database (http://leda.univ-lyon1.fr). This research has made use of the SIMBAD database, operated at CSD, Strasbourg, France.

\end{acknowledgements}

%
%

\bibliographystyle{aa}
\bibliography{biblio_QSO}


\begin{appendix} 

\section{Individual galaxies}\label{appen}

\begin{table*}
        \centering
        \caption{Basic parameters of the three nearby objects selected as models for the host galaxies of AGN at redshifts around the cosmic noon (see \S \ref{host}). }
        
        \label{tabgalaxies}
        \begin{threeparttable}

        \begin{tabular}{cccccccccccccccc} 
                \hline
    Name & D & V  & $A_B$ & $m_B$ & Type & Structure & R$_{eff}$ & log Mass & log SFR & PA & incl. & V$_{sys}$ & $\sigma$ \\
    (1) & (2) & (3) & (4) & (5) & (6) & (7) & (8) & (9) & (10) & (11) & (12) & (13) & (14) \\
\hline
                 NGC809 & 79.2 & 5340 &  0.10 &  14.59 & S0 & Ring & 3.5 & 10.6 &  0.41 & -4 & 41 & 5369 & 145 \\
              PGC055442 & 105.5 & 7026 &  0.25 &  14.76 & E/SABb\tnote{*} & Bar + ring & 3.8 & 10.9 & -0.1 & 15 & 46 & 7036 & 149\\
  NGC\,7119A & 142.3 & 9875 &  0.08 &  13.44 & SBbc & Companion & 9.9 & 11.6 & 1.4 & -45 & 57 & 9874 & 138 \\
                \hline
        \end{tabular}
        \tablefoot{ Columns correspond to: (1) galaxy name, (2) galactocentric distance in Mpc, computed assuming a Hubble flow with a Hubble constant H$_{0}$=67.8 km s$^{-1}$ Mpc$^{-1}$, (3) mean heliocentric radial velocity (c$z$) in km s$^{-1}$, (4) galactic extinction in B band, (5) total B magnitude, (6) morphological type, (7) morphological particularities reported, (8) effective radius in kpc, (9) integrated stellar mass in log M$_{\odot}$, (10) star formation rate in log M$_{\odot}$/yr, (11) photometric position angle in degrees, (12) inclination in degrees, (13) systemic velocity in km s$^{-1}$, and (14) stellar velocity dispersion in km s$^{-1}$. Data for columns (3) to (7) are as provided by HyperLeda$^{\ref{ledacat}}$. Data for columns (8) to (12) are from \citealt{laura2018}. Data for columns (13) and (14) are estimates of this work (see text), with rms errors always $\leq$10 km s$^{-1}$ in both parameters.\\}
        \tablefoottext{*}{Classified by HyperLeda as elliptical, the MUSE data shows the presence of a clear spiral structure \citep{laura2018}.}  
  \end{threeparttable}
\end{table*}

In this work, we selected three galaxies observed in the AMUSING survey \citep{Galbany16} as prototypes for the host galaxies of AGNs at the desired cosmic epoch (see \S \ref{host}). This appendix is focused on showing the main characteristics of these galaxies. Table \ref{tabgalaxies} presents the basic data for them. Details on the data processing are provided in \citet{Galbany16}. 

For each object, we obtained a 'total' spectrum (see Fig. \ref{integrated_spectra_MUSE}) by adding up the spectra of all spaxels in the MUSE data cubes (see \S \ref{host}
) inside a projected circular aperture of a diameter equal to the R$_{eff}$. We estimated the systemic velocity and stellar velocity dispersion (see Table \ref{tabgalaxies}) applying the {\scshape{ppxf}} code \citep{cappellari2004,cappellari2017} on these spectra in the rest-frame 4800-5300 \AA \ wavelength range, with the emission-lines masked. We also present here the stellar velocity fields and velocity dispersion maps extracted from the MUSE data cubes. We measured the stellar kinematics using {\scshape{ppxf}} through the GIST pipeline \citep{2019Bittner}. We applied the Voronoi binning scheme of \cite{cappellari2003} to spatially bin the data to homogenise the S/N of the analysed spectra to $\sim$8 in this spectral range. The spaxels with an estimated S/N below 1 were discarded. These S/N thresholds are the same as those selected for the AGN hosts at different redshifts. We used the MILES model library as a template \citep{vazdekis2015} at a spectral resolution of 2.51 \AA \ \citep{falcon2011}. The MILES library provides the smallest difference between the spectral resolution of the stellar templates and the MUSE instrument resolution \citep[$\sim$2.85 \AA \ at 0.5 $\mu$m,][]{guerou2017}. Figures \ref{stellar_NGC809}, \ref{stellar_PGC055442}, and \ref{stellar_NGC7119N} show the stellar velocity and velocity dispersion maps for NGC\,809, PGC\,055442, and NGC\,7119, respectively.

\subsection{NGC\,809}
\begin{figure}
\includegraphics[width=6.5cm, angle=270]{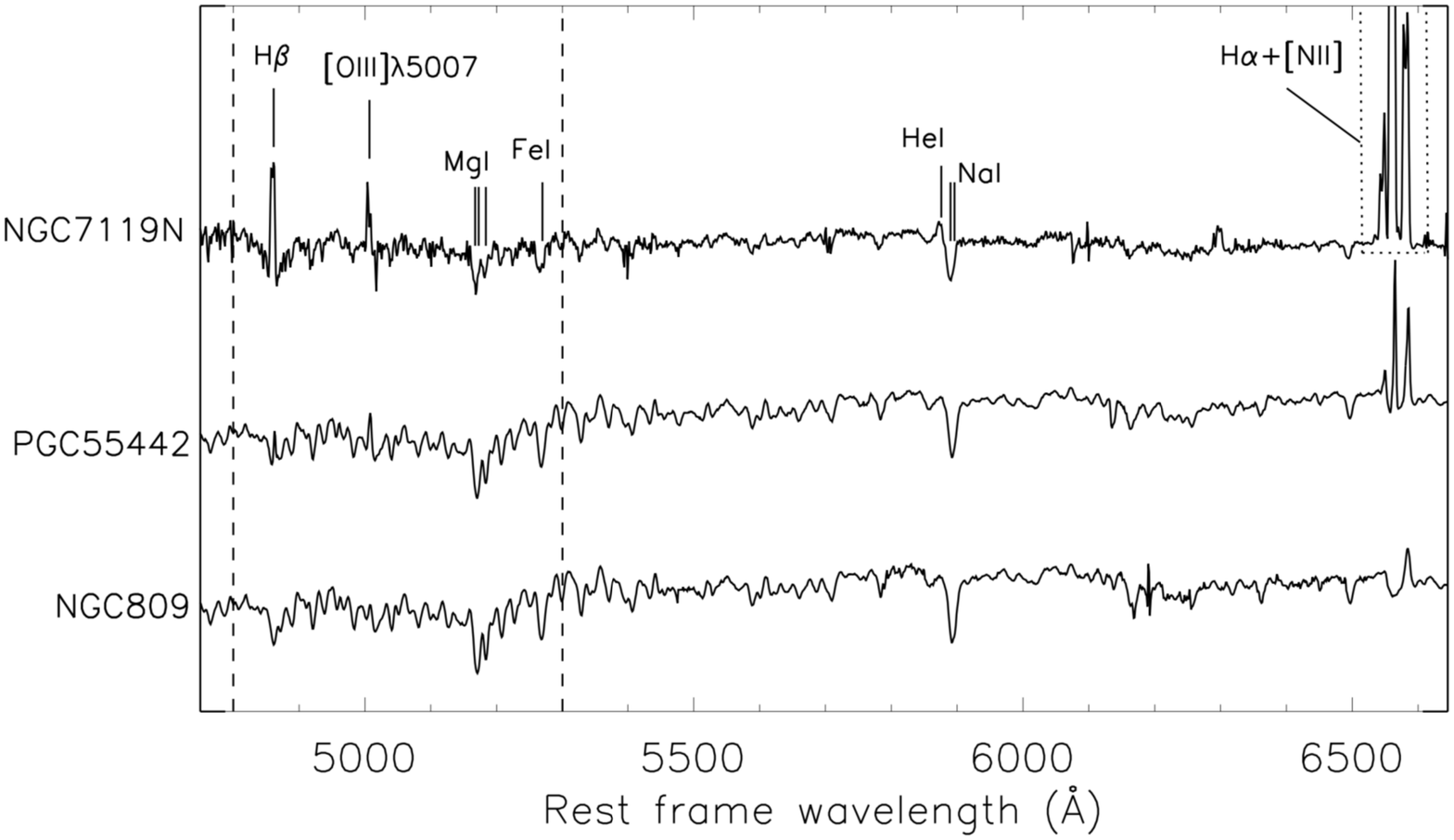}
  \caption{Total spectra of the three nearby galaxies picked as host for our simulations of AGN+host observations with HARMONI. Spectra were normalised to the average intensity in the plotted wavelength range. The horizontal axis corresponds to rest-frame wavelengths. The spectra are shown in logarithmic scale and arbitrarily shifted on the vertical axis. Dashed-vertical lines indicate the spectral range included in the HARMONI simulations. Labels mark the position of several emission (i.e. H$\beta$, [\ion{O}{iii}]$\lambda$5007, HeI$\lambda5876$, [NII]$\lambda$6543, H$\alpha$, and [NII]$\lambda$6584) and absorption lines (i.e. MgI$\lambda$5163, 5172, 5183, FeI$\lambda$5269, and NaI$\lambda$5889,5895). Dotted rectangle indicates the zoomed-in image of a region in Fig. \ref{zoom_NGC7119N_spectrum} for the NGC\,7119A spectrum.}
    \label{integrated_spectra_MUSE}
\end{figure}

\begin{figure*}
\includegraphics[width=\textwidth]{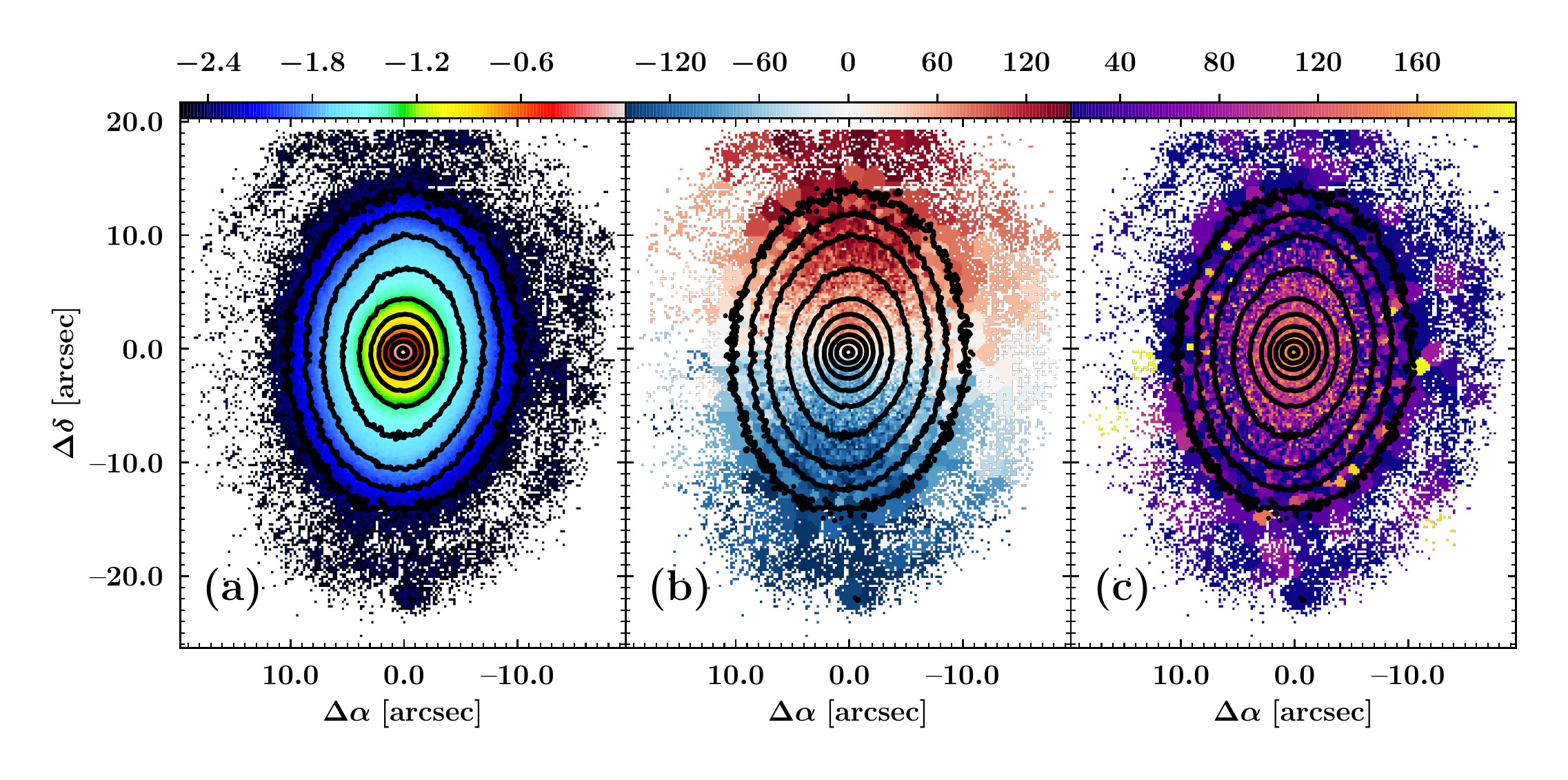}
  \caption{(a) Filter-band image of NGC\,809 obtained by collapsing the MUSE data cube over the rest-frame spectral range 5100-5230 \AA. (b) NGC\,809 line-of-sight stellar velocity field. (c) Stellar velocity dispersion map of NGC\,809. We plot all the MUSE spaxels with S/N>1 in the 5100-5230 \AA \ rest-frame. Limits of the colour bar are [-150,150] km s$^{-1}$ and [20,200] km s$^{-1}$ for velocity (panel (b)) and velocity dispersion (panel (c)), respectively. These are the same limits as those selected for the AGN hosts in Fig. \ref{resolved_kinematics_NGC809}. Contours (in black) correspond to the stellar continuum in steps of 0.25 mag.}
    \label{stellar_NGC809}
\end{figure*}

\begin{figure*}
\includegraphics[width=\textwidth]{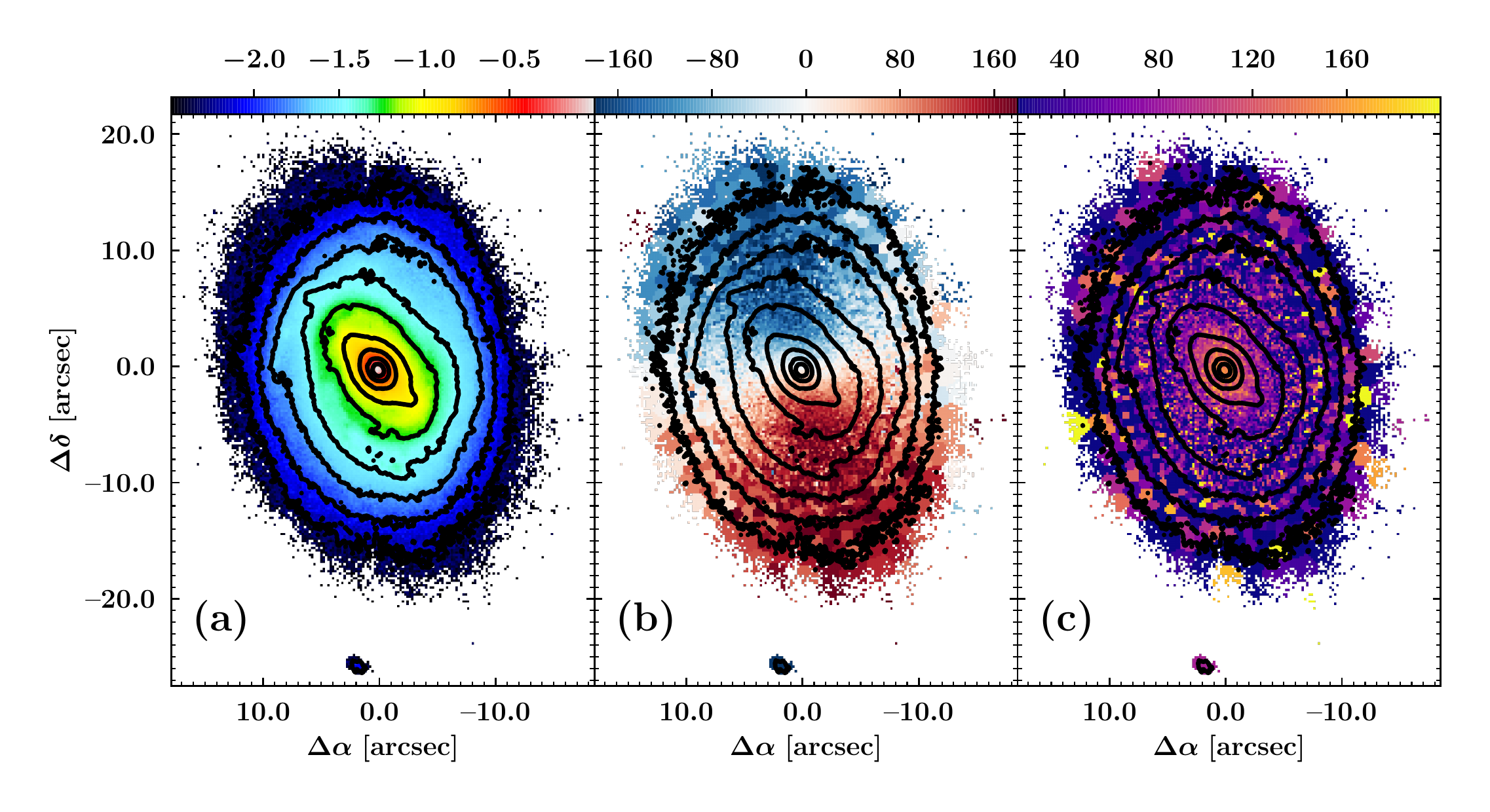}
  \caption{Same as Fig. \ref{stellar_NGC809}, but for PGC\,055442. Limits of the colour bar are [-180,180] km s$^{-1}$ and [20,200] km s$^{-1}$ for velocity (panel (b)) and velocity dispersion (panel (c)), respectively. These are the same limits as those selected for the AGN hosts in Fig. \ref{resolved_kinematics_PGC055442}.}
    \label{stellar_PGC055442}
\end{figure*}

\begin{figure*}
\includegraphics[width=\textwidth]{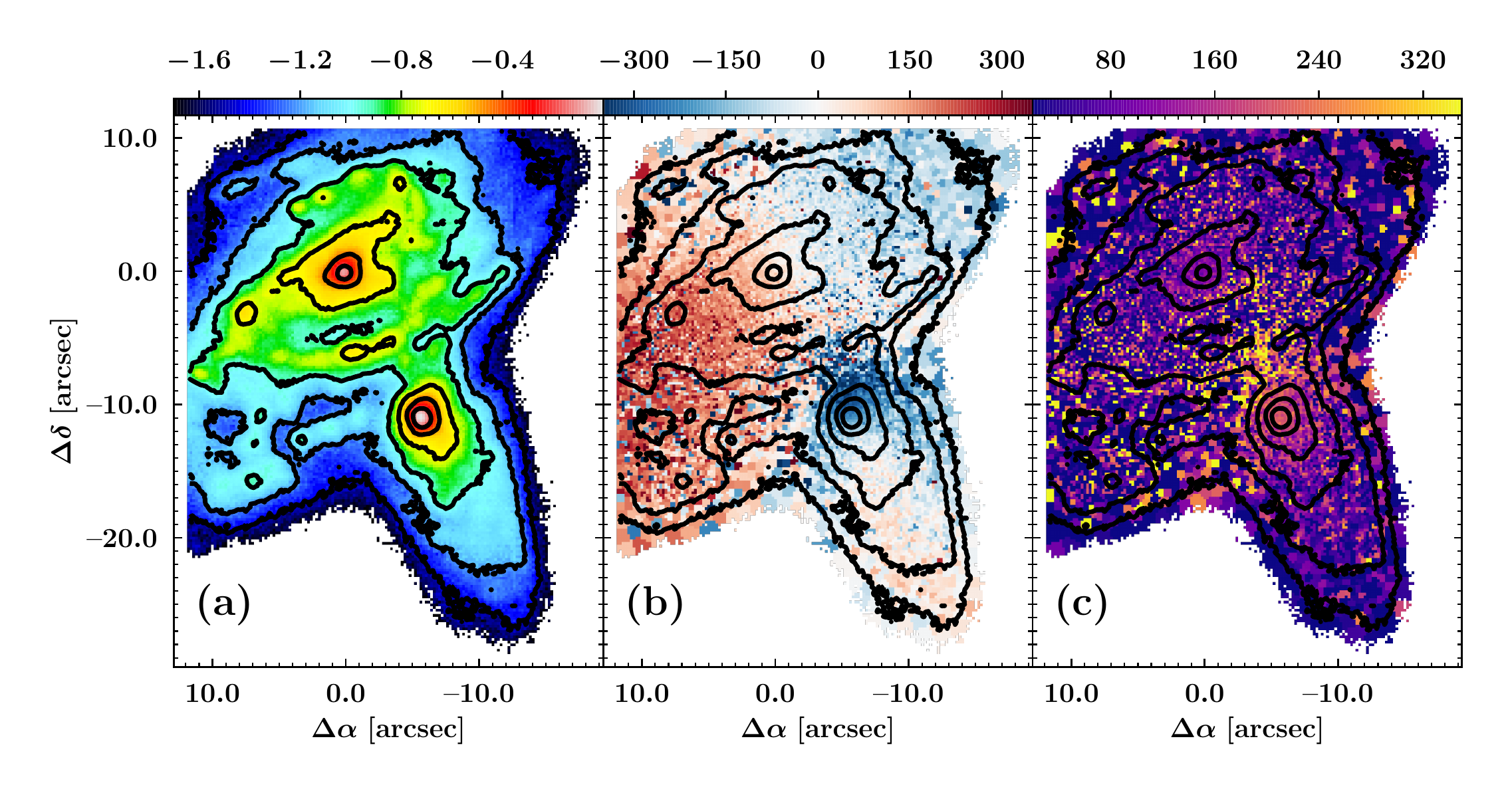}
  \caption{Same as Fig. \ref{stellar_NGC809}, but for NGC\,7119A. Limits of the colour bar are [-350,350] km s$^{-1}$ and [20,350] km s$^{-1}$ for velocity (panel (b)) and velocity dispersion (panel (c)), respectively. These are the same limits as those selected for the AGN hosts in Fig. \ref{resolved_kinematics_NGC7119N}.}
    \label{stellar_NGC7119N}
\end{figure*}

NGC\,809   is   a   lenticular   galaxy   showing   two   bright UV and optical rings at $\sim$3.5 kpc ($\sim$10 arcsec) and $\sim$9 kpc (25.5 arcsec) in which star formation activity has ceased \citep{2019proshina}. The integrated spectrum of NGC\,809 (see Fig. \ref{integrated_spectra_MUSE}) is dominated by the stellar continuum in the $4800-6600$ \AA\ spectral range. Emission line features are not evident in this range and only [NII]$\lambda6584$ can be visually identified. 
The stellar velocity field of NGC\,809 presents a clear rotation pattern, with the kinematic major axis well-aligned with the photometric position angle (Table \ref{tabgalaxies}). The peak-to-peak amplitude of the velocity field is $\sim$$267$ km s$^{-1}$ and the kinematic centre is in agreement with the optical nucleus position (coordinate origin in panels of Fig. \ref{integrated_spectra_MUSE}). The global kinematic position angle (PA$_{k}$) is -7$\pm$3 degrees. The stellar velocity dispersion peaks ($\sim$$160$ km s$^{-1}$) at the optical nucleus decreasing outwards.  

 \begin{figure}
 \centering
\includegraphics[trim={0cm 0 0 0},clip,width=6.5cm, angle=270]{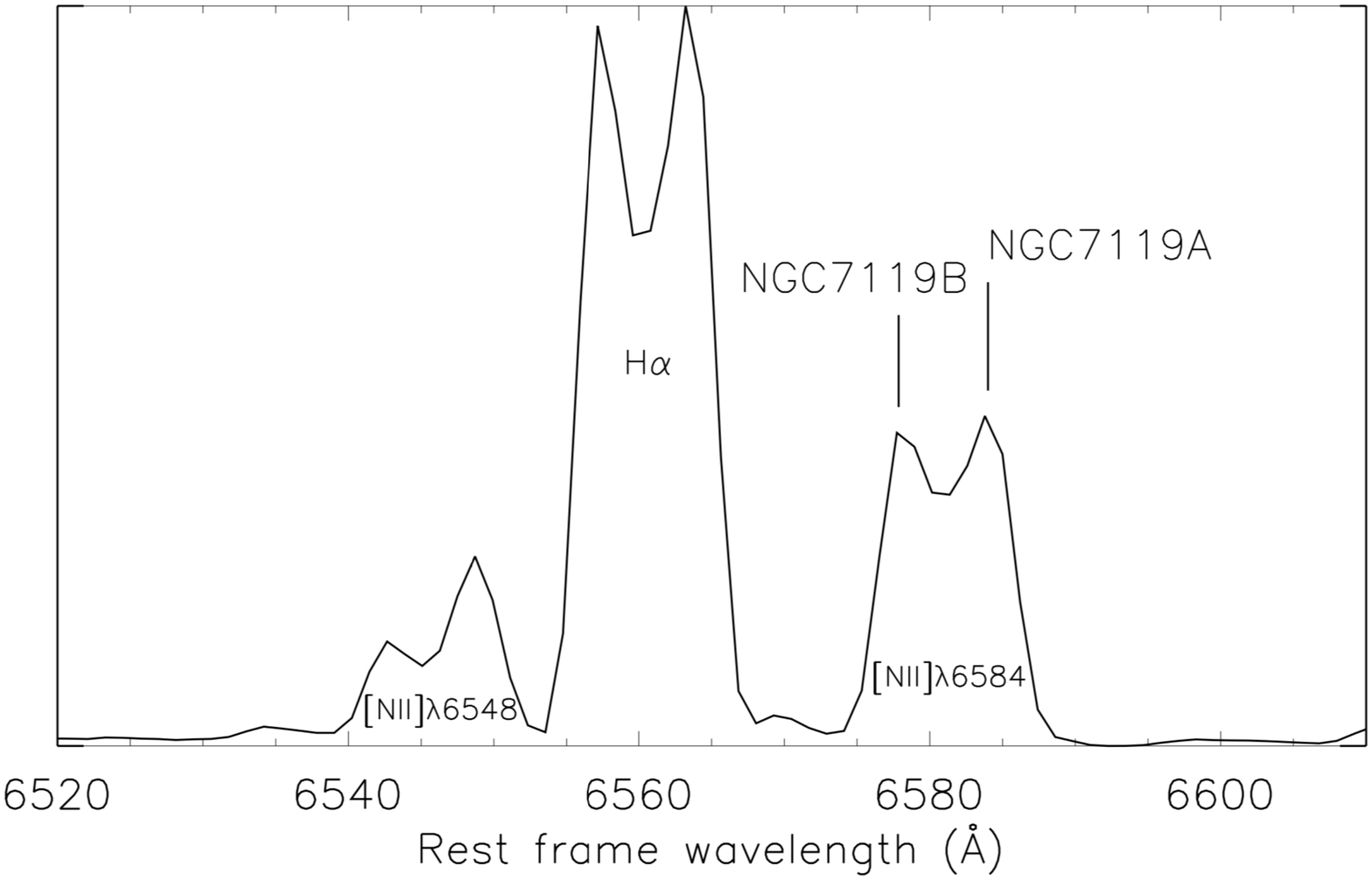}
  \caption{H$\alpha$+[NII]$\lambda\lambda6548,6584$ emission line profiles observed in the integrated spectrum of NGC\,7119A. Double-peak profiles are the result of the  contribution from its close companion, NGC7119B. }
    \label{zoom_NGC7119N_spectrum}
\end{figure}

\subsection{PGC\,055442}

PGC\,055442 is an early-type barred spiral galaxy at redshift $\sim$0.02358. It is morphologically mis-classified as an elliptical galaxy by HyperLeda$^{\ref{ledacat}}$, which is evident from the spiral structure in the H$\alpha$-filter images recovered from MUSE data (see Fig. D1 in \citealt{laura2018}).
The spectrum of PGC\,055442 (see Fig. \ref{integrated_spectra_MUSE}) is dominated by the stellar continuum, although [NII]$\lambda6548$, H$\alpha,$ and [NII]$\lambda6584$ emission lines are clearly identified. The H$\beta$ absorption feature is not evident because of an emission peak at its position. A weak [\ion{O}{iii}]$\lambda$5007 emission line also seems to be present. 
The filter-band image in Fig. \ref{stellar_PGC055442} shows the existence of a stellar bar along a position angle of about 45 degrees, offset by $\sim$$30$ degrees from the large-scale photometric axis (Table \ref{tabgalaxies}).  The stellar velocity field is consistent with a regular rotational pattern, with the major kinematic angle (PA$_{k}$=12$\pm3$ degrees) aligned with the photometric axis (Table \ref{tabgalaxies}) and a projected velocity amplitude (peak-to-peak) of $\sim$$360$ km s$^{-1}$. The stellar velocity dispersion shows relatively high values at the central region, with the peak ($\sim$$148$ km s$^{-1}$) at the optical nucleus and dropping off to an almost constant velocity dispersion of about 55 km s$^{-1}$ from 7 arcsec.

\subsection{NGC\,7119A}

NGC\,7119A is a barred spiral galaxy in an initial stage of interaction with its close companion NGC\,7119B. The systemic velocity of the system (NGC\,7119A and NGC7119B) is 9666 km s$^{-1}$ (from HyperLeda$^{\ref{ledacat}}$). The NGC\,7119B centre is located about 12 arcsec to the southeast of the NGC\,7119A nucleus, with a systemic velocity difference of $\Delta$V$\sim$-283 km s$^{-1}$ and a difference of $\sim$0.31 magnitudes in total B magnitude (from HyperLeda$^{\ref{ledacat}}$). 
The integrated spectrum of NGC\,7119A resembles those typical of spiral galaxies, displaying pronounced absorption features, bright [NII]$\lambda\lambda6548,6584$ and Balmer emission lines and faint [\ion{O}{iii}]$\lambda\lambda4959,5007$ lines. However, it shows the presence of double peaks in the emission line profiles as result of two kinematically distinct components. The reddest component is associated with the ionised gas in NGC\,7119A. The second component, at about 280 km s$^{-1}$ blueshifted, is placed at the corresponding wavelength position for the ionised gas in its close companion, NGC7119B. Figure \ref{zoom_NGC7119N_spectrum} zooms in on the wavelength range between 6520 and 6610 \AA , including only the [NII]$\lambda6548$, H$\alpha$ and [NII]$\lambda6584$ emission lines to clearly show these double-peaked profiles. The blending with NGC7119B light is unclear from the stellar absorption lines in the NGC\,7119A spectrum. The filter-band image (see Fig. \ref{stellar_NGC7119N}-left) clearly shows the spiral morphology of NGC\,7119A. Although its companion NGC\,7119B is in front of it, the continuum emission seems to be dominated by NGC\,7119A in the light-of-sight overlap region, where the prominent southeastern spiral arm of NGC\,7119A is located. The stellar velocity field of NGC\,7119A resembles a typical rotation supported spiral galaxy, with the photometric (Table \ref{tabgalaxies}) and major kinematic axes (PA$_{k}$=-35$\pm3$) aligned. However, it is distorted in the southeast due to the overlap with the NGC\,7119B stellar kinematics. The peak-to-peak velocity amplitude along the major kinematic axis is $\sim$386 km s$^{-1}$. The stellar velocity dispersion map reaches values over 400 km s$^{-1}$ in the overlapping region of NGC\,7119 system. The stellar velocity dispersion at the optical nucleus of NGC\,7119A is $\sim$116 km s$^{-1}$. There is a region with average velocity dispersion of $\sim$330 km s$^{-1}$ at about four arcsec southeast of the NGC\,7119A nucleus extending to the east and west by about 1.5 arcsec.
These high values of the velocity dispersion are most probably associated with a superposed contribution of the stellar components from NGC7119B along the lines of sight of NGC\,7119A, which is not evident in the stellar features of the integrated spectrum, but is evident in the emission lines (see Fig. \ref{zoom_NGC7119N_spectrum}).

\end{appendix}
 
\begin{appendix}
\setcounter{section}{1}
\section{Fits}
\label{appen_images}

In this appendix, we include the intensity maps, the fits to the integrated spectra within the effective radius, and the resolved stellar kinematics for the spiral galaxy PGC\,055442 and the interacting galaxy NGC\,7119A simulating the host galaxies at redshifts 0.76, 1.50, and 2.30.

\begin{figure*}
\centering
\includegraphics[trim={1cm 1.5cm 5cm 0cm},width=5.5cm]{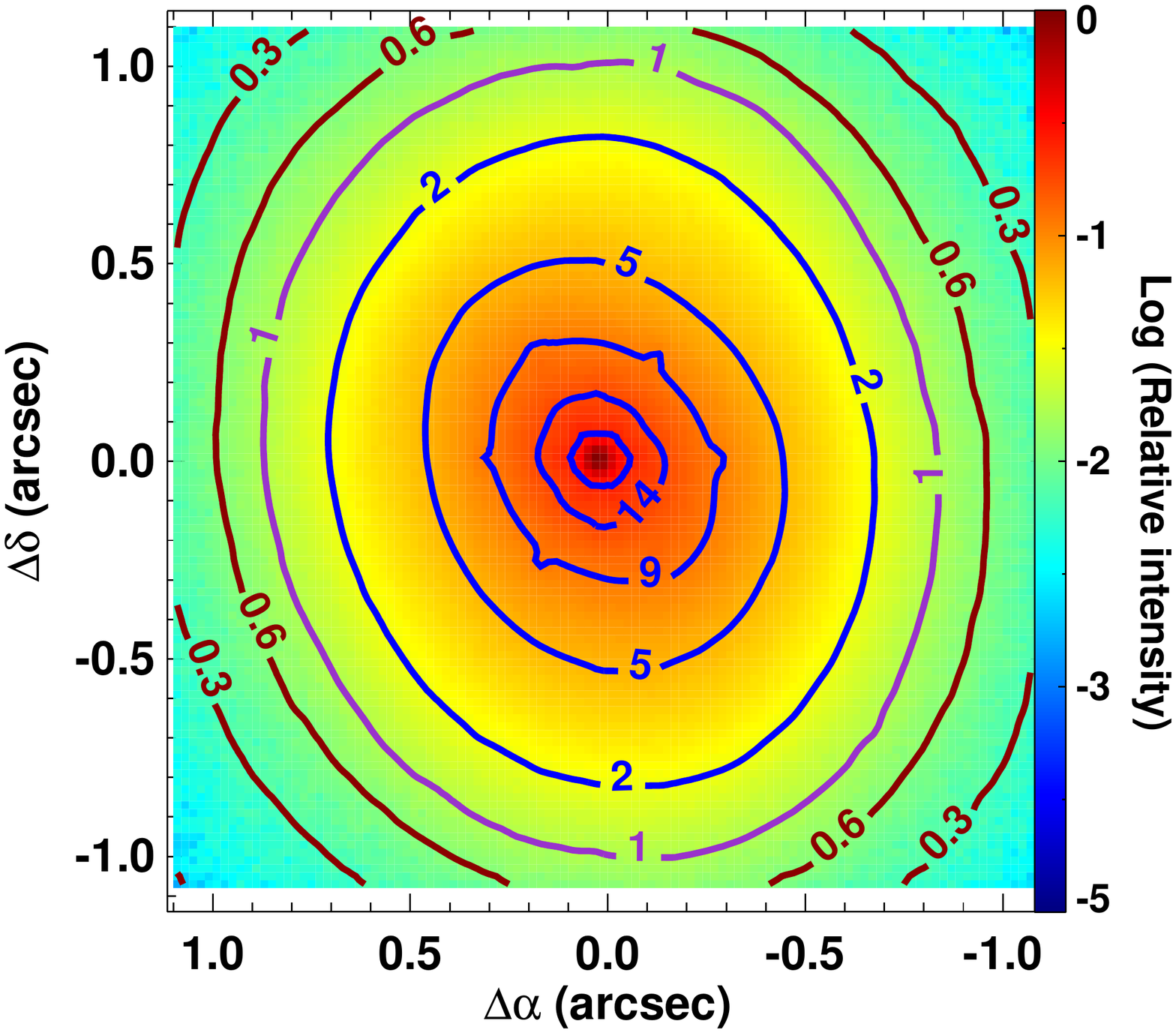}
\includegraphics[trim={1cm 1.5cm 5cm 0cm},width=5.5cm]{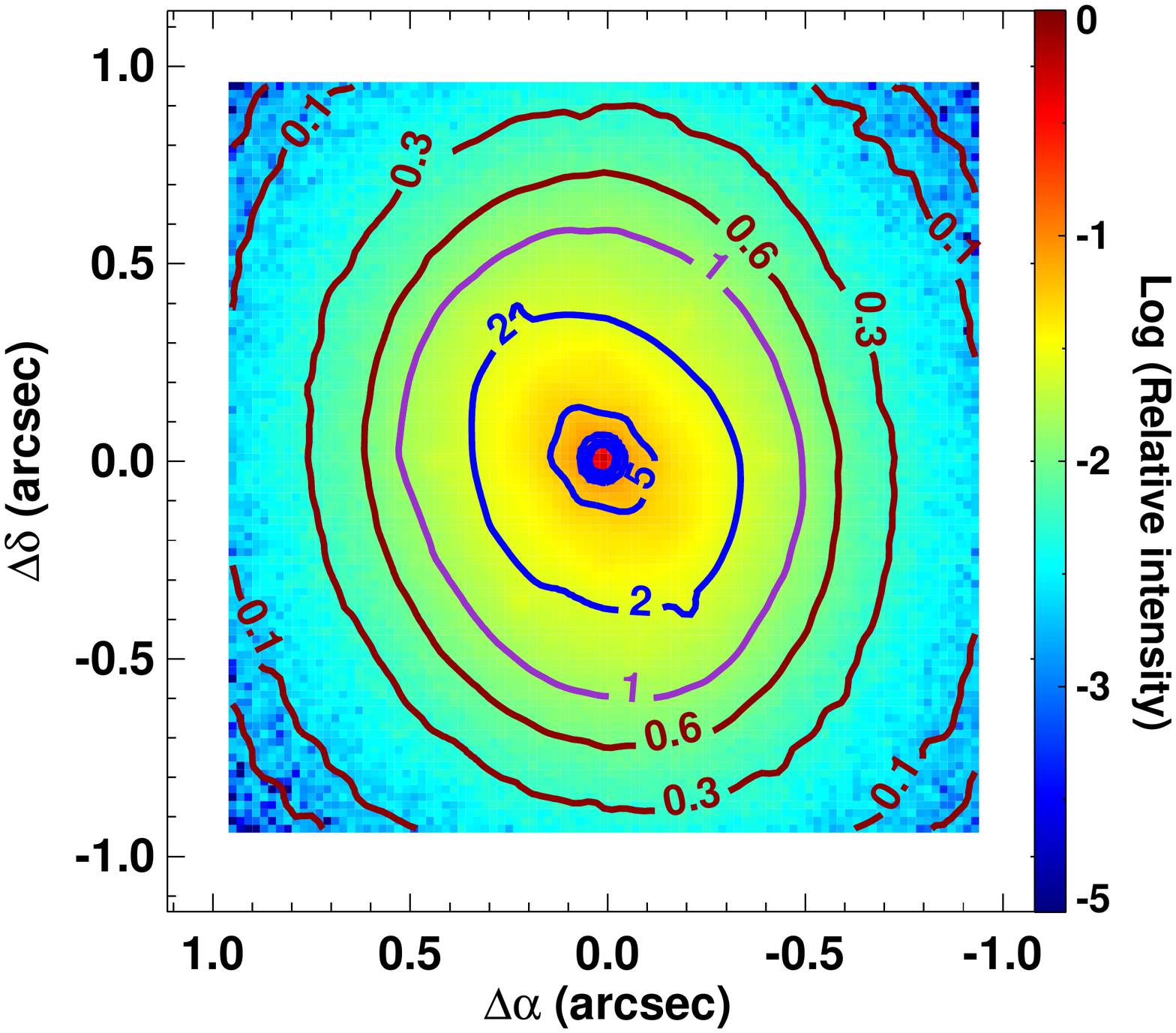}
\includegraphics[trim={1cm 1.5cm 5cm 0cm},width=5.5cm]{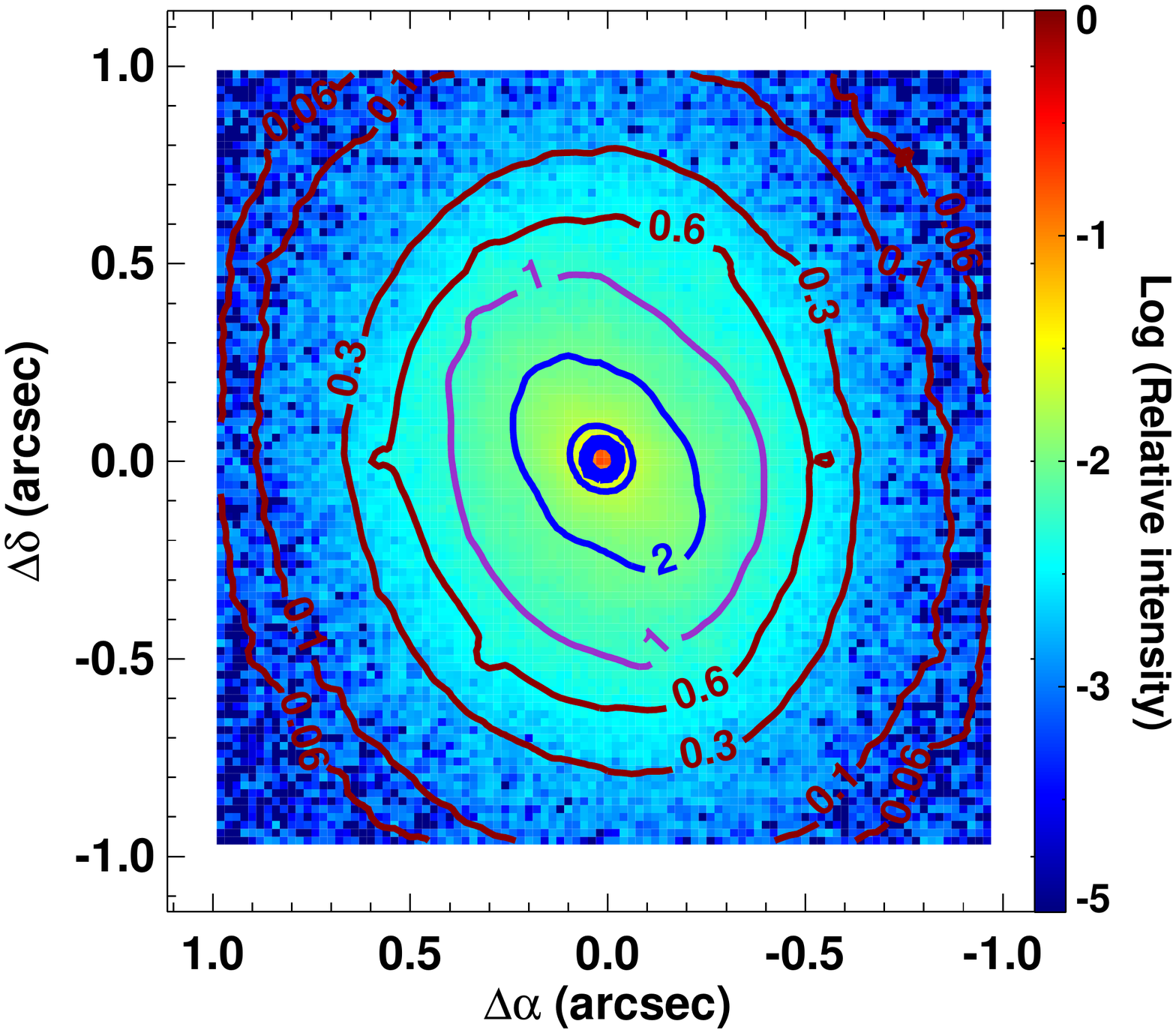}

 \includegraphics[trim={1cm 1.5cm 5cm 0cm},width=5.5cm]{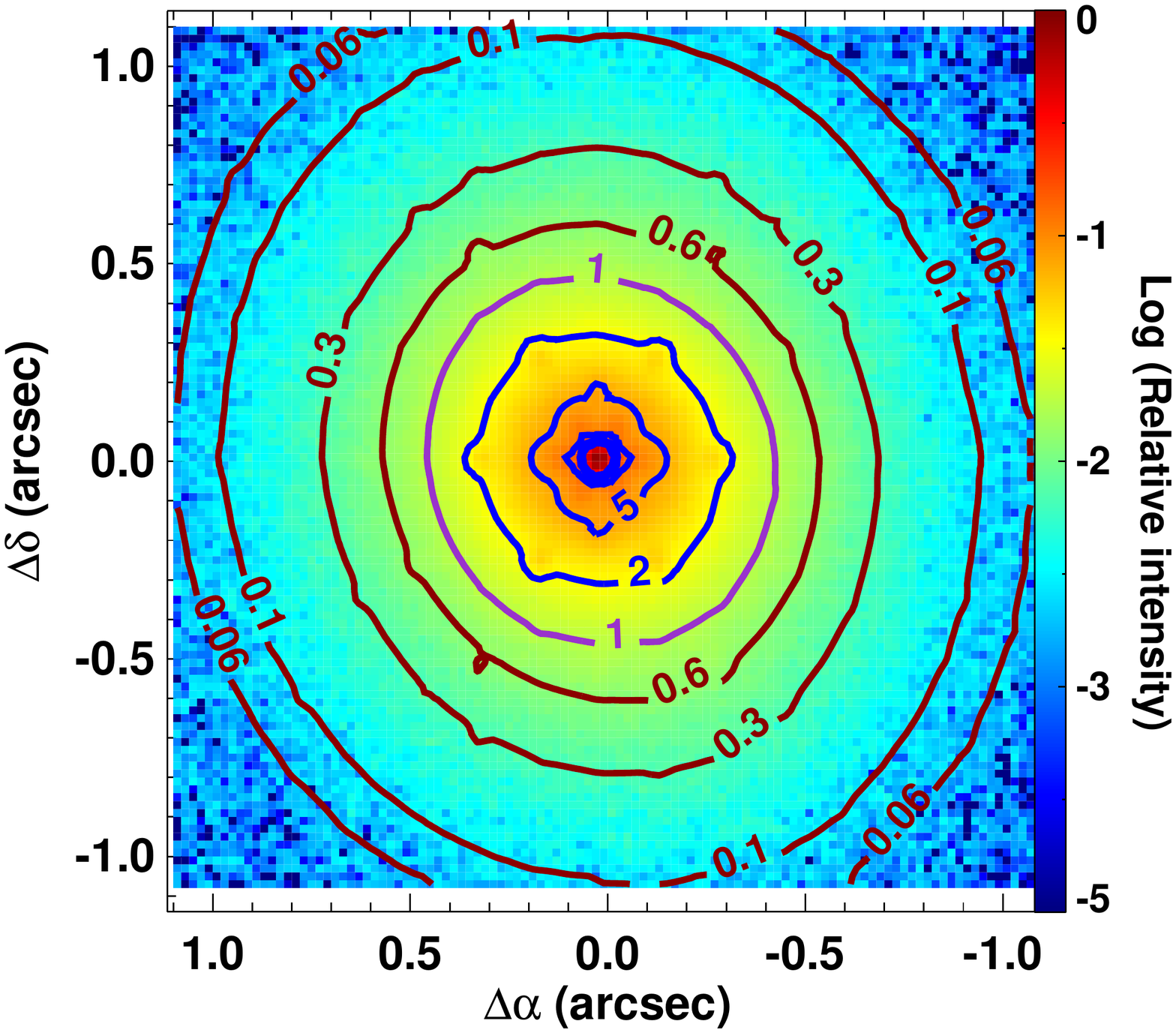}
 \includegraphics[trim={1cm 1.5cm 5cm 0cm},width=5.5cm]{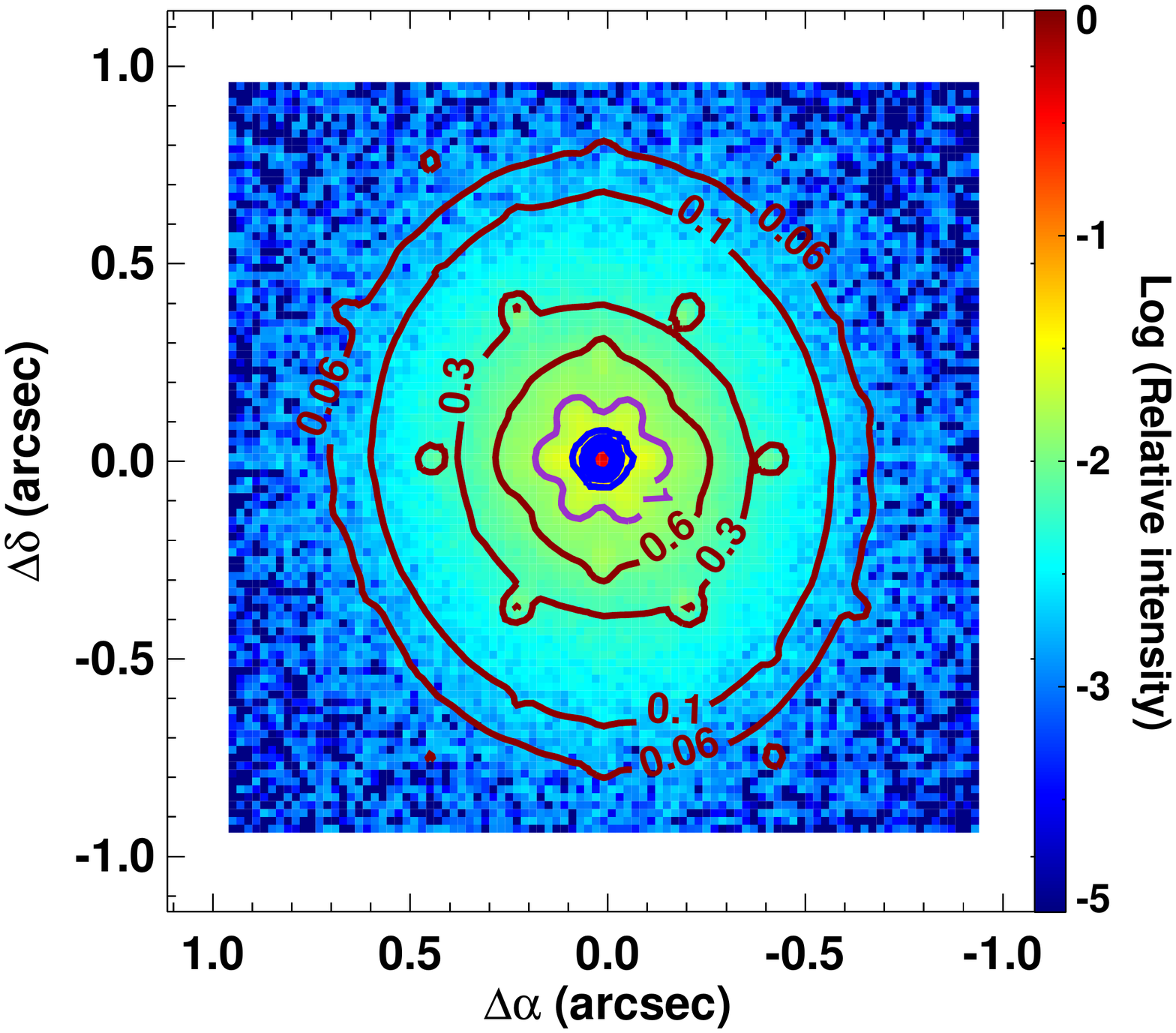}
 \includegraphics[trim={1cm 1.5cm 5cm 0cm},width=5.5cm]{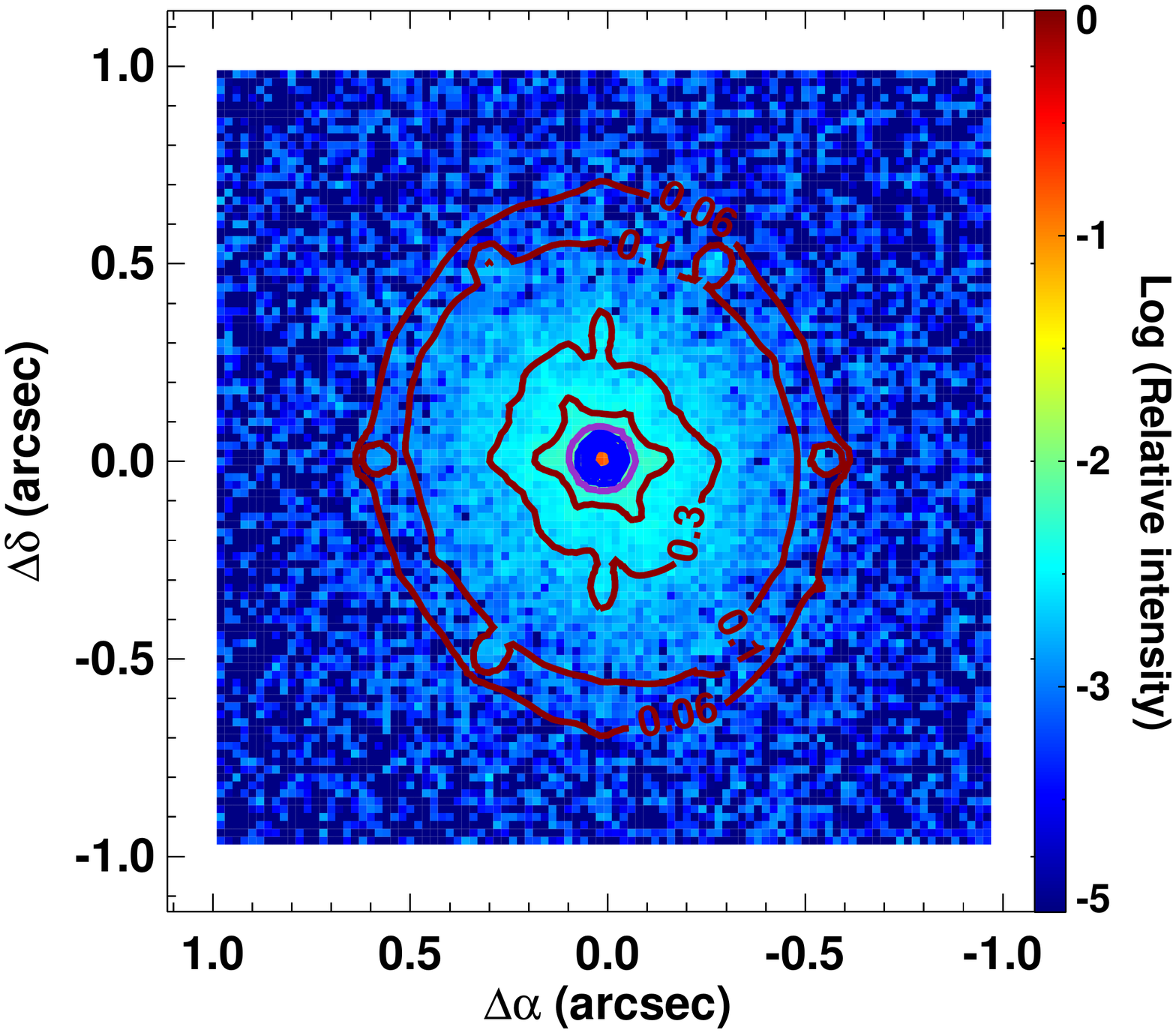}
 
  \includegraphics[trim={1cm 1.5cm 5cm 0cm},width=5.5cm]{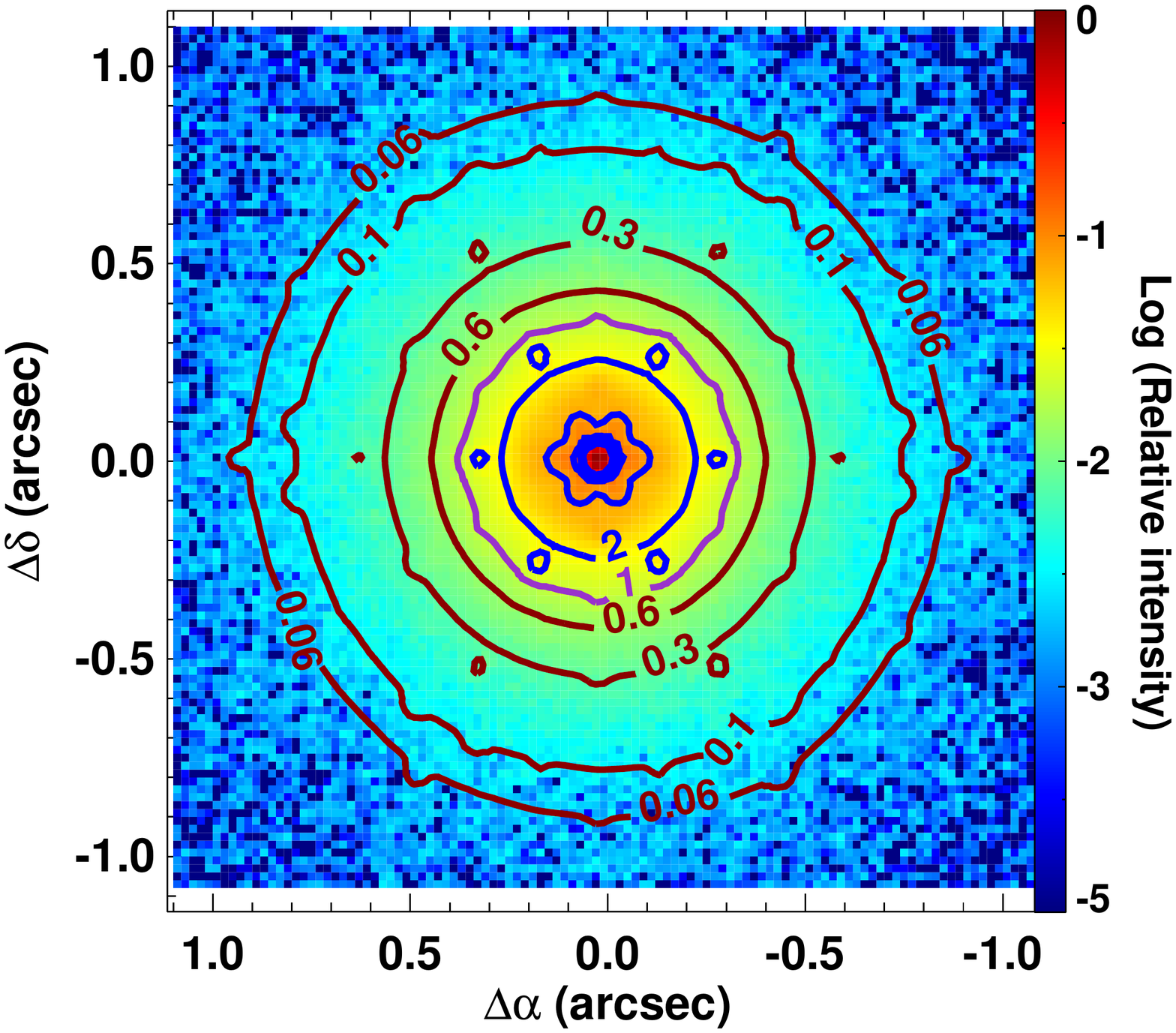}
 \includegraphics[trim={1cm 1.5cm 5cm 0cm},width=5.5cm]{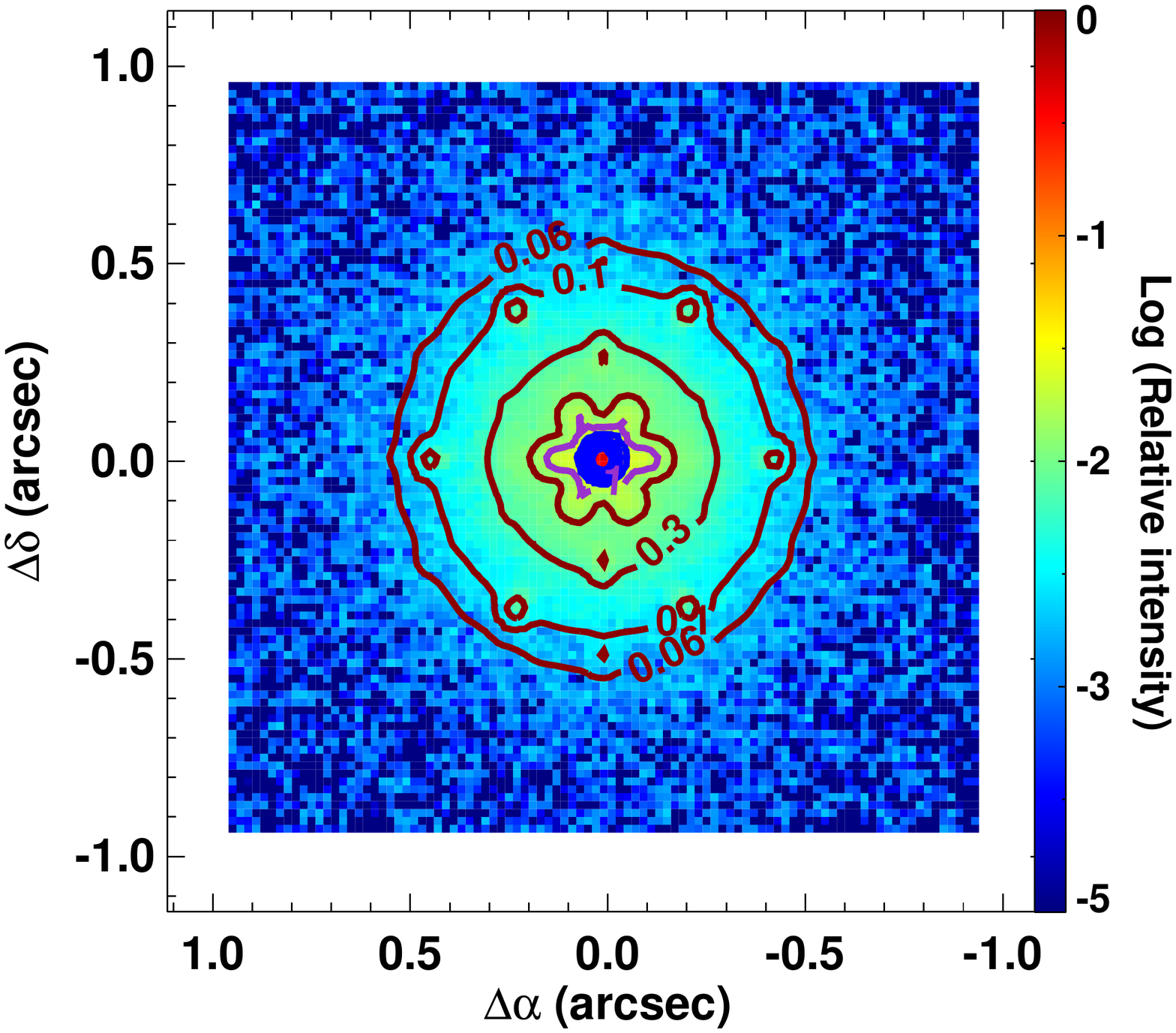}
 \includegraphics[trim={1cm 1.5cm 5cm 0cm},width=5.5cm]{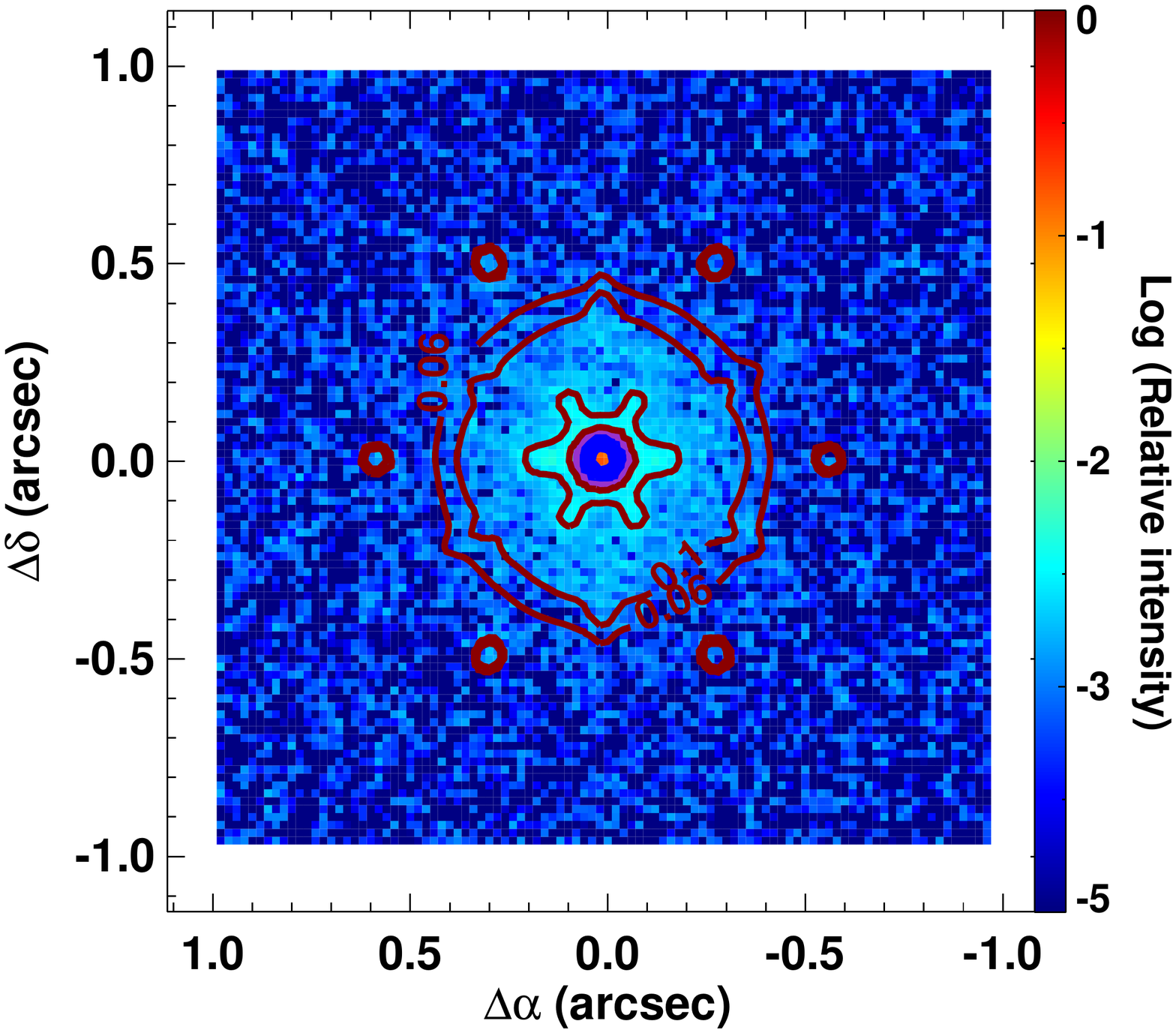}
  \caption{Same as Fig. \ref{NGC809_HbMgI_range_42}, but when considering spiral morphology (i.e. PGC\,055442) as a host.}
 \label{PGC055442_HbMgI_range_42}
\end{figure*}

\begin{figure*}
\centering
\includegraphics[trim={1cm 1.5cm 5cm 0cm},width=5.5cm]{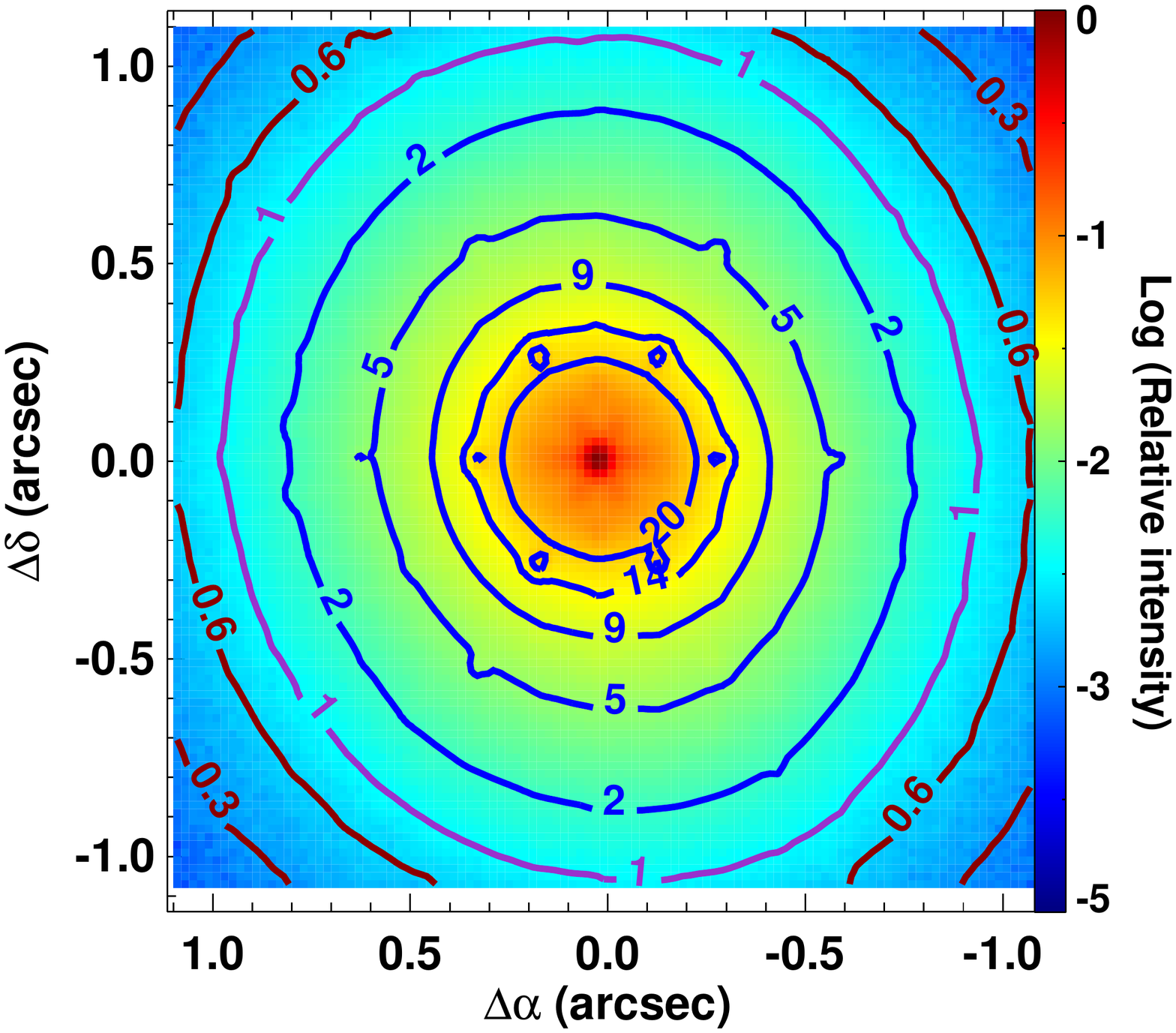}
\includegraphics[trim={1cm 1.5cm 5cm 0cm},width=5.5cm]{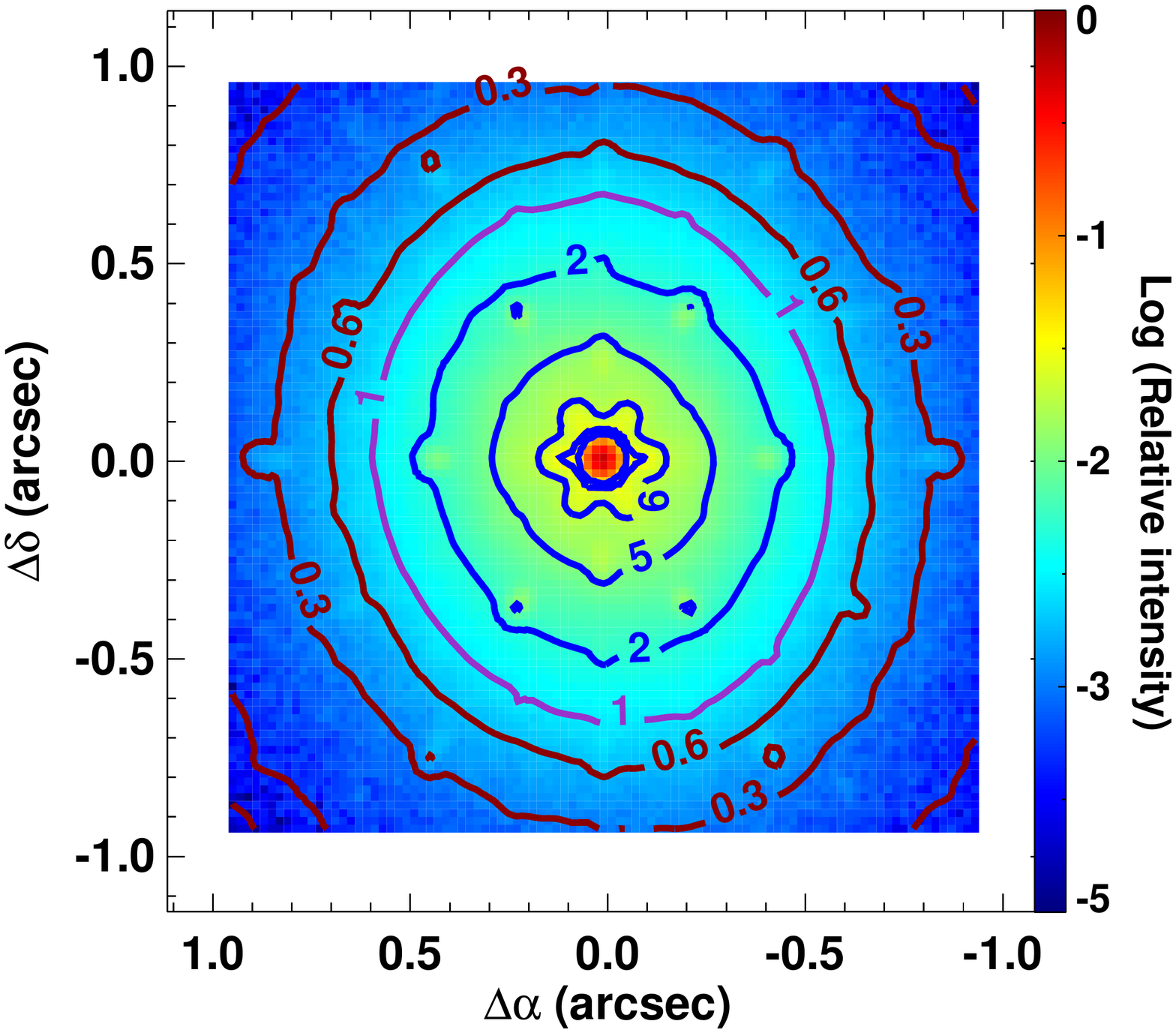}
\includegraphics[trim={1cm 1.5cm 5cm 0cm},width=5.5cm]{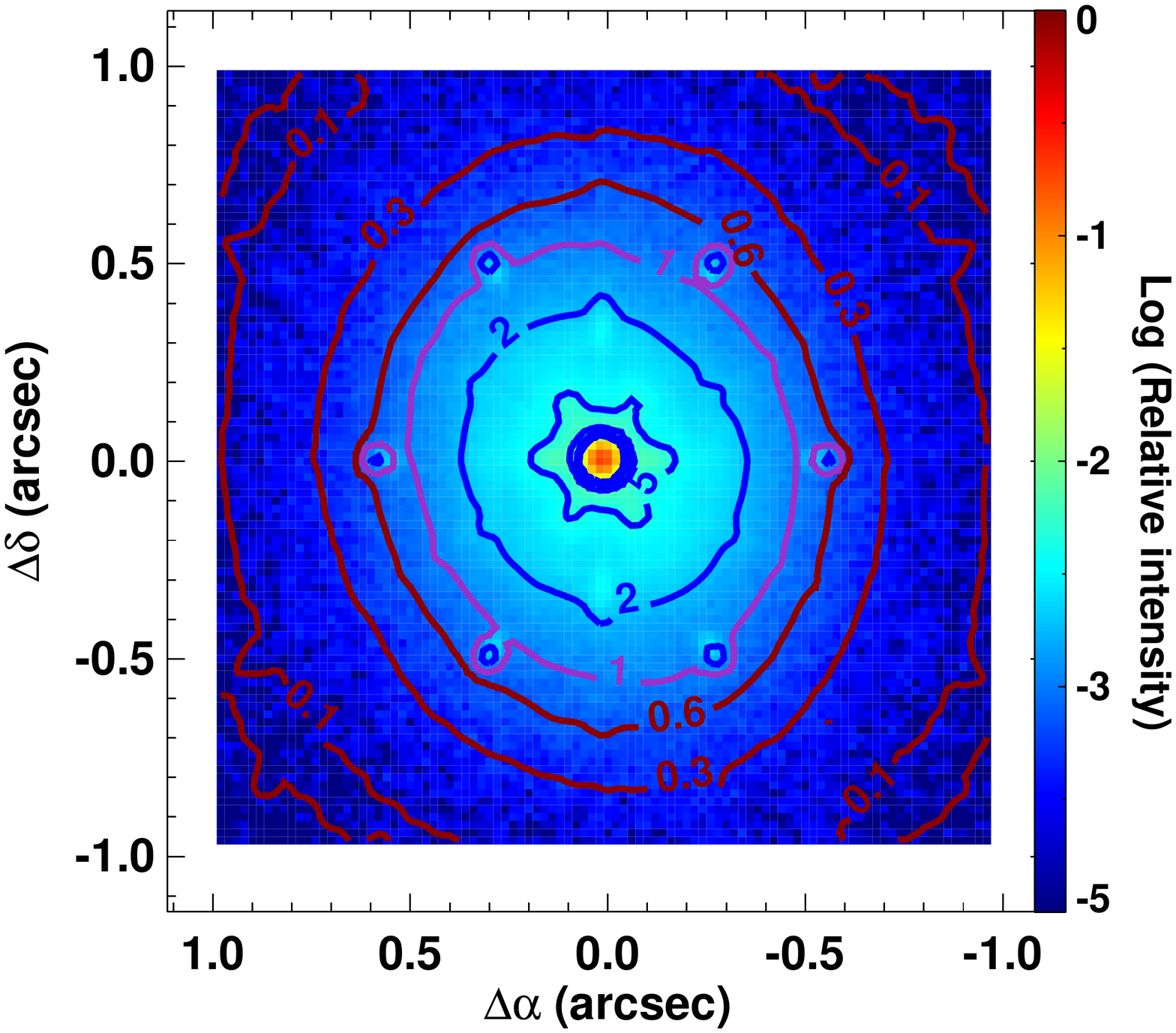}

 \includegraphics[trim={1cm 1.5cm 5cm 0cm},width=5.5cm]{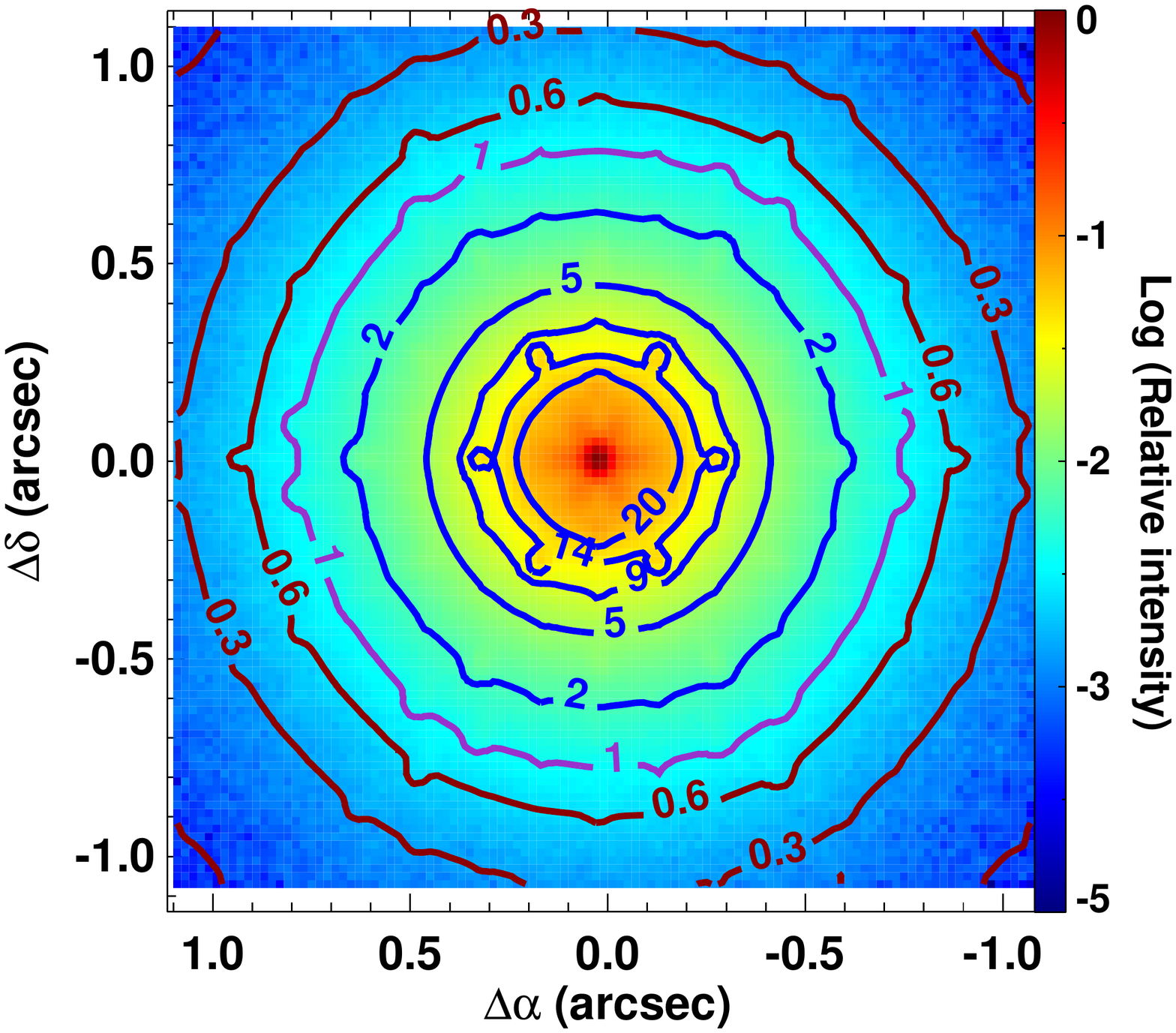}
 \includegraphics[trim={1cm 1.5cm 5cm 0cm},width=5.5cm]{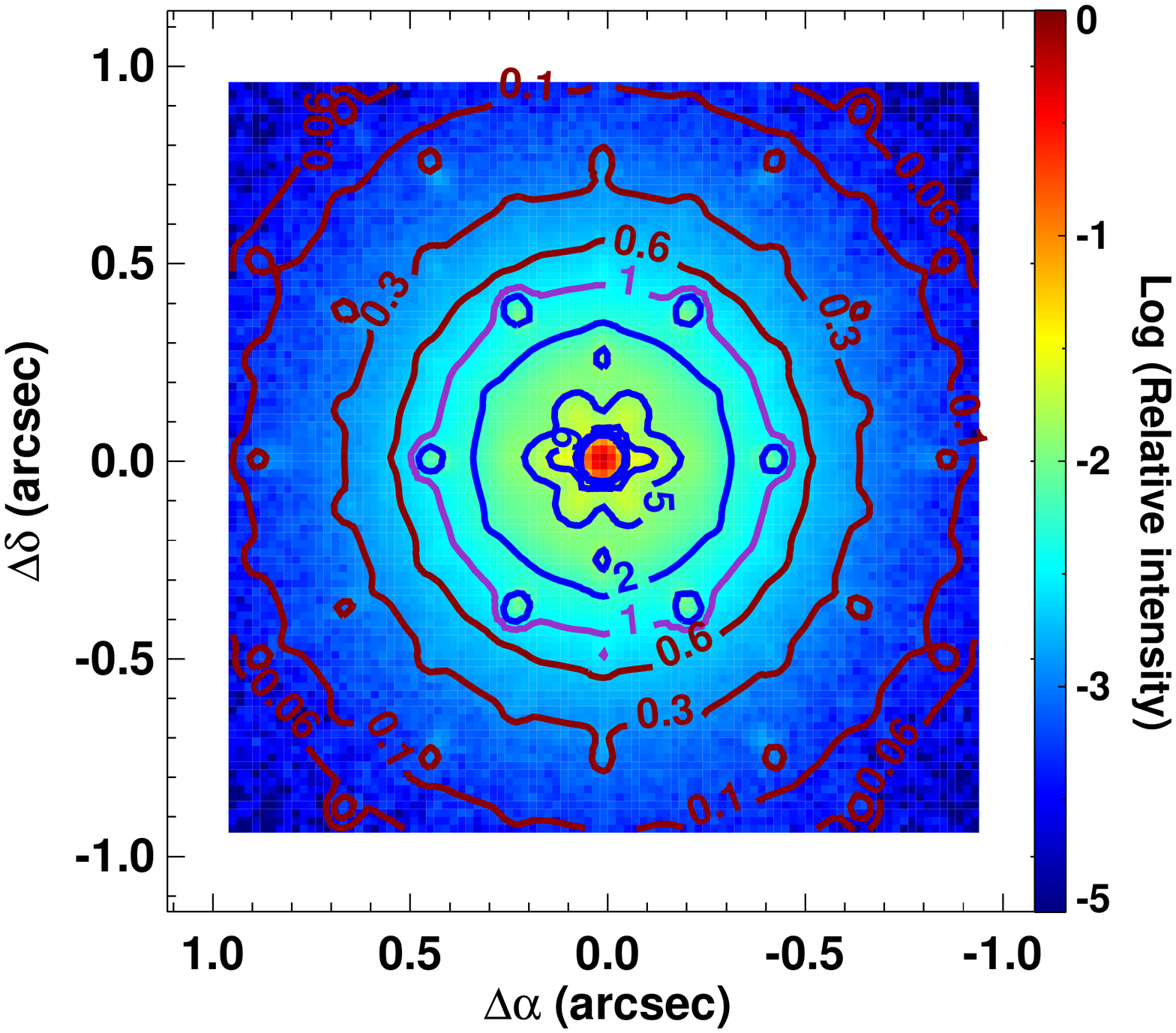}
 \includegraphics[trim={1cm 1.5cm 5cm 0cm},width=5.5cm]{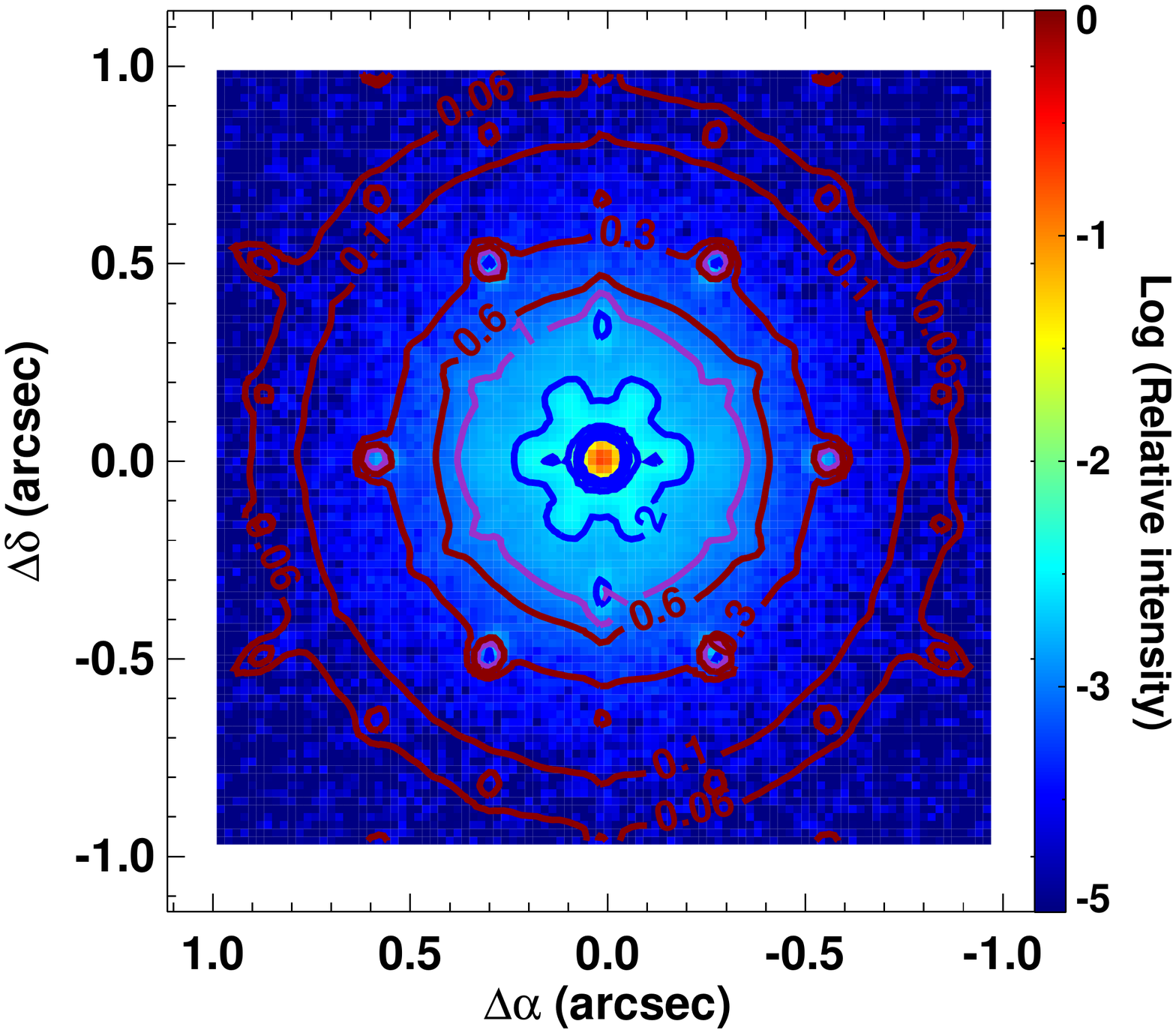}
 
  \includegraphics[trim={1cm 1.5cm 5cm 0cm},width=5.5cm]{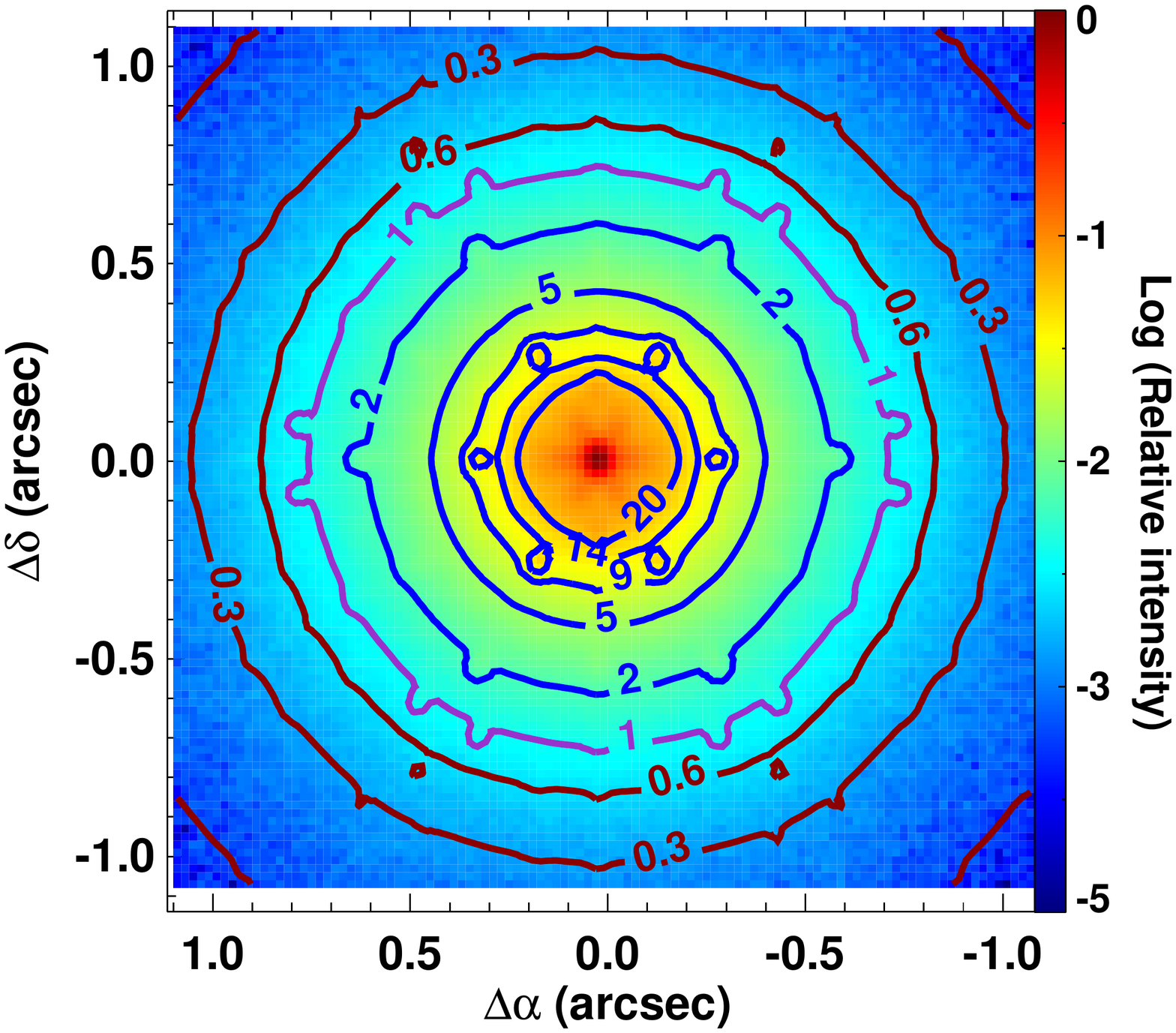}
 \includegraphics[trim={1cm 1.5cm 5cm 0cm},width=5.5cm]{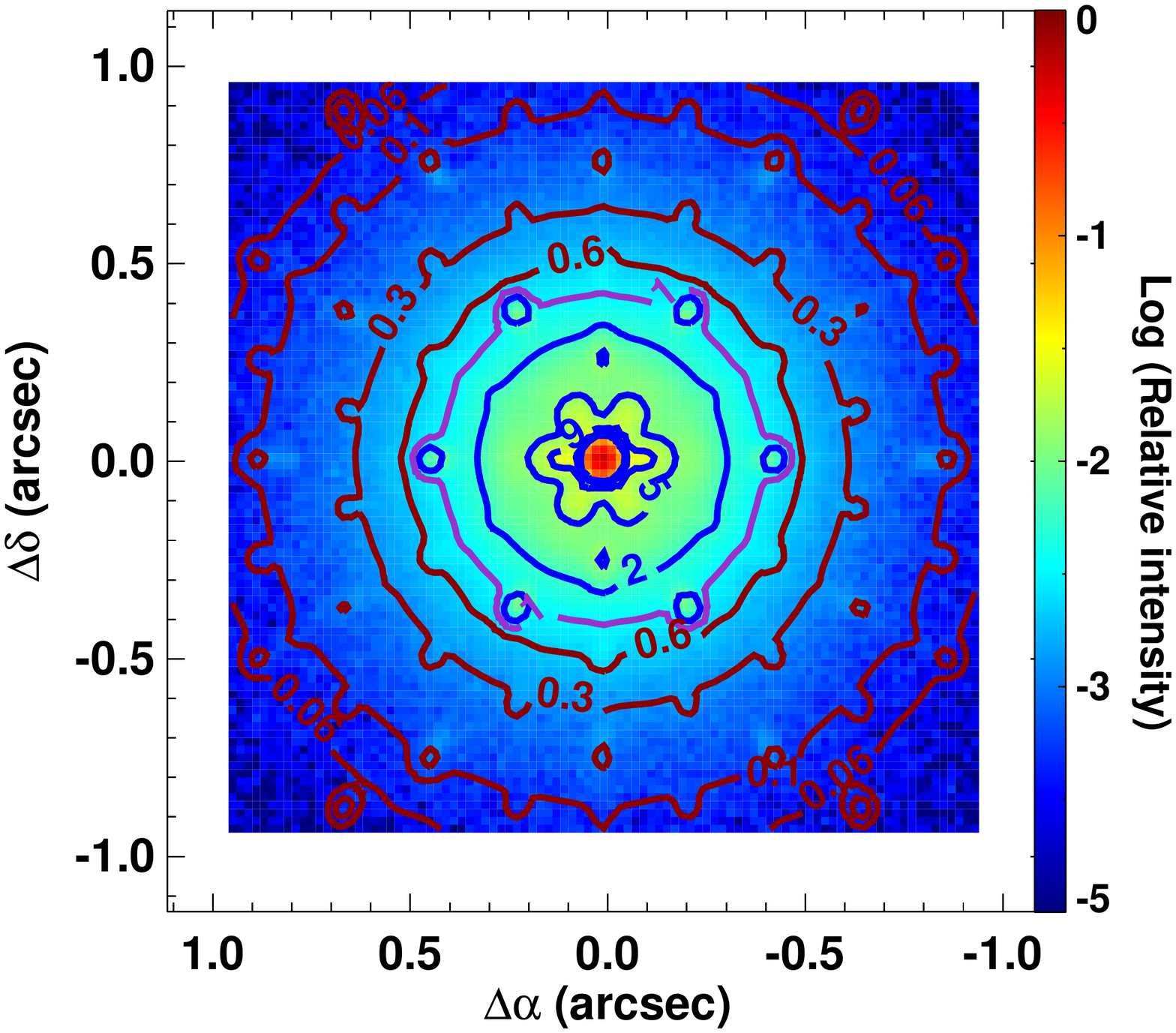}
 \includegraphics[trim={1cm 1.5cm 5cm 0cm},width=5.5cm]{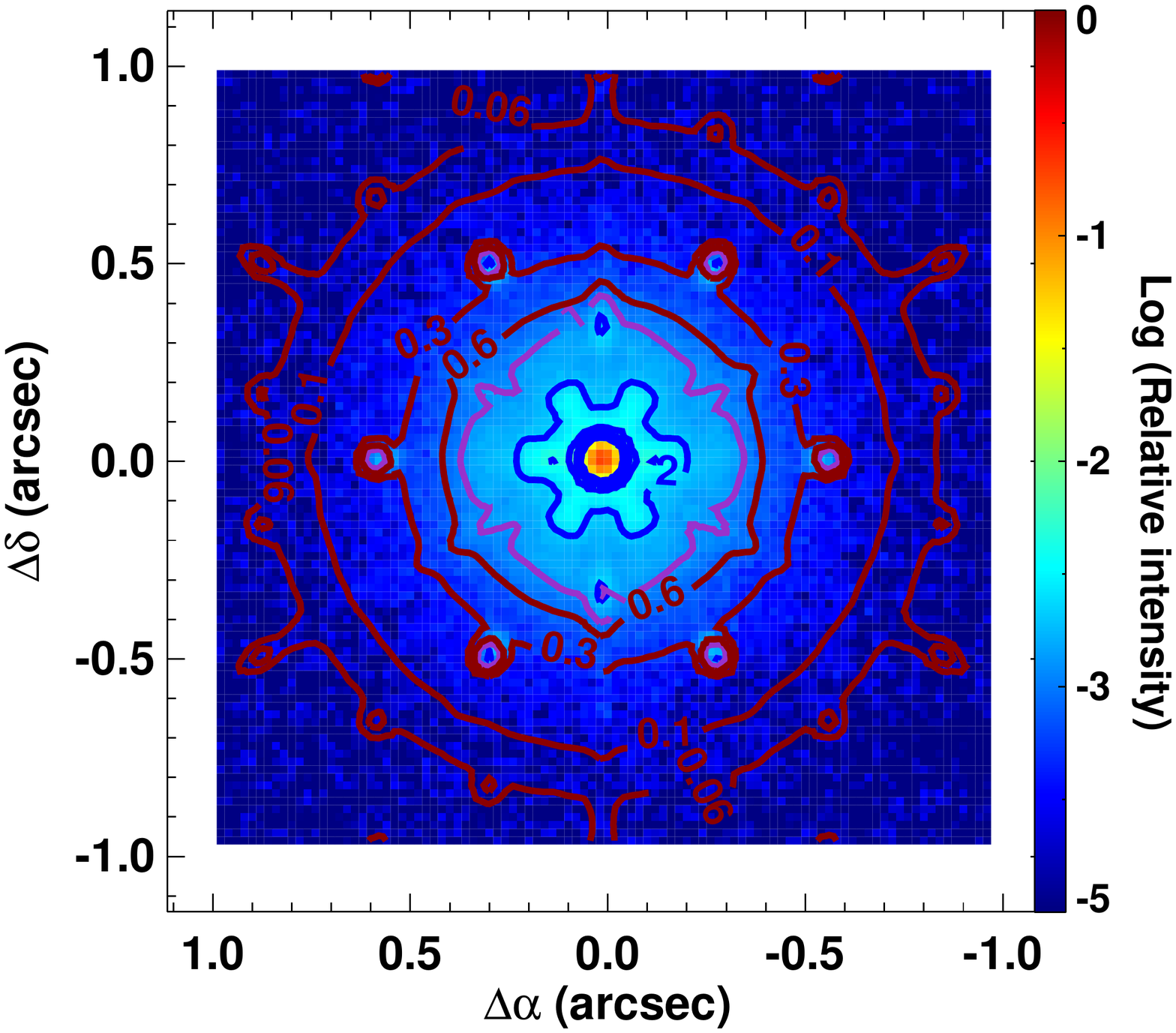}
  \caption{Same as Fig. \ref{NGC809_HbMgI_range_43}, but when considering spiral morphology (i.e. PGC\,055442) as a host.}
 \label{PGC055442_HbMgI_range_43}
\end{figure*}

\begin{figure*}
\centering
  \includegraphics[trim={1cm 1.5cm 5cm 0cm},width=5.5cm]{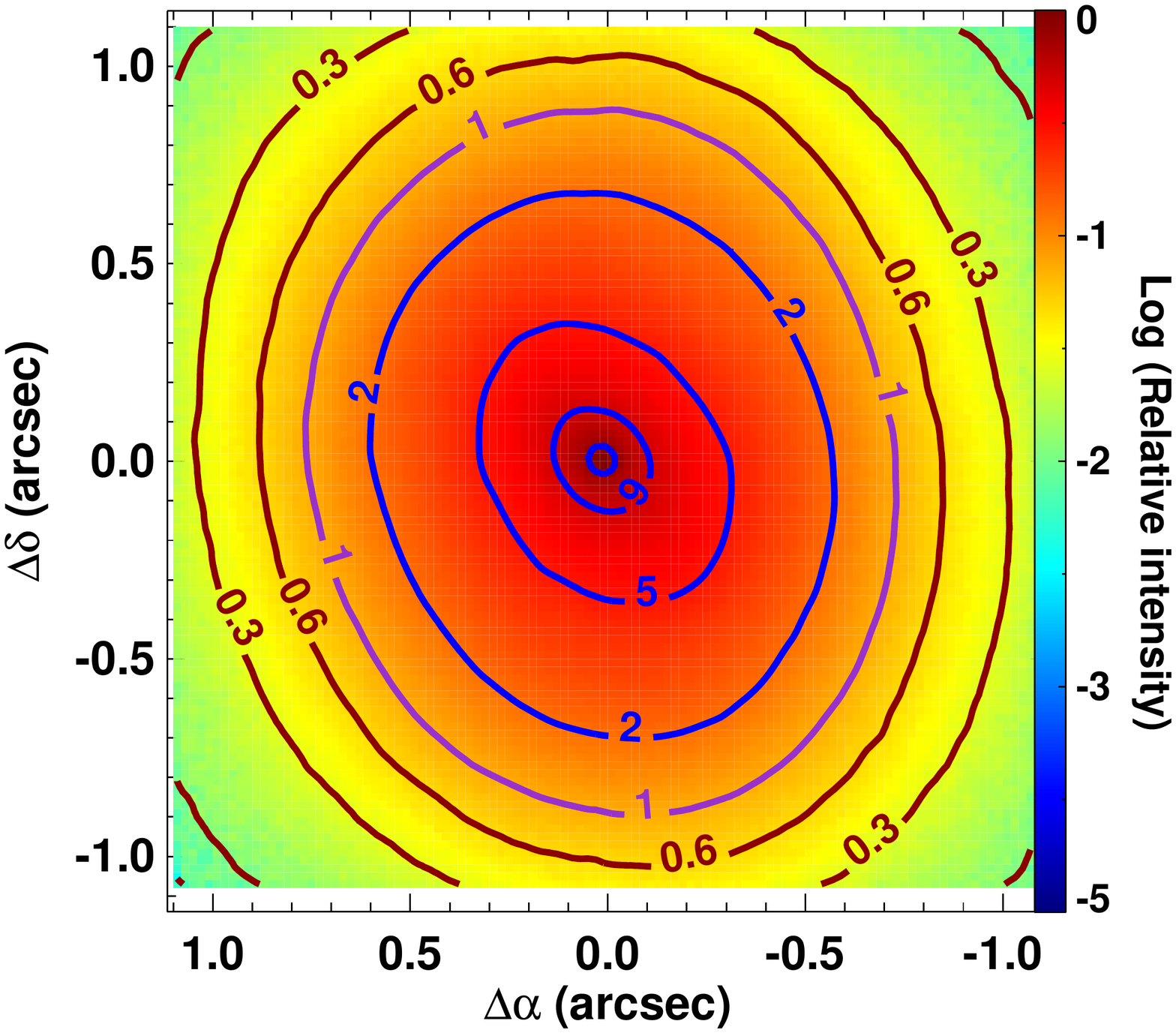}
 \includegraphics[trim={1cm 1.5cm 5cm 0cm},width=5.5cm]{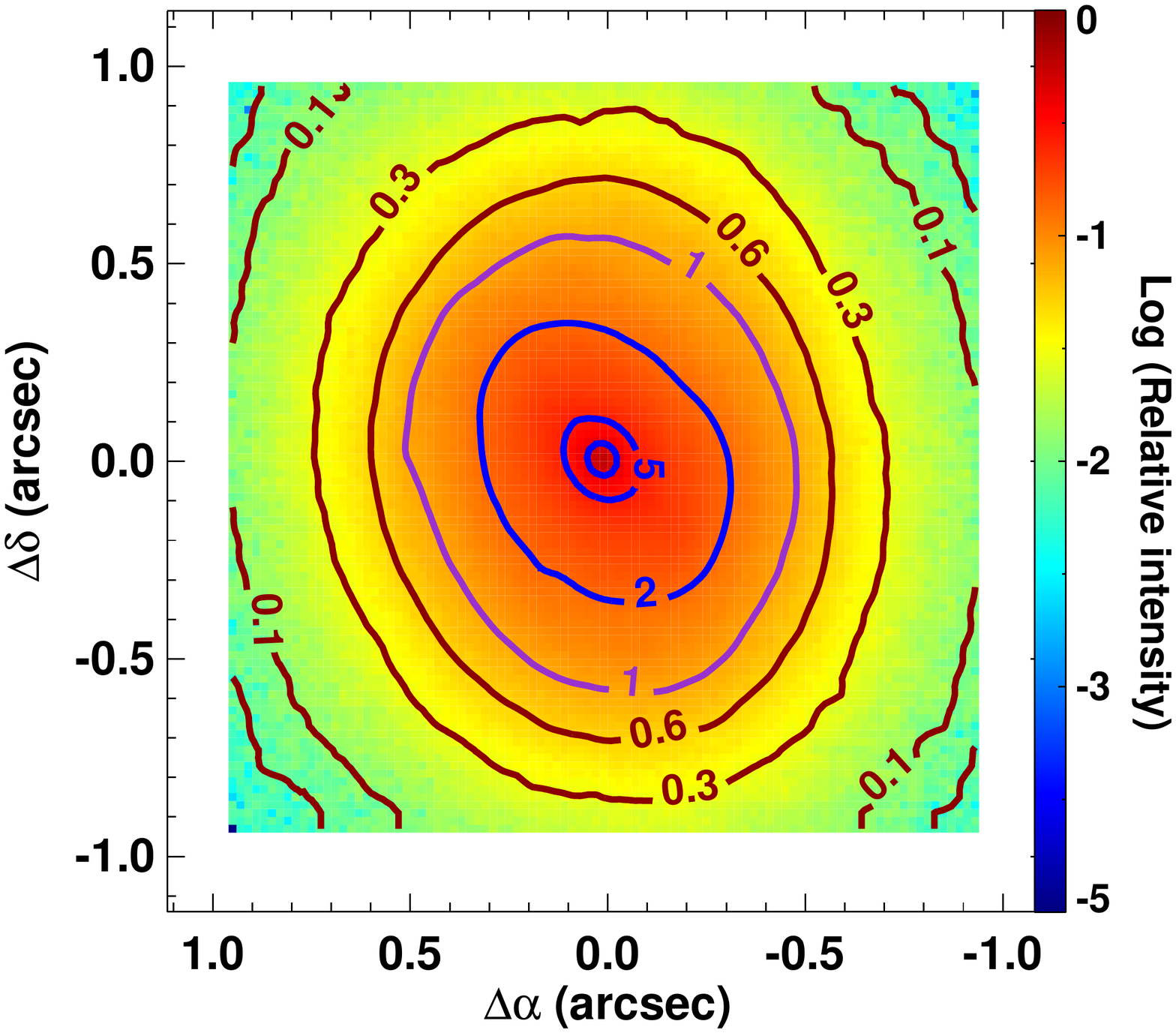}
 \includegraphics[trim={1cm 1.5cm 5cm 0cm},width=5.5cm]{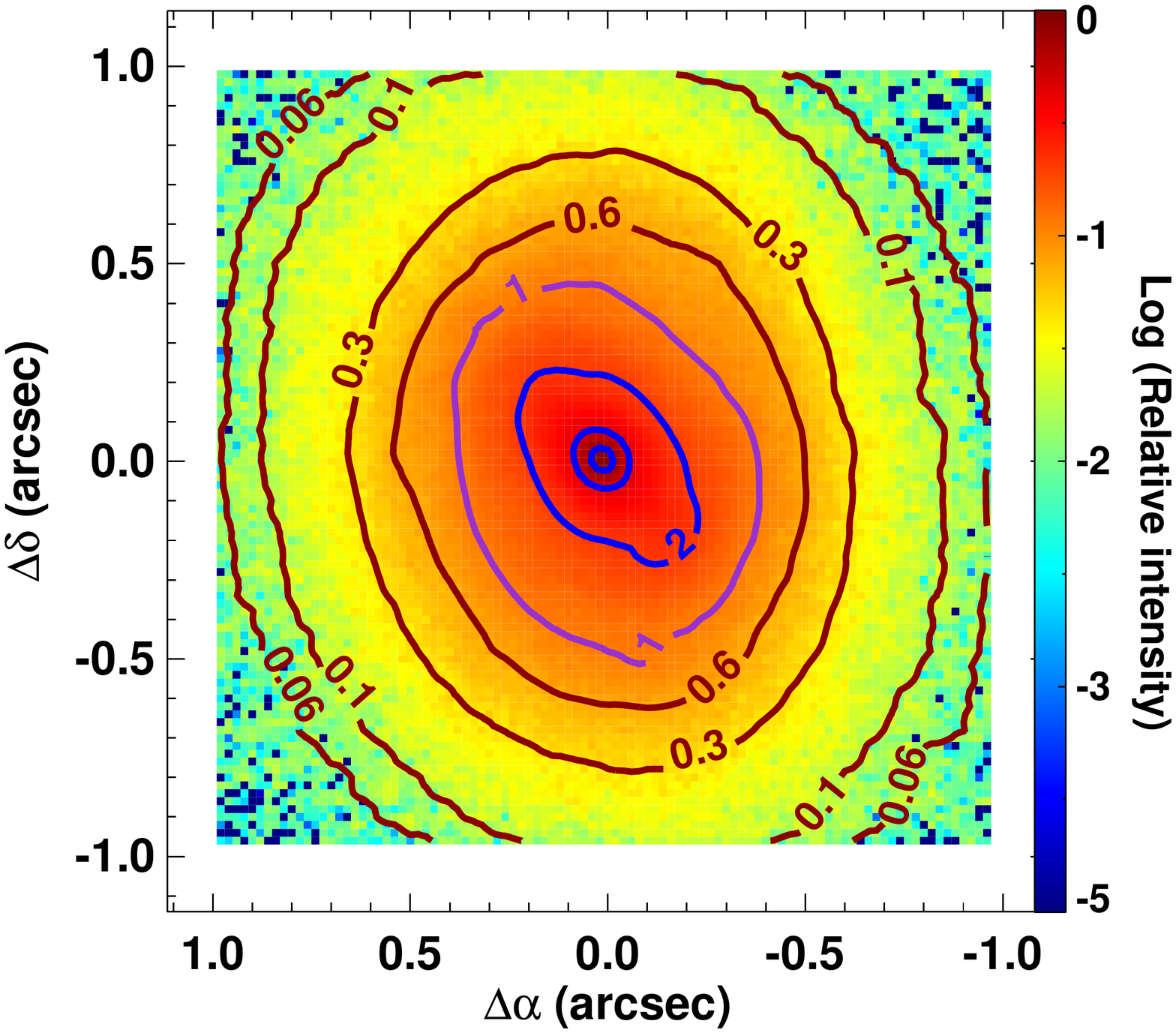}

  \includegraphics[trim={1cm 1.5cm 5cm 0cm},width=5.5cm]{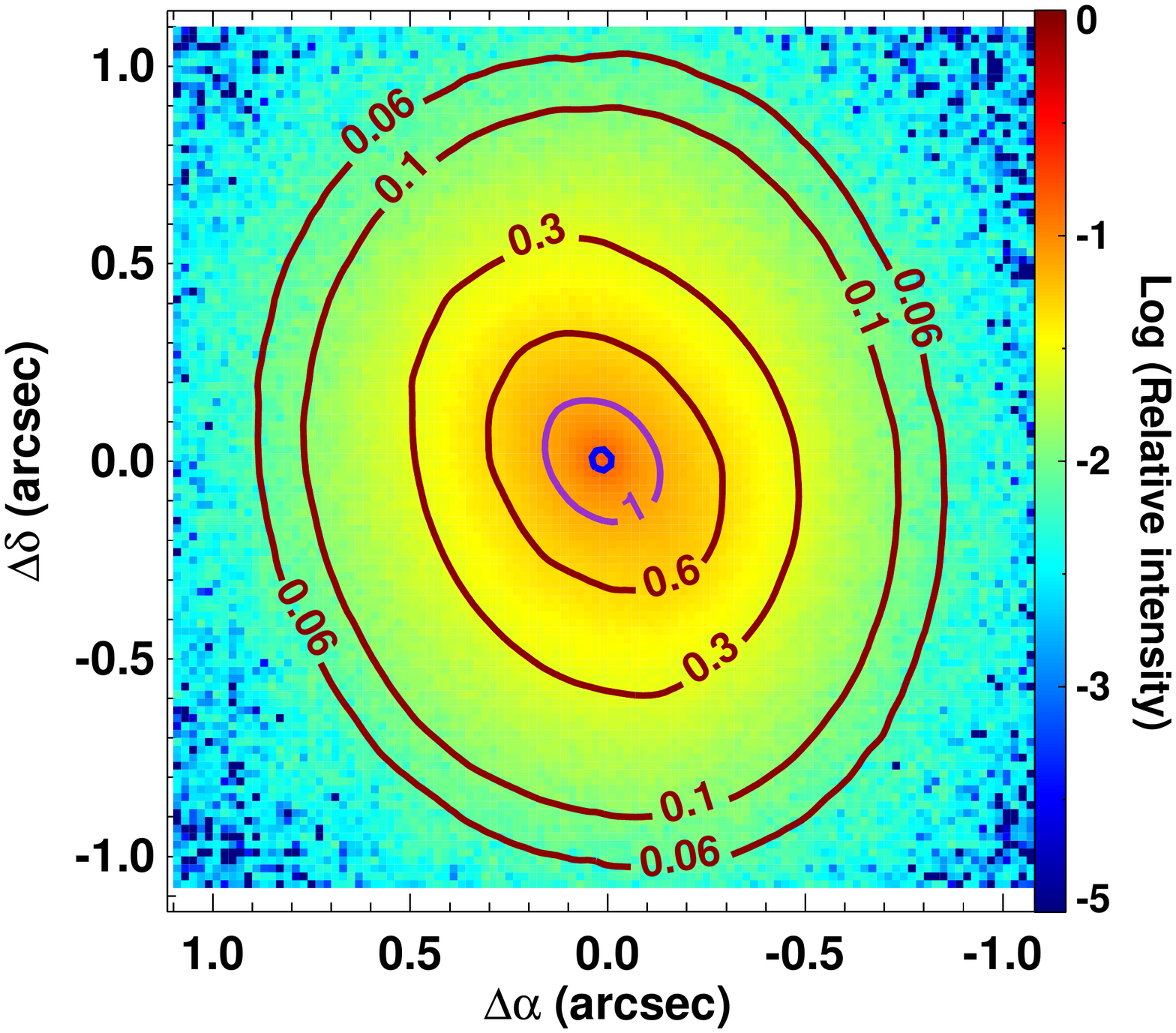}
 \includegraphics[trim={1cm 1.5cm 5cm 0cm},width=5.5cm]{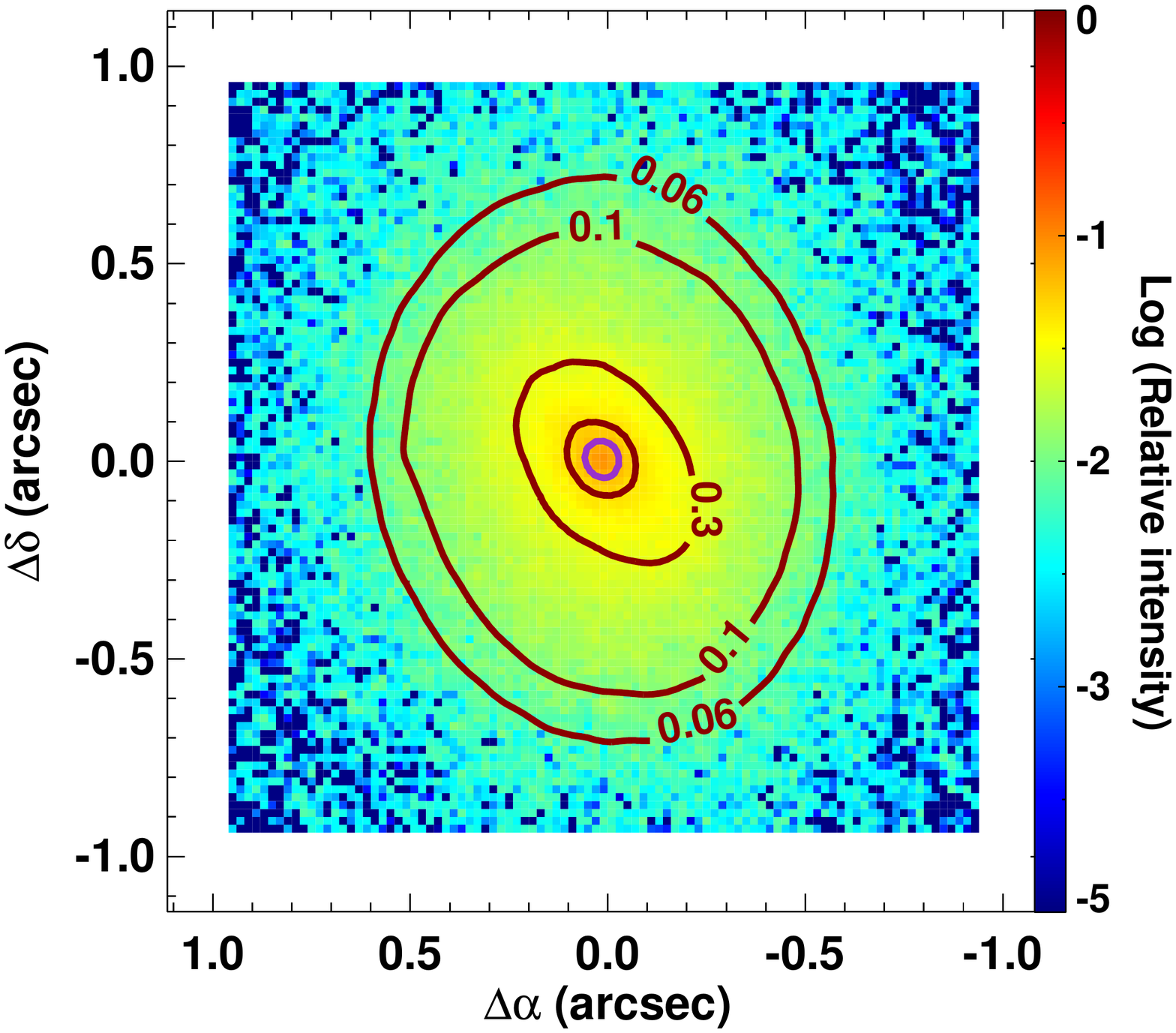}
 \includegraphics[trim={1cm 1.5cm 5cm 0cm},width=5.5cm]{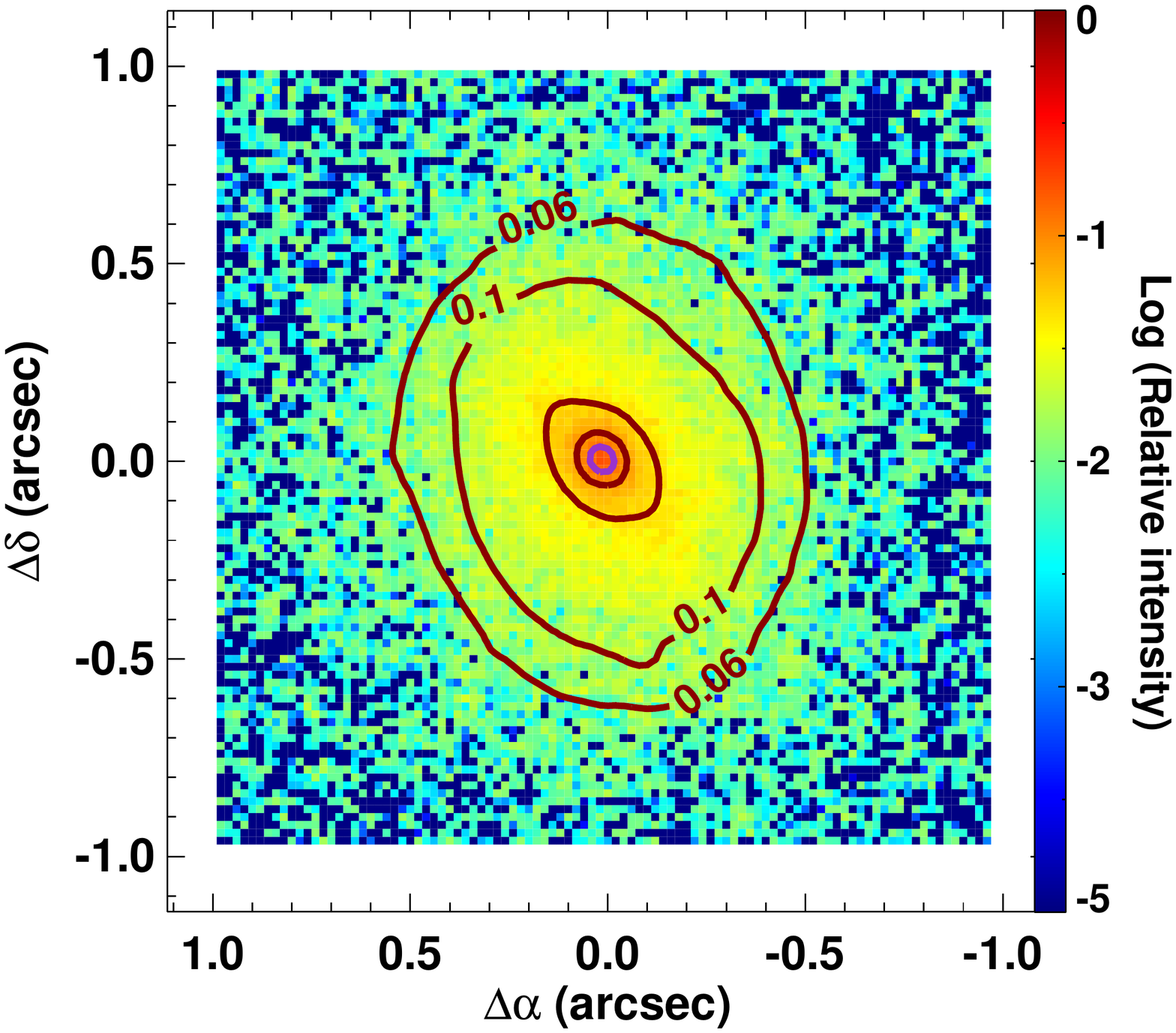}

  \includegraphics[trim={1cm 1.5cm 5cm 0cm},width=5.5cm]{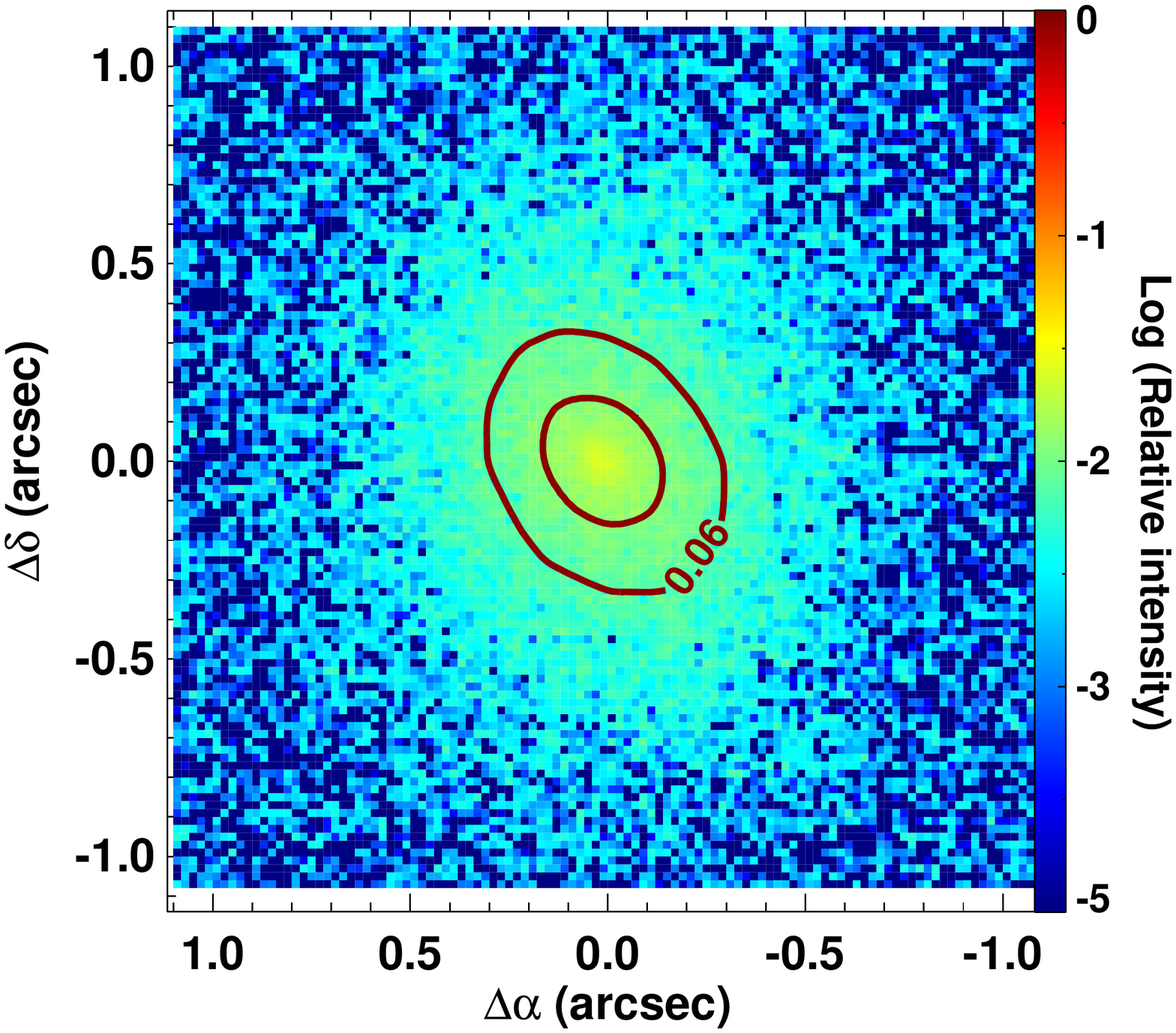}
 \includegraphics[trim={1cm 1.5cm 5cm 0cm},width=5.5cm]{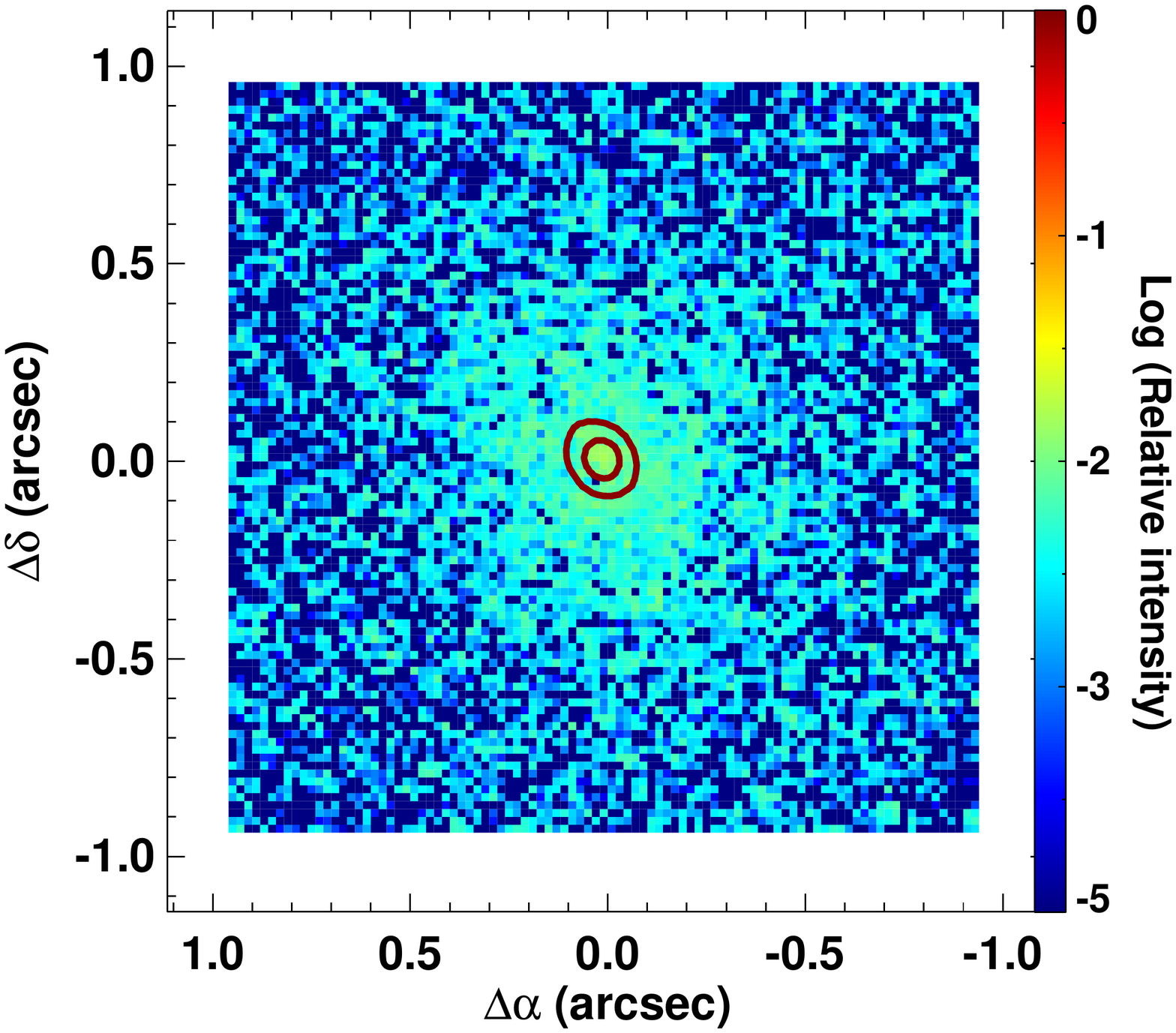}
 \includegraphics[trim={1cm 1.5cm 5cm 0cm},width=5.5cm]{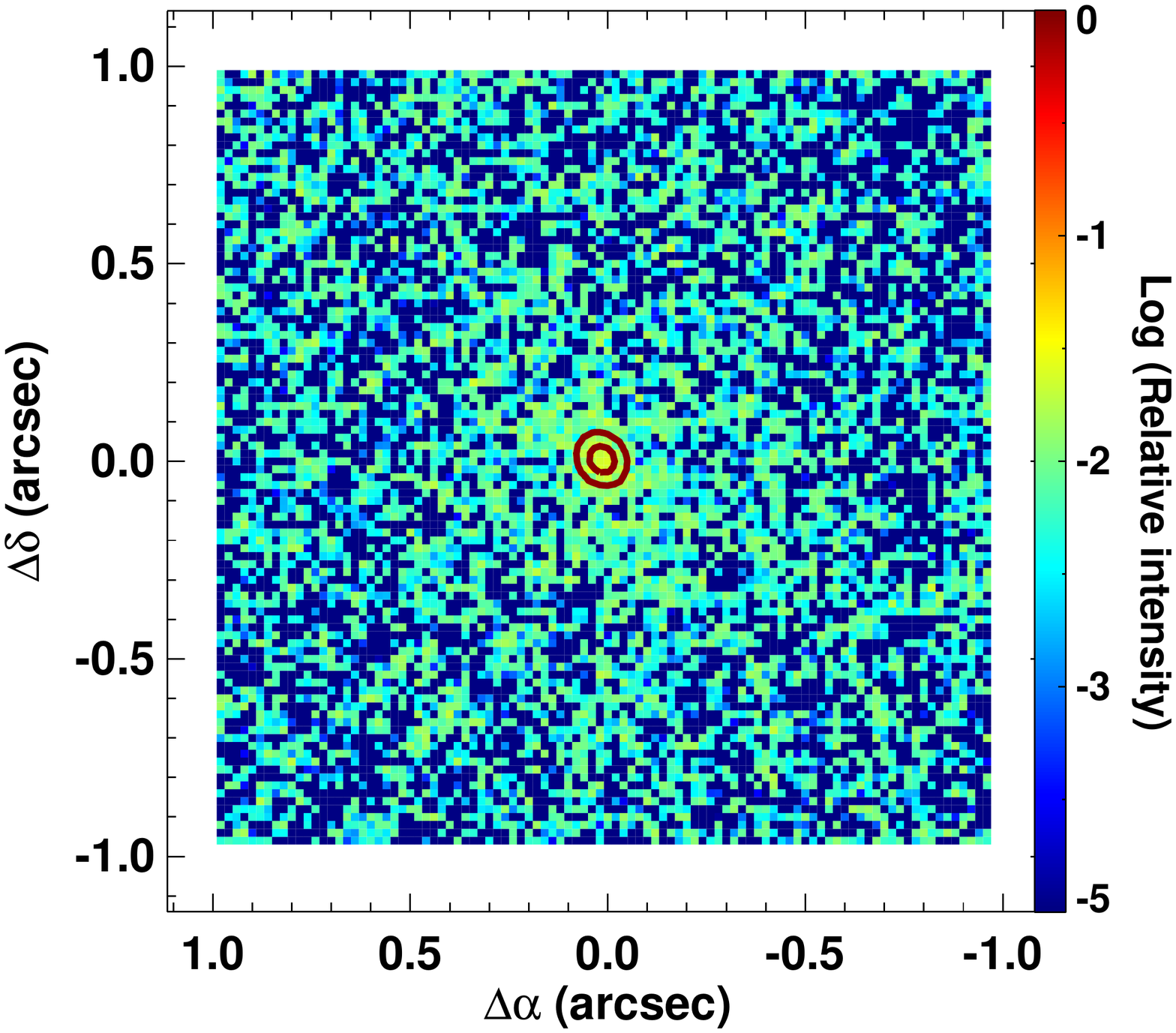}

  \caption{Same as Fig. \ref{NGC809_fullrange}, but when considering spiral morphology (i.e. PGC\,055442) as a host.}
  
 \label{PGC055442_fullrange}
\end{figure*}

\begin{figure*}
 \includegraphics[trim={0.85cm 0.cm 2cm 0},clip,width=5.75cm]{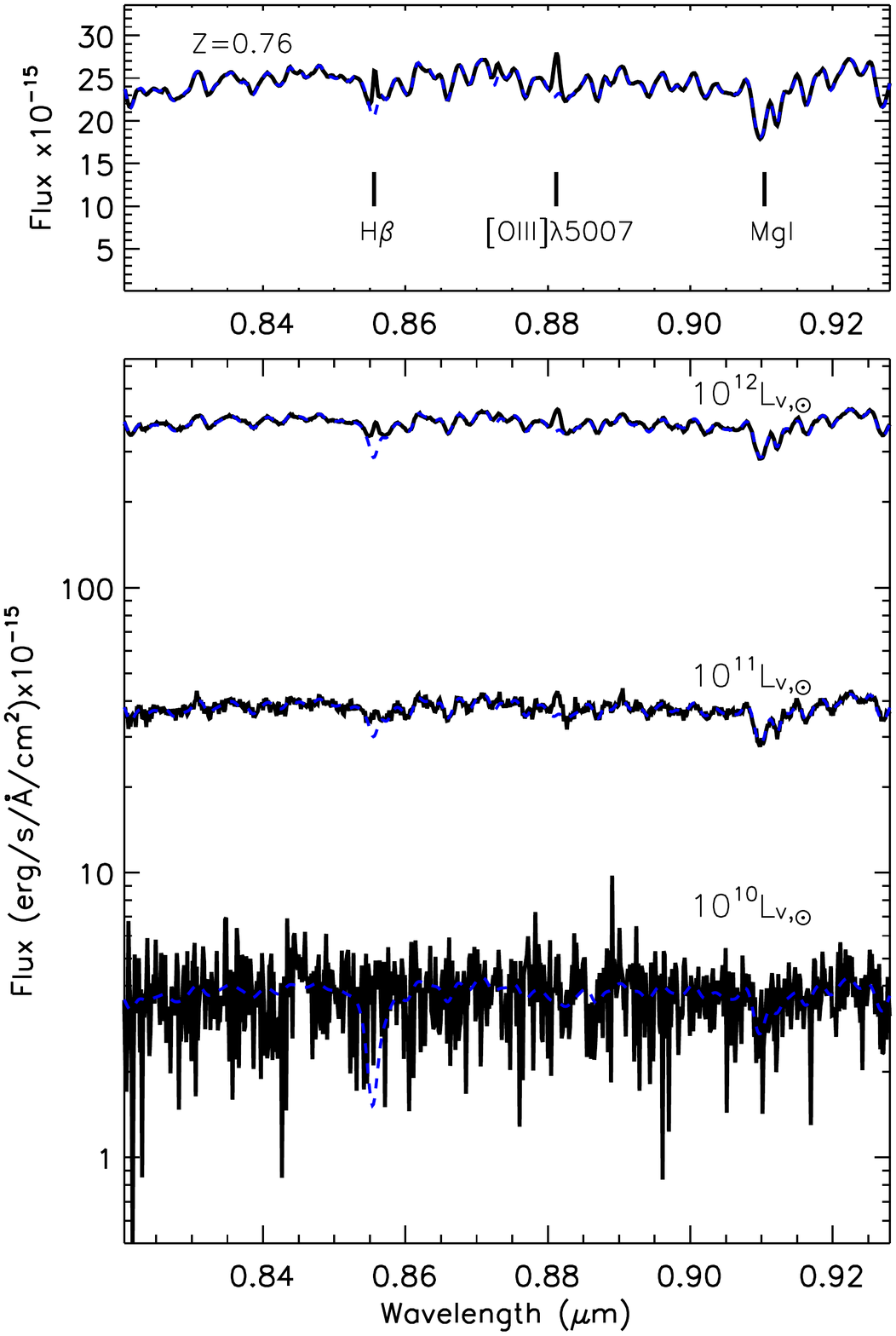}
 \includegraphics[trim={0.85cm 0.cm 2cm 0},clip,width=5.75cm]{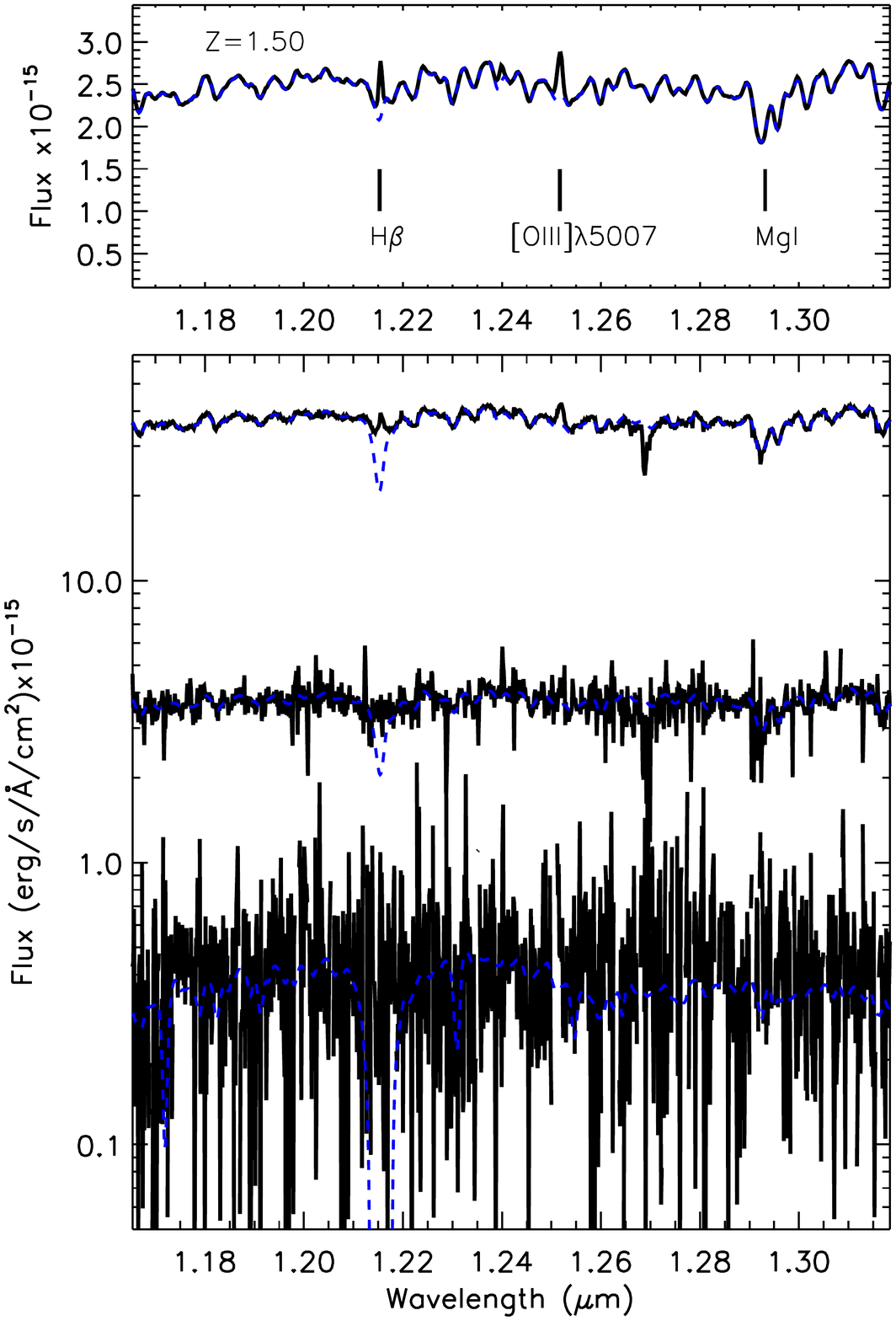}
 \includegraphics[trim={0.85cm 0.5cm 2cm 0},clip,width=5.75cm]{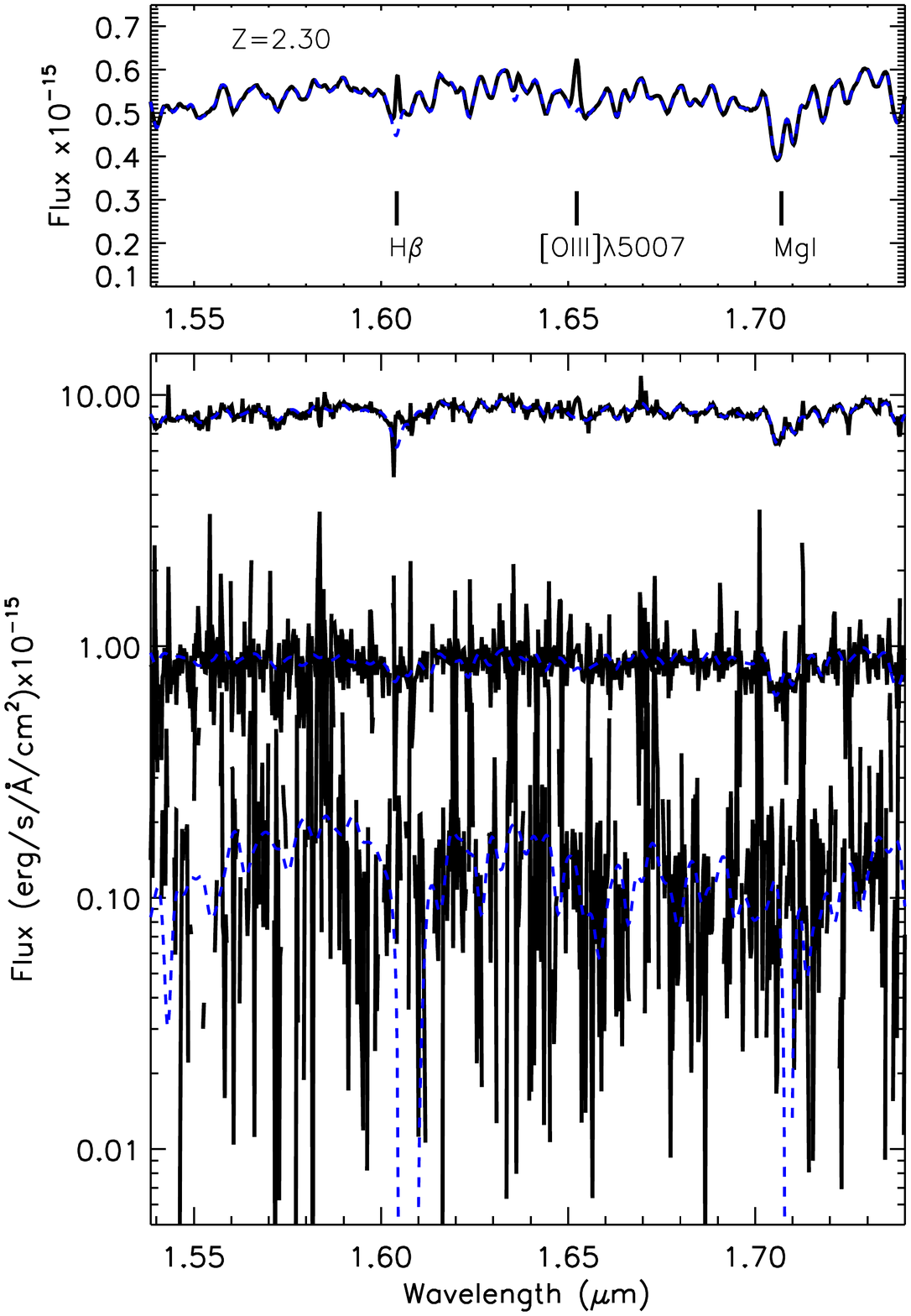}

  \caption{Same as Fig. \ref{NGC809_Reffspectra}, but for the considered spiral morphology (i.e. PGC\,055442).}

 \label{PGC055442_Reffspectra}
\end{figure*}

\begin{figure*}
\centering
 \includegraphics[trim={0.5cm 0.5cm 0.5cm 0},clip,width=4.25cm]{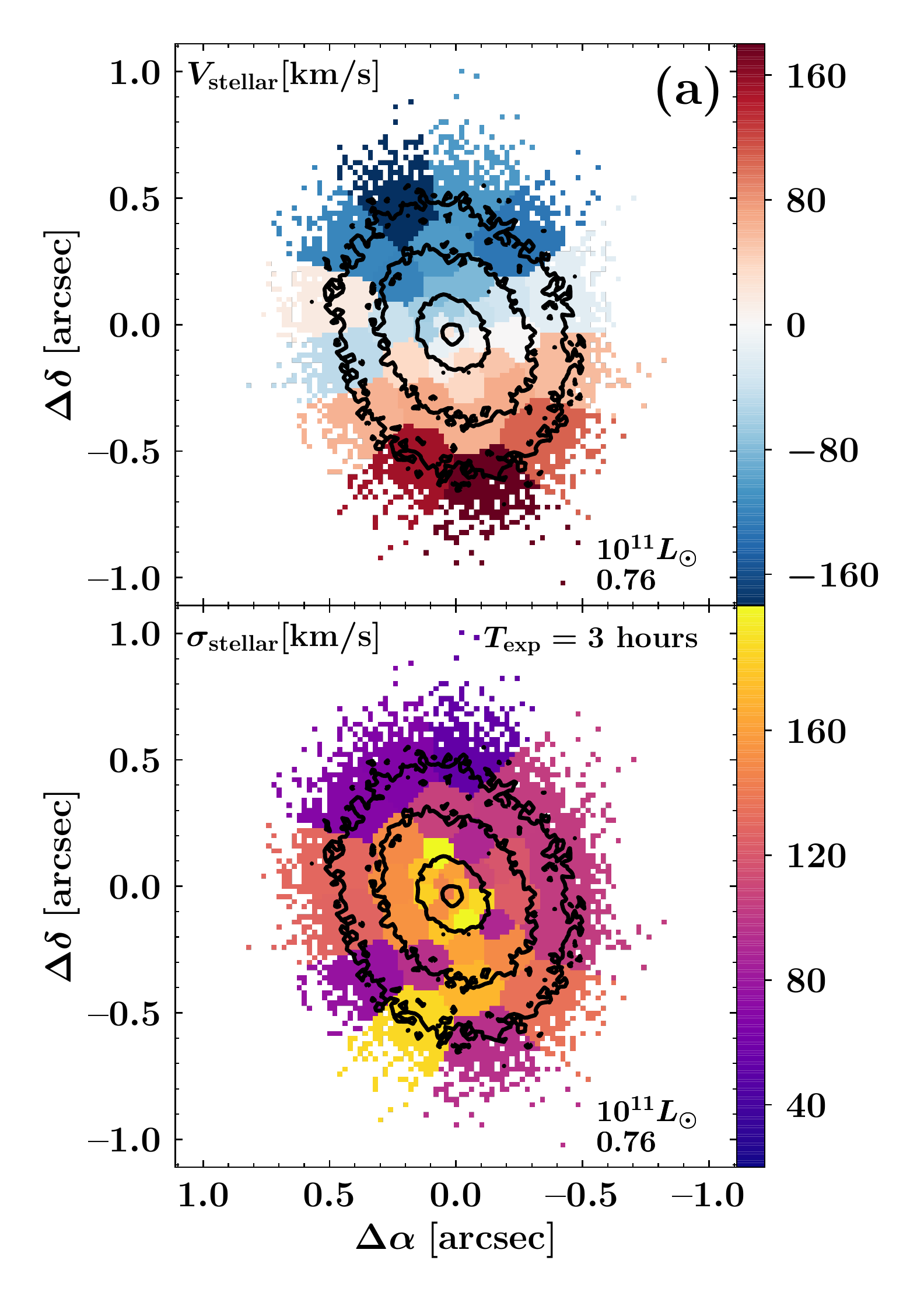}
 \includegraphics[trim={0.5cm 0.5cm 0.5cm 0},clip,width=4.25cm]{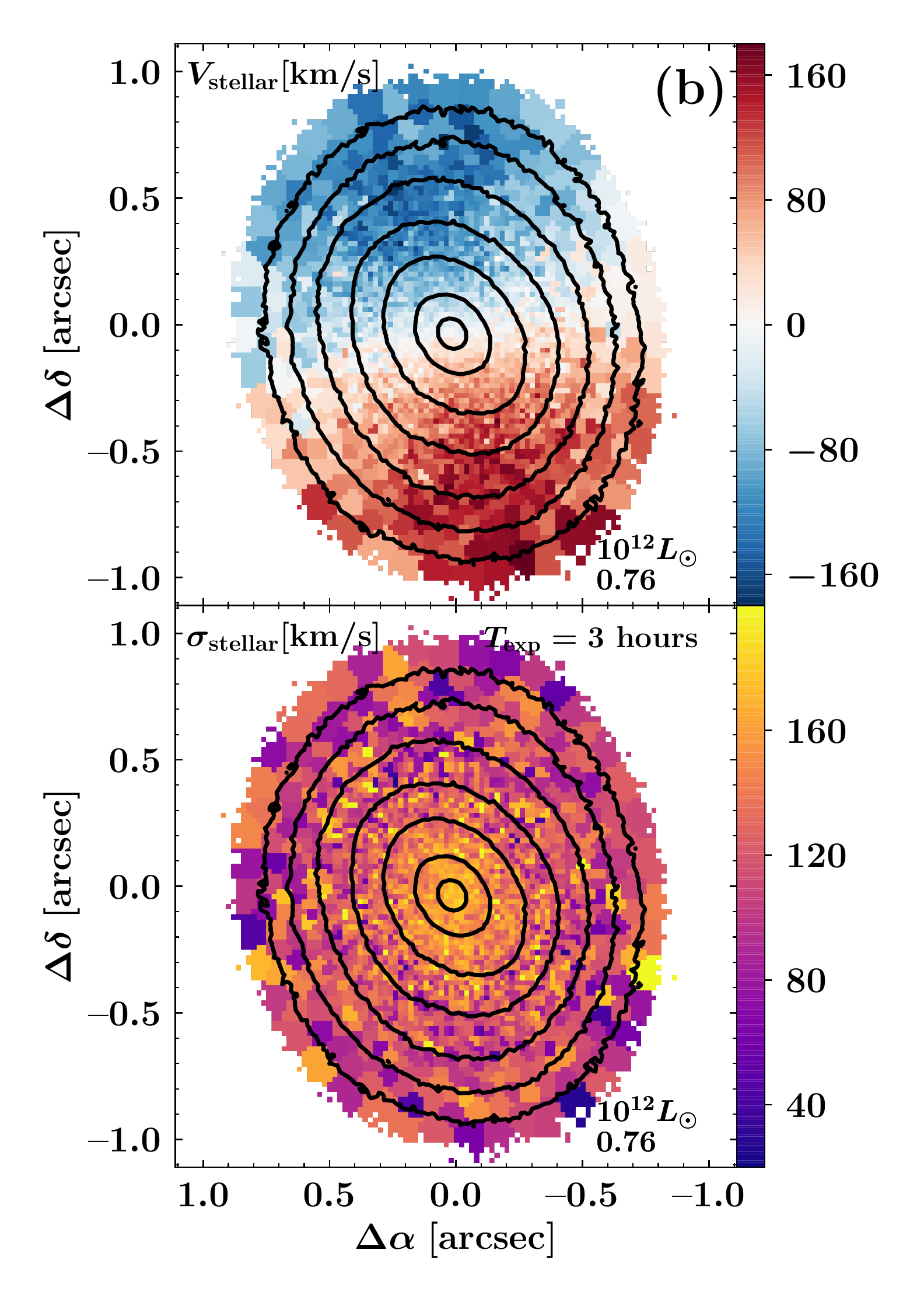}
 \includegraphics[trim={0.5cm 0.5cm 0.5cm 0},clip,width=4.25cm]{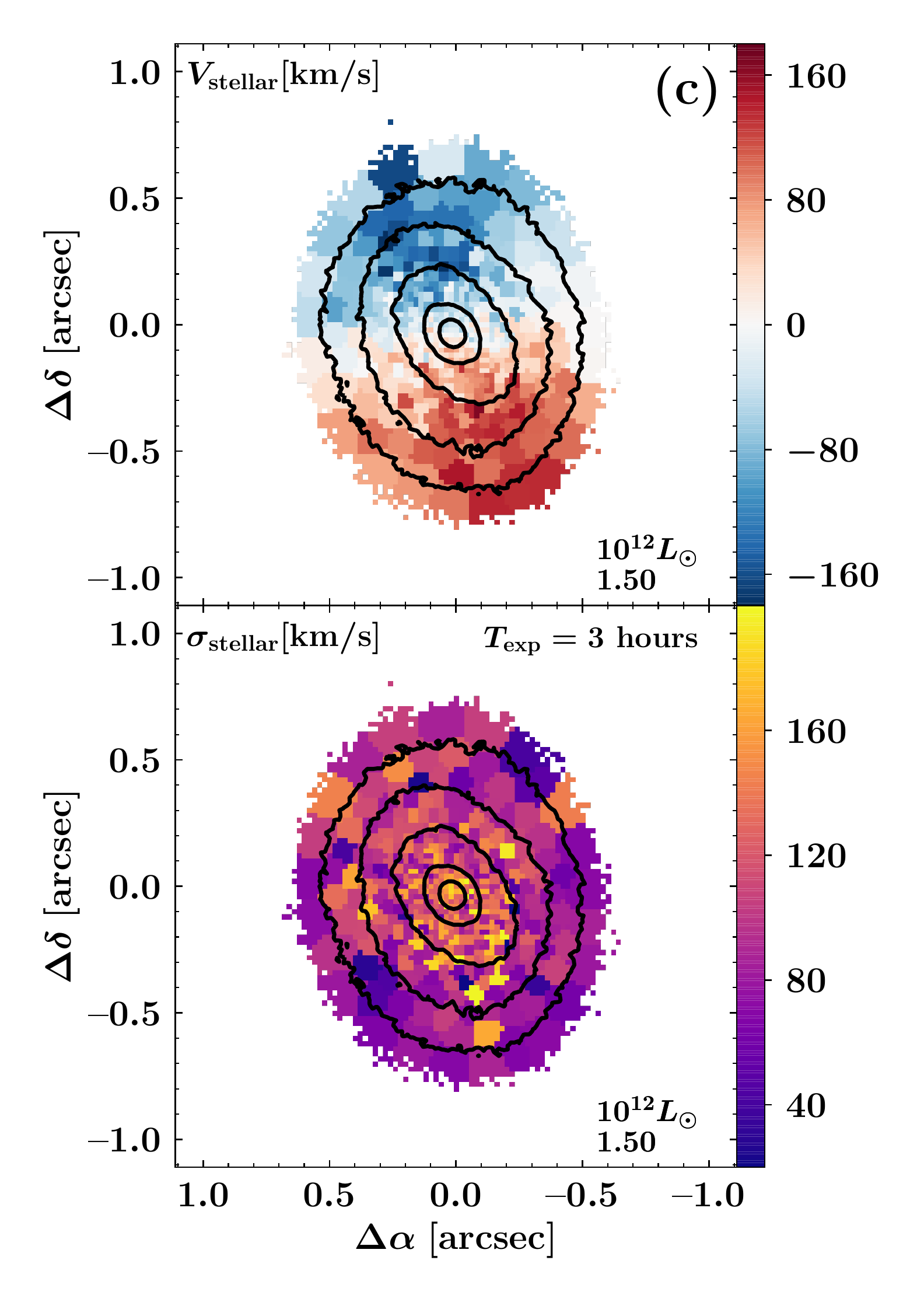}
\includegraphics[trim={0.5cm 0.5cm 0.5cm 0},clip,width=4.25cm]{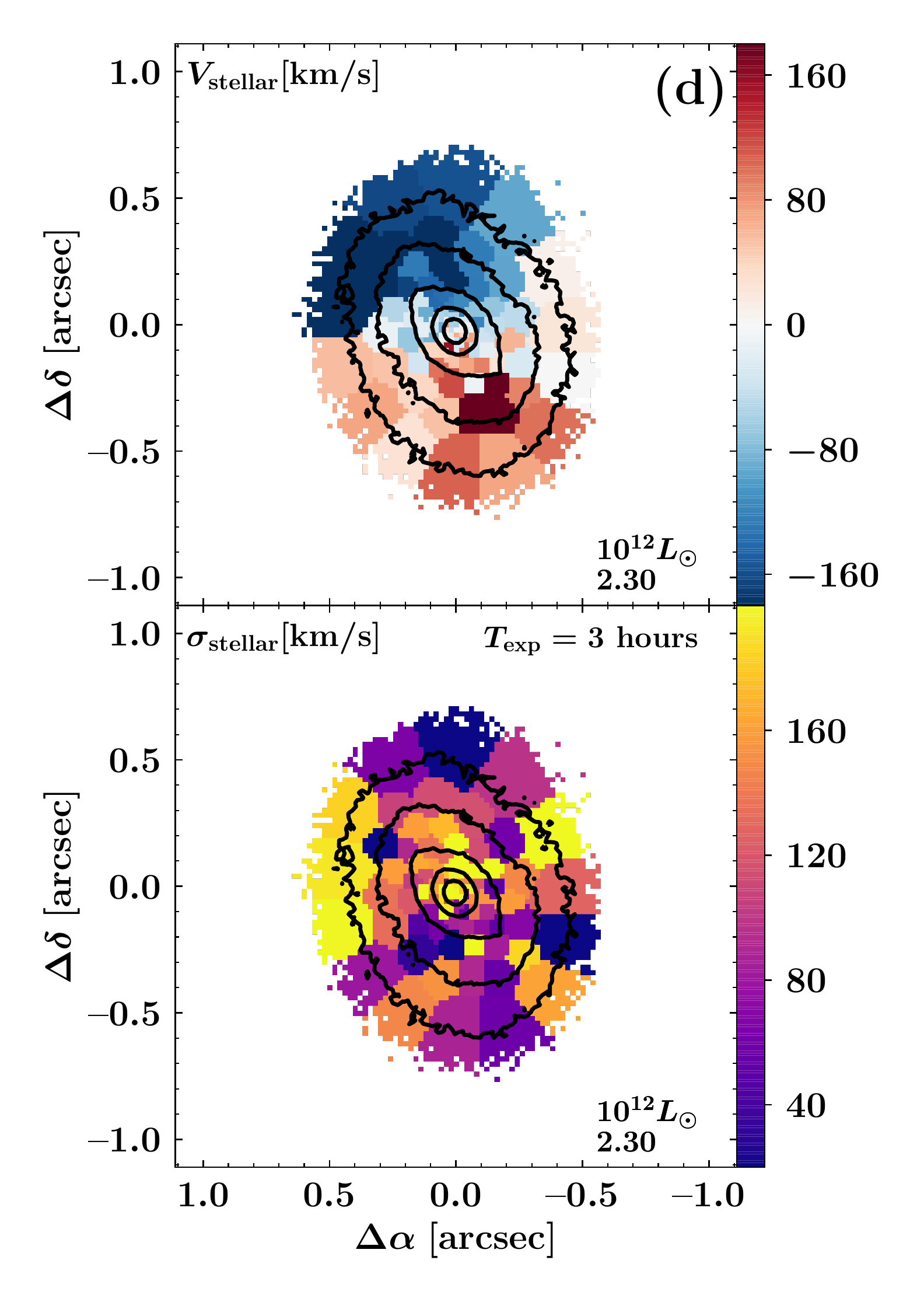}

  \caption{Same as Fig. \ref{resolved_kinematics_NGC809} (top-panels), but for a spiral host galaxy (i.e. PGC\,0055442). Colour bars are in the [-180,180] km s$^{-1}$ and [20,200] km s$^{-1}$ ranges for velocities and velocity dispersions, respectively.}

 \label{resolved_kinematics_PGC055442}
\end{figure*}

\begin{figure*}
\centering
 \includegraphics[trim={1cm 1.5cm 5cm 0cm},width=5.5cm]{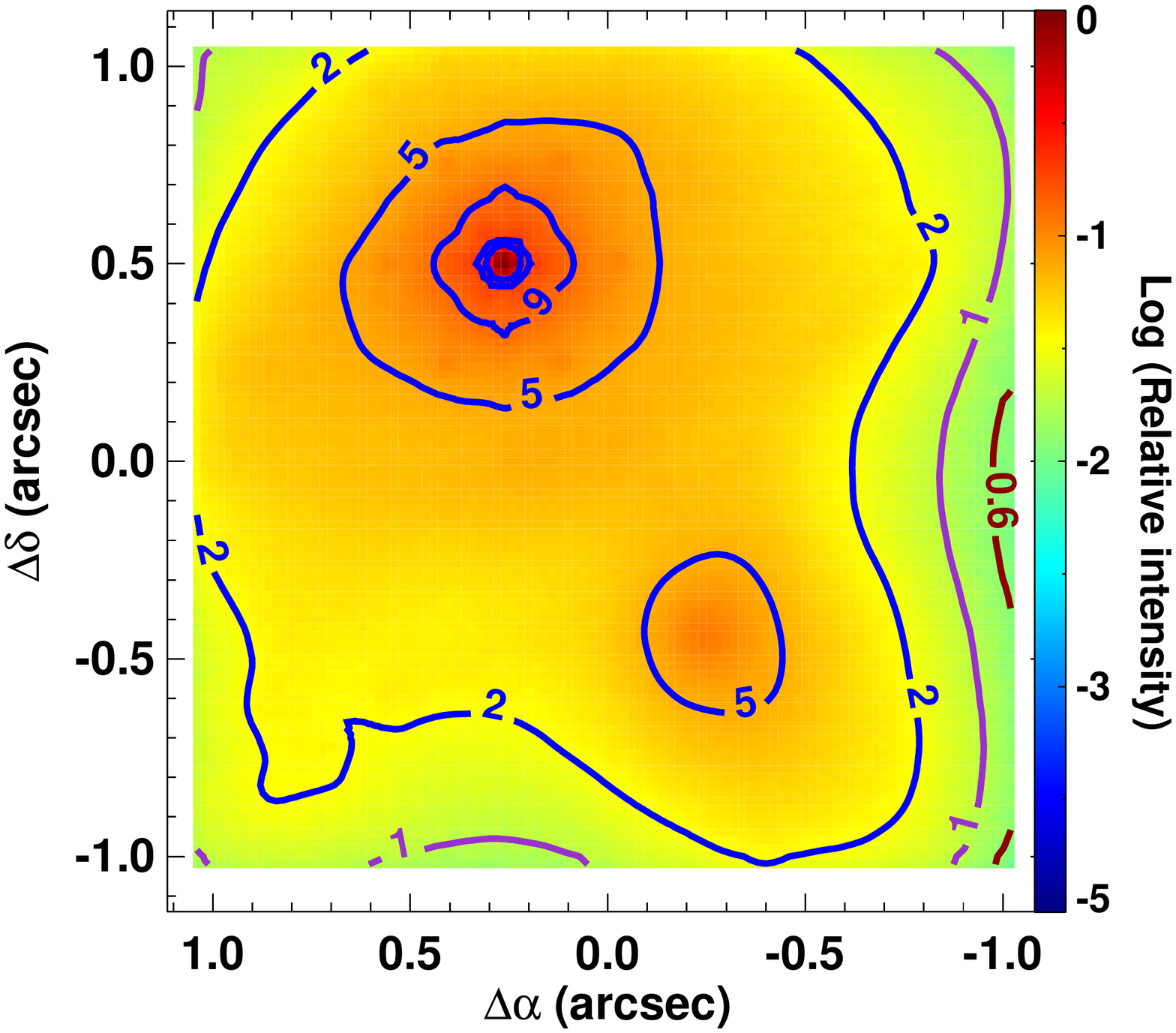}
 \includegraphics[trim={1cm 1.5cm 5cm 0cm},width=5.5cm]{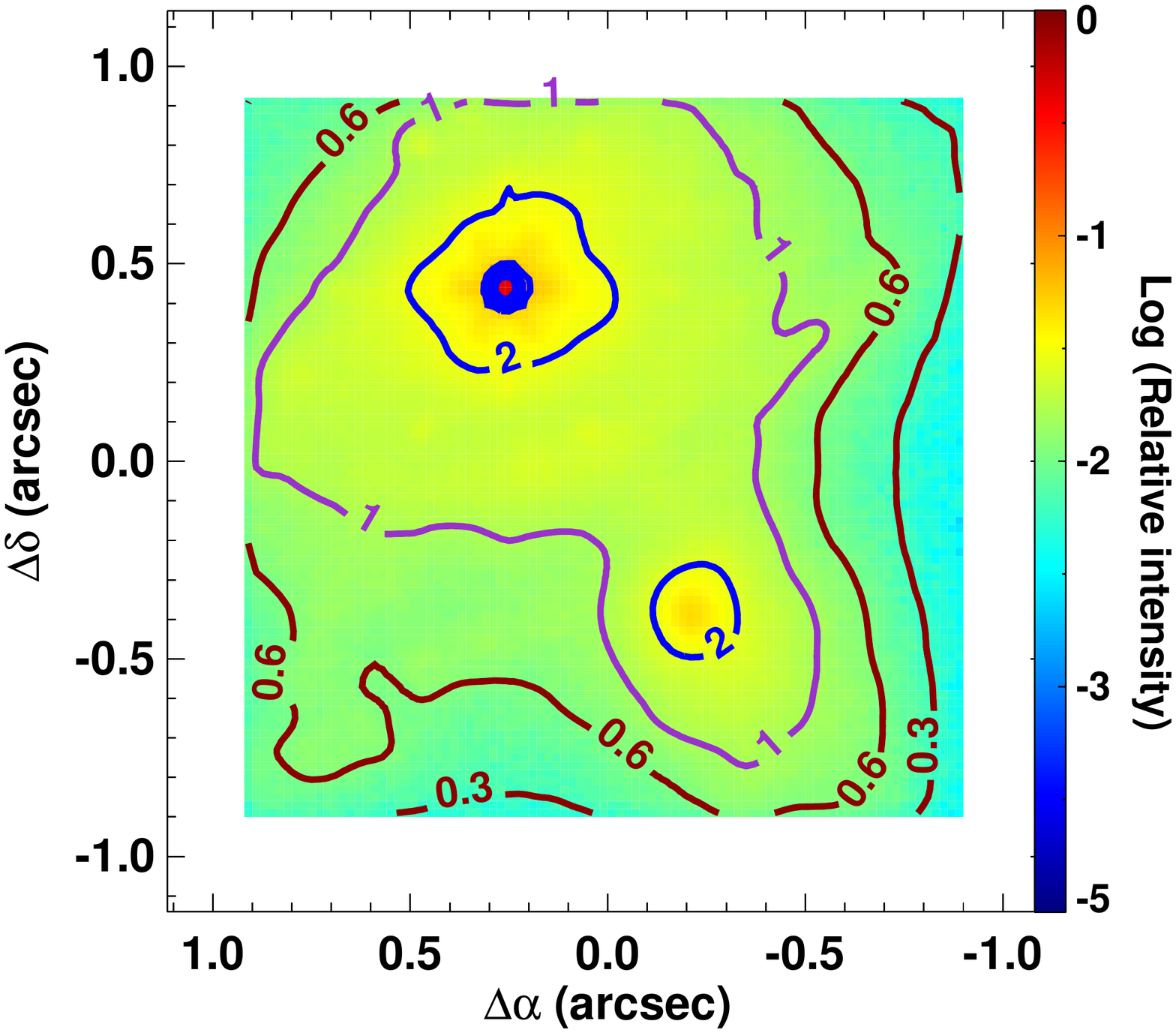}
 \includegraphics[trim={1cm 1.5cm 5cm 0cm},width=5.5cm]{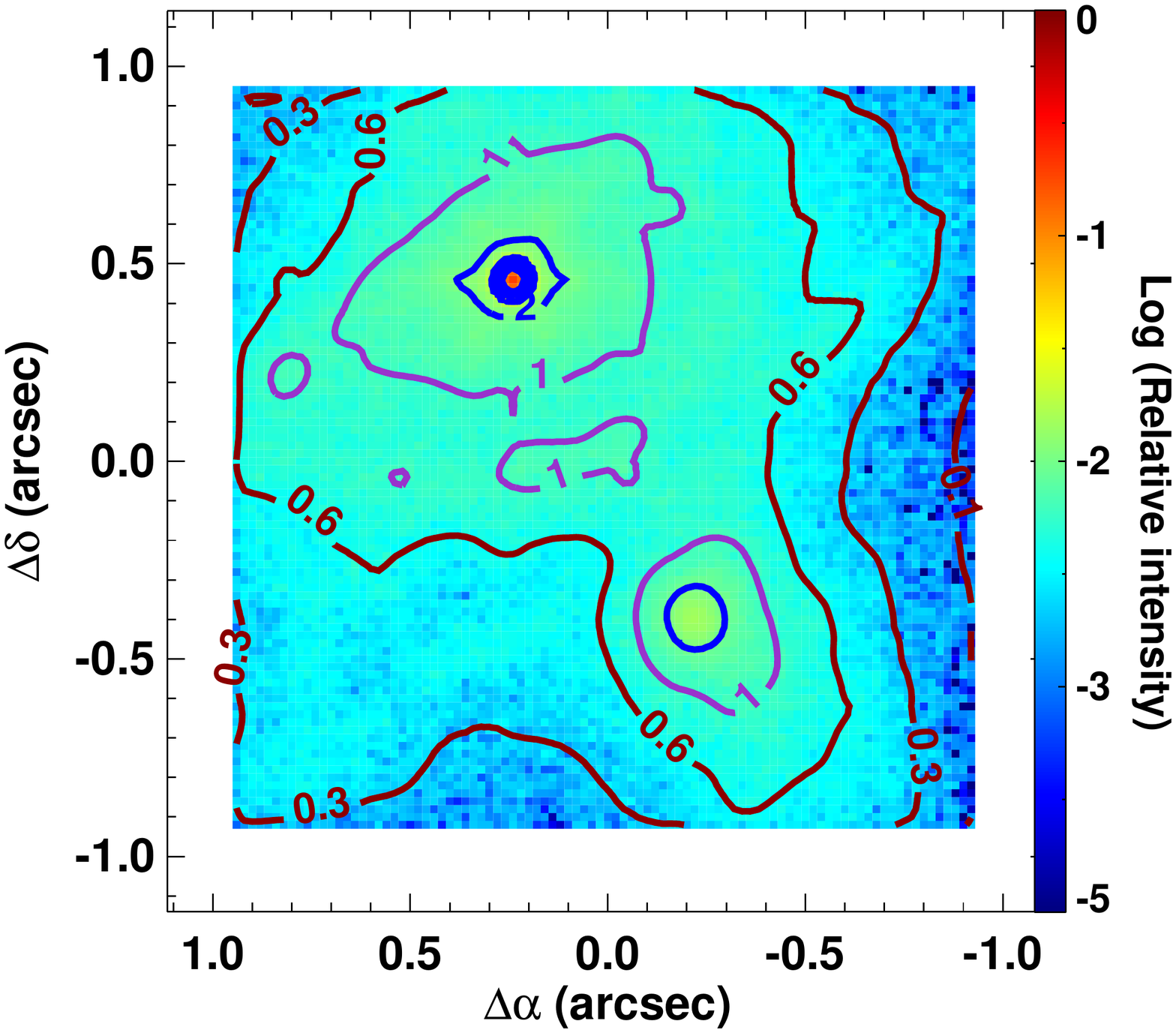}

\includegraphics[trim={1cm 1.5cm 5cm 0cm},width=5.5cm]{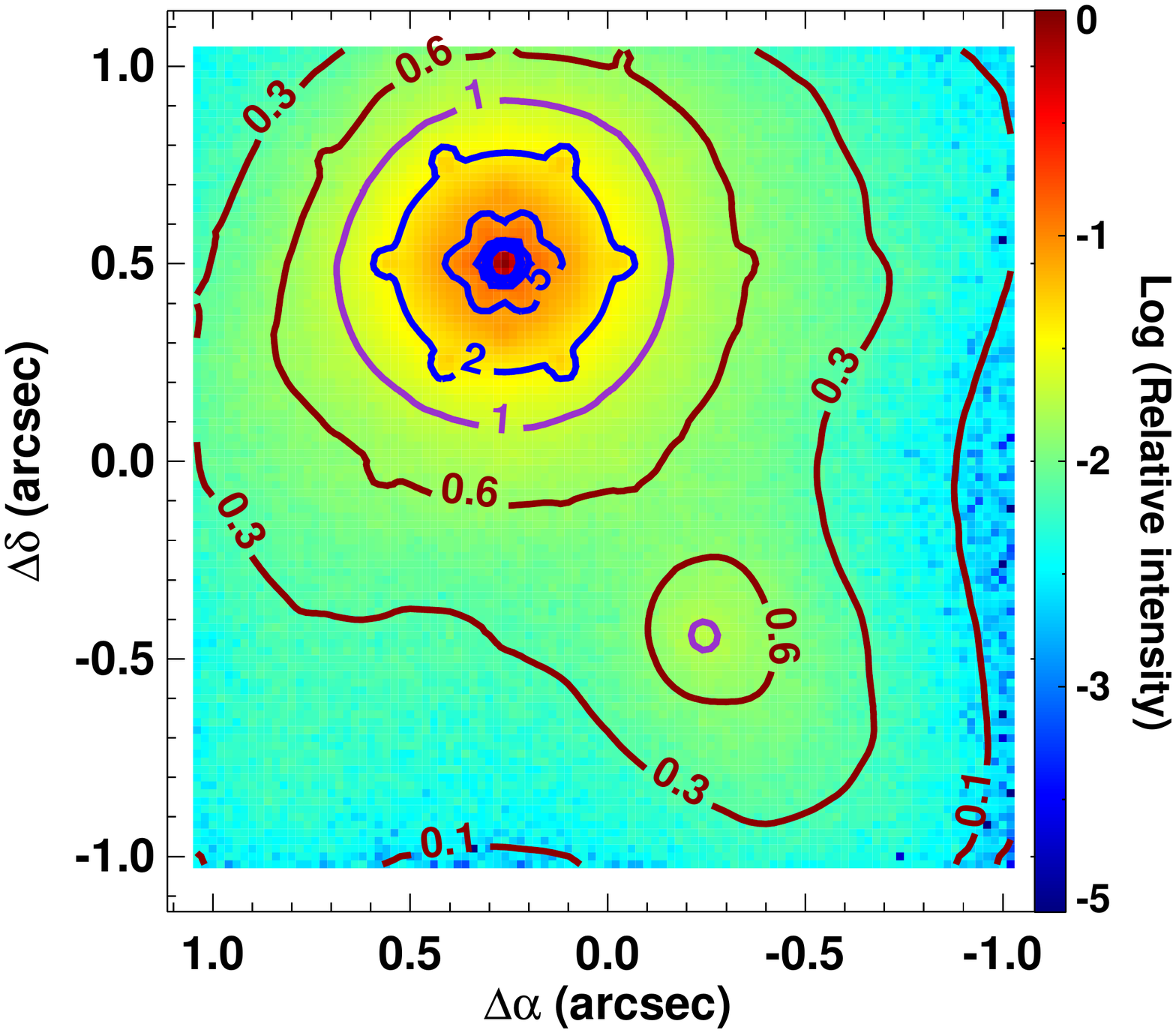}
 \includegraphics[trim={1cm 1.5cm 5cm 0cm},width=5.5cm]{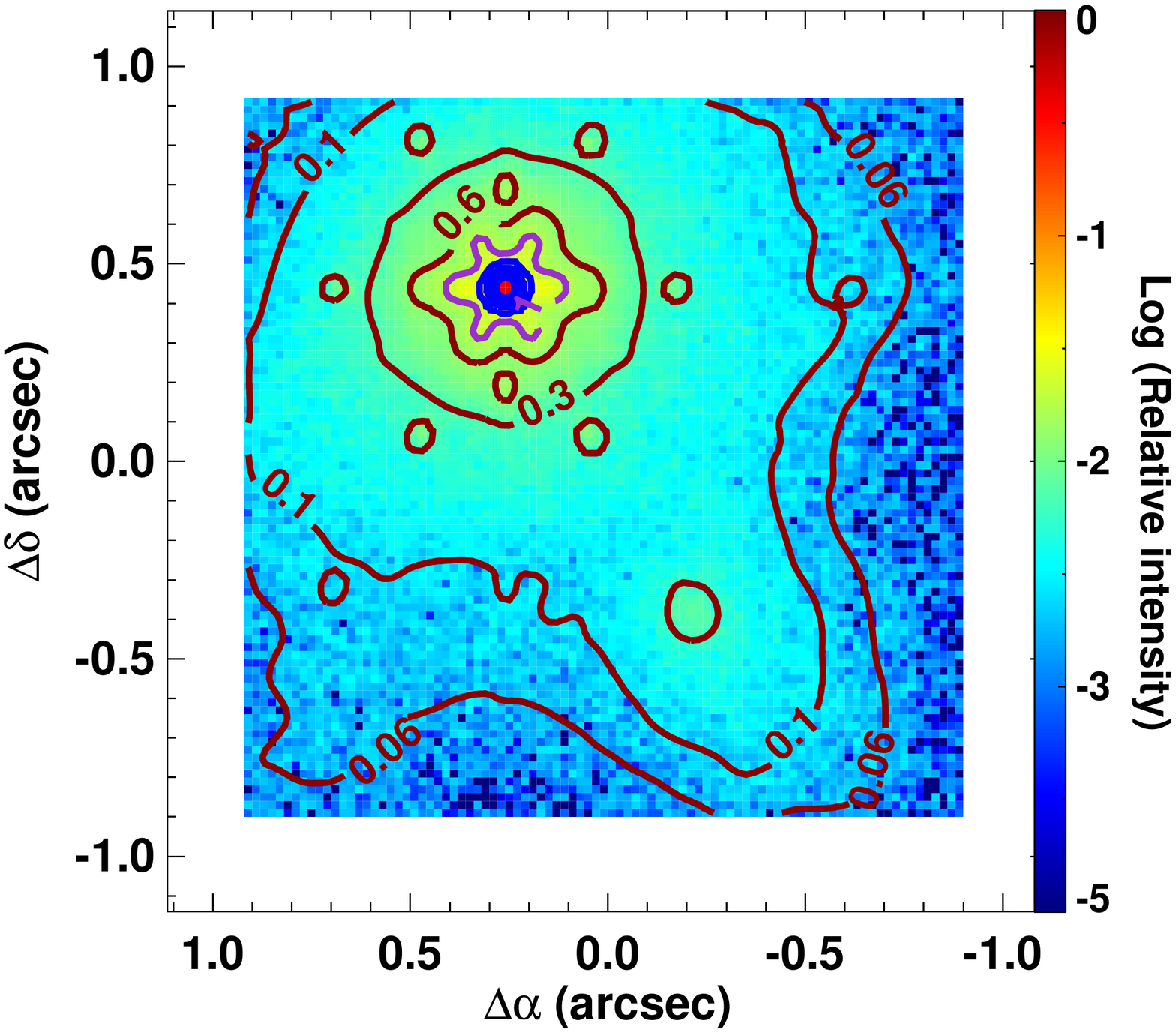}
 \includegraphics[trim={1cm 1.5cm 5cm 0cm},width=5.5cm]{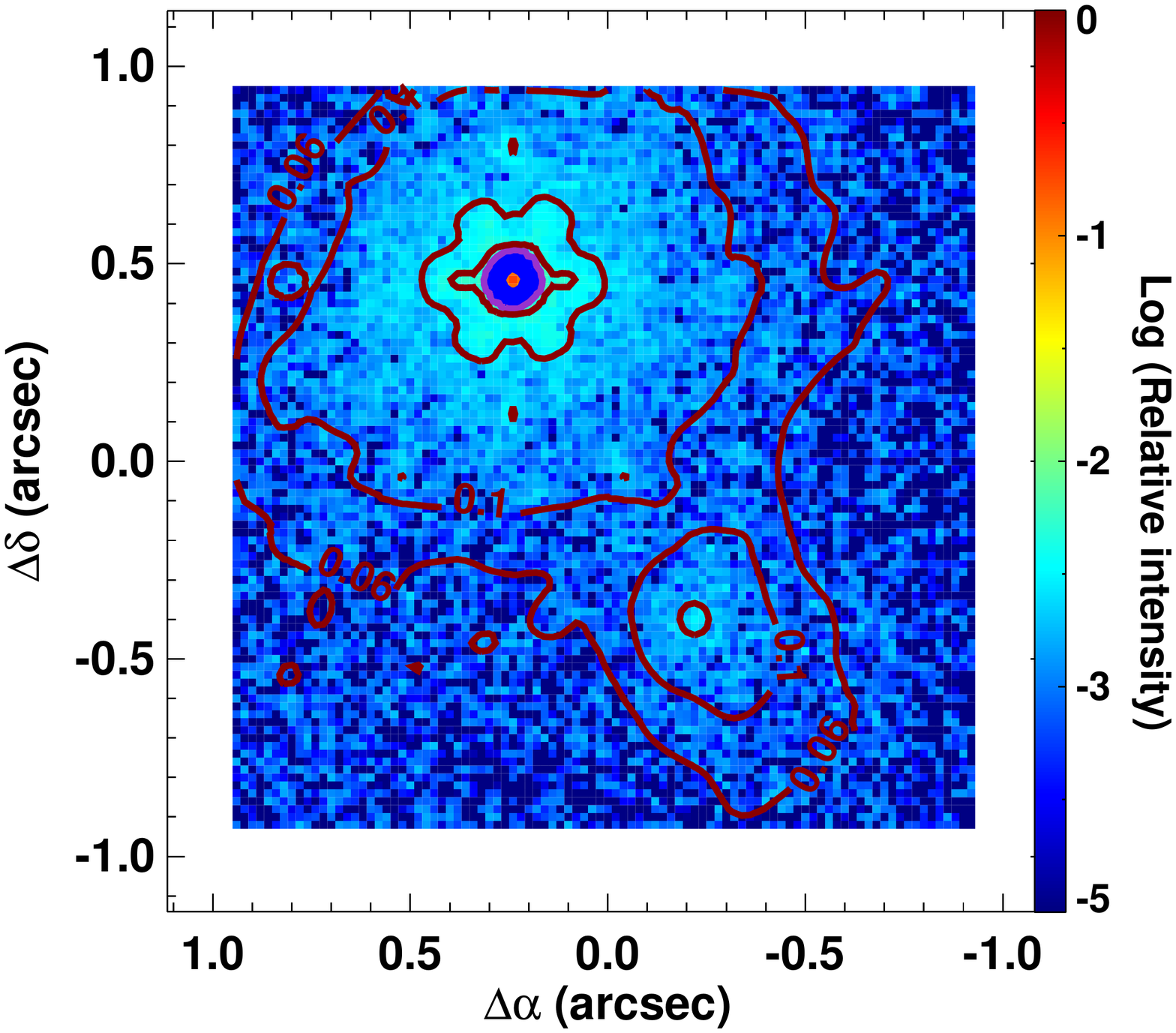}

\includegraphics[trim={1cm 1.5cm 5cm 0cm},width=5.5cm]{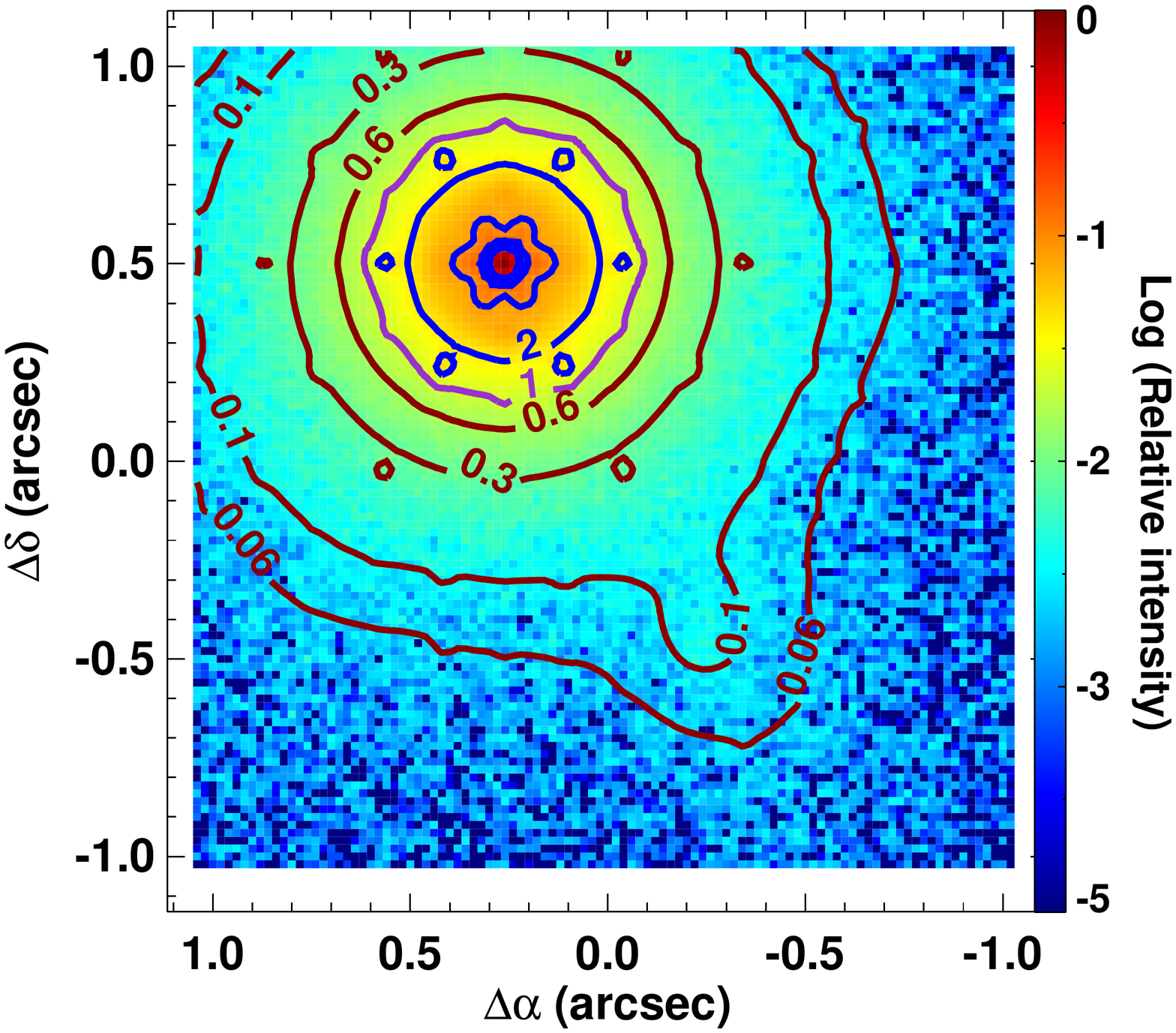}
 \includegraphics[trim={1cm 1.5cm 5cm 0cm},width=5.5cm]{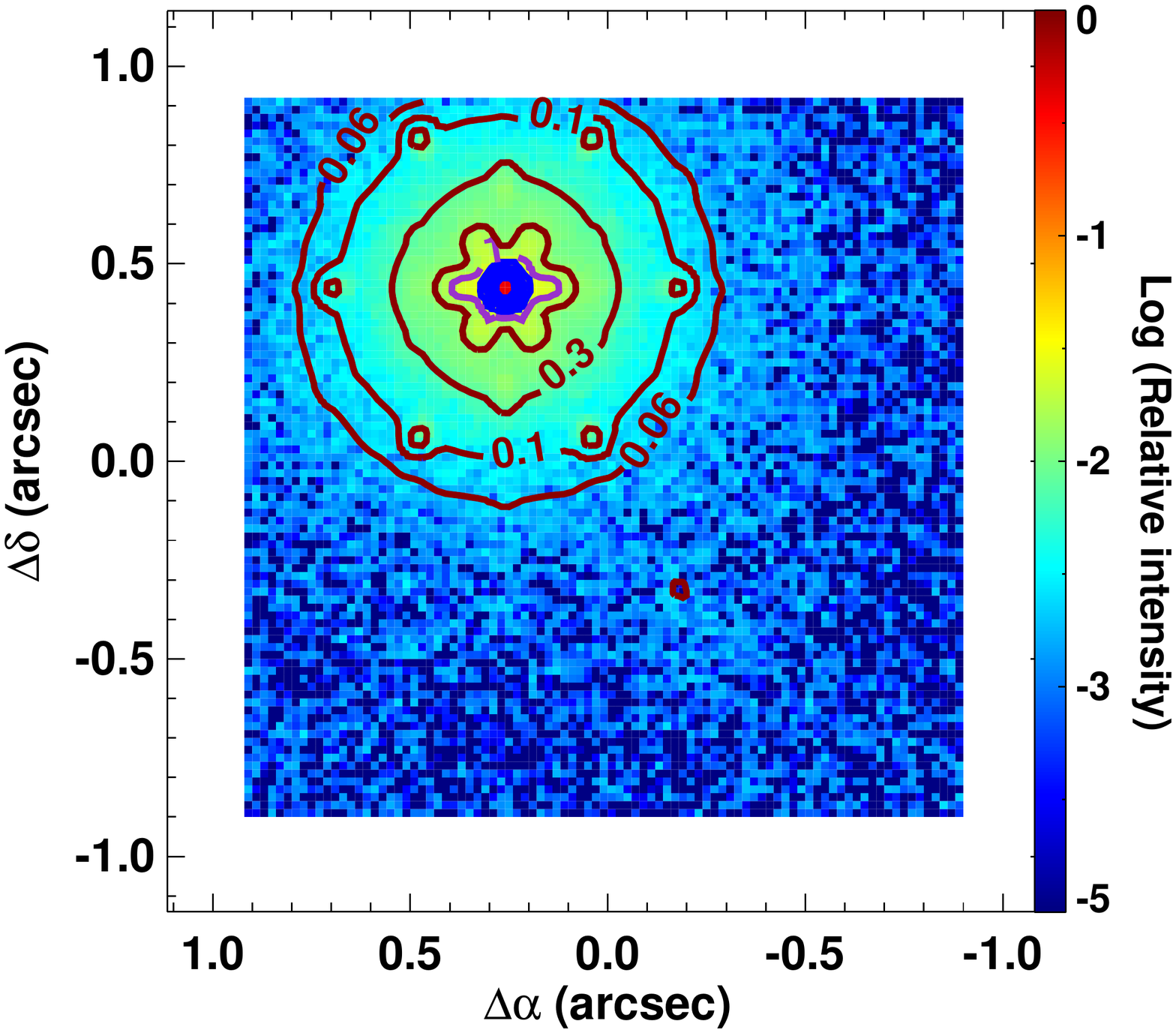}
 \includegraphics[trim={1cm 1.5cm 5cm 0cm},width=5.5cm]{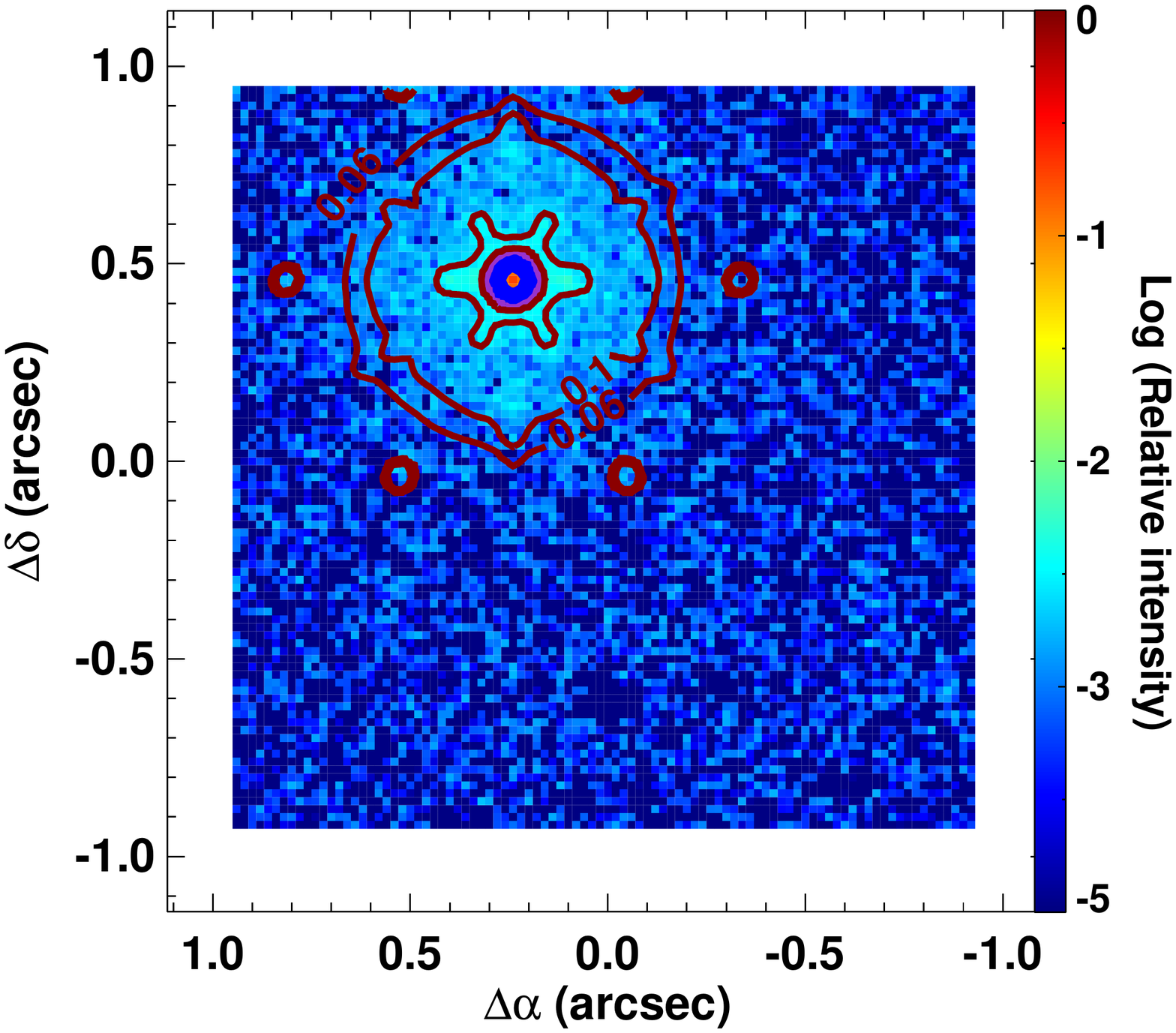}

  \caption{Same as Fig. \ref{NGC809_HbMgI_range_42}, but when considering an interacting morphology (i.e. NGC\,7119A) as a host.}

 \label{NGC7119N_HbMgI_range_42}
\end{figure*}

\begin{figure*}
\centering
 \includegraphics[trim={1cm 1.5cm 5cm 0cm},width=5.5cm]{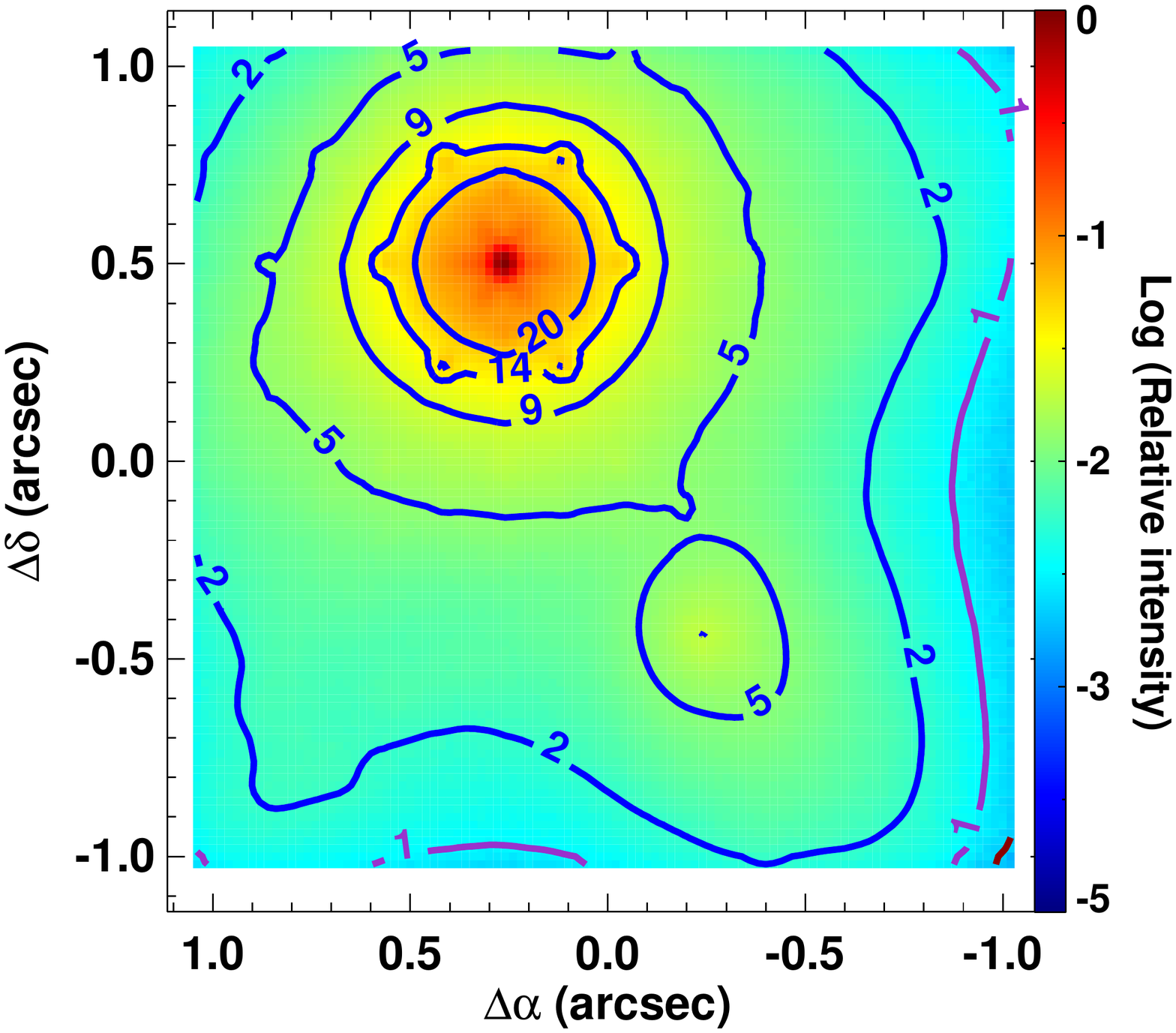}
 \includegraphics[trim={1cm 1.5cm 5cm 0cm},width=5.5cm]{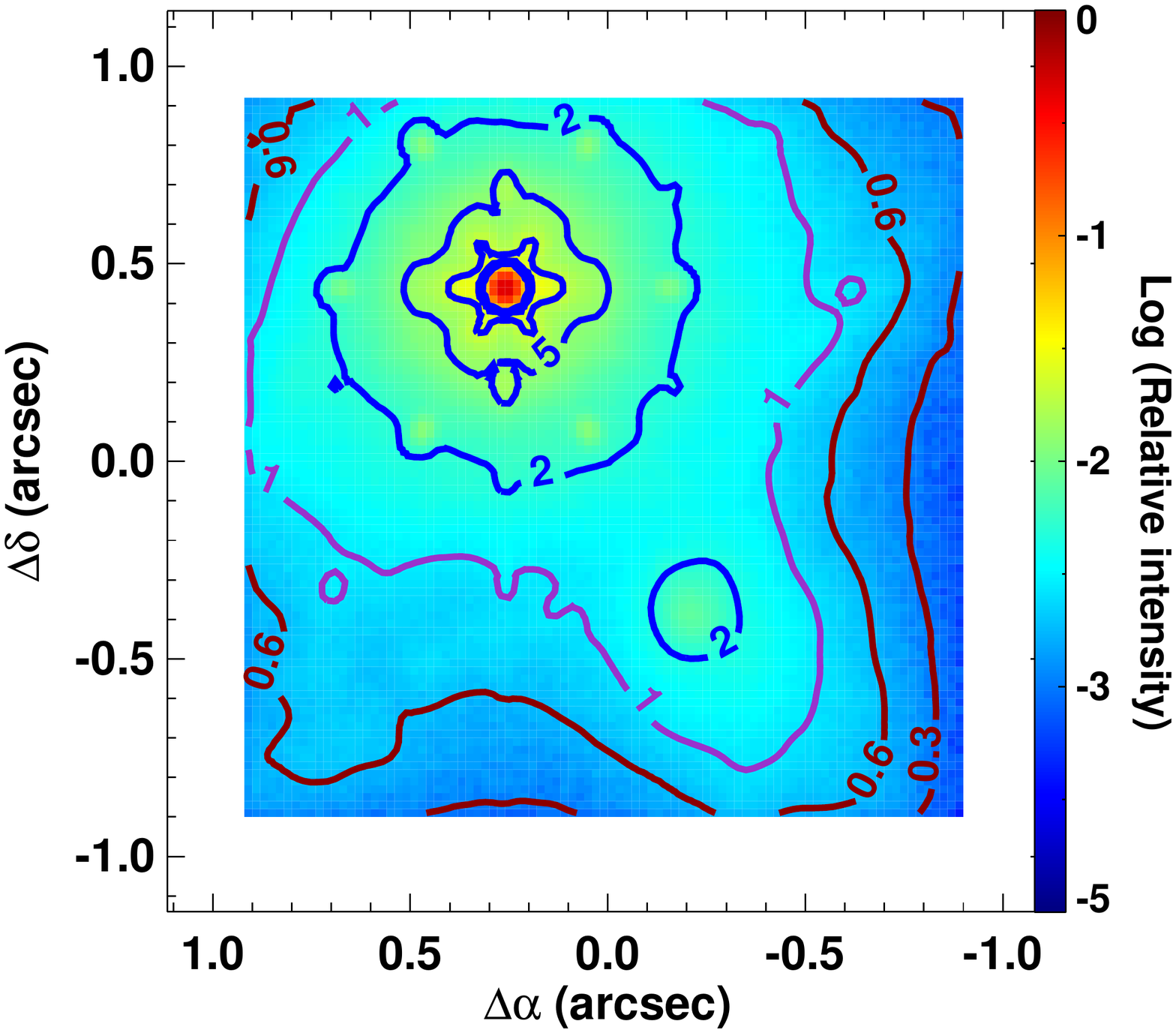}
 \includegraphics[trim={1cm 1.5cm 5cm 0cm},width=5.5cm]{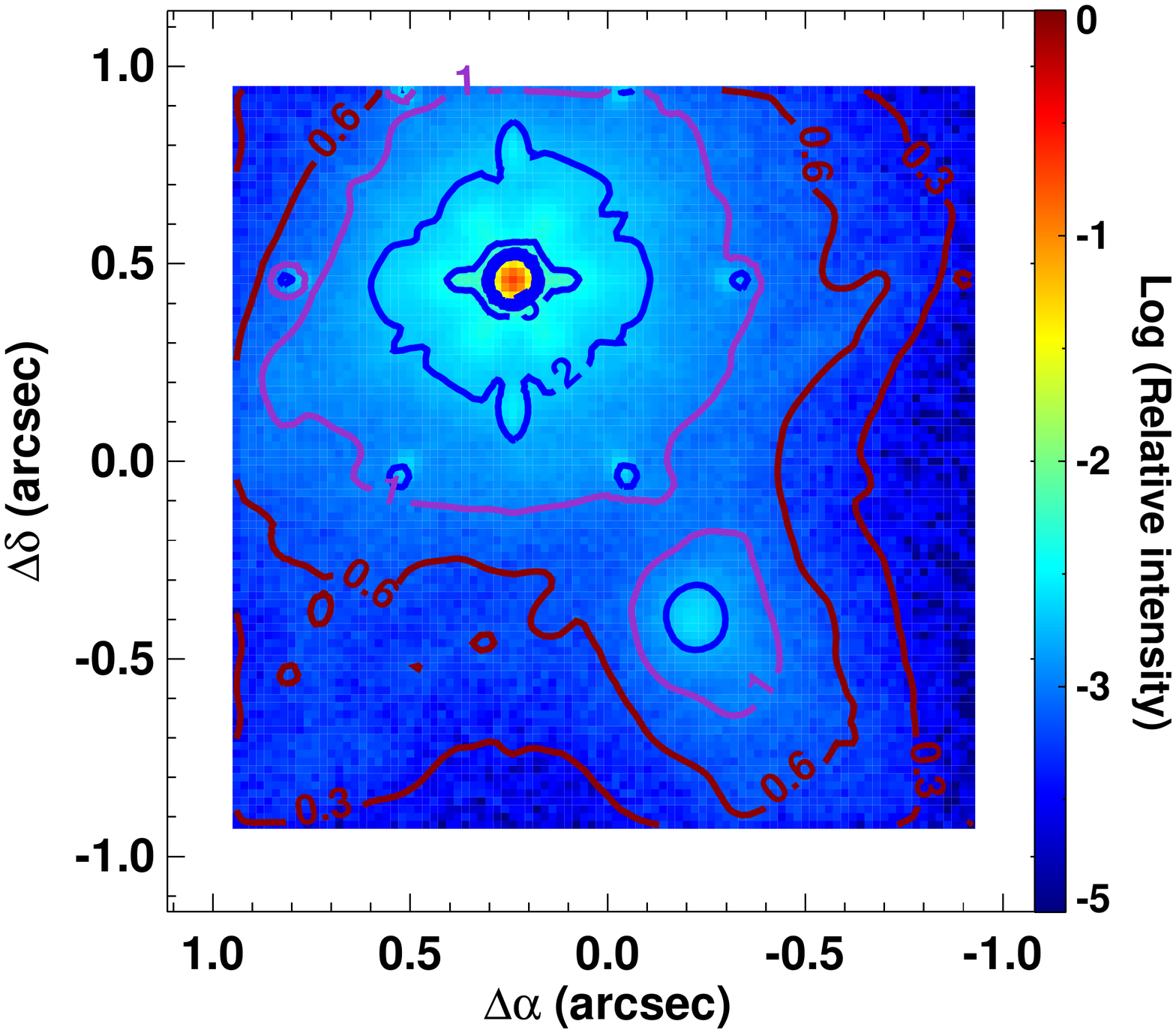}

\includegraphics[trim={1cm 1.5cm 5cm 0cm},width=5.5cm]{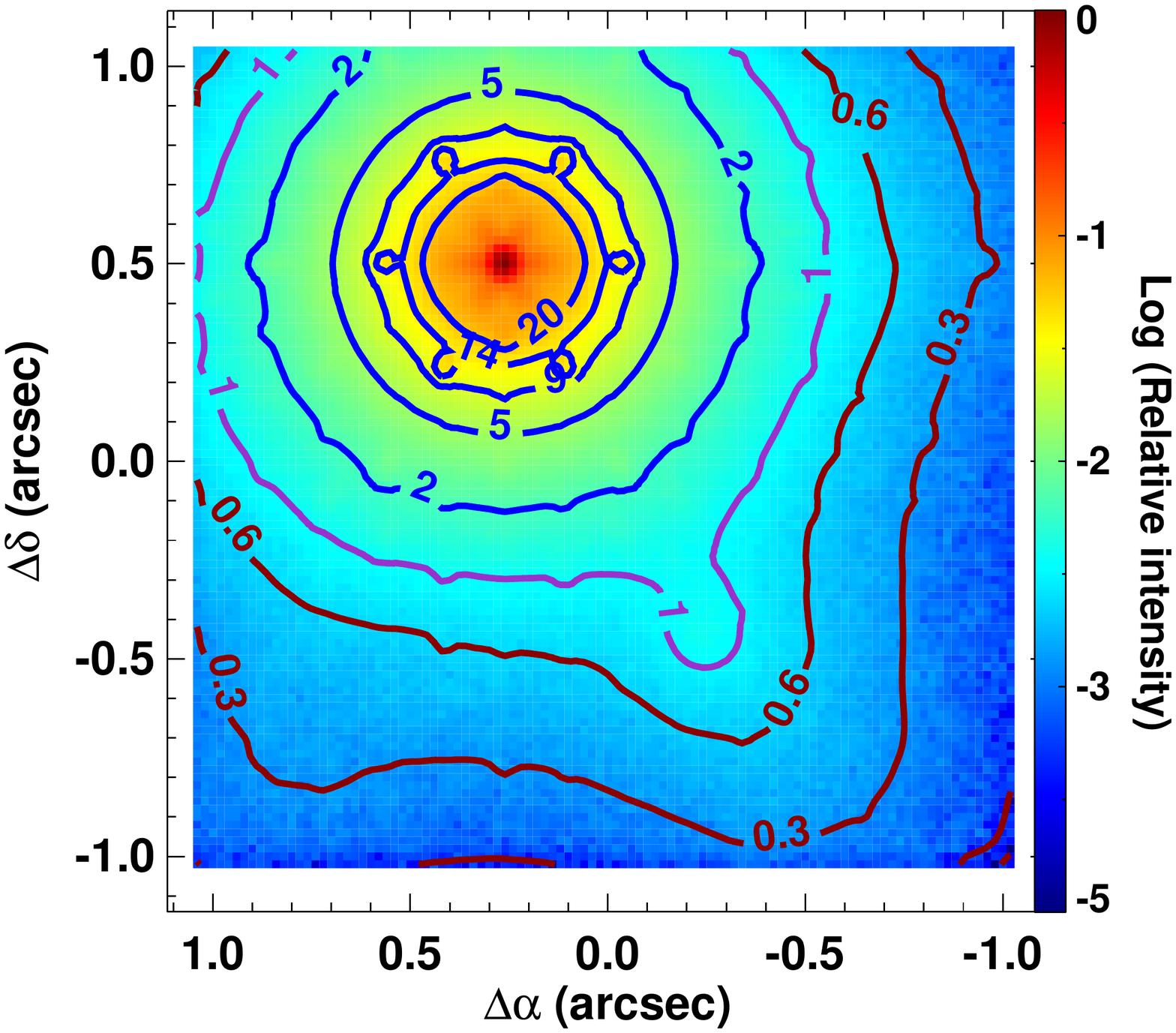}
 \includegraphics[trim={1cm 1.5cm 5cm 0cm},width=5.5cm]{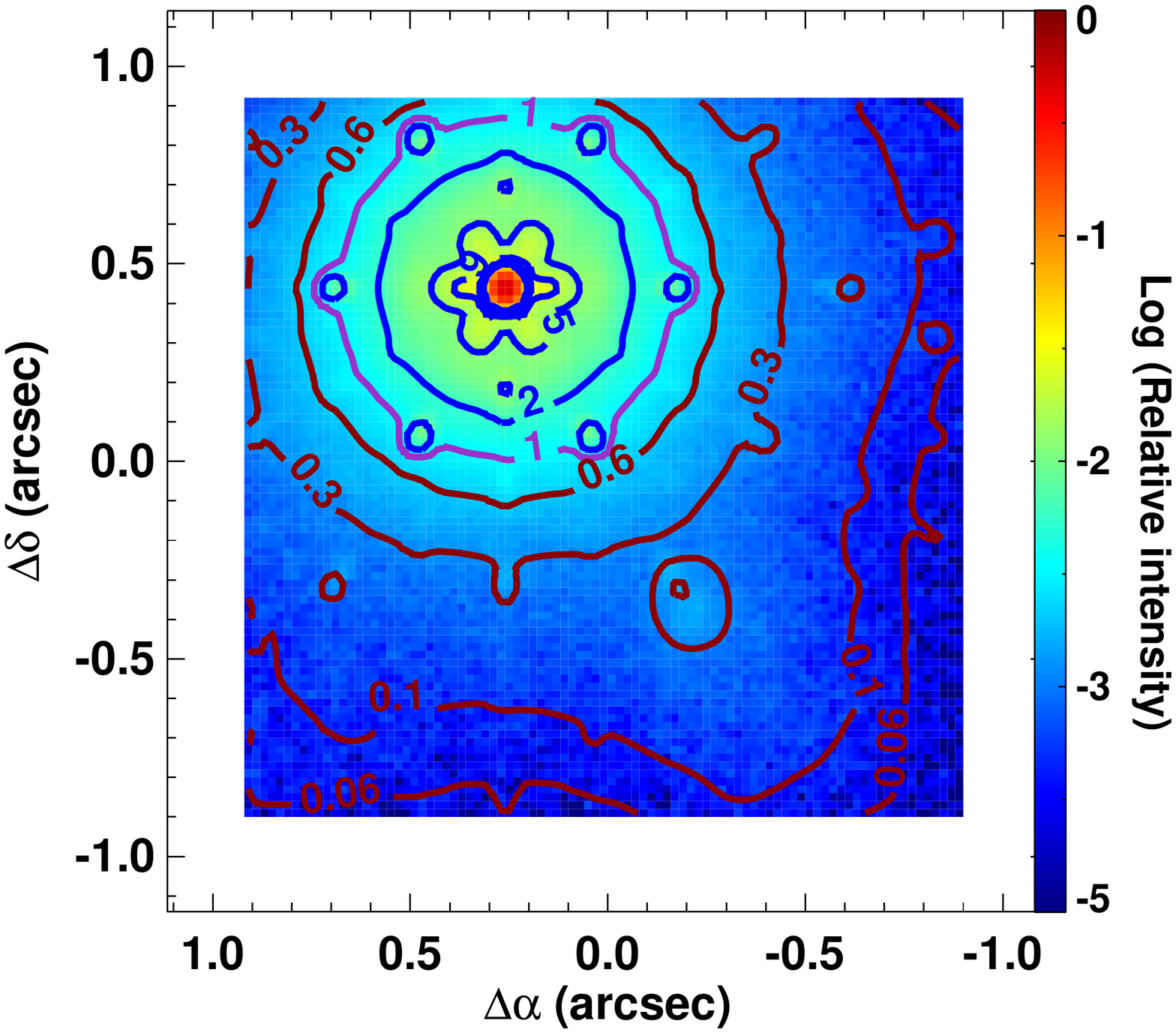}
 \includegraphics[trim={1cm 1.5cm 5cm 0cm},width=5.5cm]{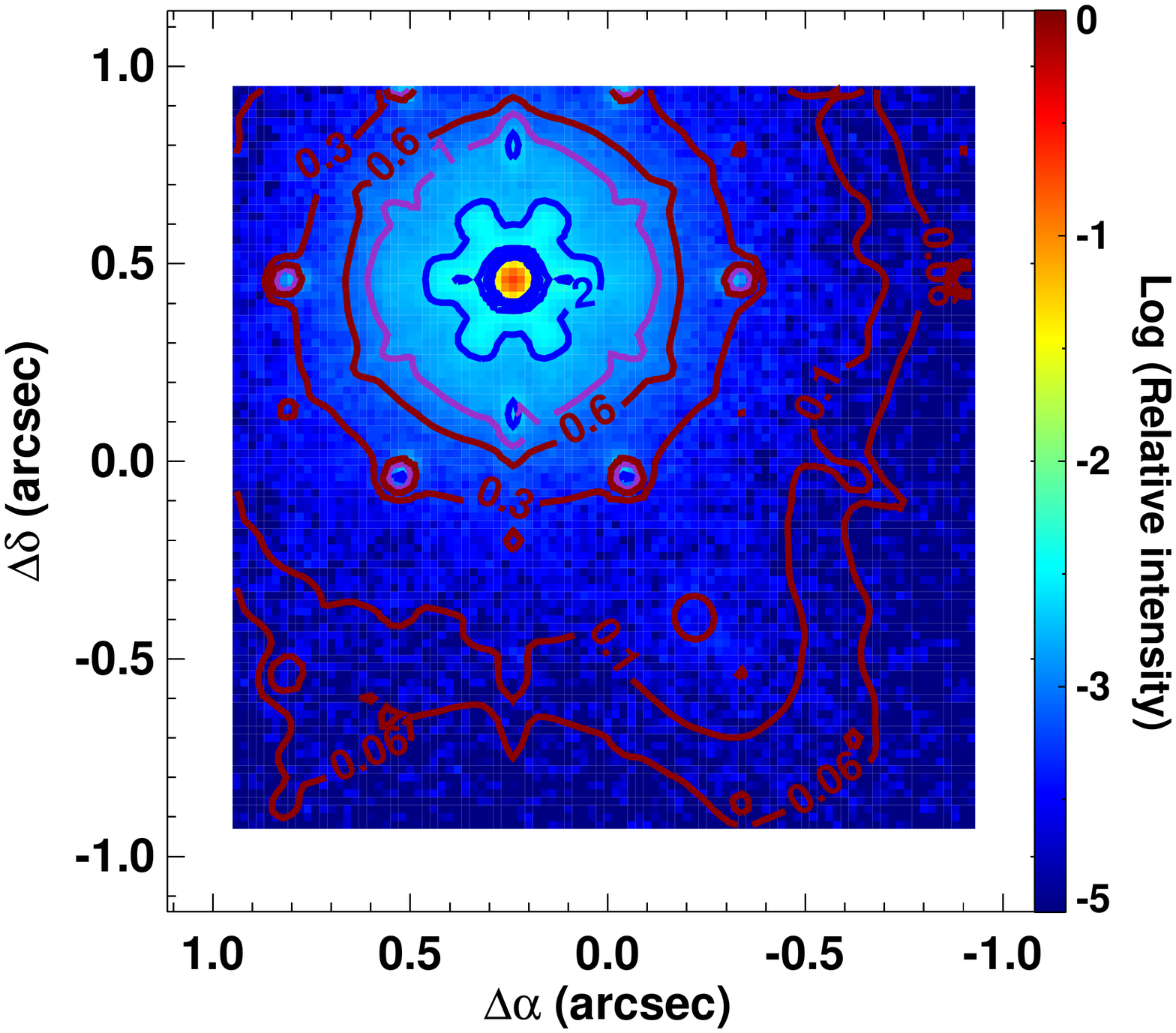}

\includegraphics[trim={1cm 1.5cm 5cm 0cm},width=5.5cm]{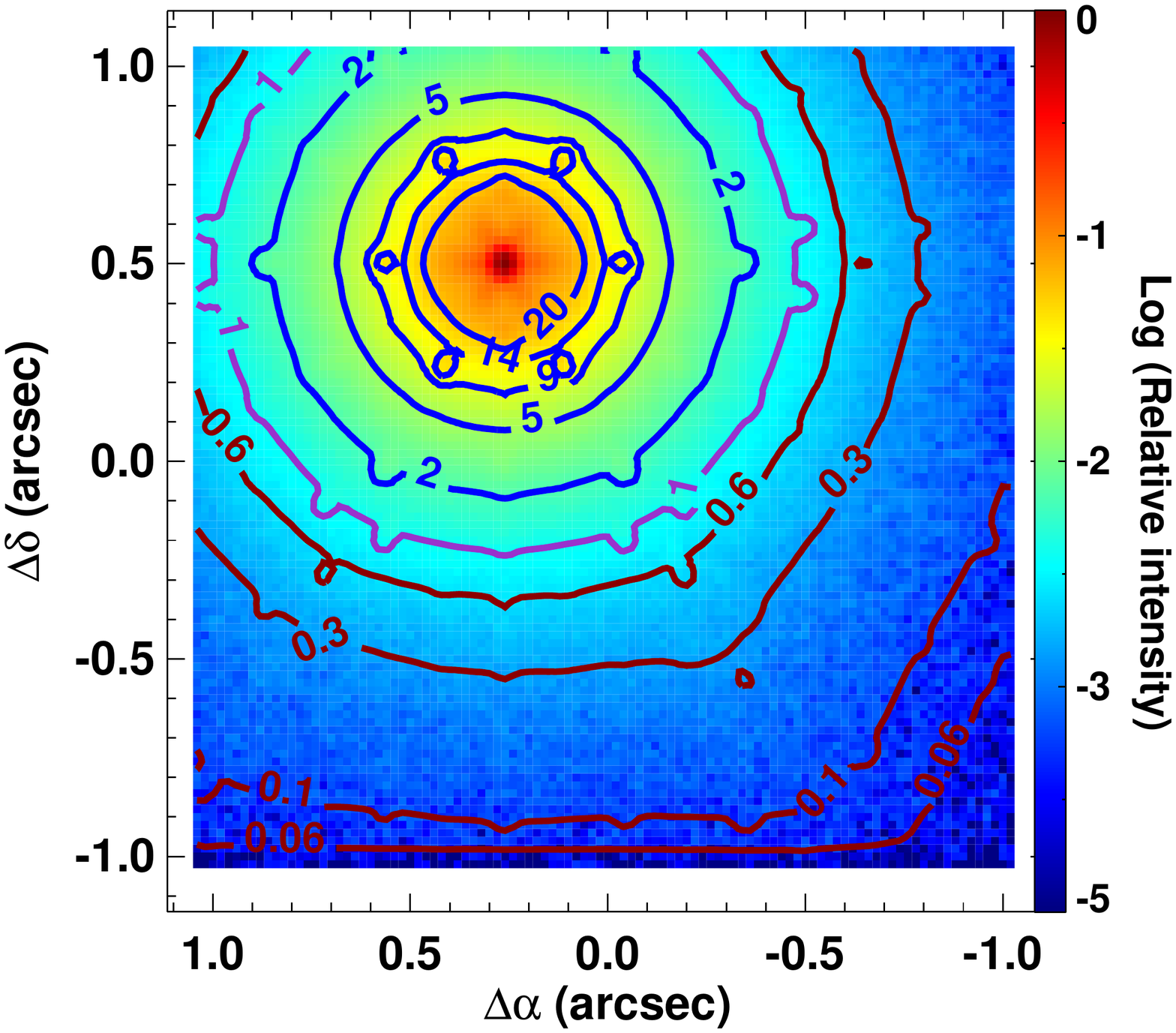}
 \includegraphics[trim={1cm 1.5cm 5cm 0cm},width=5.5cm]{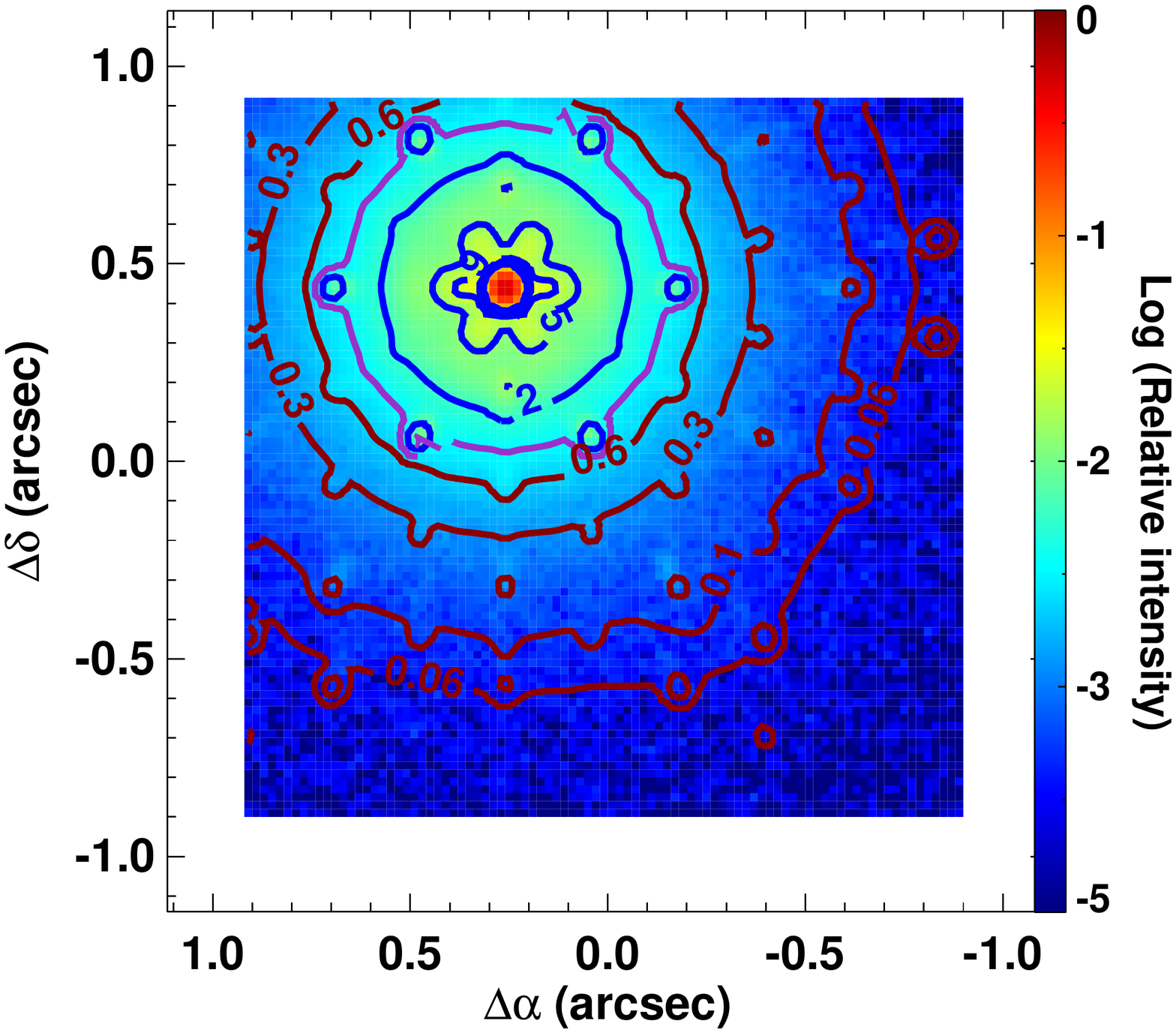}
 \includegraphics[trim={1cm 1.5cm 5cm 0cm},width=5.5cm]{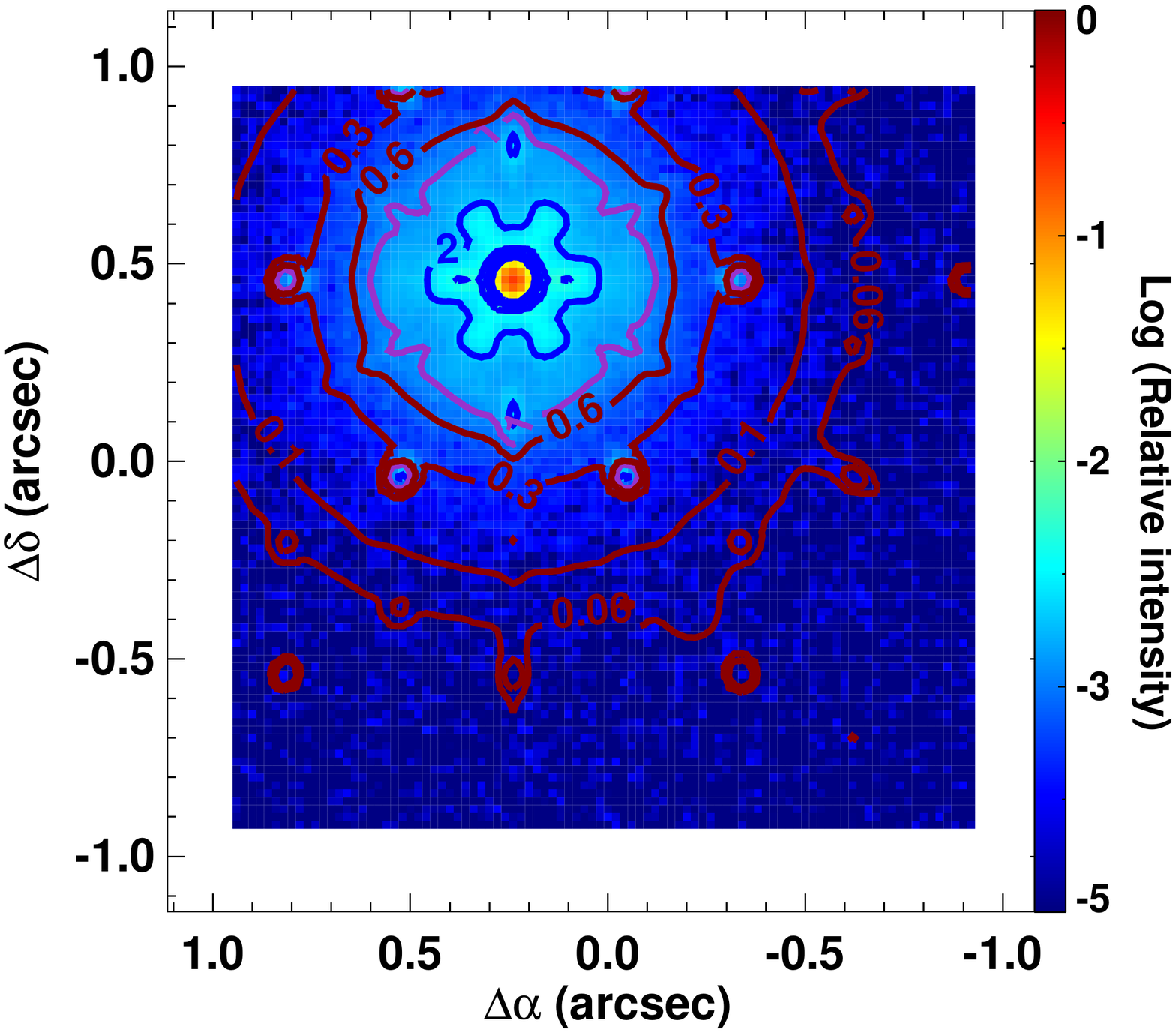}

  \caption{Same as Fig. \ref{NGC809_HbMgI_range_43}, but when considering an interacting morphology (i.e. NGC\,7119A) as a host.}

 \label{NGC7119N_HbMgI_range_43}
\end{figure*}

\begin{figure*}
\centering
 \includegraphics[trim={1cm 1.5cm 5cm 0cm},width=5.5cm]{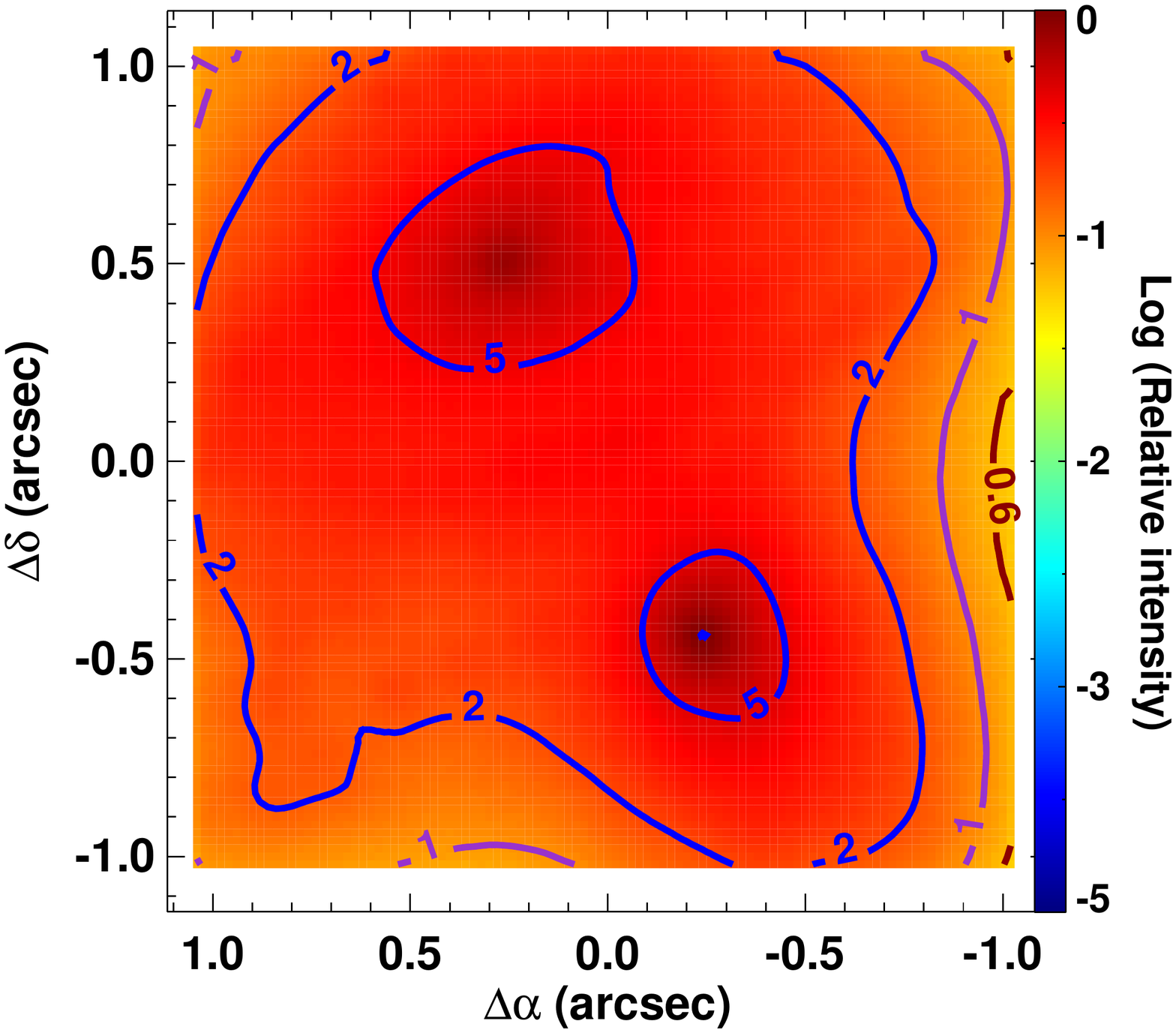}
 \includegraphics[trim={1cm 1.5cm 5cm 0cm},width=5.5cm]{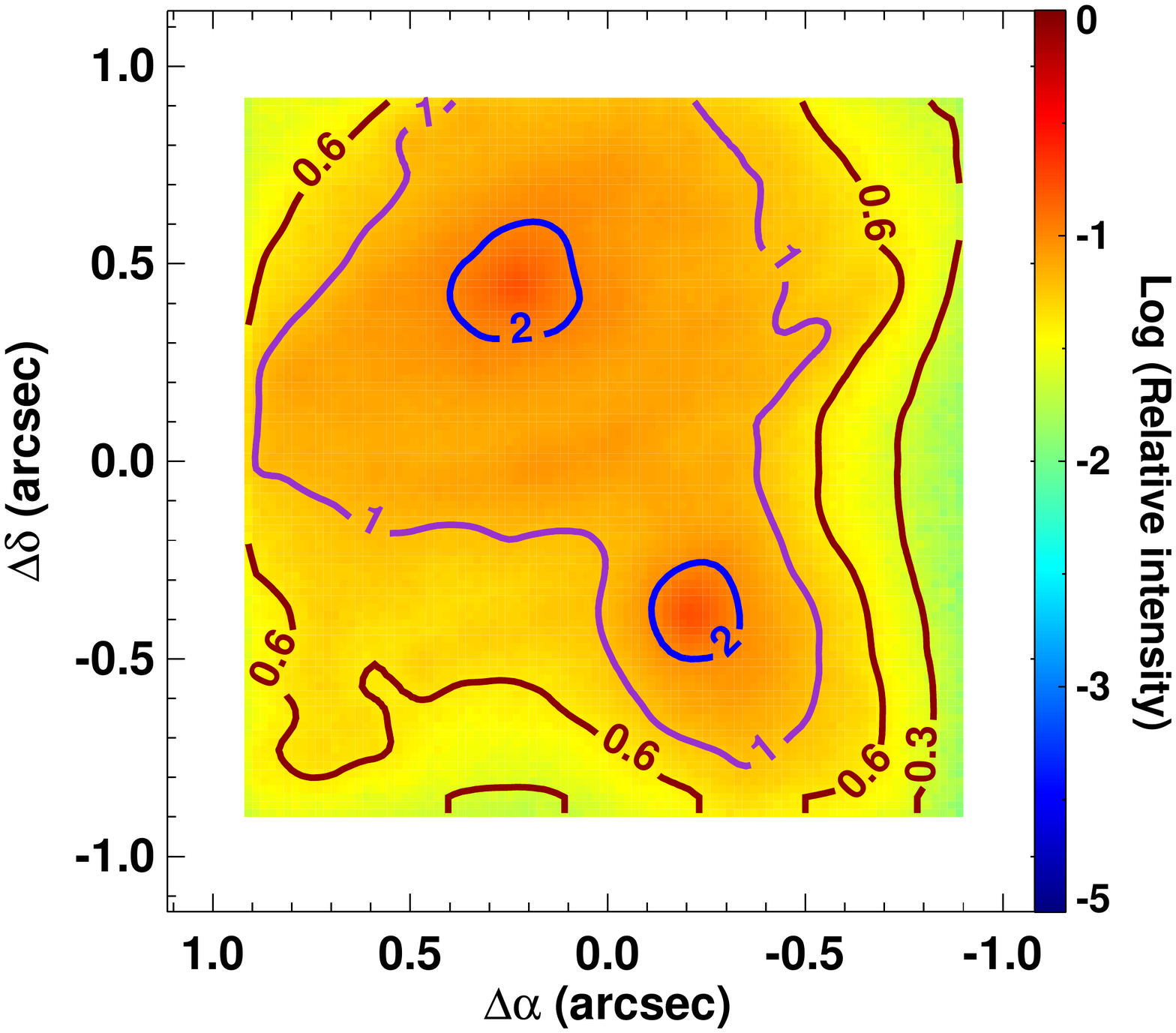}
 \includegraphics[trim={1cm 1.5cm 5cm 0cm},width=5.5cm]{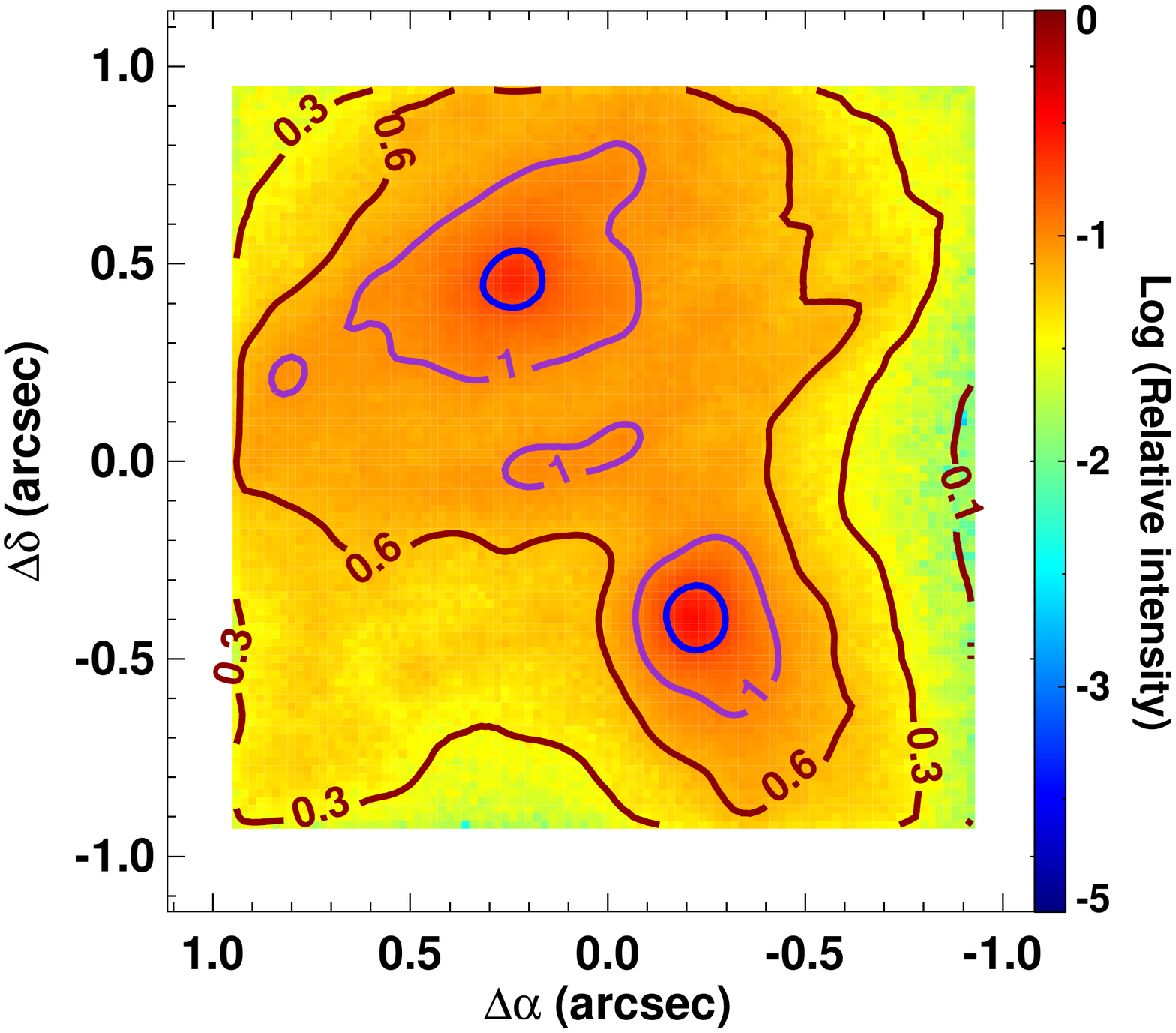}

 \includegraphics[trim={1cm 1.5cm 5cm 0cm},width=5.5cm]{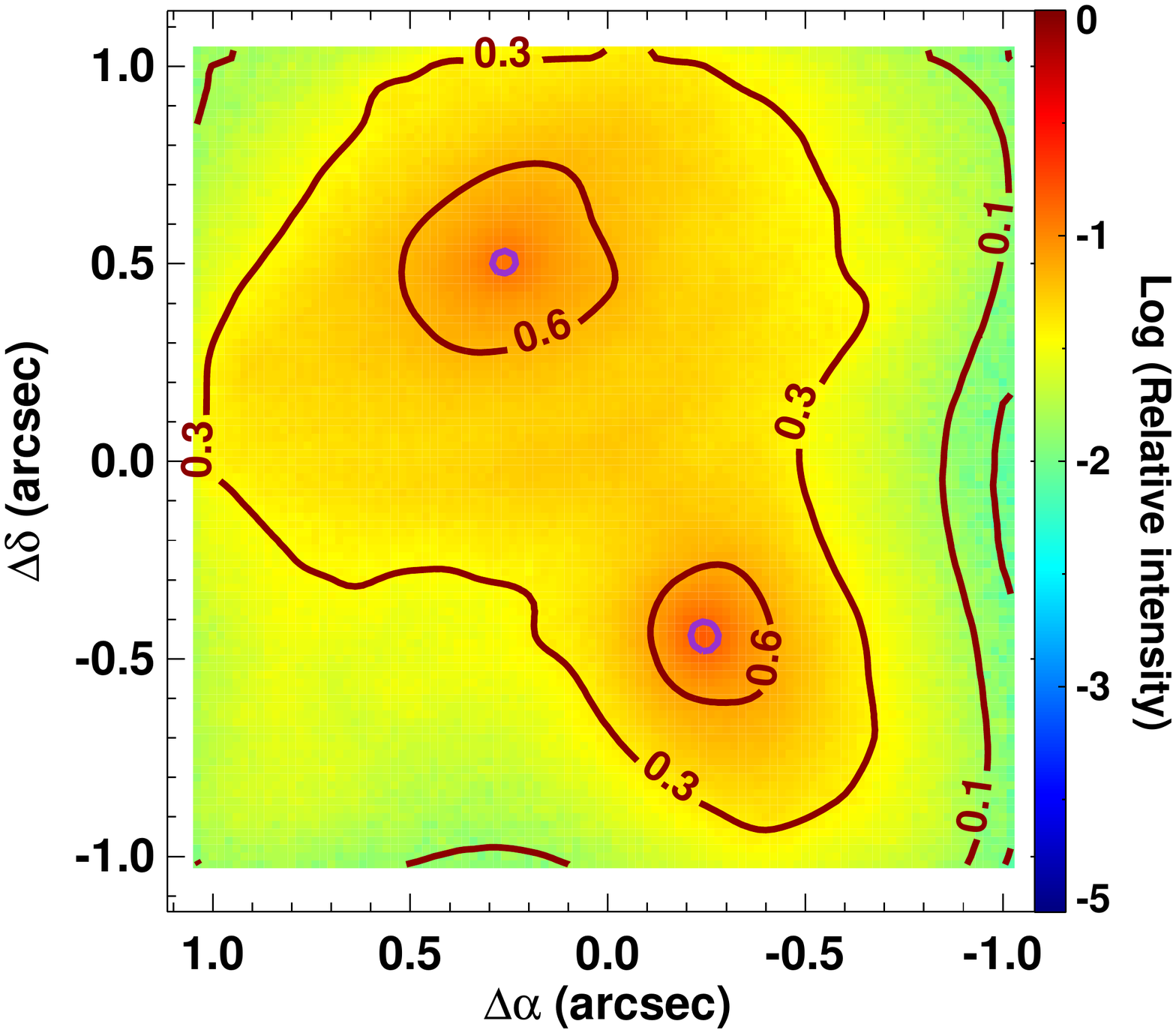}
 \includegraphics[trim={1cm 1.5cm 5cm 0cm},width=5.5cm]{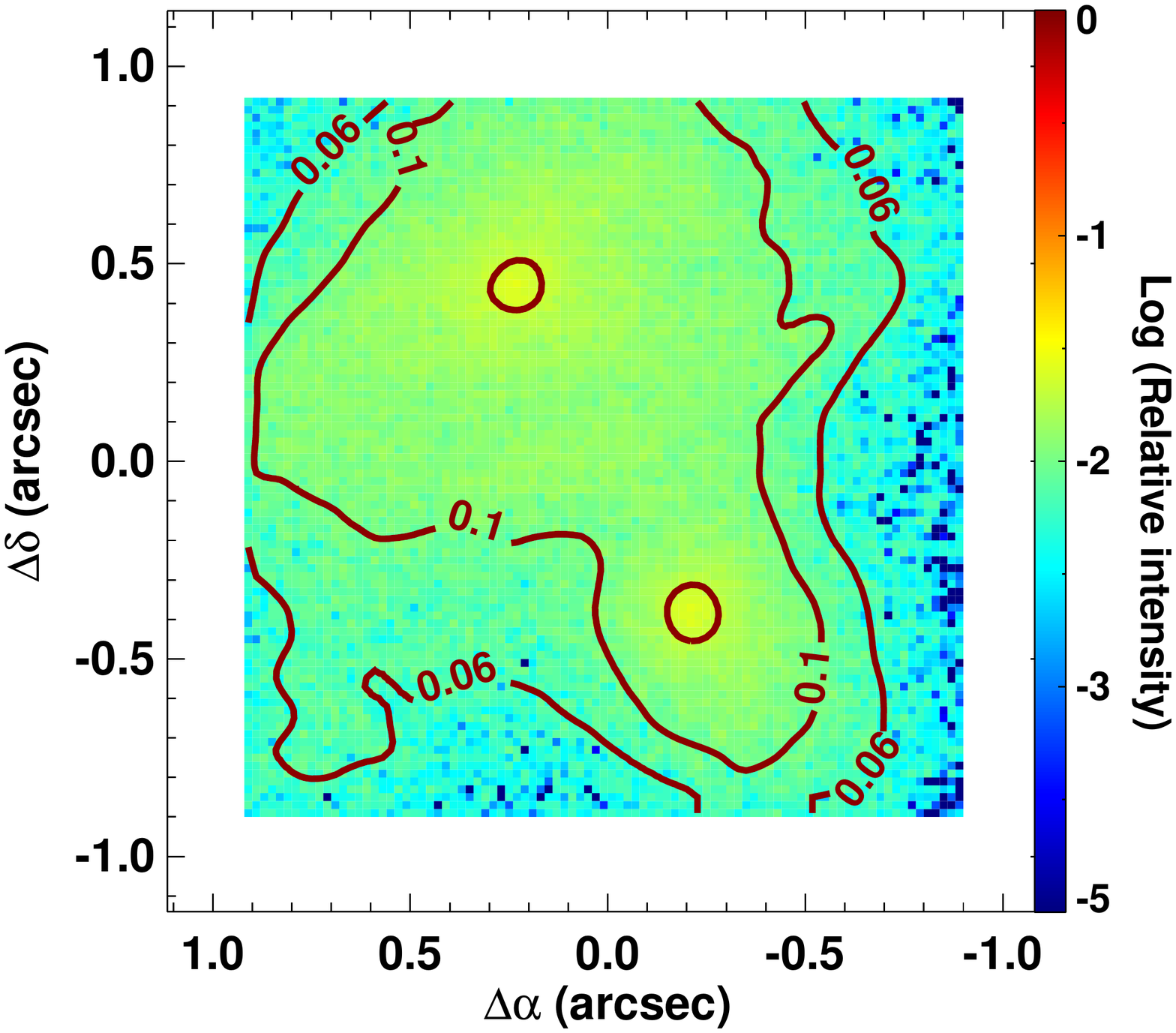}
 \includegraphics[trim={1cm 1.5cm 5cm 0cm},width=5.5cm]{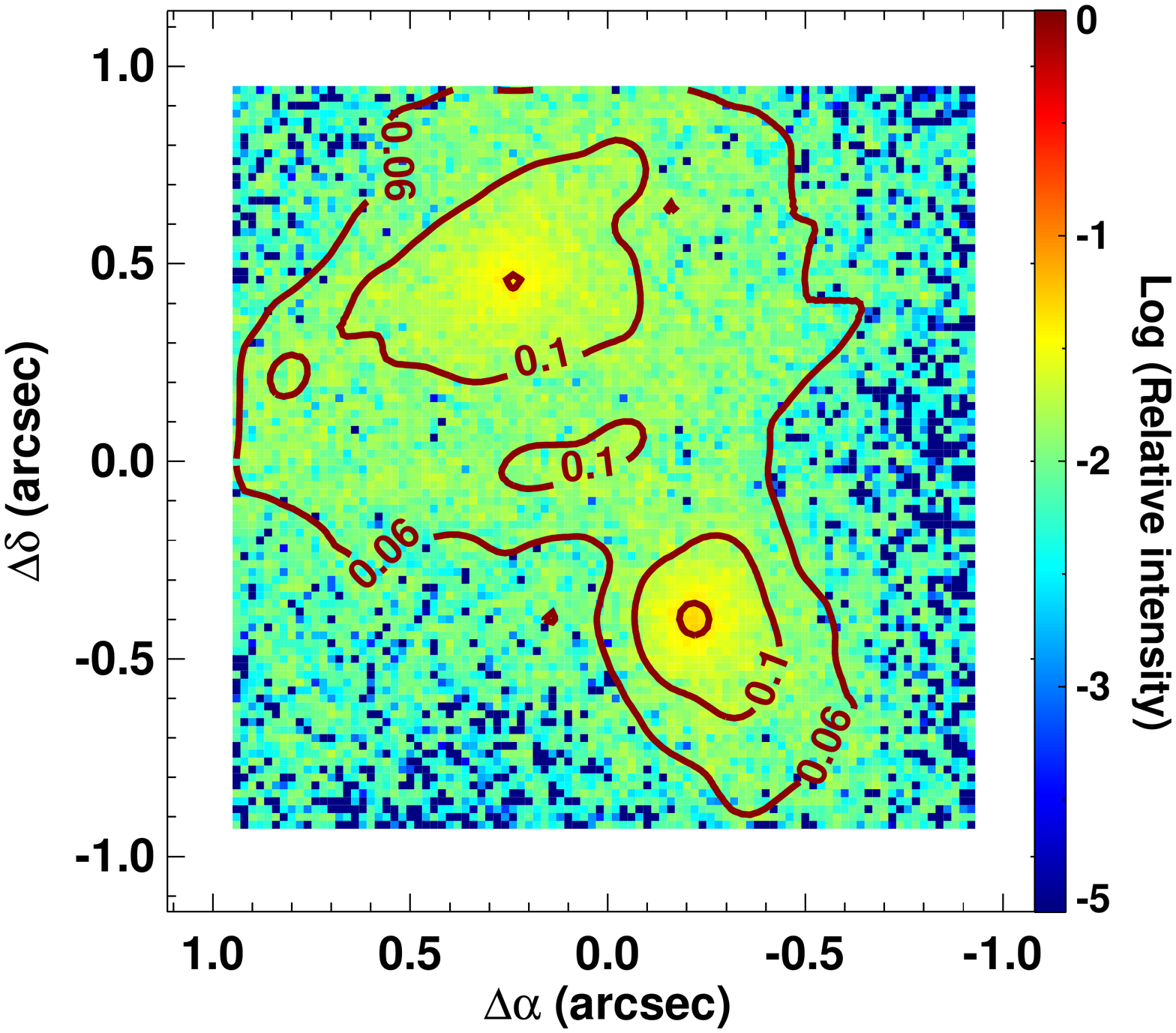}
 
 \includegraphics[trim={1cm 1.5cm 5cm 0cm},width=5.5cm]{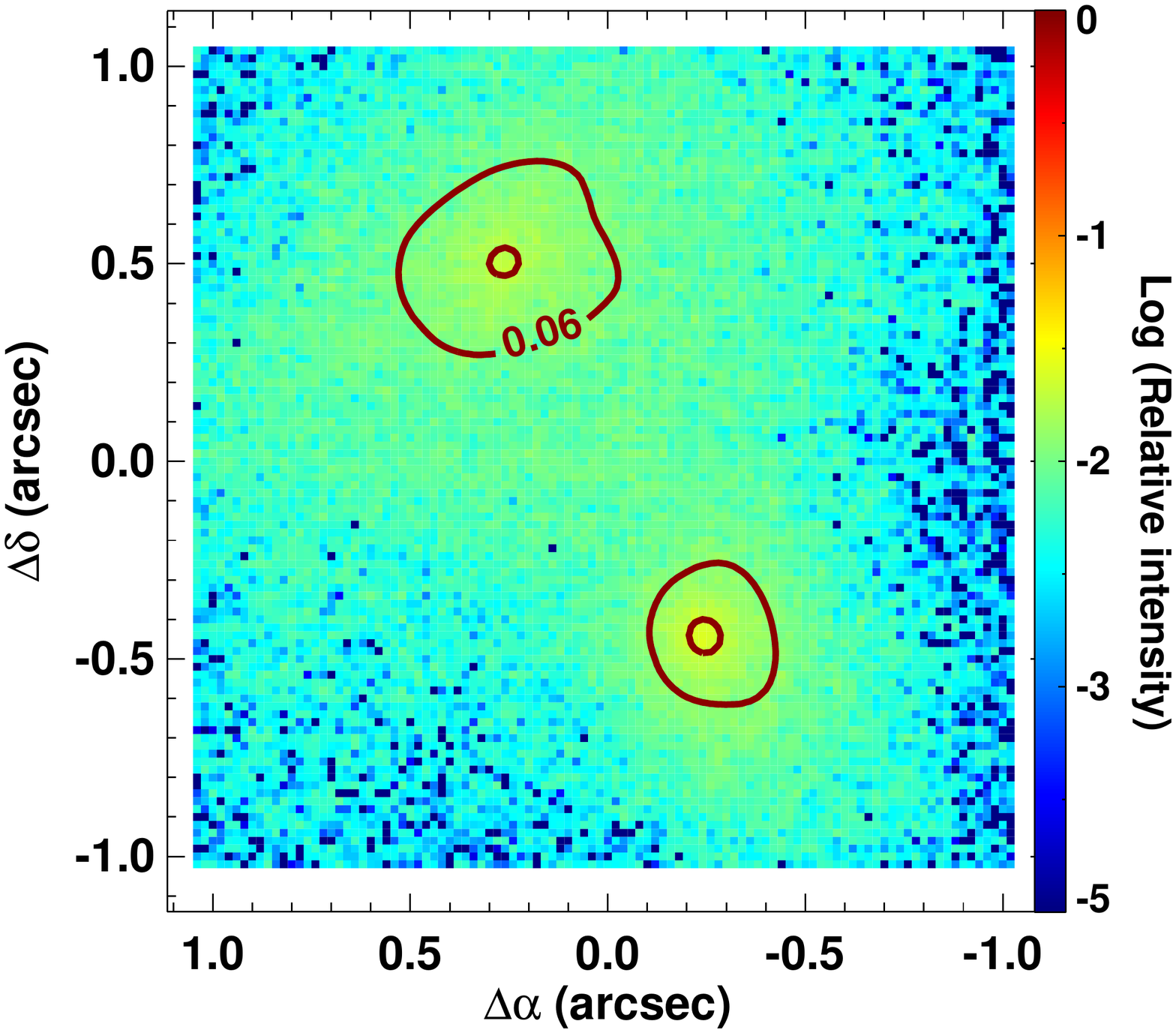}
 \includegraphics[trim={1cm 1.5cm 5cm 0cm},width=5.5cm]{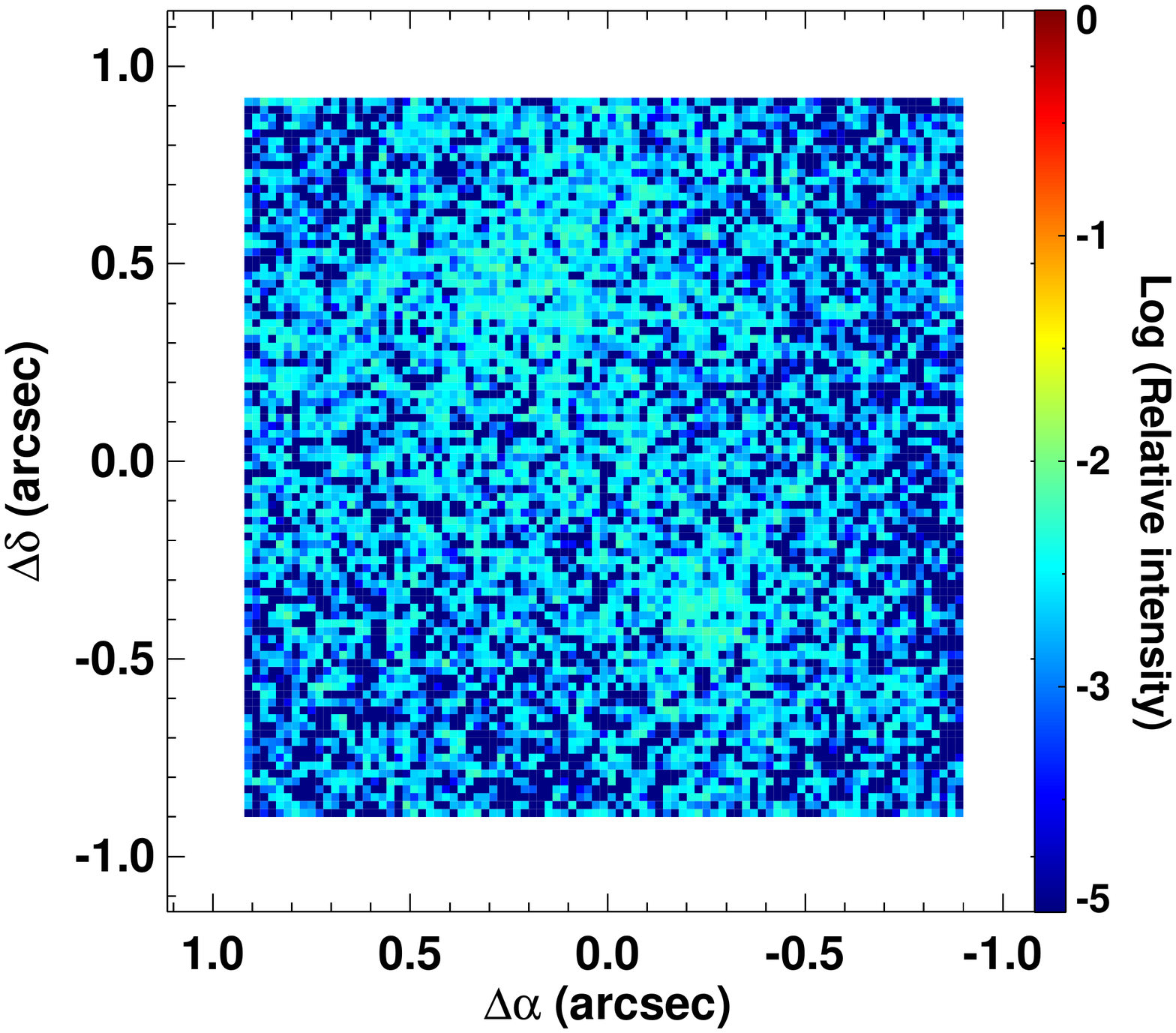}
 \includegraphics[trim={1cm 1.5cm 5cm 0cm},width=5.5cm]{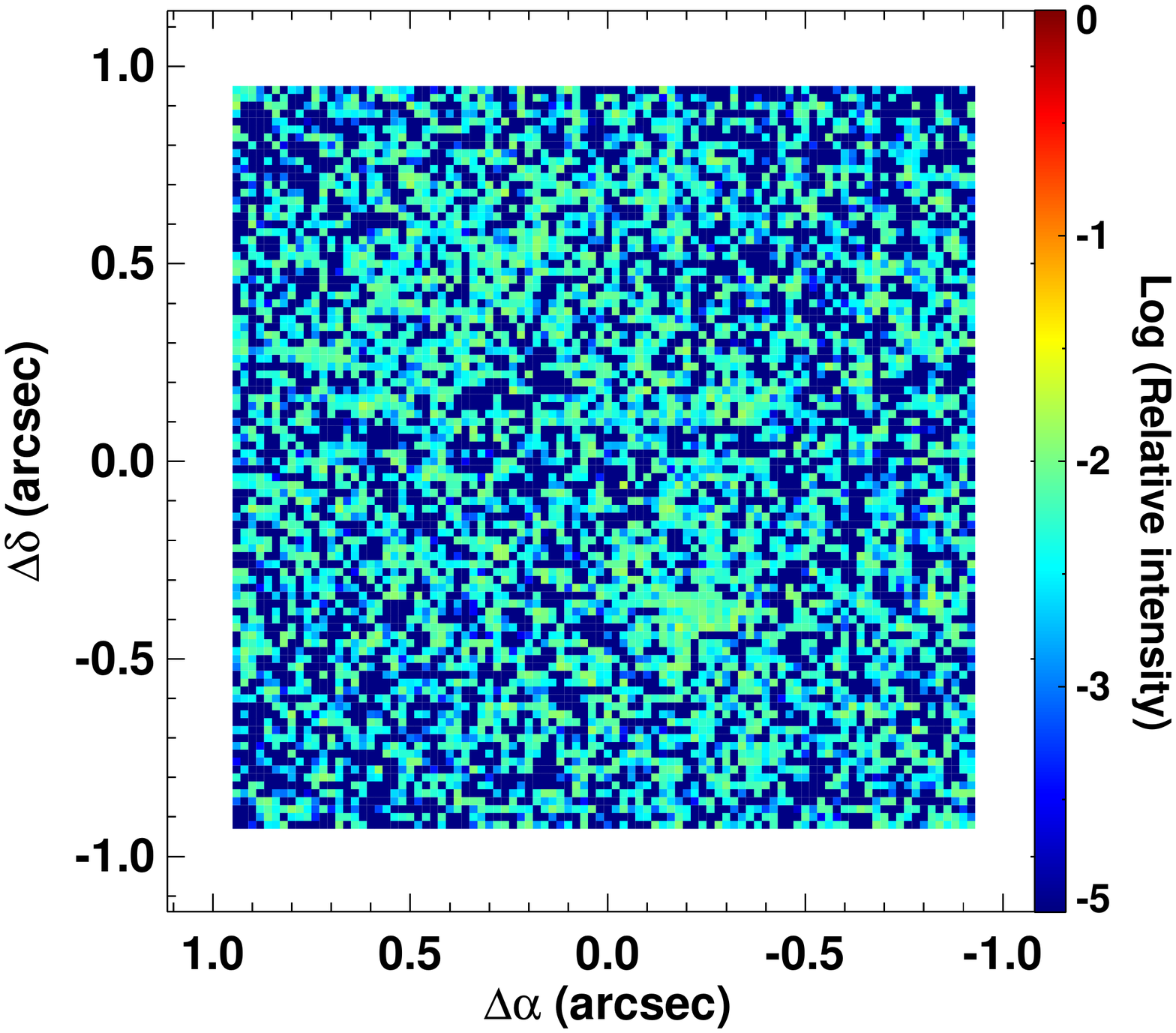}

  \caption{Same as Fig. \ref{NGC809_fullrange}, but when considering the interacting morphology (i.e. NGC\,7119A) as a host.}
  
 \label{NGC7119N_fullrange}
\end{figure*}

\begin{figure*}
 \includegraphics[trim={0.85cm 0.5cm 2cm 0},clip,width=5.75cm]{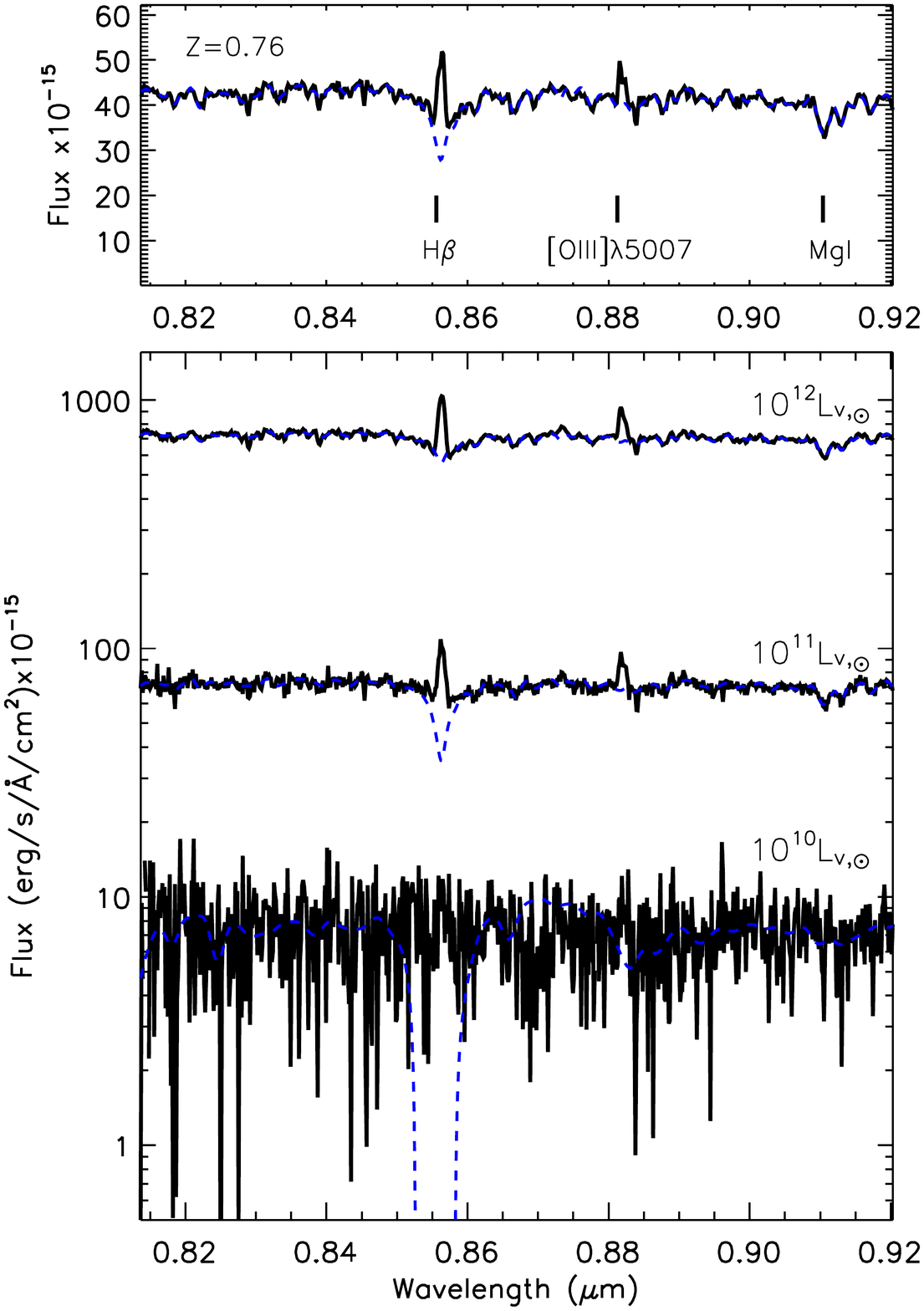}
 \includegraphics[trim={0.85cm 0.5cm 2cm 0},clip,width=5.75cm]{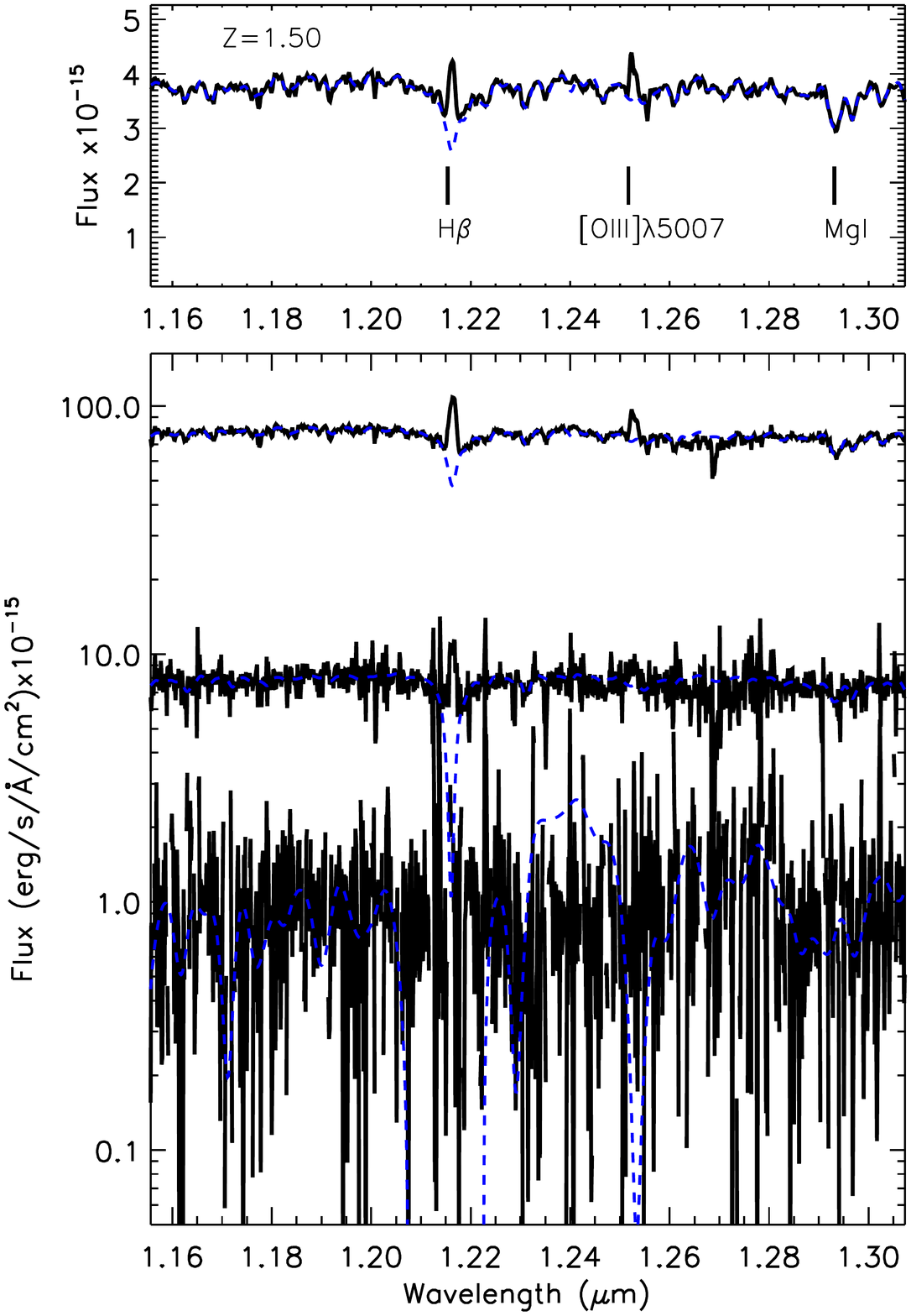}
 \includegraphics[trim={0.85cm 0.5cm 2cm 0},clip,width=5.75cm]{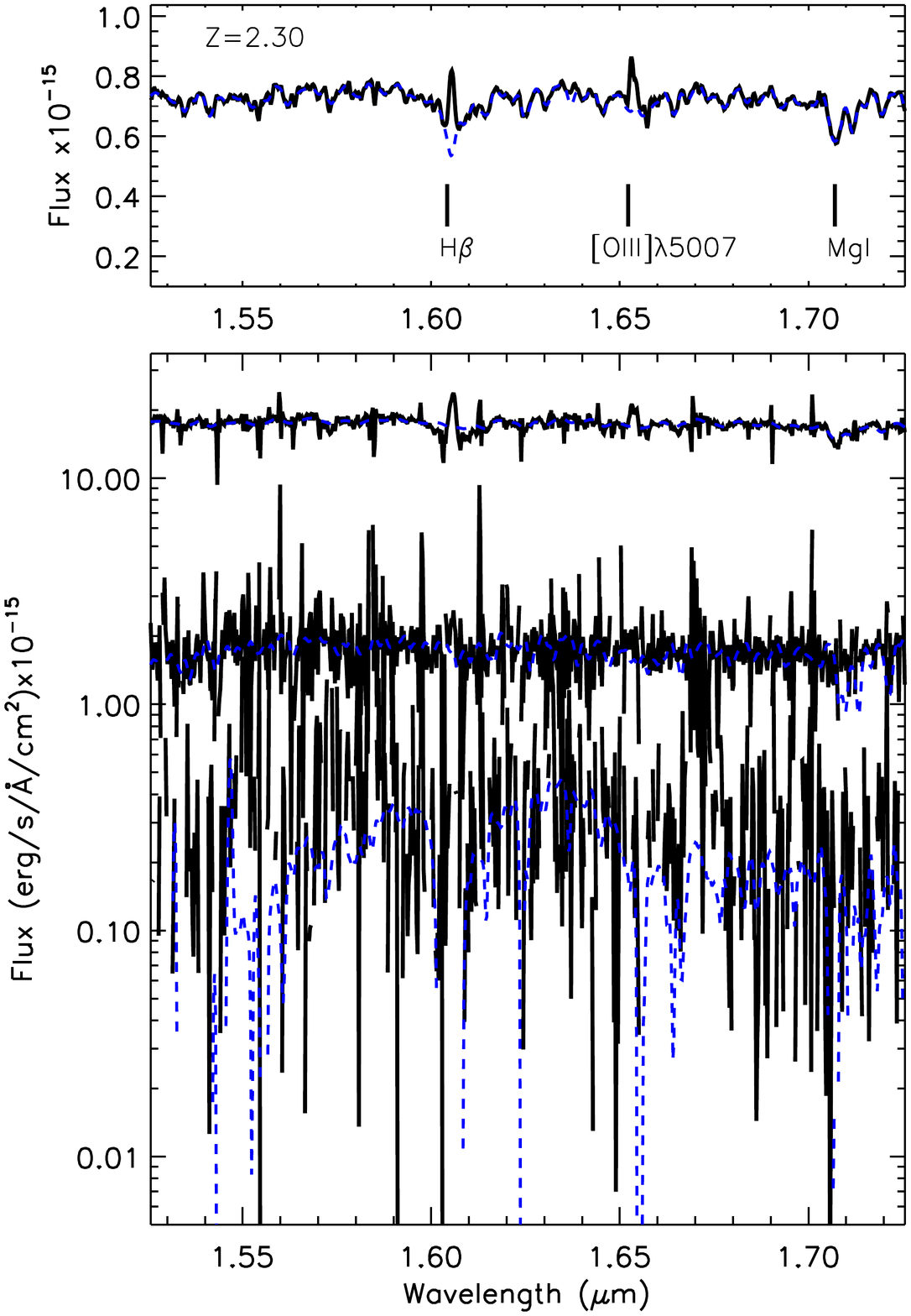}

  \caption{Same as Fig. \ref{NGC809_Reffspectra}, but for the considered interacting morphology (i.e. NGC\,7119A).}
  
 \label{NGC7119N_Reffspectra}
\end{figure*}

\begin{figure*}
\centering
 \includegraphics[trim={0.5cm 0.5cm 0.5cm 0},clip,width=4.25cm]{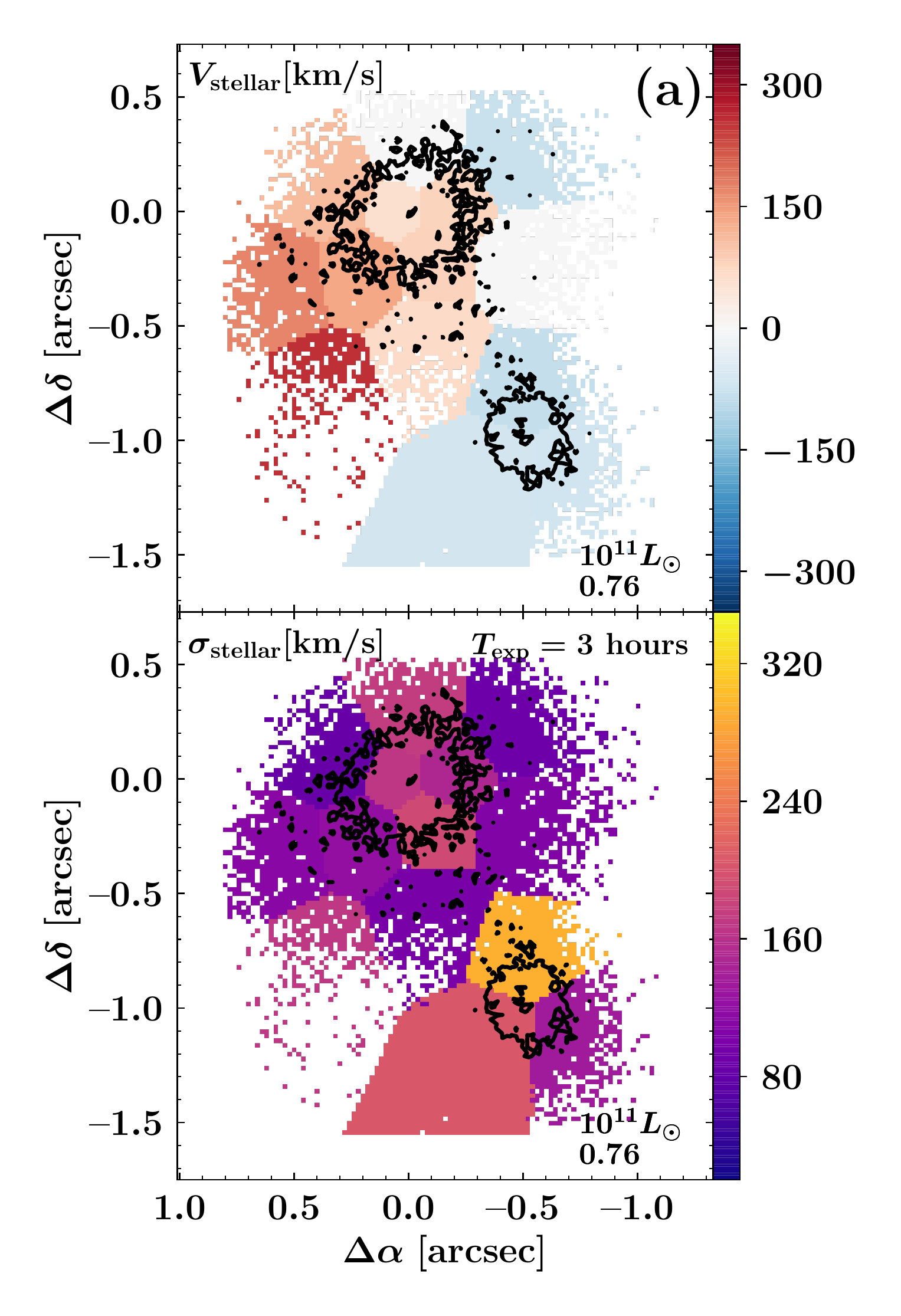}
 \includegraphics[trim={0.5cm 0.5cm 0.5cm 0},clip,width=4.25cm]{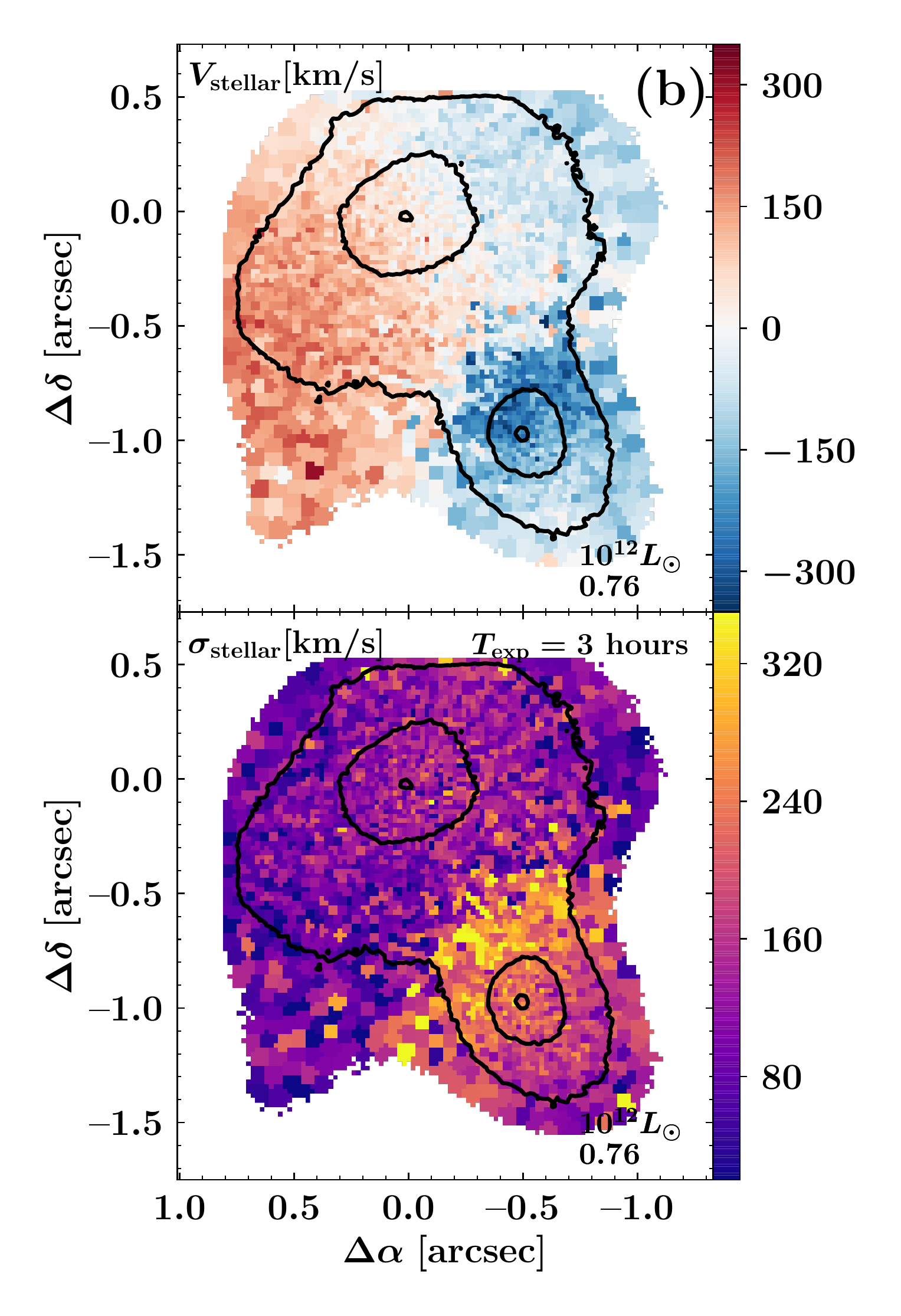}
 \includegraphics[trim={0.5cm 0.5cm 0.5cm 0},clip,width=4.25cm]{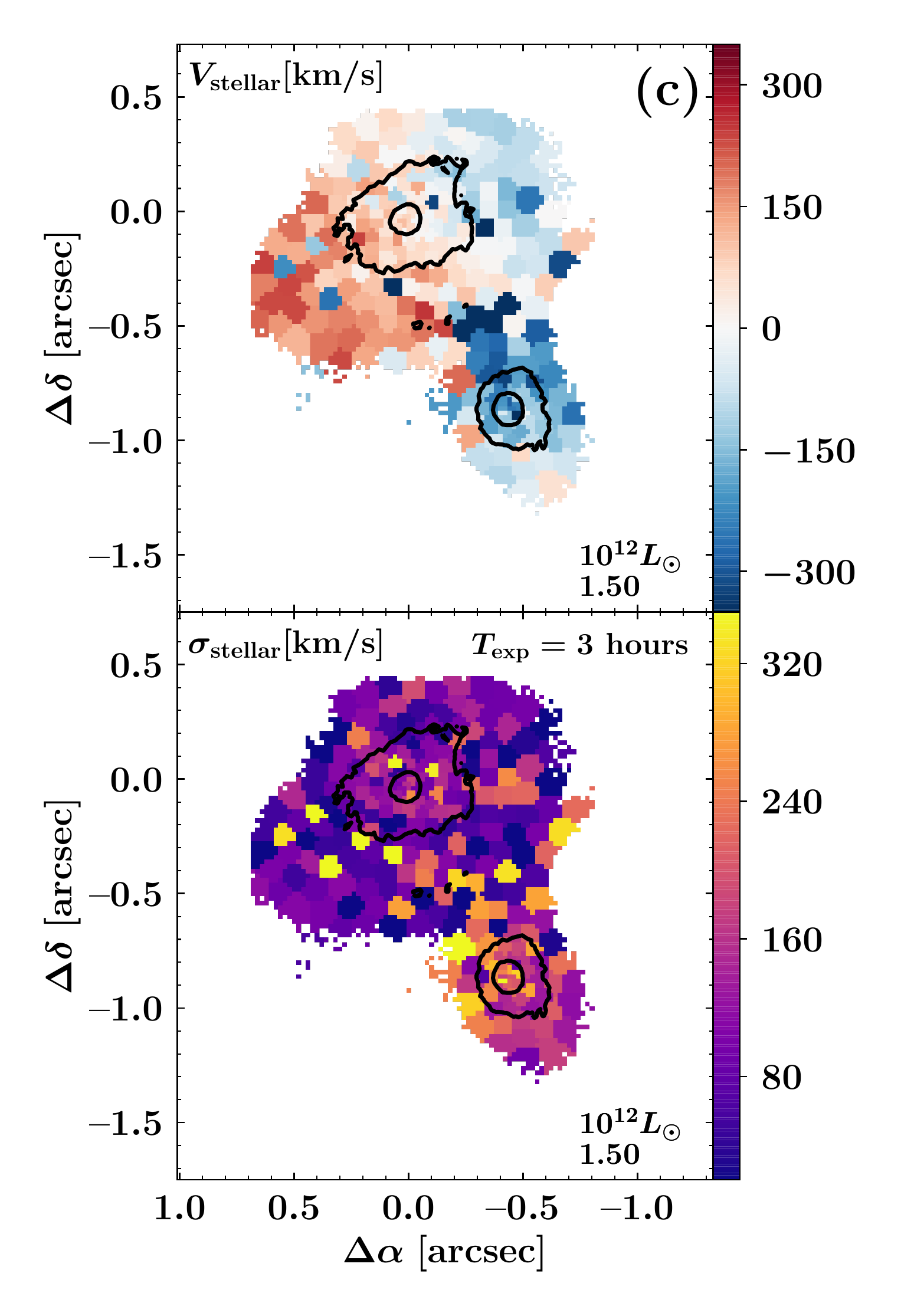}
\includegraphics[trim={0.5cm 0.5cm 0.5cm 0},clip,width=4.25cm]{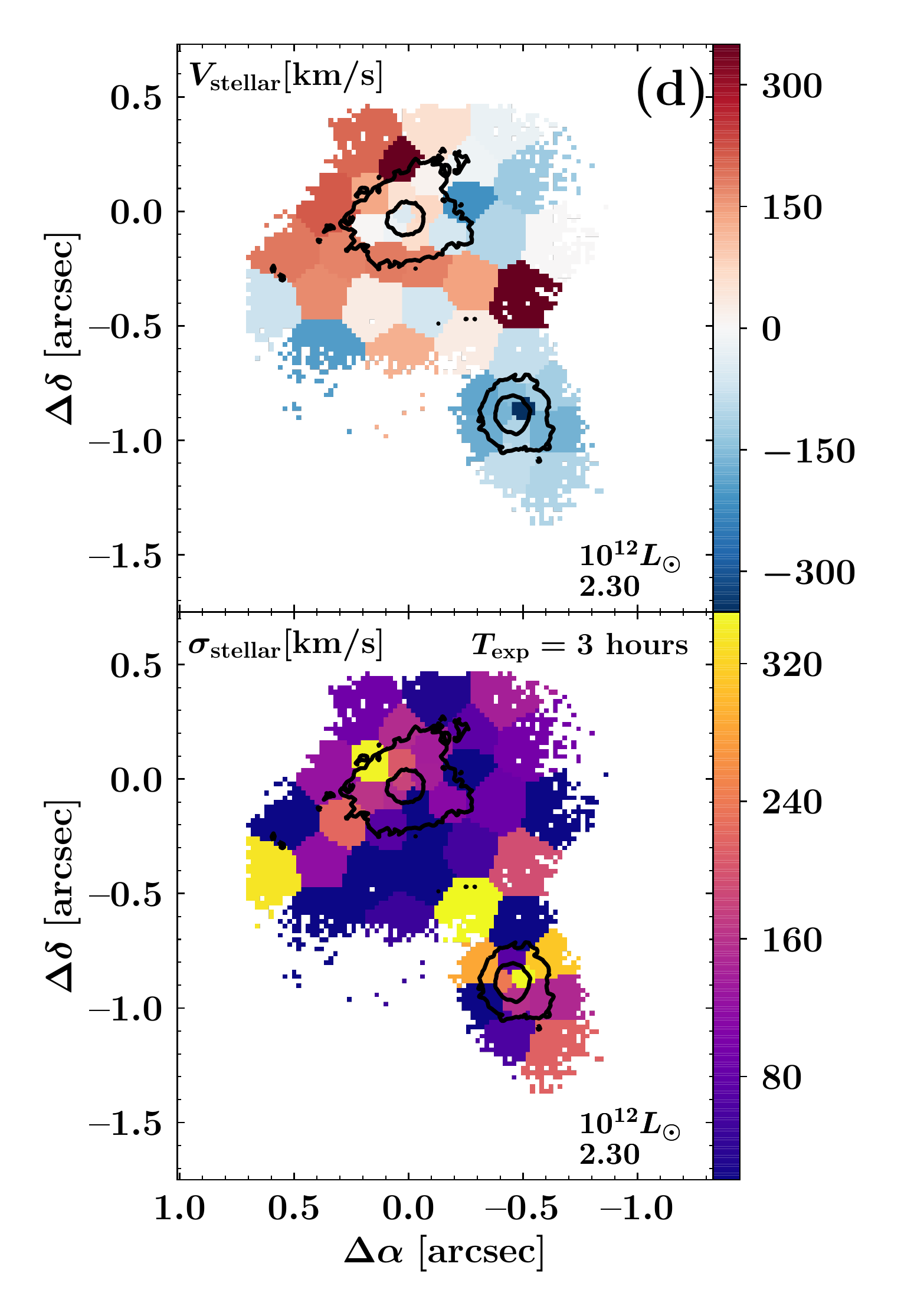}

  \caption{Same as Fig. \ref{resolved_kinematics_NGC809} (Top-panels), but for an interacting system acting as a host galaxy (i.e. NGC\,7119A). Colour bars are in the [-350,350] km s$^{-1}$ and [20,350] km s$^{-1}$ ranges for velocities and velocity dispersions, respectively.
  }
\label{resolved_kinematics_NGC7119N}
\end{figure*}

\end{appendix}

\end{document}